\documentclass[12pt]{report}
\usepackage[utf8]{inputenc}
\usepackage[T1]{fontenc}
\usepackage[french]{babel}
\usepackage{graphicx}
\usepackage{amssymb}
\usepackage{amsmath}
\usepackage{amsbsy}
\usepackage{amsfonts}
\usepackage{appendix}
\usepackage{fancyheadings}

\def\Box{\mathord{\dalemb{7.9}{8}\hbox{\hskip1pt}}}
\def\dalemb#1#2{{\vbox{\hrule height.#2pt
        \hbox{\vrule width.#2pt height#1pt \kern#1pt \vrule width.#2pt}
          \hrule height.#2pt}}}

\def\ba{\begin{eqnarray}}
\def\ea{\end{eqnarray}}
\def\be{\begin{equation}}
\def\ee{\end{equation}}

\def\gtorder{\mathrel{\raise.3ex\hbox{$>$}\mkern-14mu
             \lower0.6ex\hbox{$\sim$}}}

\def\ltorder{\mathrel{\raise.3ex\hbox{$<$}\mkern-14mu
             \lower0.6ex\hbox{$\sim$}}}

\newcommand{\ellb }{\boldsymbol{\ell }}
\newcommand{\thetab }{\boldsymbol{\theta }}

\begin{document}
\begin{titlepage}

\includegraphics[trim=1mm 1mm 1mm 1mm, clip,scale=0.3]{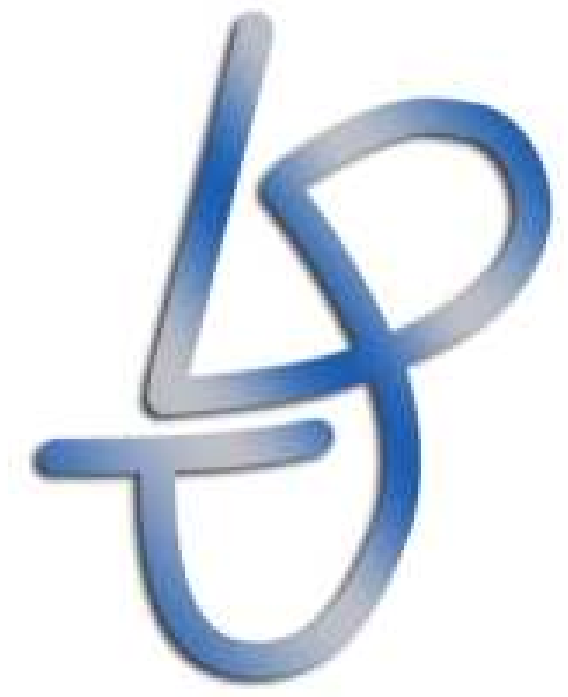}
\hfill
\includegraphics[trim=1mm 1mm 1mm 1mm, clip, scale=0.1]{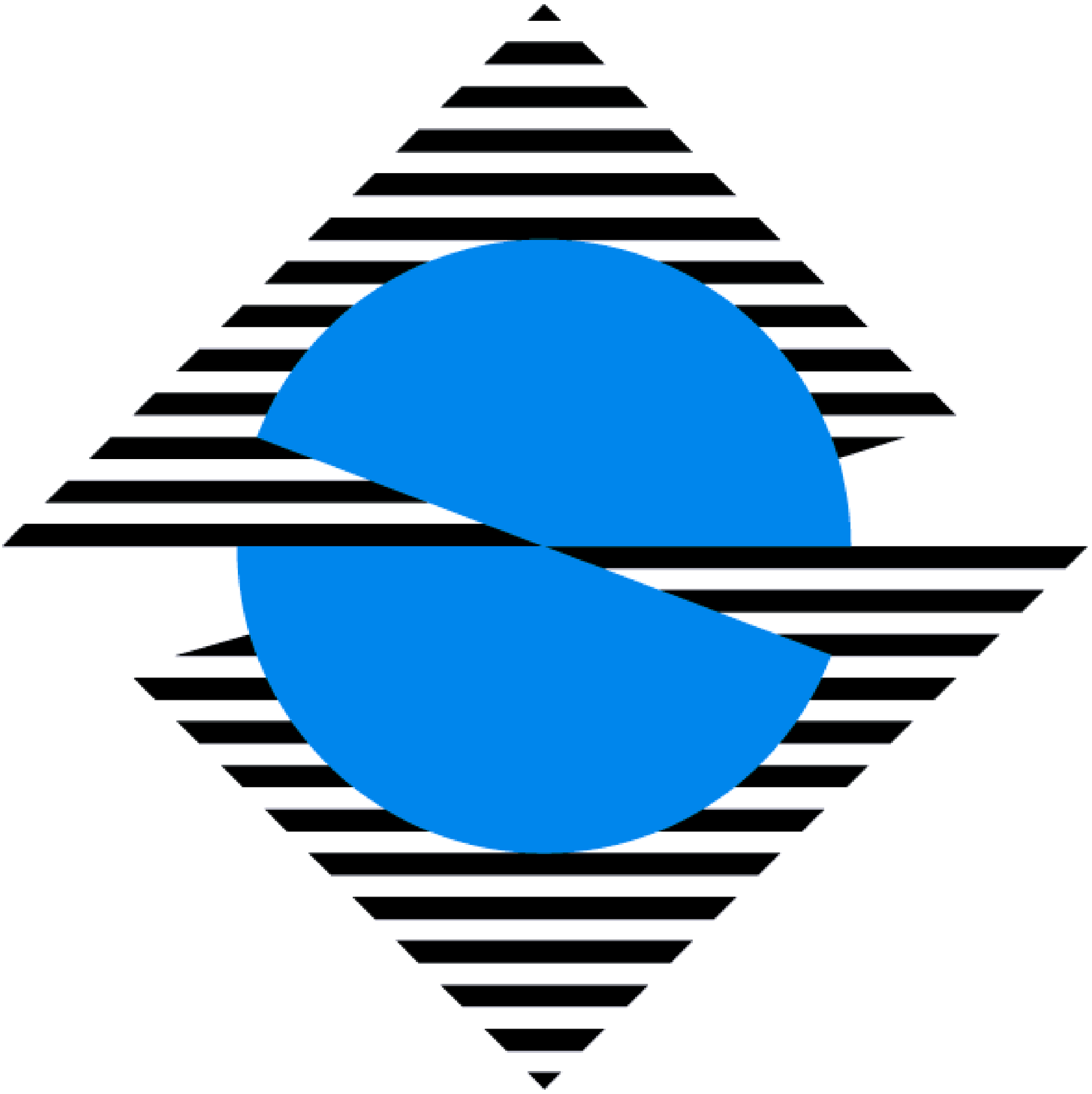}
\vspace*{2cm}
\begin{center}
\fbox{{\bf UNIVERSITÉ PARIS-SUD XI}}

\vskip1cm

{\bf THÈSE}
\vskip1cm

Spécialité: {\bf  PHYSIQUE THÉORIQUE}\\
\vskip0.3cm
Présentée\\
 pour obtenir le grade de
\vskip0.75cm
\large {\bf Docteur de l'Université Paris XI}

\vskip0.75cm

par

\vskip0.5cm

{\sc \bf Mathieu Remazeilles}

\vskip1cm

Sujet: 

\vskip0.2cm
\Large{\bf Évolution des perturbations cosmologiques dans les univers branaires}

\end{center}
\vskip0.75cm
\begin{flushleft}

Soutenue le vendredi 28 novembre 2008 devant la commission d'examen:\\
\vskip0.75cm
$\begin{array}{lll}
\mbox{MM}. & \mbox{Martin Bucher}, & \mbox{directeur de thèse},\\
   & \mbox{Brandon Carter}, & \\
   & \mbox{Jaume Garriga}, & \mbox{rapporteur},\\ 
&  \mbox{Renaud Parentani}, & \mbox{président du jury},\\
  & \mbox{Fernando Quevedo}, &\mbox{rapporteur}.
   \end{array}$

\end{flushleft}
\end{titlepage}

\thispagestyle{empty}
\vspace*{2.5cm}
~\\
\thispagestyle{empty}
\thispagestyle{empty}
\vspace*{2.5cm}
~\\
\newpage
\thispagestyle{empty}

















\begin{center}
{\bf Résumé}
\end{center}
Dans la majeure partie de cette thèse nous explorons l'évolution des
perturbations cosmologiques scalaires et tensorielles dans un Univers branaire
de type Randall-Sundrum ayant une expansion cosmologique arbitraire. Nous
adoptons un point de vue quadri-dimensionnel dans lequel les degrés de liberté
localisés sur la brane sont considérés comme un système quantique ouvert
couplé à l'environnement composé des gravitons du bulk Anti-de Sitter ($AdS$). À
cause de l'expansion non-uniforme de l'Univers, les degrés de liberté de la
brane et ceux du bulk interagissent dans le temps, faisant apparaître une
forme dissipation effective ainsi que des processus non-locaux en temps du
point de vue quadri-dimensionnel d'un observateur sur la brane. À partir des
propagateurs retardés ``nus'' sur la brane et dans le bulk nous calculons le
propagateur effectif sur la brane pour les modes ``habillés'' localisés sur la
brane en resommant les effets de rétroaction du bulk à tous les ordres dans le
couplage brane-bulk. La dissipation et la non-localité sont inscrites dans le
propagateur effectif. Nous obtenons les taux de dissipation de diverses
perturbations de matière sur la brane ainsi que du mode lié du
graviton. Ensuite nous présentons un calcul explicite de la fonction de Green
retardée covariante du graviton dans l'espace $AdS$.

Dans la dernière partie nous étudions la reconstruction du champ de lentille
gravitationnelle sur le CMB dans une cosmologie standard. Nous construisons un
estimateur statistique des champs de dilatation et de cisaillement directement
en espace réel. La méthode en espace réel développée ici est utile pour
analyser des cartes du CMB réalistes, contenant une coupure galactique et les 
nombreuses autres petites coupures excluant les points-source.

\begin{center}
{\bf Abstract}
\end{center}
In this thesis we mainly explore the evolution of both scalar and tensor cosmological perturbations  
in a Randall-Sundrum braneworld having an arbitrary expansion history. We 
adopt a four-dimensional perspective in which the localized degrees of freedom on
the brane are regarded as an open quantum system coupled to the large environment
composed of the Anti-de Sitter ($AdS$) bulk gravitons. Due to the non-uniform
expansion of the universe, the brane degrees of freedom and the bulk degrees
of freedom interact as they propagate forward in time, leading to an effective
dissipation as well as a nonlocality from the four-dimensional point of view
of an observer on the brane. Using both the "bare" retarded propagator on the
 brane and the retarded propagator in the bulk, we compute the effective propagator on the brane for the
"dressed" brane modes by resumming the bulk backreaction effects at all order
in the brane-bulk coupling. Dissipation and nonlocality are encoded into the
effective brane propagator. We find the dissipation rates of various matter
perturbations on the brane as well as of the graviton bound state. 
Next we present an explicit calculation of the covariant retarded Green
function for the graviton in $AdS$ space.  

In the last part, we investigate the reconstruction of CMB lensing in
standard cosmology. We construct a statistical estimator of the gravitational lensing dilatation and
shear fields directly in real space. The real space method developed
here is useful for analysing realistic CMB maps containing a galactic cut and
possibly numerous small excisions to exclude point sources that cannot be
reliably subtracted.

\newpage
\thispagestyle{empty}
\vspace*{2.5cm}
~\\
\thispagestyle{empty}
\thispagestyle{empty}
~\\




\vspace{24cm}
\emph{ 
Mes premiers remerciements iront à Martin Bucher pour ces trois années particulièrement enrichissantes. J'ai appris beaucoup de physique aux côtés de cet esprit brillant. C'est en ami que je tiens à lui exprimer ici toute ma gratitude pour ses encouragements, son soutien, son humour, et ses autres qualités humaines qui en font un directeur de thèse exceptionnel.}

\emph{Je tiens à remercier Brandon Carter, Jaume Garriga, Renaud Parentani et Fernando Quevedo pour m'avoir fait l'honneur de faire partie du jury de thèse, et tout particulièrement Jaume Garriga et Fernando Quevedo pour avoir accepté d'être les rapporteurs de mon manuscript de thèse.}

\emph{Je remercie également mon collaborateur Kavi Moodley qui m'a
  généreusement invité à l'Université de Durban et à l'African Institute for
  Mathematical Sciences à Cape Town. Merci aussi à Carla Carvalho pour son
  aide et son soutien.}

\emph{Un grand merci à tous les membres du LPT pour leur accueil et tout particulièrement au groupe de cosmologie: Christos Charmousis, Renaud Parentani, Bartjan van Tent, Jean Macher et Blaise Gouteraux. Merci aussi à Julien Serreau, Ana Teixeira et Robin Zegers avec qui j'ai eu le plaisir de partager mon bureau.}

\emph{Je tiens naturellement à remercier mes parents, et mes frères, pour leur soutien inconditionnel.}

\emph{Merci à Akba, Julien, Nico pour les écarts et les bons souvenirs.}

\emph{Enfin j'aurai une pensée profonde pour Claire, avec qui j'ai eu le
  bonheur de partager ma vie pendant cette thèse.}
\clearpage
\clearpage
\hspace{1cm}
\clearpage

\begin{flushleft}
\hspace{10cm}\textit{À mes parents,}\\
\hspace{10cm}\textit{à mes frères.}
\end{flushleft}


\newpage
\tableofcontents
\clearpage

\pagestyle{fancy}


\chapter*{Introduction}\label{chapter:intro}

    La possibilité que l'Univers possède plus que trois dimensions d'espace
    peut surprendre chacun à première vue puisque cela va à l'encontre de
    l'intuition  et de l'expérience quotidienne. Mais cette idée ne doit pas
    être plus surprenante que celle d'Eratosthène qui, dès l'Antiquité, soutenait que la
    Terre est ronde devant ses contemporains qui la voyaient plate parce
    qu'ils ne possédaient pas encore la technologie suffisante pour en mesurer
    sa vraie géometrie. L'existence de dimensions supplémentaires dans
    l'Univers est un scénario scientifiquement plausible, réellement pris au
    sérieux par la communauté scientifique de la physique théorique
    moderne, mais qui n'a pas encore été confirmé expérimentalement. Les
    scénarios de dimensions supplémentaires sont motivés par les
    théories de la physique des hautes énergies qui visent à unifier à très
    haute énergie, à l'échelle dite de Planck ($10^{19}$~GeV), l'interaction
    gravitationnelle avec les trois autres interactions fondamentales de la
    nature, à savoir l'interaction électromagnétique, l'interaction faible et
    l'interaction forte. Les deux grandes théories physiques du vingtième
    siècle, que sont la théorie de la relativité générale d'Einstein et la
    mécanique quantique - et la théorie quantique des champs-,
    prédisent avec une très grande précision la plupart des phénomènes
    physiques observés à ce jour, une grande partie des prédictions théoriques ayant
    été confirmées expérimentalement avec succès. La théorie de la relativité
    générale offre un cadre géométrique rigoureux pour décrire l'interaction
    gravitationnelle et les mécanismes de l'Univers aux échelles
    macroscopiques allant du dixième de millimètre jusqu'aux distances
    intergalactiques. Elle a également bouleversé notre vision de l'espace et
    du temps puisqu'elle unifie le temps aux dimensions spatiales révélant
    ainsi que notre Univers est un espace-temps à $(3+1)$ dimensions pouvant
    de surcroît être courbées. La théorie quantique des champs utilise un
    formalisme cohérent pour décrire le comportement microscopique
    des particules élémentaires du Modèle Standard de la physique des
    particules (quarks, leptons, bosons de jauge,~...) et
    leurs interactions électromagnétique, faible et forte à travers les
    théories de jauge $U(1)$, $SU(2)$ et $SU(3)$ respectivement. Mais ces deux
    grandes théories reposent sur des formalismes mathématiques différents et peu
    compatibles, la première faisant appel à la géométrie différentielle et la
    seconde aux  algèbres non-commutatives. Pourtant un cadre mathématique
    cohérent doit  pouvoir unifier ces deux théories pour prétendre décrire
    complètement la physique de phénomènes très énergétiques, incluant des
    effet quantiques en plus des effets gravitationnels, tels que la formation
    de trous noirs dans l'Univers ou la cosmologie primordiale de l'Univers
    quelques fractions de secondes après le Big-Bang. La théorie des supercordes est le
    candidat le plus prometteur dans cette quête d'unification et de
    réalisation d'une théorie quantique de la gravité \cite{string}. Cette nouvelle théorie,
    encore en développement, suppose que les constituants
    élémentaires de la matière ne sont plus les particules ponctuelles de la
    théorie quantique des champs mais des objets étendus unidimensionnels,
    à savoir des cordes minuscules de la taille de Planck ($10^{-33}$
    cm). Chaque état quantique de vibration des cordes correspond à une
    particule du Modèle Standard de la physique mais le succès de la théorie
    des cordes repose en partie sur le fait que le spectre quantique des
    cordes contient également un état de spin $2$ sans masse qui correspond
    exactement au graviton, la particule quantique qui doit véhiculer
    l'interaction gravitationnelle. Ce succès vers l'unification s'accompagne
    cependant d'un nouveau bouleversement dans notre compréhension de la
    nature de l'espace-temps : un des résultats majeurs de la théorie est que
    cette dernière exige, pour des raisons de cohérence mathématique, que les
    cordes évoluent dans un espace-temps à $(9+1)$ dimensions. L'Univers
    pourrait donc posséder $6$ dimensions spatiales supplémentaires
    "cachées". La découverte complémentaire que les théories de cordes comprennent également des
    membranes de dimensions diverses, appelées "branes", en plus des cordes
    elles-même, a motivé l'étude de la possibilité que notre Univers
    observable réside en fait sur une membrane à $(3+1)$ dimensions entourée de
    dimensions supplémentaires, pas forcément petites ni compactifiées \cite{hw2,rs2}.

    L'étude de modèles d'Univers branaires plongés dans un espace-temps à
    \emph{cinq} dimensions peut déjà nous révéler une
    phénoménologie très riche dans le cadre de la cosmologie.
    Des questions immédiates se posent dans cette
    nouvelle perspective : comment sont ``dissimulées'' les (ou la) dimensions
    supplémentaires ? Est-ce que la présence de dimensions supplémentaires
    dans l'Univers ne contredit pas les résultats expérimentalement valides
    de la physique standard quadridimensionnelle ?
    Quelles sont les modifications cosmologiques éventuelles par rapport à la cosmologie
    standard quadrimensionnelle et comment les détecter ? Dans la plupart des
    modèles branaires tous les
    champs de particules du Modèle Standard sont confinés sur une hypersurface de genre
    temps à quatre dimensions, la brane, et seuls les gravitons peuvent se
    propager dans les dimensions supplémentaires de l'espace-temps total
    appelé ``bulk''. De cette fa\c{c}on la physique du Modèle Standard à
    quatre dimensions se trouve préservée sur la brane. Les signatures de la présence
    éventuelle de dimensions supplémentaires dans l'Univers ne peuvent donc
    être appréhendées que par l'étude du
    comportement de la gravitation et de la cosmologie. Une fa\c{c}on,
    alternative aux modèles branaires, de s'affranchir du
    problème des dimensions supplémentaires est de supposer qu'elles sont
    compactifiées, à la Kaluza-Klein, à très petite échelle de sorte
    qu'aucun instrument de mesure n'ait pu les détecter jusqu'à
    aujourd'hui. Cependant les scenarios branaires, qui autorisent pour
    certains d'entre eux la présence
    d'au moins une dimension supplémentaire non-compacte de grande taille voire infinie,
    offrent par conséquent la possibilité plus optimiste de pouvoir détecter cette dimension
    supplémentaire, au moins de fa\c{c}on indirecte. Si la dimension supplémentaire est
    infinie on peut objecter que la gravité risque de trop se diluer dans cette dimension
    supplémentaire et que la force de gravitation de Newton doit apparaitre
    beaucoup plus faible que prévue aux observateurs localisés sur l'Univers-brane à
    quatre dimensions ; et cela contredirait les mesures expérimentales de la
    loi de Newton qui confirment que le comportement de la force de gravitation est
    quadridimensionnel puisque son intensité diminue comme l'inverse de
    la distance au carré, au moins jusqu'aux petites distances de $0.1$ mm. Cependant dans leurs travaux
    de 1999 \cite{rs2}, Randall et Sundrum ont résolu ce paradoxe en
    proposant un modèle d'Univers branaire cohérent où la dimension supplémentaire
    est non-compacte et infinie mais préserve néanmoins un comportement 
    quadridimensionnel pour la gravitation sur la brane à basse énergie
    (distances $> 0.1$ mm), du point de vue d'un observateur
    ``confiné'' sur la brane. Dans le modèle de Randall-Sundrum, l'Univers est une
    brane de géométrie Minkowski à quatre dimensions plongée dans un espace-temps à
    cinq dimensions de courbure négative, c'est-à-dire de géométrie Anti-de
    Sitter ($AdS$). L'idée astucieuse de ce modèle est que la courbure négative de la
    dimension supplémentaire empêche la gravité de se diluer complètement dans
    cette dimension, de telle sorte que la gravité reste localisée de fa\c{c}on
    dynamique sur la brane à quatre dimensions. De cette fa\c{c}on le
    comportement de la gravité standard à quatre dimensions reste
    effectivement conservé sur la brane à basse énergie.  Bien-sûr, la brane
    étant ici plate et vide de matière, ce
    premier modèle est pauvre du point de vue de la cosmologie. L'étape suivante
    est d'étudier la cosmologie dans le cas d'un Univers-brane ayant un contenu de
    matière et une histoire d'expansion cosmologique, c'est-à-dire une brane
    de géométrie Friedmann-Robertson-Walker plongée dans l'espace anti-de
    Sitter. Un outil efficace sinon indispensable pour évaluer les
    conséquences cosmologiques de ces modèles d'Univers reste la théorie
    des perturbations cosmologiques.

    L'Univers constitue le plus grand laboratoire pour étudier les processus
    de la physique des hautes énergies et la cosmologie observationnelle est devenue à
    partir de la fin du vingtième siècle une science de précision extrême à
    travers les relevés de supernovae, les observations des structures à
    grande échelle de l'Univers, et surtout la cartographie des anisotropies
    de température et de polarisation du Fond Diffus Cosmologique (CMB, en
    anglais, pour Cosmic Microwave Background) grâce aux satellites COBE en
    1992 puis WMAP en 2003. Le rayonnement de photons du Fond Diffus nous
    parvient depuis la surface de dernière diffusion, c'est-à-dire au moment du
    découplage entre la matière et les photons, 380 000 ans après le Big-Bang
    (redshift $z\sim 1100$) quand l'Univers ne faisait qu'un millième de sa taille
    actuelle et lorsque sa température était de 3000 K. C'est donc l'image la
    plus ancienne de notre Univers accessible aux observations. Ce rayonnement
    du ciel est un rayonnement micro-onde de type corps noir dont la température est
    aujourd'hui de 2.7 K à cause de l'expansion de l'Univers mais il présente néanmoins des anisotropies relatives, de l'ordre de
    $10^{-5}$, qui sont le reflet des inhomogénéités initiales de l'Univers, 
    responsables de la formation ultérieure des grandes structures de l'Univers
    par attraction gravitationnelle. Ces inhomogénéités résultent elles-même
    de fluctuations quantiques intrinsèques aux échelles subhorizon lors de
    l'époque de l'Inflation de l'Univers. La
    théorie des perturbations  cosmologiques permet de relier
    les perturbations inhomogènes initiales aux anisotropies du Fond Diffus
    Cosmologique. Les spectres d'anisotropies de température et de
    polarisation ont été calculés dans le cadre quadridimensionnel du Modèle
    Standard de la cosmologie et reproduisent dans un certain intervalle de
    précision les cartes du CMB observées. Dans le cadre d'une cosmologie
    branaire du type Randall-Sundrum, en présence d'une dimension
    supplémentaire ayant une structure Anti-de
    Sitter, on ne peut exclure que cette cinquième dimension ait un rayon de
    courbure aussi important qu'un dixième de millimètre. L'enjeu majeur pour
    la cosmologie branaire est de calculer l'évolution des perturbations
    cosmologiques dans un Univers branaire en expansion quelconque afin de comparer les
    résultats obtenus aux données du Fond Diffus Cosmologique. C'est dans ce
    cadre que s'inscrit la majeure partie de ce travail de thèse.

    La théorie des perturbations cosmologiques standard quadridimensionnelle
    est bien connue depuis 1963 grâce aux travaux pionniers de Lifshitz et Khalatnikov
    \cite{lifshitz} suivis en 1980 du formalisme invariant de jauge
    développé par Bardeen \cite{bardeen}. Selon le
    principe cosmologique, l'Univers apparait homogène et isotrope à grande échelle
    dans les trois directions spatiales. De ce fait les équations d'Einstein
    linéarisées pour les perturbations peuvent être diagonalisées par
    transformation de Fourier dans l'espace tridimensionnel et chaque mode de
    Fourier évolue selon une équation aux derivées ordinaires (EDO) en temps.
    Dans le cadre de la cosmologie branaire, la théorie des perturbations cosmologiques demeure
    bien plus compliquée parce que la présence de la brane, en
    tant que défaut topologique, brise l'isotropie de l'espace-temps à cinq
    dimensions, de sorte que les équations de perturbations ne sont plus des
    EDO en temps mais des équations aux dérivées partielles (EDP) en temps et
    en la cinquième dimension. Des résultats analytiques ont cependant été
    obtenus pour des Univers-branes vides à symétrie maximale, telle
    qu'une brane Minkowski ou une brane purement (Anti-)de Sitter. Mais dans
    le cas d'une brane ``cosmologique'' avec un contenu de matière et ayant
    une expansion arbitraire dans le temps, les équations linéarisées
    satisfaites par les perturbations cosmologiques ne
    sont plus séparables, de sorte que peu de prédictions quantitatives
    statisfaisantes ont été obtenues pour l'évolution des perturbations
    cosmologiques dans un Univers branaire en expansion. Cependant ce premier
    problème reste d'ordre technique et peut donc a priori être abordé de
    façon numérique. Un second problème, d'ordre plus
    fondamental ou physique, est celui des conditions initiales dans
    Anti-de Sitter :  en plus d'être specifiées sur
    les degrés de liberté localisés sur la brane, les conditions initiales doivent être spécifiées sur une
    infinité de degrés de liberté dans le bulk $AdS$. Mais il n'y a pas de
    conditions initiales naturelles dans l'espace $AdS$. En cosmologie standard
    l'existence de conditions initiales naturelles est résolue par les
    scenarios d'Inflation et ce succès repose sur la structure causale de la
    géometrie de fond de Sitter ($dS$) qui tend a effacer les irrégularités
    initiales. Mais dans le scénario branaire de Randall-Sundrum la géométrie
    de fond est $AdS$, dont la structure causale fait que l'amplitude des
    irregularités initiales est conservée au cours du temps, empêchant ainsi de
    générer l'homogénéité et l'isotropie de l'Univers observées
    aujourd'hui. Si les difficultés techniques peuvent être écartées par des
    approches numériques, le problème des conditions initiales persiste
    puisque les schémas numériques ont besoin de spécifier des conditions
    initiales particulières dans le bulk pour faire évoluer les
    perturbations. Dans ce travail de thèse nous abordons le problème de
    l'évolution des perturbations cosmologiques dans un Univers branaire en
    expansion de façon analytique: on utilise une approche effective
    quadri-dimensionnelle où on ``trace'' sur l'information manquante 
    portée par les degrés de liberté du bulk, que l'on englobe dans le
    propagateur retardé de $AdS$, ce qui évite de définir un vide initial dans
    le bulk. Nous calculons le propagateur retardé effectif sur
    la 
    brane en resommant les effets de rétroaction du bulk, contenus dans le
    propagateur retardé du bulk $AdS$, à tous les ordres dans le couplage
    brane-bulk. L'objectif suivi dans ce travail de thèse est d'obtenir des
    résultats quantitatifs 
    quant à l'ordre de magnitude des effets cosmologiques dûs à la
    présence de dimensions supplémentaires, cela  dans un modèle d'Univers
    branaire en expansion arbitraire et du type Randall-Sundrum. Au lieu de
    chercher des solutions exactes aux équations de perturbations
    cosmologiques, nous nous sommes intéressés davantage au rôle des gravitons
    du bulk dans leur interaction avec les degrés de liberté localisés sur la
    brane. Dans le cas d'une expansion \emph{non-uniforme} de l'Univers, les
    excitations quantiques de degrés de liberté localisés sur la brane peuvent engendrer
    l'émission de gravitons dans le bulk ; ceux-ci peuvent alors s'échapper vers l'infini
    futur, faisant apparaitre, du point de vue quadridimensionnel d'un
    observateur  localisé sur la brane, une forme de \emph{dissipation}. Les
    gravitons émis peuvent aussi être réfléchis ou diffractés dans le bulk
    inhomogène Anti-de Sitter à cause de la courbure, puis
    réabsorbés par la brane, engendrant une sorte de \emph{non-localité} du
    point de vue d'un observateur sur la brane. Ces processus de dissipation
    et de non-localité sont ``codés'' dans le propagateur $AdS$ du
    graviton et par conséquent dans le propagateur effectif sur la brane, 
    et constituent des signatures de la présence de
    dimensions supplémentaires. La majeure partie de cette thèse est consacrée
    au calcul des phénomènes dissipatifs en cosmologie branaire. Le dernier
    chapitre de cette thèse traite un sujet différent concernant un
    travail effectué en cosmologie standard : il s'agit d'un travail en
    progrès sur la reconstruction en espace réel des effets de lentilles gravitationnelles
    qui contaminent les spectres primordiaux du CMB. Ce mémoire de thèse est
    organisé comme suit :
    \begin{itemize}
    \item Au chapitre \ref{chapter:motiv} nous donnons les motivations physiques
    de la présence de dimensions supplémentaires dans l'Univers et introduisons
    les modèles d'Univers branaires.
    \item Au chapitre \ref{chapter:rs} nous présentons le modèle de
      Randall-Sundrum
    ainsi que ses conséquences cosmologiques, ce travail de thèse
    s'inscrivant dans ce modèle.
    \item Au chapitre \ref{chapter:pert} nous présentons la théorie des
    perturbations cosmologiques branaires dans le formalisme de Mukohyama et nous discutons quels sont les
      problèmes et les difficultés rencontrées dans le calcul de l'évolution
      des perturbations cosmologiques dans un Univers branaire en expansion.
    \item Au chapitre \ref{chapter:dissip} nous exposons l'approche que nous
    avons adoptée pour étudier l'interaction brane-bulk et évaluer l'évolution des
      perturbations cosmologiques dans un Univers branaire en expansion.
      Nous montrons comment des processus de dissipation et
      de non-localité sont ``codés'' dans le propagateur du bulk Anti-de
      Sitter et nous calculons les taux de
    dissipation de divers degrés de liberté localisés sur la brane. Nos
    résultats que nous exposons dans ce
    chapitre ont donné lieu à la publication d'un article \cite{moi}:
    ``\emph{Dissipation and nonlocality in a general expanding braneworld
    universe}'', Mathieu Remazeilles, Phys. Rev. D79:043523 (arXiv:0807.4238 [hep-th]).
    \item Au chapitre \ref{chapter:adsprop} nous calculons le propagateur
    retardé exact du graviton dans l'espace Anti-de Sitter de fa\c{c}on
      covariante. Ce travail n'a pas été soumis à publication.
    \item Au chapitre \ref{chapter:cmb} nous terminons sur un tout autre sujet concernant un travail effectué
      en cosmologie standard sur les effets de lentilles gravitationnelles qui
      contaminent les spectres du rayonnement du Fond Diffus Cosmologique. Il
      s'agit de construire un estimateur statistique optimal du potentiel de lentille
      en \emph{espace réel} (par opposition à l'espace de Fourier tel qu'il a été
    fait dans la littérature) à partir des
      spectres observés des anisotropies de température et de
      polarisation, cela dans le but de pouvoir nettoyer les spectres
      primordiaux des effets de lentilles gravitationnelles tout en tenant
    compte des éventuelles coupures et excisions du ciel (points source, Voie
    Lactée, ...). Les résultats de
    chapitre sont en progrès et doivent bientôt donner lieu à la
      publication d'un article : ``\emph{CMB lensing reconstruction in real
    space}'',
    Martin Bucher, Kavilan Moodley and Mathieu Remazeilles.
    \end{itemize}
    La convention de sommation d'Einstein sur les indices sera souvent
    utilisée :
    $A_\mu A^{\mu}\equiv \sum_{i = 0}^{d-1}\sum_{j = 0}^{d-1} g_{ij} A^i A^j$
    pour un espace-temps à $d$ dimensions de métrique $g_{\mu\nu}$.

    La forme de ce mémoire de thèse traduit une volonté de
    présenter de fa\c{c}on claire et concise
    quels sont les enjeux et les difficultés du calcul des perturbations
    cosmologiques dans un Univers branaire en expansion et quelle a éte mon
    approche pour les résoudre. Il ne s'agit en aucun cas de rappeler les
    bases de la théorie de la relativité générale ou les résultats de la
    cosmologie  standard que le lecteur intéressé pourra trouver dans 
    les nombreux ouvrages pédagogiques de qualité déjà existant sur le sujet
    et cités dans la bibliographie à la fin de ce manuscript.


\chapter{Dimensions supplémentaires et Univers
  branaires}\label{chapter:motiv}

L'existence de dimensions spatiales supplémentaires dans l'Univers a d'abord
été proposée dans la théorie de Kaluza-Klein en 1921. En introduisant une
cinquième dimension dans l'espace-temps, Kaluza \cite{kk1} puis Klein
\cite{kk2} ont remarqué qu'on
pouvait de cette fa\c{c}on unifier l'interaction gravitationnelle et
l'interaction électromagnétique. En effet les équations d'Einstein de la relativité générale
à \emph{cinq} dimensions peuvent se décomposer, lorsqu'on les projette sur
l'espace-temps usuel à quatre dimensions, en les équations d'Einstein à
quatre dimensions de la
relativité générale, les équations de Maxwell décrivant
l'interaction électromagnétique à quatre dimensions plus une équation de
Klein-Gordon pour un nouveau champ scalaire appelé dilaton. L'absence de
détection de la cinquième dimension a été expliquée par le fait que cette
dimension supplémentaire devait être compactifiée autour du cylindre
$R^4\times S^1$ de rayon suffisamment petit de fa\c{c}on à échapper
aux instruments de mesure. La théorie de Kaluza-Klein est ensuite tombée un
peu dans l'oubli avec l' engouement des physiciens provoqué par les
découvertes de la mécanique quantique et des théories de jauges $U(1)$, $SU(2)$ et $SU(3)$
au cours du vingtième siècle, ces théories quantiques de champs décrivant avec succès les
interactions électromagnétique, faible et forte entre les particules
élémentaires ainsi que grand nombre d'expériences subatomiques. Cependant les
phénomènes gravitationnels extrêmes existant dans l'Univers incluent à la fois des
effets quantiques et des effets gravitationnels. C'est le cas pour la
formation des trous noirs ou le comportement de
la physique proche du Big-Bang. Dans l'objectif de comprendre la physique de
ces phénomènes, la construction d'une théorie
quantique de la gravitation s'est avérée indispensable. Dans cette quête
d'une unification de la gravité avec les autres interactions fondamentales, la question de la
dimensionnalité de l'espace-temps a resurgi à travers la théorie des cordes
supersymétriques, ou supercordes, qui reste le meilleur candidat à l'unification. En
rempla\c{c}ant la représentation ponctuelle des particules de la théorie
quantique des champs par une représentation unidimensionnelle sous forme de
cordes élémentaires, cette théorie est à même de réconcilier la relativité générale
avec les théories quantiques de champs, mais au prix d'une reconsidération de
la dimensionnalité de l'espace-temps.  En
effet la quantification de la théorie des cordes n'est pertinente d'un point de
vue mathématique que si les cordes évoluent dans un espace-temps à
$(9+1)$ dimensions. Cette découverte, associée à la découverte que la théorie
des cordes contient des solutions classiques correspondant à des membranes,
appelées ``D-branes'', de dimensions diverses,  en plus des cordes elles-même,
a inspiré l'idée que notre Univers observable réside sur une brane à $(3+1)$ 
dimensions plongée dans un espace-temps de plus haute dimension. Nous exposons
dans ce chapitre les motivations physiques issues de
la théorie des cordes qui ont inspiré les scénarios branaires et nous
présentons quelques modèles branaires.

\section{Deux résultats fondamentaux de la théorie des cordes}\label{sec:string}

Nous déduisons dans cette section deux résultats majeurs de la théorie des
cordes \cite{string} qui ont motivé l'étude de modèles d'Univers
branaire : la présence du graviton dans le spectre des états quantiques de la
corde et la dimensionnalité de l'espace-temps. Nous nous restreignons à
l'étude de la corde bosonique libre pour appréhender ces deux propriétés.

\subsection{Le spectre quantique des cordes contient le graviton}\label{subsec:graviton}

En théorie quantique des champs la particule ponctuelle relativiste libre
se propage dans l'espace-temps le long d'une ligne d'Univers de longueur
minimale (géodésique), paramétrisée par le temps propre $\tau$ et à l'aide des coordonnées
d'espace-temps notées $x^{\mu}(\tau)$. Suivant le principe de moindre action,
les équations du mouvement de la particule sont obtenues en minimisant l'action
proportionnelle à la longueur de la ligne d'Univers. La corde
de la théorie des cordes n'étant que la généralisation bidimensionnelle de la
particule, elle se propage dans l'espace-temps le long d'une \emph{feuille}
d'Univers d'aire minimale, paramétrisée par deux
variables $\sigma=(\sigma_0$, $\sigma_1$) et à l'aide des coordonnées
d'espace-temps notées $X^{\mu}(\sigma_0,\sigma_1)$. Le choix naturel pour
l'action de la corde bosonique est l'action de Nambu-Goto proportionnelle à
l'aire de la feuille d'Univers, ou plutôt son homologue Lorentzien, $S = T\int d^2\sigma \sqrt{\vert
  \det g_{\mu\nu}\partial_aX^\mu\partial_bX^\nu\vert}$. En utilisant la
contrainte (\ref{qqq:tcorde}) calculée plus bas on montre aisément que cette
action est équivalente à l'action de Polyakov
\ba\label{qqq:ng}
S & = &  -{T\over 2}\int d^2\sigma \sqrt{\vert
  h\vert}h^{ab}(\sigma)g_{\mu\nu}(X)\partial_a X^{\mu}\partial_b X^{\nu}
\ea
où la tension $T$ de la corde a la dimension d'une masse au carré et
où il faut distinguer la métrique intrinsèque $h_{ab}(\sigma)$ de la feuille d'Univers
bidimensionnelle ($a,b=0,1$) de la métrique $g_{\mu\nu}(X)$ de l'espace-temps
physique (ou espace cible) à $d$ dimensions ($\mu,\nu=0,...,d-1$). Notons que si
l'espace cible est Minkowskien on peut interpréter cette théorie d'un autre
point de vue comme étant la gravité classique à deux dimensions, décrite par
la métrique $h_{ab}$, en interaction avec $d$ champs scalaires $X_{\mu}$ ; le
terme scalaire de Ricci de l'action d'Einstein-Hilbert n'ayant aucune raison
d'être présent dans l'action parce que
c'est un invariant topologique à deux dimensions et en ce sens un terme non
dynamique\footnote{Notons de plus que l'on peut ajouter un terme de
  ``constante cosmologique'' dans l'action à deux dimensions qui contraint à respecter
  l'invariance relativiste de l'aire de la feuille d'Univers.}.  La minimisation de l'action
(\ref{qqq:ng}) par rapport au champ métrique $h_{ab}$ fournit une équation de
contrainte sur l'annulation du tenseur énergie-impulsion de la feuille d'Univers
\ba\label{qqq:tcorde}
T_{ab} & = &  - {1\over\sqrt{\vert
  h\vert}}{\delta S\over\delta h^{ab}} = \partial_a X^{\mu}\partial_b
X_{\mu}-{1\over 2}h_{ab}h^{cd}\partial_c X^{\mu}\partial_d
X_{\mu} = 0.
\ea
La feuille
d'Univers de la trajectoire de la corde est conformément plate, c'est-à-dire
conforme à l'espace Minkowski, comme toute
variété bidimensionnelle. En effet l'action est invariante sous les
difféomorphismes à deux dimensions, ainsi que sous transformation conforme
de Weyl de la métrique de la feuille d'Univers, $h_{ab}\rightarrow
\Lambda(\sigma_0,\sigma_1)h_{ab}$.
L'invariance sous les difféomorphismes nous laisse la liberté de choisir la jauge telle que la métrique
bidimensionnelle soit conforme à Minkowski $h_{ab} = e^{\phi}\eta_{ab}$, où
$\eta_{ab}$ est la métrique Minkowskienne, car
$\phi$ disparaît de l'action.
Nous considérons pour la suite que l'espace-temps physique de fond est
Minkowskien afin de
simplifier le raisonnement et nous notons $\sigma_0 = \tau$, $\sigma_1 = \sigma$, alors
l'action (\ref{qqq:ng}) se simplifie comme
\ba\label{qqq:action}
S & = &  -{T\over 2}\int d\sigma d\tau \left[{X'}^2-\dot{X}^2\right],
\ea
où un ``prime'' dénote une dérivation par rapport à $\sigma$ et un ``point''
une dérivation par rapport au temps $\tau$.
Dans le choix de jauge conforme la minimisation de l'action (\ref{qqq:action}) par
rapport à $X^\mu$ conduit à l'équation du mouvement de la corde qui n'est
autre qu'une équation d'onde libre bidimensionnelle, et (\ref{qqq:tcorde})
conduit à deux équations de contraintes appelées conditions de Virasoro :
\ba\label{qqq:eomcorde}
X''-\ddot{X} & = & 0, \qquad \dot{X}\cdot X' = 0, \qquad {X'}^2+\dot{X}^2 = 0.
\ea
Il est nécessaire de choisir des conditions de bord pour la corde. Si la
corde est une corde \emph{fermée} il est naturel d'imposer une condition de périodicité
sur le paramètre spatial $\sigma$ tel que $X^{\mu}(\sigma,\tau) =
X^{\mu}(\sigma+\pi,\tau)$. Si la corde est une corde \emph{ouverte}, chacune
de ses extrémités peut satisfaire des conditions de bord de type Neumann,
${X'}^\mu(0,\tau) = {X'}^\mu(\pi,\tau) = 0$, ou peut aussi satisfaire des
conditions de bord de type Dirichlet, $\dot{X}^\mu(0,\tau) = \dot{X}^\mu(\pi,\tau)
= 0$. Les conditions de Dirichlet imposent à la corde ouverte d'avoir ses
deux extrémités fixées dans une dimension spatiale, c'est-à-dire que la corde
se trouve localisée dans l'espace-temps sur une hypersurface de genre temps et de codimension un où
sont attachées ses extrémités. Cette hypersurface de l'espace-temps est
appelée une D-brane où ``D'' vient de Dirichlet, ou encore une $(d-2)$-brane car elle
comporte $(d-2)$ dimensions spatiales. Son action est une généralisation à $(d-1)$
dimensions de celle de la corde qui n'est autre qu'une $1$-brane. Comme nous allons le voir, le spectre
quantique de la corde ouverte contient les états particulaires du Modèle
Standard tel que le photon alors que le spectre de la corde fermée contient
l'état du graviton. Les ingrédients sont donc réunis pour concevoir l'idée que
les particules du Modèle Standard ``seraient'' des cordes ouvertes contraintes de
se déplacer sur une D-brane de codimension un, représentant l'Univers
observable, qui est entourée d'une dimension supplémentaire où seuls les
gravitons, qui ``seraient'' des cordes fermées, peuvent se déplacer. La
solution générale des équations du mouvement (\ref{qqq:eomcorde}) pour la
corde ouverte de type Neumann est donnée par
\ba\label{qqq:open}
X^\mu & = & x^\mu+\ell_s^2p^\mu\tau+i\ell_s\sum_{n\neq 0}{1\over
  n}\alpha_n^\mu e^{-in\tau} \cos n\sigma,
\ea
où $\ell_s = (\pi T)^{-1/2}$ est la longueur fondamentale de la 
corde\footnote{La taille caractéristique des cordes peut être
calculée par analyse dimensionnelle en fonction de la constante de gravitation
$G$, de la vitesse de la lumière $c$ et de la constante de Planck~$\hbar$~:~
$\ell_s~=~\sqrt{\hbar G/c^3}~\sim~10^{-33}$~cm. Cette taille minuscule, dite
de Planck, justifie que les cordes
puissent être représentée à plus basse énergie par les particules ponctuelles
de la théorie quantique des champs.} et
$x^\mu$ et $p^\mu$ correspondent respectivement à la position et à l'impulsion
du centre de masse de la corde. Pour la
corde fermée, la solution est une su\-per\-position de modes gauches G et droits
D,\\
\noindent
$X^\mu(\tau,\sigma)~=~X_D^\mu(\tau-\sigma)+X_G^\mu(\tau+\sigma)$, avec
\ba\label{qqq:closed}
X_D^\mu & = & {1\over 2}x^\mu+{1\over 2}\ell_s^2p^\mu(\tau-\sigma)+{1\over 2}i\ell_s\sum_{n\neq 0}{1\over
  n}\alpha_n^\mu e^{-in(\tau-\sigma)},\cr
X_G^\mu & = & {1\over 2}x^\mu+{1\over 2}\ell_s^2p^\mu(\tau+\sigma)+{1\over 2}i\ell_s\sum_{n\neq 0}{1\over
  n}\tilde{\alpha}_n^\mu e^{-in(\tau+\sigma)}.
\ea

Nous pouvons quantifier la corde de fa\c{c}on canonique, comme en théorie
quantique des champs, en élevant au rang d'opérateurs les variables conjuguées
$X^\mu$ et $P^\mu~=~\delta~S~/~\delta~\dot{X}_\mu~=~T~\dot{X}^\mu$ et en imposant
des relations de commutation canoniques à temps égaux :\\
\noindent
$\left[X^\mu(\tau,\sigma),P^\nu(\tau,\sigma')\right] =
 i\eta^{\mu\nu}\delta(\sigma-\sigma')$ et
$\left[P^\mu(\tau,\sigma),P^\nu(\tau,\sigma')\right] =
\left[X^\mu(\tau,\sigma),X^\nu(\tau,\sigma')\right] = 0$. Par conséquent les
coefficients des modes de la corde deviennent eux-même des opérateurs de création,
$(\alpha_m^\mu)^\dagger = \alpha_{-m}^\mu$, et
d'annihilation, $\alpha_m^\mu$, qui satisfont les relations de commutation suivantes
\ba\label{qqq:comm}
\left[\alpha_m^\mu,\alpha_n^\nu\right] & = & m\delta_{m+n}\eta^{\mu\nu},
\qquad \left[x^\mu,p^\mu\right] = i \eta^{\mu\nu}
\ea
pour la corde ouverte. L'état du vide $\vert 0,k^\nu\rangle$ est annihilé
selon $\alpha_m^\mu\vert 0,k^\nu\rangle =
0$ ($m > 0$) et l'opérateur impulsion du centre de masse satisfait $p^\mu\vert
0,k^\nu\rangle = k^\mu\vert 0,k^\nu\rangle$. Les états excités sont créés
selon $\vert m_1,\mu_1,...,m_k,\mu_k;k^\nu\rangle =
\alpha_{-m_1}^{\mu_1}...\alpha_{-m_k}^{\mu_k}\vert 0;k^\nu\rangle$. Des
relations similaires sont obtenues pour les opérateurs $\tilde{\alpha}_n$ de
la corde fermée. Les contraintes classiques de Virasoro sont
représentées au niveau quantique par les opérateurs de Virasoro
\ba\label{qqq:virasoro}
L_0 & = & \sum_{n=1}^{\infty}\alpha_{-n}^\mu\alpha_{n,\mu}+{\alpha_0^2\over
  2}-a, \qquad L_m = {1\over 2}\sum_{n=-\infty}^{\infty}\alpha_{m-n}^\mu\alpha_{n,\mu}
\ea
où on a défini $\alpha_0^\mu = \ell_s p^\mu$ et où $a$ est une constante
due à l'ambiguité dans l'ordonnement normal des opérateurs puisque $\alpha_{-n}^\mu$ et $\alpha_{n}^\mu$
ne commutent plus dans $L_0$. On peut vérifier que les contraintes
classiques de Virasoro $(\dot{X}\pm X')^2 = 0$ sont \emph{équivalentes} à $L_n
= 0$ pour tout $n$. La contrainte $L_0\vert phys \rangle = 0$
permet d'obtenir le spectre de masse des états \emph{physiques} $\vert
phys\rangle$, c'est-à-dire
sur la couche de masse, selon
\ba\label{qqq:openm}
M^2 = -p^2 & = & {2\over
  \ell_s^2}\left(\sum_{n=1}^{\infty}\langle phys\vert\alpha_{-n}\cdot\alpha_{n}\vert
phys\rangle-a\right) = {2\over
  \ell_s^2}\left(N-a\right)
\ea
où $N$ est la valeur propre de l'opérateur nombre d'excitations  $\hat{N} =
\sum_{n=1}^{\infty}\alpha_{-n}\cdot\alpha_{n}$. Le spectre de la corde ouverte
démarre donc avec
\begin{itemize}
\item Le vide $\vert 0,k^\nu\rangle$ :  $N = 0$ et $M^2= -2a/\ell_s^2$
  qui est un \emph{tachyon} si $a > 0$.
\item Le premier état excité est un état bosonique vectoriel,
  $\alpha_{-1}^\mu\vert 0,k^\nu\rangle$ : $N = 1$,
  $M^2 = 2(1-a)/\ell_s^2$. Cet état s'identifie au \emph{photon} si sa
  masse est nulle, c'est à dire si $a = 1$.
\end{itemize}
Nous constatons donc que le choix $ a = 1$ dans la théorie des cordes
bosoniques fait que le spectre de la corde ouverte contient le photon du
Modèle Standard, c'est-à-dire la particule de jauge $U(1)$ de spin 1 et à
$(d-2)$ degrés de liberté véhiculant l'interaction électromagnétique. Cependant le vide est tachyonique ce qui traduit une
instabilité du vide (masse imaginaire, impulsion de genre espace) et donc une
incohérence à priori de la théorie quantique. En réalité cette instabilité tachyonique
disparaît pour la théorie complète, dite des supercordes, qui inclue la présence
de cordes fermioniques en plus des cordes bosoniques. L'action des supercordes
est intrinsèquemnt supersymétrique (symétrie bosons-fermions), ce qui a
pour effet d'éliminer le tachyon. Regardons maintenant le spectre de la corde
fermée bosonique. Les relations de commutation supplémentaires pour les modes
gauches de la corde fermée sont similaires à celles des modes droits
$\left[\tilde{\alpha}_m^\mu,\tilde{\alpha}_n^\nu\right]  =
m\delta_{m+n}\eta^{\mu\nu}$, $\left[\tilde{\alpha}_m^\mu,\alpha_n^\nu\right]  =
0$, où on note $\tilde{\alpha}_0^\mu =
\alpha_0^\mu = (1/2)\ell_s p^\mu$. Les états quantiques de la corde fermée
sont crées à l'aide des opérateurs de création selon $\Pi_{m=1}^i\Pi_{m'=1}^j
\alpha_{-n_m}^{\mu_m}\tilde{\alpha}_{-n_{m'}}^{\mu_{m'}}\vert
0,0,k^\nu\rangle$. Les conditions de Virasoro deviennent au niveau quantique $L_m\vert
0,0,k^\nu\rangle = \tilde{L}_m\vert
0,0,k^\nu\rangle = 0$. L'annulation de l'opérateur $(L_0-\tilde{L}_0)$ impose
l'égalité $N = \tilde{N}$ entre les nombres de particules générées par les
modes droits et gauches. L'annulation supplémentaire de l'opérateur
$(L_0+\tilde{L}_0)$ fournit le spectre de masse suivant pour la corde fermée
\ba\label{qqq:closedm}
M^2 & = & {2\over
  \ell_s^2}\sum_{n=1}^{\infty}\left(2N-2a\right).
\ea
Le vide  de la corde fermée bosonique est encore tachyonique si $a > 0$. Le
premier état excité $\vert \Omega_{\mu\nu}\rangle =
\alpha_{-1}^\mu\tilde{\alpha}_{-1}^\nu\vert 0,0,k^\rho\rangle$ ($N = 1$) peut se décomposer selon
\begin{itemize}
\item un état de spin 0 appelé dilaton, $\vert\Phi\rangle = \vert
  \Omega_\mu^\mu\rangle$, de masse $M^2 = {4\over \ell_s^2}(1-a)$,
\item un état tensoriel antisymétrique, $\vert B_{\mu\nu}\rangle = \left(\vert
    \Omega_{\mu\nu}\rangle-\vert\Omega_{\mu\nu}^\dagger\rangle\right)/2$, de masse $M^2 = {4\over \ell_s^2}(1-a)$,
\item un état tensoriel symétrique et sans trace, $\vert
  g_{\mu\nu}\rangle$, de masse $M^2 = {4\over \ell_s^2}(1-a)$.
\end{itemize}
En faisant le choix $ a = 1$ comme pour la corde ouverte, on constate
le spectre de la  corde fermée contient un état de spin 2 sans
masse, symétrique et sans trace,  état qui s'identifie naturellement au
\emph{graviton}, le champ de métrique de l'espace-temps. Le tachyon de la
corde fermée disparait en théorie des supercordes lorsqu'on inclue les
superpartenaires fermioniques dans la théorie. Le calcul des premiers états
quantiques du spectre révèle que la théorie des supercordes engendre, parmi
une infinité de particules, le photon, boson de jauge
véhiculant l'interaction électromagnétique, ainsi que le graviton, boson de
jauge devant véhiculer l'interaction gravitationnelle\footnote{On peut montrer que
la théorie des cordes contient également les bosons de jauge $SU(2)$ de
l'interaction faible et les gluons de l'interaction forte mais cela va
au-delà de l'objectif de cette section.}.  En contenant le champ
du graviton dans son spectre, la théorie des
supercordes se présente donc comme une théorie quantique qui doit contenir la
théorie de la relativité générale à basse énergie.

\subsection{L'espace-temps possède  (9+1)  dimensions}\label{subsec:xtradim}

Le choix de la jauge conforme, $h_{ab} = e^\phi\eta_{ab}$, adoptée
précédemment ne fixe pas totalement la jauge pour l'action de Polyakov puisque
$\phi$ peut-être choisi arbitrairement. Il reste une symétrie de jauge
résiduelle liée aux transformations conformes de la feuille d'Univers. On peut
utiliser cette liberté résiduelle pour fixer davantage la jauge. Introduisons pour cela les coordonnées du cône de lumière pour la feuille
d'univers, $\sigma_\pm = \tau\pm\sigma$,  et pour
l'espace-temps $X^\pm = \left(X^0\pm X^{d-1}\right)$, les coordonnées
transverses $X^i$, $i=1,...d-2$, demeurant inchangées. L'action
(\ref{qqq:action}) puis l'équation du mouvement deviennent dans ces coordonnées
\ba\label{qqq:actionlum}
S & = & \int d\sigma_+ d\sigma_-\partial_+ X\cdot\partial_- X, \qquad \partial_+\partial_- X = 0.
\ea
La symétrie
résiduelle de l'action (\ref{qqq:actionlum}) correspond à la possibilité de
reparamétrisations arbitraires de la feuille d'Univers $\sigma_+\rightarrow
f(\sigma_+)$, $\sigma_-\rightarrow g(\sigma_-)$. Comme les solutions de l'équation
du mouvement sont elles-même la somme d'une fonction de $\sigma_+$ et
d'une fonction de $\sigma_-$, l'invariance résiduelle nous autorise à fixer
la jauge simple, dite du cône de lumière, pour un degré de liberté :
\ba\label{qqq:lightgauge}
X^+(\sigma,\tau) & = &x^+ + \ell_s^2p^+\tau.
\ea
Les contraintes de Virasoro classiques, $2\ell_s^2p^+\left(\dot{X}^-\pm
  {X^-}'\right) = \left(\dot{X}^i\pm {X^i}'\right)^2$, où $i=1,...,d-2$,
permettent d'obtenir $X^- = x^-
+\ell_s^2p^-\tau+i\ell_s^2\sum_{n\neq 0} {\alpha_n^-\over
  n}e^{-in\tau}\cos(n\sigma)$ en fonction des $X^i$ :
\ba\label{qqq:alphamoins}
\alpha_n^- & = & {1\over p^+}\left({1\over 2}\sum_{i = 1}^{d-2}\sum_{m =
    -\infty}^{\infty}:\alpha^i_{n-m}\alpha_{i,m}: - a\delta_n\right),
\ea
où on a noté $\alpha_0^- = \ell_s^2 p^-$ et où $a$ est une constante due à l'ambiguité d'ordonnement normal des
opérateurs quantiques de création et d'annihilation $\alpha_{-m}^i$ et
$\alpha_{m}^i$. L'encadrement entre deux points dénote le produit en
ordre normal des opérateurs. On en déduit le spectre de masse des
états quantiques $M^2 = 2p^+p^- - p^ip_i =
\langle phys\vert
2p^+{\alpha_0^-\over\ell_s^2}-{1\over\ell_s^2}\sum_{i=1}^{d-2}\alpha_0^i\alpha_0^i\vert
phys\rangle = 2(N-a)$ où $N$ est la
valeur propre de l'opérateur nombre d'excitations $\hat{N} = \sum_{m =
  1}^{\infty} \alpha^i_{-m}\alpha_{i,m}$. Le premier état excité $\vert
\alpha_{-1}^i \rangle$ possède $(d-2)$ degrés de liberté et  correspond au
photon sans masse si $a = 1$ de nouveau. Cette condition se
trouve être en fait indispensable pour satisfaire l'invariance de Lorentz de la théorie
puisque le premier état excité se transforme sous Lorentz comme un vecteur du
groupe $SO(d-2)$ qui correspond au groupe de symétrie d'une particule
vectorielle \emph{sans masse}. Dans la jauge du cône de lumière on peut calculer cette fois directement
la constante d'ordonnement normal $a$, d'après la relation
\ba\label{qqq:formule}
{1\over 2}\sum_{i = 1}^{d-2}\sum_{n =
  -\infty}^{+\infty}:\alpha_{-n}^i\alpha_n^i: & = & {1\over 2}\sum_{i = 1}^{d-2}\sum_{n =
  -\infty}^{+\infty}\alpha_{-n}^i\alpha_n^i - {d-2\over 2}\sum_{n =
  1}^{\infty} n.
\ea
On en déduit $a = -{d-2\over 2}\sum_{n = 1}^{\infty} n$. Cette série divergente
peut être régularisée au moyen de la fonction Zeta de Riemann selon $\sum_{n
  = 1}^{\infty} n = \zeta(-1) = -1/12$ de telle sorte que
\ba\label{qqq:avsd}
a & = & {d-2\over 24}.
\ea
Les conditions d'existence du photon et du graviton et la condition d'invariance de Lorentz
dans la théorie des cordes se sont traduits par la condition $a = 1$. Combinée à
l'expression explicite (\ref{qqq:avsd}), cette condition impose à son tour la condition
que l'espace-temps de la théorie des cordes bosoniques possède
exactement $d = 25+1$ dimensions. De fa\c{c}on alternative la quantification de la
théorie par intégrale fonctionnelle à la Fadeev-Popov  aurait montré qu'une anomalie
quantique apparait dans l'algèbre des opérateurs de Virasoro (incluant les
champs de \emph{ghost}). La condition $d =
26$ est alors nécessaire pour annuler l'anomalie quantique et ainsi rendre la théorie cohérente au niveau
quantique.  La théorie des
cordes bosoniques n'est cependant pas satisfaisante d'un point de vue physique
puisqu'elle ne contient pas de fermions. Une quantification analogue de la version supersymétrique de la
théorie montre que les supercordes doivent en réalité évoluer dans un
espace-temps à $9+1$ dimensions. Enfin précisons qu'il existe en fait cinq théories des
supercordes différentes mais duales entre elles qui ne sont que des limites particulières
d'une seule et unique théorie fondamentale, appelée ``théorie M'' \cite{hw1}, qui impose à l'espace-temps de posséder
finalement $10+1$ dimensions.

Deux alternatives, qui ne s'excluent pas l'une et l'autre, ont été proposées pour
expliquer l'absence de détection des six (ou sept) dimensions supplémentaires
de l'espace-temps : ces dimensions peuvent soit être compactifiées à la
Kaluza-Klein sur une échelle suffisament petite pour échapper à la résolution
des instruments de mesure, soit être non-compactes mais orthogonales à une
brane à quatre dimensions sur laquelle réside notre Univers observable.

\section{Modèles d'Univers branaires}\label{sec:brane}

\begin{figure}
  \begin{center}
\includegraphics[width=13.5cm]{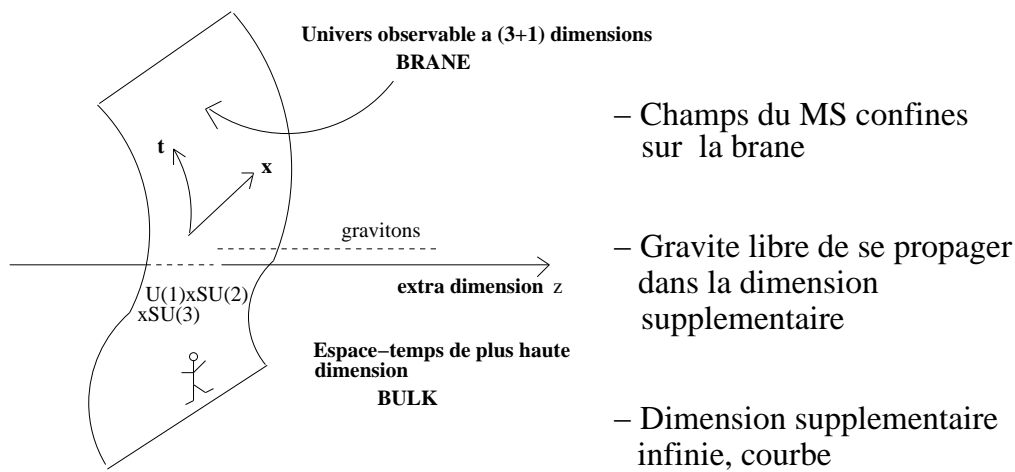}
 \end{center}
\caption{
Description schématique d'un Univers branaire entourée d'\emph{une} dimension
 supplémentaire infinie. Les particules du Modèle Standard sont contraintes de se déplacer dans les quatre
dimensions de la brane. Seuls les gravitons
peuvent s'échapper dans la dimension supplémentaire.}
\label{fig:schemabrane}
\end{figure}

Nous avons vu que la théorie des cordes englobe toutes les théories
de jauge y compris la gravitation à travers la présence du graviton dans son
spectre. En contrepartie la quantification cohérente de cette théorie d'unification implique que
l'espace-temps possède des dimensions supplémentaires. De plus les champs du
Modèle Standard, décrits par des cordes ouvertes, peuvent être contraints de
se déplacer sur des membranes (D-branes) de codimension 1. Seuls les
gravitons, décrits par des cordes fermées, peuvent se propager dans la
dimension supplémentaire. Ces conséquences majeures de la théorie des cordes
ont été la première source d'inspiration dans la construction de modèles
branaires, considérant que notre Univers observable réside
en fait sur une brane à quatre dimensions plongée dans un espace-temps de
dimension supérieure (Fig. \ref{fig:schemabrane}). 
La physique du Modèle
Standard conserve son comportement quadridimensionnel puisque tous les
champs du Modèle Standard sont confinés sur la brane. La gravitation et
la cosmologie doivent en revanche être affectées par la présence de
dimensions supplémentaires puisque les gravitons peuvent se déplacer dans
toutes les dimensions du bulk.

Dans cette section nous présentons dans les grandes lignes quelques premiers modèles branaires
dont on pourra comparer les caractéristiques avec le modèle de Randall-Sundrum
présenté de manière détaillée au chapitre \ref{chapter:rs}.

\subsection{Modèle de Ho$\check{r}$ava-Witten}\label{subsec:hw}

\begin{figure}
  \begin{center}
\includegraphics[width=8.5cm]{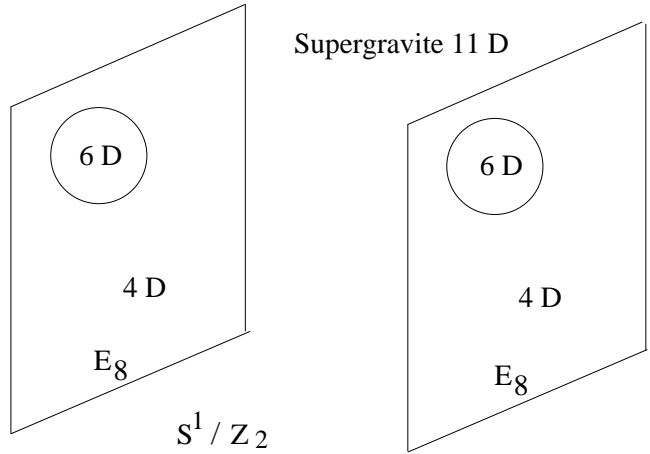}
 \end{center}
\caption{
Modèle de Ho$\check{r}$ava-Witten. L'espace-temps de la théorie M à 11 dimensions
est décomposé en deux branes à 10 dimensions, situées aux points fixes de
l'orbifold $S^1/Z^2$, dont 6 dimensions sont compactifiées sur une variété de
Calabi-Yau et (3+1) dimensions sont non-compactes. Une théorie de jauge $E_8$
est confinée sur chaque brane. La supergravité se propage
dans les 11 dimensions.}
\label{fig:witten}
\end{figure}

Ce modèle \cite{hw1, hw2} a été construit pour trouver des dualités qui relient les cinq versions perturbatives
de la théorie des supercordes à la théorie M. On ne connait pas encore
 complètement la structure non-perturbative de la théorie M mais on sait que
 cette théorie fondamentale contient des $p$-branes de dimension $(p+1)$
 diverses et qu'elle admet la supergravité\footnote{Théorie avec supersymétrie locale qui induit la
 présence de la gravité.} à 11 dimensions comme limite de
 basse énergie. En 1996, Ho$\check{r}$ava et Witten ont exprimé la théorie M, ou sa
 limite de basse énergie, la supergravité à 11 dimensions, sur la variété
 quotient $\mathcal{M} = R^{10}\times S^1/Z_2$, c'est-à-dire que la onzième dimension est
 compactifiée sur un cercle $S1$ ayant une symétrie miroir $Z_2$ et donc deux
 points fixes, le terme
 mathématique pour $S_1/Z_2$ étant ``orbifold''. En faisant cela ils ont
 constaté que, dans la limite d'un petit rayon de compactification,  la théorie M
 est décrite par la théorie des cordes hétérotiques $E_8\times E_8$ à 10
 dimensions en
 couplage fort, qui est une
 des cinq versions de la théorie des cordes. Les deux points fixes de
 l'orbifold $S_1/Z_2$ constituent des sous-espaces branaires à $(9+1)$ dimensions sur
 chacun desquels est confinée une théorie de jauge $E_8$ afin d'éliminer les
 anomalies de jauge et gravitationnelle dans la théorie. On
 peut encore compactifier à la Kaluza-Klein six dimensions spatiales sur une
 variété dite de Calabi-Yau, $X$, parce que ce type de variété compacte conserve la
 supersymétrie de la théorie des cordes, réduisant ainsi le nombre de
 dimensions non-compactes à $(3+1)$ sur chaque brane fixe de l'orbifold :
 $\mathcal{M} = R^4\times X\times S^1/Z_2$. En supposant que l'échelle de
 compactification des six dimensions sur la variété de Calabi-Yau est bien
 plus petite que le rayon de l'orbifold, on obtient le scénario d'un monde à
 cinq dimensions $\mathcal{M} \approx R^4\times S^1/Z_2$ où se propage la
 gravité et contenant deux univers
 branaires à $(3+1)$ dimensions sur lesquels sont confinées les interactions
 de jauge (Fig.\ref{fig:witten}). De part et d'autre de chaque brane l'espace est symétrique dans la
 cinquième dimension d'après la symétrie $Z_2$. Le modèle ``cordiste'' d'Horava-Witten a
 inspiré le modèle de Randall-Sundrum présenté plus loin, ainsi que le besoin
 de comprendre la phénoménologie et la cosmologie branaire à \emph{cinq}
 dimensions \emph{non compactes}.

\subsection{Modèle ADD}\label{subsec:add}

On peut se poser la question de savoir pourquoi l'interaction gravitationnelle
entre deux particules est d'intensité beaucoup plus faible que les trois autres interactions de jauge du Modèle Standard. En effet l'échelle d'énergie à laquelle les interactions de jauge du Modèle Standard
sont d'intensité équivalente est donnée par l'échelle électrofaible $m_{EW}~
\sim~ 1~ TeV$, alors que l'échelle d'énergie à laquelle l'intensité de la gravité
serait aussi importante que celle des interactions de jauge est donnée par
l'échelle de Planck $M_p~ =~ G_N^{-1/2}~ \sim~ 10^{19}~ GeV$, où $M_p$ est la masse
de Planck et $G_N$ la constante de gravitation de Newton. Ceci constitue le
problème de \emph{hiérarchie} en physique des particules : pourquoi l'échelle
de Planck de l'interaction gravitationnelle est-elle si éloignée de l'échelle
électrofaible des trois autres interactions de jauge ($m_{EW}/M_p \sim 10^{-16}
\ll 1$) ?

Le modèle d'Arkani-Hamed, Dimopoulos et Dvali (ADD) \cite{add}, publié en 1998, propose
une solution au problème de hiérarchie. Dans ce modèle l'Univers
réside sur une brane Minkowski à quatre dimensions entourée de $n$ dimensions
supplémentaires compactes de taille $R$ ($R$ est le rayon de
compactification), donnant à  l'espace-temps total à $(4+n)$ dimensions
la topologie $\mathcal{M} = R^4\times S^n$. Les champs de jauge du Modèle
Standard sont confinés sur la brane, seule l'interaction gravitationnelle peut
se diluer dans les dimensions compactes. L'échelle fondamentale de la
gravitation est ainsi donnée par la masse de
Planck à $(4+n)$ dimensions $M_{p(4+n)}$ et la masse de Planck à quatre
dimensions n'est qu'une échelle effective sur la brane. En supposant que
l'échelle de Planck à $(4+n)$ dimensions est du même ordre de grandeur que
l'échelle électrofaible, $M_{p(4+n)}\sim m_{EW}\sim TeV$, on élimine ainsi tout
problème de hiérarchie. L'importance de la masse de Planck  à quatre
dimensions $M_p\sim
10^{19} GeV$,  mesurée par un observateur localisé sur l'Univers-brane,
s'explique donc par l'importance de la taille $R$ et du nombre $n$ des dimensions supplémentaires.


En appliquant le théorème de Gauss en
$(4+n)$ dimensions, on trouve que le potentiel Newtonien d'interaction
gravitationnelle entre deux masses test $m_1$, $m_2$ séparées d'une distance $r
\ll R$ vaut
\ba
V(r) \stackrel{r \ll R}{\sim} {m_1 m_2\over M_{p(4+n)}^{n+2}}{1\over r^{n+1}},
\ea
illustrant que la gravité se dilue dans les $n$ dimensions
supplémentaires. Par contre si les masses sont séparées d'une distance $r \gg
R$ alors elles ne ressentent plus la présence des dimensions supplémentaires
compactes et subissent un champ gravitationnel quasi quadridimensionnel :
on obtient le comportement habituel du potentiel gravitationnel en $1/r$
\ba
V(r) \stackrel{r \gg R}{\sim} {m_1 m_2\over M_{p(4+n)}^{n+2}R^n}{1\over r}.
\ea
La constante de Planck effective à quatre dimensions $M_p$ est donc reliée à la
à la constante de Planck fondamentale de plus haute dimension  selon $M_p^2 =
M_{p(4+n)}^{n+2} R^n$. En posant $M_{p(4+n)}^{n+2}~\sim~
m_{EW}~\sim~TeV$, la résolution du problème de hiérarchie exige donc que la
taille des dimensions supplémentaires compactes soit
\ba
R\sim 10^{-17+30/n} \mbox{ cm}.
\ea
Le cas $n = 1$ implique des modifications de la gravité de Newton usuelle
jusqu'à des distances $R \sim 10 ^{13}$ cm, c'est-à-dire de l'ordre des
distances caractéristiques du système
solaire. Ce cas est bien-sûr exclu par toutes les observations astronomiques. Par
contre, le comportement quadridimensionnel en $1/r$ de la loi de Newton n'ayant
jamais été testé à des distances inférieures au millimètre, cela laisse
entrevoir la possibilité que l'Univers est une brane entourée d'au moins deux
dimensions supplémentaires. Le cas $n = 2$ est donc le plus intéressant car
il implique l'existence de dimensions compactes de ``grande''\footnote{L'échelle
  $R\sim 0.1$ mm est ``grande'' par rapport aux autres échelles du problème
  telles que l'échelle électrofaible.} taille $R\sim
0.1$ mm et donc des modifications de la loi de Newton potentiellement détectables à cette
 ``grande'' distance. L'idée de ce modèle a été de profiter du fait que les trois
 interactions de jauge du Modèle Standard ont été testée aux échelles du TeV
 alors que la loi de la gravité n'a jamais été testée en dessous de $0.1$
 mm. Cela autorise la gravité à se propager dans un espace de plus haute
 dimension tout en contraignant les interactions de jauge à se propager sur une
 brane à quatre dimensions.

La présence de dimensions supplémentaires entraine d'autres conséquences
phénoménologiques, notamment l'existence de gravitons massifs comme nous
allons le voir. De
fa\c{c}on générale les gravitons
sans masse (ou ondes gravitationnelles) se déplacent à la vitesse de la lumière dans un espace-temps à
$(4+n)$ dimensions et correspondent aux fluctuations
transverses et sans trace de la métrique. Les equations d'Eintein linéarisées
pour ces perturbations de métrique se réduisent dans un espace-temps plat
(Minkowskien) à une équation d'onde scalaire dans le vide
($\Box_{M^4}+\Box_{S^n} = 0$), c'est-à-dire dans
l'espace des impulsions :
\ba
p^2 & = & 0 = -p_0^2+p_1^2+p_2^2+p_3^2+m^2,
\ea
ce qui entraine l'apparence, du point de vue quadridimensionnel, d'une masse
effective $m^2 = \Box_{S^n} h_{ab}$ pour les gravitons $h_{ab}$. Dans le modèle ADD l'espace-temps total n'est pas
Minkowskien puisque compactifié mais de la même fa\c{c}on en projetant les
équations linéarisées du graviton à $(4+n)$ dimensions sur l'espace
Minkowskien à quatre dimensions on obtient l'équation du mouvement d'un
graviton massif à quatre dimensions. Les dimensions supplémentaires étant
compactifiées sur un cercle de rayon $R$, le spectre de masse des gravitons
est discret et les masses sont séparées de $\Delta m \sim 1/R$. Cette tour
discrète de gravitons massifs est appelée tour de Kaluza-Klein (KK) en
référence au modèle du même nom qui partage cette propriété due aux dimensions
supplémentaires. Plus $R$ est petit plus le premier graviton massif est lourd
et peut avoir échappé aux détecteurs des accélérateurs de particules. Hormis
les modifications de la loi de Newton à petite distance, la
détection de gravitons KK dans les accélérateurs (\emph{e.g} au Large Hadron Collider) constituerait une autre signature
forte de la présence de dimensions supplémentaires.

\subsection{Modèle RS 1}\label{subsec:rs1}

\begin{figure}
  \begin{center}
\includegraphics[width=10.5cm]{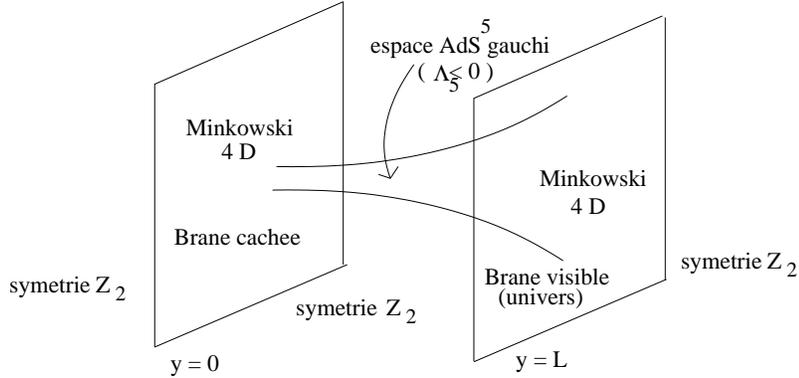}
 \end{center}
\caption{
Modèle de Randall-Sundrum (RS 1) à deux branes. L'espace-temps Anti-de Sitter entre les
branes, situées aux points fixes de l'orbifold, a une constante cosmologique
négative. L'espace est $Z_2$-symétrique de part et d'autre de chaque
brane. L'Univers réside sur la brane visible en $y = L$.}
\label{fig:rs1model}
\end{figure}

Le modèle ADD avec $n = 2$ dimensions supplémentaires élimine certes la hierarchie entre l'échelle de Planck $M_p$ et
l'échelle électrofaible $m_{EW}$ mais introduit une nouvelle hiérarchie entre
l'échelle électrofaible et l'échelle de compactification $R^{-1}\sim 10^{-3} eV$. De plus le
fait que $R$ soit ``grand'' dans le modèle ADD autorise la présence de gravitons KK
\emph{légers}, ce qui semble incompatible avec l'absence de leur détection dans
les accélérateurs actuellement en fonctionnement. Randall et Sundrum ont
proposé un scénario alternatif pour résoudre le problème de
hiérarchie en utilisant seulement une seule dimension supplémentaire, courbée mais
non-compacte \cite{rs1}. Dans ce premier modèle de Randall-Sundrum (modèle appelé RS
1), deux branes Minkowski de genre temps à quatre dimensions sont plongées dans une portion de
l'espace Anti-de Sitter à \emph{cinq} dimensions ($AdS^5$) où la cinquième
dimension possède de plus la structure orbifold $S^1/Z_2$ :
\ba\label{qqq:adsmetric}
ds^2_{bulk~AdS^5} & = & g_{ab}dX^adX^b = e^{-2\vert y \vert/\ell}\eta_{\mu\nu}dx^\mu dx^\nu+dy^2.
\ea
$\eta_{\mu\nu}$ est la métrique Minkowski à $(3+1)$ dimensions et $\ell$
est le rayon de courbure l'espace $AdS^5$.
La première brane, située sur le point fixe $y = 0$ de
l'orbifold, est cachée tandis que l'autre brane, parallèlement positionnée à distance finie
dans la dimension supplémentaire sur l'autre point fixe $y = L$, constitue
l'Univers visible. L'espace $AdS$ entre les deux branes a donc une constante
cosmologique négative $\Lambda_5 = -6/\ell^2$ et est $Z_2$-symétrique de part et d'autre de chaque brane
$y\leftrightarrow -y$ , $L+y\leftrightarrow L-y$ (Fig. \ref{fig:rs1model}). 
Cet édifice branaire
correspond vraiment à une solution exacte des équations d'Einstein à cinq
dimensions, qui conserve l'invariance de Poincaré à quatre dimensions sur
chaque brane à
la seule condition que les deux branes possède une tension de signe opposé $\pm
\sigma$ qui soit ajustée selon $\sigma = 3M_5^3/(4\pi\ell)$, où $M_5$ est l'échelle
fondamentale de Planck à cinq dimensions, afin d'annuler la constante
cosmologique effective sur les branes Minkowski. Le modèle RS parait plus
satisfaisant du point de vue géométrique que le modèle ADD puisque c'est une
solution des équations d'Einstein et que les branes
apparaissent cette fois comme des objets géométriques auto-gravitants,
influen\c{c}ant la géométrie du bulk. Le point faible est que cet
ajustement fin des paramètres n'a pas encore de justification
fondamentale. Par contre le modèle RS s'insère très bien dans la théorie
effective à cinq dimensions du scénario
cordiste d'Horava-Witten (voir paragraphe \ref{subsec:hw}). À partir de
(\ref{qqq:adsmetric}) on relie
aisément le volume $5D$ avec le volume $4D$ ainsi que les scalaires de courbure de Ricci
$5D$ et $4D$, de telle fa\c{c}on que l'action d'Einstein-Hilbert inclue l'action
effective à $4D$ selon :
\ba\label{qqq:volrs1}
M_5^3\int d^5 X\sqrt{\vert g\vert}R & \supset & M_5^3\int_0^L dy
e^{-2y/\ell}\int d^4 x \sqrt{\vert g^{(4)}\vert} R^{(4)},
\ea
entrainant la relation suivante entre l'échelle fondamentale  et l'échelle de
Planck effective $M_p$ : $
M_p^2 = M^3_5\ell\left[1-e^{-2L/\ell}\right]$.
Si l'on veut discuter le problème de hiérarchie du point de vue de la brane
visible supportant notre Univers il convient de renormaliser à $1$ le
facteur conforme de la métrique induite sur la brane visible située à $y = L$,
en modifiant le facteur de gauchissement\footnote{``warp factor'' en
  anglais.} dans (\ref{qqq:adsmetric}) selon $e^{-2y/\ell}\rightarrow
e^{-2(y-L)/\ell}$. Dès lors la masse de Planck effective mesurée en $y = L$
sur la brane visible vaut
\ba\label{qqq:rshierarchy}
M_p^2 & = & M^3_5\ell\left[e^{2L/\ell}-1\right].
\ea
En supposant que l'échelle fondamentale soit le TeV et $M_5,\ell\sim m_{EW}\sim TeV$,
 le problème de hiérarchie $M_p\sim 10^{19} GeV$ se resoud donc en supposant
 que les branes sont distantes de $L\sim 37 \ell$. On constate que le modèle
 RS 1 n'introduit pas de nouvelle hiérarchie contrairement au modèle ADD. De
 plus la petite taille de la dimension supplémentaire $L~\sim~ TeV^{-1}$ répond de
 manière plus satisfaisante aux contraintes expérimentales que le modèle $ADD$
 puisque les gravitons KK massifs sont cette fois beaucoup plus lourds avec
 des masses de départ de l'ordre du TeV. La distance entre les branes correspond à la
 valeur moyenne dans le vide d'un champ quantique scalaire (appelé le module),
 ce dernier doit être stabilisé si l'on veut que la hiérarchie soit
 respectée au niveau quantique. Notons qu'un mécanisme de stabilisation a été proposé par Goldberger et Wise
 \cite{stab} en introduisant un champ scalaire dans le bulk.

Je ne m'étendrai pas davantage sur le vaste problème de hiérarchie ni sur la phénoménologie des modèles
 branaires puisque je vais désormais me concentrer à partir du chapitre
 suivant sur la gravitation et la cosmologie dans le modèle de Randall-Sundrum
à \emph{une} brane (modèle appelé RS 2). Dans le modèle RS 2, l'Univers réside cette
fois sur la brane en $y = 0$ et l'autre brane a été repoussée à l'infini
$L\rightarrow\infty$ de sorte qu'il n'y a plus qu'une seule brane dans ce modèle.


\chapter{Le modèle de Randall-Sundrum}\label{chapter:rs}

Randall et Sundrum ont amené un point de vue nouveau sur la cosmologie et la
gravitation en proposant en 1999 un modèle branaire simple et élégant
\cite{rs2} où la gravité standard est localisée de fa\c{c}on dynamique sur la brane quadridimensionnelle
 malgré la présence d'une dimension supplémentaire non-compacte et
 infinie. Dans ce modèle (dit RS 2), notre Univers observable réside sur une
 hypersurface à $(3+1)$ dimensions, une $3-$brane, de géométrie
 Minkowskienne plongée dans un espace-temps Anti-de Sitter à cinq dimensions
 ($AdS^5$). De fa\c{c}on qualitative l'idée générale du modèle de
 Randall-Sundrum est que la courbure du bulk Anti-de Sitter entraine en fait une
 ``compactification'' effective de la cinquième dimension empêchant ainsi la
 gravité $4D$ de se diluer dans cette dimension supplémentaire. En
 conséquence le comportement $4D$ de la gravité d'Einstein reste préservé sur
 la brane, au moins aux grandes distances correspondant aux échelles où la
 force de gravité a été testée expérimentalement. Dans ce chapitre on présente
 quantitativement et qualitativement les conséquences gravitationnelles et
 cosmologiques du modèle branaire de Randall-Sundrum.

\section{Localisation dynamique de la gravité}\label{sec:bound}

\begin{figure}
  \begin{center}
\includegraphics[width=8.5cm]{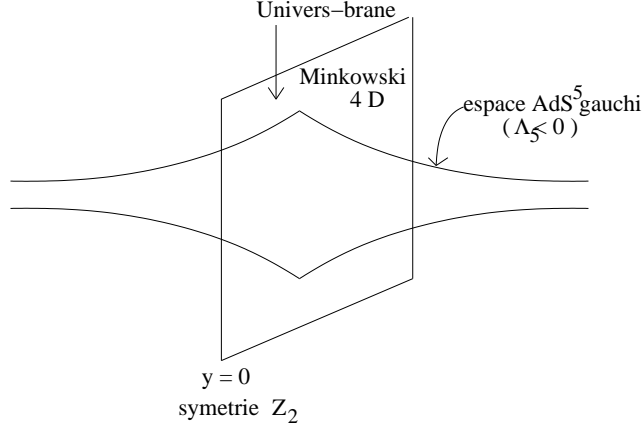}
 \end{center}
\caption{
Modèle de Randall-Sundrum (RS 2) à une brane. L'espace-temps Anti-de Sitter à
cinq dimensions est gauchi et symetrique de part et d'autre de la brane
Minkowski à quatre dimensions.}
\label{fig:rs2model}
\end{figure}

Dans le modèle de Randall-Sundrum l'Univers à $(3+1)$ dimensions
est une brane Minkowki de tension $\sigma$ plongée dans une portion de l'espace Anti-de Sitter $AdS^5$ à cinq dimensions de rayon de
courbure $\ell$ :
\ba\label{qqq:metricrs}
ds^2 & = & e^{-2\vert y \vert/\ell}\eta_{\mu\nu}dx^\mu dx^\nu+dy^2,
\ea
où $\eta_{\mu\nu}$ est la métrique de Minkowski à $(3+1)$ dimensions (Fig. \ref{fig:rs2model}). La
métrique de l'espace $AdS^5$ dans les coordonnées (\ref{qqq:metricrs}) induit
effectivement une métrique quadridimensionnelle de géométrie
Minkowski sur la brane située en $y = 0$. Le facteur de gauchissement contient la valeur absolue
$\vert y \vert$ parce que l'espace
$AdS$ est supposé $Z_2$-symétrique de part et d'autre de la brane (symétrie
miroir $y\leftrightarrow -y$), en référence à la symétrie orbifold du modèle
de cordes de Ho$\check{r}$ava-Witten (paragraphe \ref{subsec:hw}). Cette construction
branaire (\ref{qqq:metricrs}) est la solution métrique $g_{ab}$ des équations
d'Einstein à cinq dimensions :
\ba\label{qqq:einsteinrs}
R_{ab}-{1\over 2}g_{ab}R & = & -\Lambda_5 g_{ab} -\kappa^2 \delta(y)\sigma
\eta_{\mu\nu} \delta_a^\mu\delta_b^\nu,
\ea
où $a,b=0,...,4$ et $\mu,\nu = 0,...,3$, $\kappa^2 = 8\pi/M_5^3$, et où le tenseur énergie-impulsion
(tension) quadridimensionnel de la brane est localisé en $y = 0$. Enfin la
tension $\sigma$ est ajustée à la constante cosmologique du bulk $\Lambda_5$, afin qu'il n'y ait pas de constante cosmologique
effective sur la brane Minkowski ($\Lambda_4 = 0$), selon :
\ba\label{qqq:finetuning}
\Lambda_5 = {-6\over\ell^2}, \qquad \sigma = {3M^3_5\over 4\pi\ell}.
\ea
La géométrie de l'espace $AdS$ (\ref{qqq:metricrs}) est non-factorisable et
gauchie.

\subsection{Equation du mouvement du graviton dans le bulk $AdS^5$}\label{subsec:eomrs}

La théorie des perturbations de la métrique permet de calculer le spectre des gravitons (ou ondes
gravitationnelles d'un point de vue classique) apparaissant dans le modèle de
Randall-Sundrum. Les ondes gravitationnelles correspondent aux fluctuations
inhomogènes et anisotropes de la métrique de l'espace-temps, il s'agit donc
d'étudier l'évolution des perturbations de la métrique à l'ordre linéaire à
travers les équations d'Einstein linéarisées. Pour linéariser les équations
d'Einstein il est pratique
de se placer dans le système de coordonnées conformes de Poincaré, $z=\ell
e^{y/\ell}$, pour décrire
le bulk $AdS^5$
\ba\label{qqq:poincare}
ds^2 & = & \ell^2 {-dt^2+d\mathbf{x}_{3}^2+dz^2 \over z^2}.
\ea
Dans ces coordonnées, la métrique $AdS$ est conforme à la métrique Minkowski et
$z$ est la dimension supplémentaire. Le calcul des perturbations, $h_{AB} \ll
1$, de la métrique (\ref{qqq:poincare})
du bulk $AdS$
\ba\label{qqq:poincarepert}
d\tilde{s}^2 & = & {\ell^2 \over z^2}\left(\eta_{AB}+h_{AB}\right)dx^A dx^B
\ea
se ramène ainsi d'une certaine fa\c{c}on au calcul, plus simple, des
perturbations de l'espace plat Minkowskien. Notons  $\tilde{g}_{AB} =
\eta_{AB}+h_{AB}$ la métrique plate perturbée et $g_{AB} =
e^{-W(z)}\tilde{g}_{AB}$ la métrique $AdS$ perturbée (\ref{qqq:poincarepert})
avec le facteur conforme $W(z) = \ln(z^2/\ell^2)$. La perturbation linéaire du
tenseur d'Einstein, $\tilde{G}_{AB}\equiv
\tilde{R}_{AB}-(1/2)\tilde{g}_{AB}\tilde{R}$, dans l'espace Minkowski perturbé est
triviale (voir par exemple dans le livre de Robert Wald \cite{wald})
\ba\label{qqq:deltagplat}
\delta \tilde{G}_{AB}& = &
\partial^C\partial_{(B}h_{A)C}-\frac{1}{2}\partial^C\partial_Ch_{AB}-\frac{1}{2}\partial_A\partial_Bh^C_C\cr
&
&-\frac{1}{2}\eta_{AB}(\partial^C\partial^Dh_{CD}-\partial^C\partial_Ch^D_D),
\ea
où $(A,B)$ signifie la symétrisation sur les indices. De plus les tenseurs
d'Einstein de deux espaces-temps conformes entre eux sont liés selon (voir par
exemple dans le livre de Wald \cite{wald})
\ba\label{qqq:einsteinconf}
G_{AB}& = &
\tilde{G}_{AB}+\frac{3}{2}[\frac{1}{2}\tilde{\nabla}_AW\tilde{\nabla}_BW
+\tilde{\nabla}_A\tilde{\nabla}_BW\cr
& & -\tilde{g}_{AB}(\tilde{\nabla}_C\tilde{\nabla}^CW
-\frac{1}{2}\tilde{\nabla}_CW\tilde{\nabla}^CW)]
\ea
où $\tilde{\nabla}$ est la dérivée covariante selon la métrique
$\tilde{g}_{AB}$ de Minkowski perturbé.
A partir de (\ref{qqq:deltagplat}) et (\ref{qqq:einsteinconf}) on en déduit la perturbation linéaire du tenseur d'Einstein dans l'espace $AdS$
\ba\label{eqnarray:deltag}
\delta G_{AB} & = &
\partial^C\partial_{(B}h_{A)C}-\frac{1}{2}\partial^C\partial_Ch_{AB}-\frac{1}{2}\partial_A\partial_Bh^C_C
\cr
                                      &   &-\frac{1}{2}\eta_{AB}(\partial^C\partial^Dh_{CD}-\partial^C\partial_Ch^D_D)
\cr
                                      &   &
                                      -\frac{3}{2}[\frac{1}{2}\eta^{CD}(\partial_Ah_{BC}+\partial_Bh_{AC}-\partial_Ch_{AB})\partial_DW
                                      \cr
                                      &   &
                                      +h_{AB}\partial_C\partial^CW+\eta_{AB}(\frac{1}{2}\partial_Ch^D_D\partial^CW\cr
                                      &   &-\partial_Ch^{CD}\partial_DW-h^{CD}\partial_C\partial_DW)\cr
                                      &   &
                                      -\frac{1}{2}(h_{AB}\partial_CW\partial^CW-\eta_{AB}h^{CD}\partial_CW\partial_DW)].
\ea
Il s'ensuit que les équations d'Einstein linéarisées dans le bulk $AdS^5$, $\delta G_{AB} = -(6/\ell^2)h_{AB}$, sont
\ba\label{qqq:lineinstein}
& 0 = \frac{1}{2}[\partial^C\partial_{B}h_{AC}+\partial^C\partial_{A}h_{BC}-\Box_5h_{AB}-\partial_A\partial_Bh^C_C \nonumber\\
 &-\eta_{AB}(\partial^C\partial^Dh_{CD}-\Box_5h)]-\frac{6}{z^2}\eta_{AB}h_{zz} \cr
 &-\frac{3}{2z}[\partial_Ah_{zB}+\partial_Bh_{zA}-\partial_zh_{AB}+\eta_{AB}(\partial_zh-2\partial_Ch^C_z)].
\ea
 Comme la théorie de la relativité générale à cinq dimensions est invariante
 sous les cinq reparamétrisations locales des coordonnées d'espace-temps
 $x^A\rightarrow x^A+(\ell^2/z^2)\epsilon^A(x^B)$, $h_{AB}\rightarrow
 h_{AB}-\partial_A\epsilon_B-\partial_B\epsilon_A+2\eta_{AB}(\epsilon^z/z)$,
 on peut fixer la jauge
\ba\label{qqq:jauge1}
h_{Az} & = & 0
\ea
qui élimine cinq composantes des quinze composantes de la perturbation
$h_{AB}$. Cependant la jauge n'est pas complètement fixée \cite{der1}
puisqu'on peut trouver cinq fonctions arbitraires $c_\mu(x^\nu),
d(x^\nu)$ dépendant des coordonnées 4D ($\mu,\nu=0,...,3$) telles que les
reparamétrisations locales suivantes des coordonnées d'espace-temps,
$\epsilon^z = zd$, $\epsilon_\mu = (1/2)z^2\partial_\mu d+c_\mu$ ne modifient
pas la jauge $h_{Az}  =  0$ mais seulement les composantes transverses de la
perturbation selon
\ba\label{qqq:jaugeres}
h_{\mu\nu} &\rightarrow &
h_{\mu\nu}+z^2\partial_\mu\partial_\nu d-\partial_\mu
c_\nu-\partial_\nu c_\mu-2\eta_\mu\nu d.
\ea
Dans la jauge (\ref{qqq:jauge1}) on peut décomposer l'équation du mouvement (\ref{qqq:lineinstein}) selon
\ba\label{qqq:eomsss}
\partial_{\rho\sigma}h^{\rho\sigma}-\Box_4h+\frac{3}{z}\partial_zh
& = & 0, \qquad \partial_z(\partial_\rho h^\rho_\mu-\partial_\mu h) =
0\mbox{,}\cr
{\partial}^2_zh-\frac{1}{z}\partial_z h & = & 0, \cr
\Box_4h_{\mu\nu}+\partial^2_zh_{\mu\nu}-\frac{3}{z}\partial_zh_{\mu\nu}
& = & \partial_{(\mu|\rho} h^\rho_{\nu)}-\partial_{\mu\nu}
h+\frac{\eta_{\mu\nu}}{z}\partial_zh
\ea
où les indices sont élevés avec la métrique plate 4D $\eta_{\mu\nu}$, $h =
h^\mu_\mu$ est la trace et $\Box_4 = \partial_\mu\partial^\mu$. Les trois
premières équations de (\ref{qqq:eomsss}) sont non-dynamiques et leur résolution
donne $h = -(1/6)z^2\partial_\mu D^\mu+C$ et $\partial_\rho h^\rho_\mu =
-1/6\partial_\mu\partial_\rho D^\rho+\partial_\mu C+D_\mu$, où
$C(x^\nu)$ et $D_\mu(x^\nu)$ sont cinq fonctions arbitraires des coordonnées
4D. On peut fixer complètement la jauge en choisissant les reparamétrisations
résiduelles $c_\mu$, $d$ telles qu'elles satisfont $\Box_4d =
-(1/6)\partial_\mu D^\mu$ et
$\partial_\mu\partial_\rho c^\rho+\Box_4 c_{\mu}-2\partial_{\mu}d =
-\partial_\mu C-D_\mu$. En faisant ce choix on annule ainsi $h$ et
$\partial_\rho h^\rho_\mu$. Cette jauge complètement fixée est la \emph{jauge
  de Randall-Sundrum} (RS), transverse et sans trace :
\ba\label{qqq:jaugers}
h_{Az} & = & 0, \qquad h = 0, \qquad \partial_\rho h^\rho_\mu = 0,
\ea
qui réduit les composantes de la perturbation métrique aux cinq degrés de
libertés physiques du graviton $5D$. Dans la jauge RS (\ref{qqq:jaugers}),
l'équation (\ref{qqq:eomsss}) du mouvement du graviton dans le bulk $AdS^5$ se
réduit à une équation d'onde de type Klein-Gordon sur chaque degré de liberté
du graviton. En ce sens les gravitons découplent et se propagent comme des
champs scalaires libres dans le bulk $AdS$ selon :
\ba\label{qqq:eompoinc}
\left(\Box_4+\partial^2_z-\frac{3}{z}\partial_z\right)h_{\mu\nu} & = & 0.
\ea
L'équation de Klein-Gordon pour un champ scalaire libre $\phi(x^A)$ dans n'importe quel système de coordonnées de métrique $g_{AB}$ s'écrit
\ba\label{qqq:equkg}
{1\over \sqrt{-g}}\partial_A \sqrt{-g} g^{AB} \partial_B \phi & = & 0.
\ea
On peut donc réecrire cette équation de Klein-Gordon dans les coordonnées
Gaussiennes normales (GN) initiales (\ref{qqq:metricrs}) du modèle de Randall-Sundrum
 et faire de plus le changement de variable $h_{\mu\nu} =
e^{2\vert y\vert \over\ell}\bar{h}_{\mu\nu}$ afin d'observer l'évolution du
champ réel de perturbation de métrique : $d\tilde{s}^2 = \left[e^{-2\vert y
    \vert/\ell}\eta_{\mu\nu}+\bar{h}_{\mu\nu}\right]dx^\mu
dx^\nu+dy^2$. L'équation du mouvement des gravitons physiques
$\bar{h}_{\mu\nu}$ devient dans les coordonnées GN $Z_2$-symétriques :
\ba\label{qqq:eomrs}
\left[e^{2\vert y\vert /\ell}\Box_4+\partial^2_y-\frac{4}{\ell^2}+\frac{4}{\ell}\delta(y)\right]\bar{h}_{\mu\nu} & = & 0.
\ea
La fonction $\delta$ de Dirac provient de la symétrie $Z_2$ et correspond aux 
conditions de bord suivantes pour le graviton sur la brane : 
\ba
\left(\partial_y
+{2\over\ell}\right)\bar{h}_{\mu\nu}\vert_{y = 0}
 & = &  0.
\ea

On peut retrouver l'équation du mouvement du graviton (\ref{qqq:eomrs}) en
 linéarisant de fa\c{c}on indépendante l'équation d'Einstein \emph{dans le bulk} et
 les conditions de jonction d'Israel \emph{sur la brane}. Les conditions de
 jonction d'Israel sont l'analogue en relativité générale des conditions de
 saut en électromagnétisme qui indiquent la variation du champ
 électromagnétique à la traversée d'une surface contenant des charges et des
 courants. Les conditions d'Israel découlent des équations d'Einstein, une fois
 prise en compte la matière présente sur la brane, et elles indiquent les
 variations du champ de métrique à la traversée de la brane contenant le
 tenseur énergie-impulsion $T_{\mu\nu}$ selon : 
\ba
K_{\mu\nu}(y =
 0^+)-K_{\mu\nu}(y = 0^-)  & = &
 -\kappa^2\left[T_{\mu\nu}-\frac{1}{3}Tg^{(4)}_{\mu\nu}\right],
\ea
 où
 $g^{(4)}_{\mu\nu}$ est la métrique induite sur la brane et $K_{\mu\nu}$ est
 la courbure \emph{extrinsèque} de la brane, c'est-à-dire à la
 courbure de la brane induite par immersion dans l'espace $AdS$ de dimension
 supérieure. Dans des coordonnées GN, la courbure extrinsèque est simplement
 donnée par la dérivée normale à la brane de la métrique
 induite \cite{wald} : $K_{\mu\nu} =
\frac{1}{2}\partial_y g^{(4)}_{\mu\nu}$. De plus, ici il n'y a pas de matière sur
 la brane, de sorte que le tenseur énergie-impulsion correspond simplement à la
 tension de la brane $T_{\mu\nu} = \sigma g^{(4)}_{\mu\nu}$. En tenant compte
 enfin de la symétrie $Z_2$, $\partial_y g^{(4)}_{\mu\nu}(y=0^+) = -\partial_y
 g^{(4)}_{\mu\nu}(y=0^-)$, la linéarisation des conditions de jonction
 d'Israel selon $g^{(4)}_{\mu\nu} = e^{-2\vert y
 \vert/\ell}\eta_{\mu\nu}+\bar{h}_{\mu\nu}$ entraine
\ba\label{qqq:israel}
\partial_y \left(e^{-2\vert y \vert/\ell}\eta_{\mu\nu}+\bar{h}_{\mu\nu}\right)
 & = & -{\kappa^2\over 3}\sigma\left(e^{-2\vert y
 \vert/\ell}\eta_{\mu\nu}+\bar{h}_{\mu\nu}\right), \qquad (y = 0).
\ea
Cela se traduit par les conditions de bord (\ref{eqnarray:vcb}) du graviton sur la
brane, complémentaires de l'équation du mouvement (\ref{eqnarray:cb}) dans le bulk :
\ba
\left(\partial_y +{2\over\ell}\right)\bar{h}_{\mu\nu} & = & 0, \qquad (y = 0),\label{eqnarray:vcb}\\
\left(e^{2\vert y\vert /\ell}\Box_4+\partial^2_y-\frac{4}{\ell^2}\right)\bar{h}_{\mu\nu} & = & 0, \qquad \mbox{(bulk)}.\label{eqnarray:cb}
\ea
On constate qu'on peut encapsuler la condition de bord de (\ref{eqnarray:vcb}) dans
 l'équation du bulk (\ref{eqnarray:cb}) de fa\c{c}on $Z_2$-symétrique et obtenir une écriture
 "compacte" de l'équation du mouvement du graviton correspondant exactement à
 la forme (\ref{qqq:eomrs}) précédemment obtenue.

Les indices tensoriels du champ de graviton n'apportant aucune information sur
la propagation, nous utiliserons désormais une notation de type champ scalaire
pour le graviton : $\bar{h}_{\mu\nu}(t,\mathbf{x}, y) =
\bar{\Phi}(t,\mathbf{x}, y)e_\mu\otimes e_\nu$. Notons enfin que les cinq
degrés de liberté du gravitons 5D peuvent se décomposer par ``projection'' sur la
brane 4D en deux degrés de liberté tensoriels transverses et sans trace (ondes
gravitationnelles), deux degrés de liberté vectoriels sans divergence
(appelés ``gravi-photon'') et un degré de liberté scalaire (appelé ``gravi-scalaire'').

\subsection{Spectre des gravitons 5D}\label{subsec:spectrers}

L'équation du mouvement (\ref{qqq:eomrs}) est séparable, de telle sorte qu'on peut
chercher les modes $m^2$ du graviton $5D$, $\bar{\Phi}(t,\mathbf{x}, y) =
\bar{\psi}(t,\mathbf{x})\bar{\phi}(y)$, selon
\ba\label{qqq:separationrs}
\Box_4 \bar{\psi}(t,\mathbf{x}) & = & m^2 \bar{\psi}(t,\mathbf{x}),\cr
\left[-\partial^2_y+\frac{4}{\ell^2}-\frac{4}{\ell}\delta(y)\right] \bar{\phi}(y)& = & m^2e^{2\vert y\vert /\ell}\bar{\phi}(y).
\ea
Du point de vue $4D$ d'un observateur confiné sur la brane, ces modes $5D$
correspondent à des excitations de spin $2$ massives, c'est-à-dire des
gravitons massifs de type Kaluza-Klein (KK).

Un moyen intuitif de cerner la physique du Modèle de Randall-Sundrum est
d'analyser le problème de mécanique quantique analogue. En se pla\c{c}ant dans
les coordonnées conformes $z = sgn(y)\ell\left(e^{\vert y
  \vert/\ell}-1\right)$ et en changeant d'échelle le champ de graviton selon
$\phi(y) = \bar{\phi}(y)e^{\vert y \vert/(2\ell)}$, l'équation du mouvement des
gravitons (\ref{qqq:separationrs})  dans le bulk prend la
forme d'une équation de type Schrödinger
\ba\label{qqq:schrors}
\left[-\partial^2_z+V(z)\right]\phi(z) & = & m^2\phi(z),
\ea
où la brane est située en $z = 0$ et le potentiel de Schrödinger vaut
\ba\label{qqq:potentialrs}
V(z) & = & {15\over 4\left(\vert z\vert +\ell\right)^2}-{3\over\ell}\delta(z).
\ea
\begin{figure}
  \begin{center}
\includegraphics[width=6.5cm]{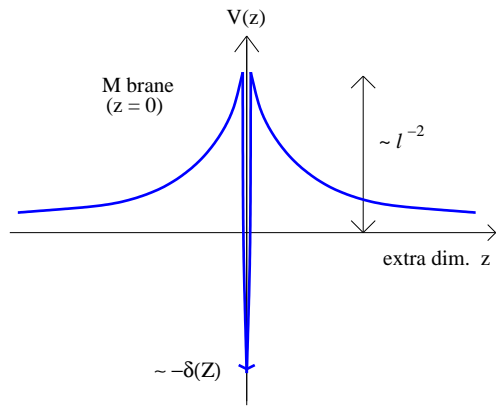}
 \end{center}
\caption{
Potentiel-volcan pour une brane Minkowski plongée dans Anti-de Sitter.}
\label{fig:volcanoM}
\end{figure}
La forme du potentiel (Fig. \ref{fig:volcanoM}) dans lequel évoluent les gravitons nous renseigne sur le
spectre possible des gravitons $5D$ : la présence de la brane créée un puit de
potentiel infini $-\delta(z)$ alors que la courbure du bulk $AdS$ forme une
barrière de potentiel décroissante comme l'inverse au carré de la distance à
la brane. L'allure du potentiel qui en résulte est celle d'un "potentiel-volcan" qui
autorise la présence d'un unique état lié, le mode zéro $m^2 = 0$ du graviton
$5D$, accompagné d'un continuum d'états $m^2 > 0$ libres de se déplacer dans
la cinquième dimension. L'état lié signifie que la fonction d'onde
correspondante du mode zéro a son support localisé près de la brane, autour
de $z = 0$\footnote{Dans un langage plus mathématique, l'état lié est un état
  discret normalisable.}. Du point de vue $4D$ d'un observateur sur la brane,
l'état lié du graviton s'identifie naturellement au graviton sans masse
standard de la théorie effective à $4D$ alors que les autres modes 
s'identifient à une infinité continue de gravitons KK massifs. Le
 potentiel tend asymptotiquement vers zéro quand $z\rightarrow 0$, ce qui
 justifie l'absence de gap d'énergie entre le mode lié et le continuum KK dont
 les masses possibles sont $m^2 > 0$ et qui indique que les modes massifs se
 comportent asymptotiquement comme des ondes planes dans le bulk. La présence
 de ce continuum de gravitons KK massifs est le signe de la taille infinie de
 la dimension supplémentaire de structure $AdS$. Nous tenons ici le résultat
 essentiel du modèle de Randall-Sundrum : malgré la présence d'une dimension
 supplémentaire non-compacte et infinie, la gravité standard $4D$ (donnée par
 le mode de masse nulle) reste \emph{localisée} sur la brane et ne fuit pas
 dans la cinquième dimension. Quant aux gravitons massifs libres et parasites,
 leur interaction avec la brane est supprimée à basse énergie grâce à la barrière de
 potentiel.  Ainsi la théorie de la relativité générale quadridimensionnelle 
reste préservée en partie sur la brane, à basse énergie.

\begin{figure}
  \begin{center}
\includegraphics[width=7.5cm]{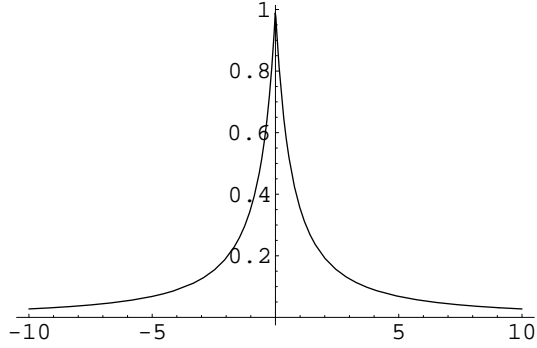}
 \end{center}
\caption{
Fonction d'onde du mode zéro du graviton : état lié à la brane. Sur la figure
$\ell \equiv 1$.}
\label{fig:lie}
\end{figure}

Concrètement la fonction d'onde (Fig. \ref{fig:lie}) du mode zéro lié du graviton est 
\ba\label{qqq:bsrs}
\phi_0(z) & = & {\ell^{5/2}\over \left(\vert z\vert +\ell\right)^{3/2}},
\ea
et les fonctions d'onde des modes libres des gravitons KK massifs sont
\ba\label{qqq:kkrs}
\phi_m(z) & = & N_m\left(\vert z\vert+\ell\right)^{1/2}\left[Y_2\left(m\left(\vert z\vert +\ell\right)\right)+{4\over\pi (m\ell)^2}J_2\left(m\left(\vert z\vert +\ell\right)\right)\right].
\ea
 Cette combinaison particulière des fonctions de Bessel satisfait les
 conditions de bord de $\phi$ inscrites dans la fonction $\delta$ du potentiel
 (\ref{qqq:potentialrs}) : $\left(\partial_{z}+{3\over
   2\ell}\right)\phi(z)\vert_{z = 0} = 0$. La normalisation du mode zero, qui
 est un état discret normalisable, est imposée par
 $\langle\phi_0\vert\phi_0\rangle = 1$. On peut trouver les coefficients de
 normalisation $N_m$ puisque les modes continus satisfont la condition
 d'orthonormalité $\langle\phi_m\vert\phi_{m'}\rangle = \delta(m-m')$. Le
 produit scalaire de Klein-Gordon correspondant est donné par
 $\langle\phi_m\vert\phi_{m'}\rangle = 2\int_0^\infty {dz\over z^3}
 \phi_m^*(z)\phi_{m'}(z)$. On remarquera que la fonction d'onde du mode lié
 correspond également à la limite $m\rightarrow 0$ de la fonction d'onde des
 modes massifs.

La courbure du bulk $AdS$ réalise en fait une sorte de "compactification
effective" puisque, malgré une dimension supplémentaire infinie, le volume
effectif supplémentaire est fini :
\ba\label{qqq:voleffrs}
R_{eff} = \int_0^\infty \sqrt{-g} dy & = & \int_0^\infty e^{-4y/\ell} dy = {\ell\over 4}.
\ea

\subsection{Modifications de la loi de gravitation de Newton}\label{subsec:gt}

Même si la gravité se comporte de fa\c{c}on quadridimensionnelle du point de
vue d'un observateur sur la brane à basse énergie, elle n'est pas purement
locale (\emph{i.e} confinée en $y = z = 0$) mais \emph{localisée} près de la
brane avec une extension
typique $\ell$ dans la cinquième dimension. La présence de modes massifs de
gravitons KK doit provoquer des corrections pentadimensionnelles à la gravité
standard à haute énergie (ou aux petites distances). Il est d'un intérêt cosmologique
d'estimer ces corrections pour savoir si leur magnitude est suffisamment
importante pour être accessible aux observations et aux expériences. Ces
corrections de plus haute dimension constituent la signature des dimensions
supplémentaires. Dans ce paragraphe  on évalue les corrections
pentadimensionnelles au potentiel Newtonien crée par une source
gravitationnelle sphérique purement localisée sur la brane. Le calcul
relativiste a été effectué par Garriga et Tanaka \cite{gt} en 2000.

On choisit temporairement la jauge dite Gaussienne normale (GN) au lieu de la jauge
RS afin de faciliter le
calcul des conditions de jonction d'Israel sur la brane en présence d'une
source de matière. Dans la jauge GN, par définition, la brane reste fixée en $\tilde{y} = 0$ dans les
coordonnées GN (\ref{qqq:metricrs}) malgré les
perturbations de la géométrie, c'est-à-dire
\ba\label{qqq:jaugegn}
\tilde{h}_{A\tilde{y}} = 0, \qquad\mbox{pour tout $A$},
\ea
où le ``tilde'' est utilisé pour distinguer les variables dans la jauge GN des
variables dans la jauge RS (qui seront dénotées par une ``barre'' pour les perturbations
de métrique et sans rien pour les coordonnées de fond).

À l'ordre linéaire dans les
perturbations et dans la jauge GN, les équations de jonction d'Israel
(\ref{qqq:israel}) $Z_2$-symétriques deviennent, en présence de matière sur la brane,
\ba\label{qqq:israelgt}
\partial_{\tilde{y}} \left(e^{-2\vert \tilde{y}
    \vert/\ell}\eta_{\mu\nu}+\tilde{h}_{\mu\nu}\right) & = &
-\kappa^2\left[T_{\mu\nu}-{1\over 3}e^{-2\vert \tilde{y}
    \vert/\ell}\eta_{\mu\nu}T+{\sigma\over 3}\left(e^{-2\vert \tilde{y}
    \vert/\ell}\eta_{\mu\nu}+\tilde{h}_{\mu\nu}\right)\right], \qquad
(\tilde{y} = 0^+)
\ea
où $T_{\mu\nu}$ est le tenseur énergie-impulsion de la source de matière
isolée sur la brane et
$T$ est sa trace. Comme la tension de la brane est ajustée selon $\sigma =
3M^3_5/(4\pi\ell)$ et $\kappa^2 = 8\pi/M^3_5$, ces conditions de bord se
réduisent dans la jauge GN à
\ba\label{qqq:cbgt}
\left(\partial_y +{2\over\ell}\right)\tilde{h}_{\mu\nu} & = &
-\kappa^2\left[T_{\mu\nu}-{1\over 3}e^{-2\vert \tilde{y}
    \vert/\ell}\eta_{\mu\nu}T\right], \qquad
(\tilde{y} = 0^+).
\ea

\begin{figure}
  \begin{center}
\includegraphics[width=3.5cm]{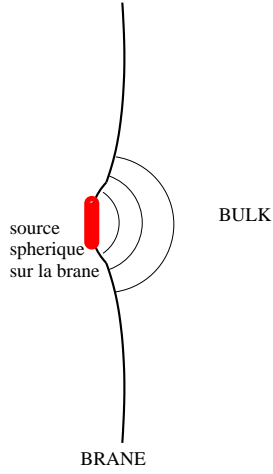}
 \end{center}
\caption{
``Brane bending'' : dans la jauge RS, une source de matière sphérique sur la
brane engendre un champ gravitationnel dans le bulk tout en créant un recul de
la brane (schema tiré de \cite{gt}).}
\label{fig:garriga}
\end{figure}

Dans la jauge RS on a aussi la condition $\bar{h}_{Ay} = 0$. La
reparamétrisation $x^A = \tilde{x}^A-\xi^A$ la plus génerale qui conserve cette condition
\ba\label{qqq:gnrs}
\xi^y & = & \hat{\xi}^y(x^\rho), \qquad \xi^\mu = {-\ell\over 2}e^{-2\vert y
  \vert/\ell}\eta_{\mu\nu} \partial_\nu
\hat{\xi}^y(x^\rho)+\hat{\xi}^\mu(x^\rho),\cr
\bar{h}_{\mu\nu} & = & \tilde{h}_{\mu\nu}-\ell\partial_\mu\partial_\nu\hat{\xi}^y-{2\over\ell}e^{-2\vert y
  \vert/\ell}\eta_{\mu\nu}\hat{\xi}^y+\partial_{(\mu}\hat{\xi}_{\nu)},
\ea
où les $\xi^A$ ne dépendent que des quatre coordonnées transverves,
permet de passer de la jauge GN ($\tilde{h}_{\mu\nu}$) à la jauge RS
($\bar{h}_{\mu\nu}$). Dans la jauge RS la brane est
légèrement déplacée en $y = -\xi^y(x^\rho)$ et les conditions de jonction sur
la brane
deviennent
\ba\label{qqq:cbrs}
\left(\partial_y +{2\over\ell}\right)\tilde{h}_{\mu\nu} & = &
-\kappa^2\Sigma_{\mu\nu} \qquad (y = -\xi^y(x^\rho)),\cr
\Sigma_{\mu\nu} & = & \left(T_{\mu\nu}-{1\over 3}e^{-2\vert \tilde{y}
  \vert/\ell}\eta_{\mu\nu}T\right)+{2\over\kappa^2}\partial_\mu\partial_\nu\xi^y,
\ea
où la source $\Sigma_{\mu\nu}$ inclue le léger déplacement de la
brane. L'écriture compacte de l'équation du mouvement dans le bulk et des conditions de
bord de fa\c{c}on $Z_2$ symétrique devient donc dans la jauge RS
\ba\label{qqq:eomgt}
\left[e^{2\vert y\vert
    /\ell}\Box_4+\partial^2_y-\frac{4}{\ell^2}+\frac{4}{\ell}\delta(y+\xi^y)\right]\bar{h}_{\mu\nu}
& = & -2\kappa^2\Sigma_{\mu\nu}\delta(y+\xi^y),
\ea
dont la solution formelle s'écrit
\ba\label{qqq:solgt}
\bar{h}_{\mu\nu} & = & -2\kappa^2\int_{y = -\xi^y(x^\rho)} d^4
x'G_R(x,x')\Sigma_{\mu\nu}(x')
\ea
où $G_R(x,x')$ est la fonction de Green retardée de l'équation du mouvement du graviton.
La jauge RS impose $h^\mu_\mu = 0$ et donc (\ref{qqq:solgt}) impose
$\Sigma^\mu_\mu = 0$, qui à son tour à travers (\ref{qqq:cbrs}) implique que le
déplacement de la brane satisfasse
\ba\label{qqq:bending}
\Box_4 \hat{\xi}^y & = & {\kappa^2\over 6} T.
\ea
La présence d'une source de matière isolée sur la brane aura donc tendance
à créer un déplacement local de la brane (``brane bending'' en anglais) dans
le bulk , $y_{brane} = -\xi^y(x^\rho)$, au niveau de la
position de la source (Fig. \ref{fig:garriga}). Signalons que la condition $\partial_\mu
\bar{h}^\mu_\nu = 0$ de la jauge RS est par conséquence aussi vérifiée d'après
l'expression de $\Sigma_{\mu\nu}$ (\ref{qqq:cbrs}).

La fonction de Green est séparable $G_R(x,x') = \sum_{m}g_{m}(x^\mu,{x'}^\mu)
u_m(y)u_m^*(y')$ avec
\ba\label{qqq:calculfg}
\left(p^2+m^2\right)\hat{g}_m(p) & = & 1 \qquad\mbox{(espace de Fourier 4D)}\cr
\left(\partial^2_y-\frac{4}{\ell^2}+m^2e^{2\vert y\vert /\ell}\right)u_m(y) &
= & 0 \qquad\mbox{(bulk)},\cr
\left(\partial_y +{2\over\ell}\right)u_m(y) &
= & 0 \qquad\mbox{(condition de bord)},
\ea
où $\hat{g}_m(p)$ est la transformée de Fourier de la fonction de Green
massive dans Minkowski 4D et $p^2 = -\omega^2+\mathbf{p}^2$. Les solutions dans le bulk ont déjà été calculées
en (\ref{qqq:kkrs})
\ba\label{qqq:kkgt}
u_m(y) & = & \sqrt{m\ell/2\over
  J_1(m\ell)^2+Y_1(m\ell)^2}\left(J_1(m\ell)Y_2\left(m\ell
e^{y/\ell}\right)-Y_1(m\ell)J_2\left(m\ell e^{y/\ell}\right)\right).
\ea
La fonction de Green retardée est donc donnée par
\be\label{qqq:greengt}
G_R(x,x')  =  -\int {d^4p\over
  (2\pi)^4}e^{ip_\mu(x^\mu-{x'}^\mu)}\left[{\ell^{-1}e^{-2y/\ell}e^{-2y'/\ell}\over \mathbf{p}^2-(\omega+i\epsilon)^2}
+\int_0^\infty  {u_m(y)u_m(y')dm\over m^2+\mathbf{p}^2-(\omega+i\epsilon)^2}\right].
\ee
Le premier terme provient du mode zéro du graviton alors que le second terme
provient de la contribution des modes KK massifs continus. 

On considère une
\emph{source stationnaire} et on définit la fonction de Green stationnaire
$G(\mathbf{x},y,\mathbf{x'},y') = \int_{-\infty}^{+\infty} dt'G_R(x,x')$, où
les coordonnées en gras couvrent les trois dimensions spatiales transverses
sur la brane. Si les deux points sont sur la brane ($y = y' = 0$, $r = \vert
\mathbf{x}-\mathbf{x'}\vert$) alors, en considérant les asymptotes des
fonctions de Bessel, on trouve que la fonction de Green se comporte comme
\ba\label{qqq:greencorr}
G(\mathbf{x},0,\mathbf{x'},0) & \approx & {-1\over 4\pi\ell r}\left[1+{\ell^2\over
    2 r^2}+ ...\right].
\ea
Le premier terme issu du mode zéro correspond à la fonction de Green de l'équation de
Poisson tridimensionnelle $\nabla^2 f = \delta(r)$ et le second terme est
correctif et provient des modes massifs du bulk.

On peut décomposer les
perturbations de métrique selon $\bar{h}_{\mu\nu}  =
\bar{h}^{(m)}_{\mu\nu}+\partial_\mu\partial_\nu \bar{h}^{(\xi)}$, où on a
volontairement séparé la partie matière de la partie déplacement dans la
source :\\
\noindent
$\bar{h}^{(m)}_{\mu\nu} = -2\kappa^2\int d^3\mathbf{x'}
G(\mathbf{x},\mathbf{x'})\left(T_{\mu\nu}-(1/3)\eta_{\mu\nu} T\right)(\mathbf{x'})$ et $\bar{h}^{(\xi)} = -4\int d^3\mathbf{x'}
G(\mathbf{x},\mathbf{x'}) \hat{\xi}^y(\mathbf{x'})$. Comme il est plus
pratique d'utiliser la jauge GN pour évaluer les perturbation de métrique sur
la brane fixée en $\tilde{y} = 0$, nous nous repla\c{c}ons dans la jauge GN. De
(\ref{qqq:gnrs}), on voit qu'on peut choisir $\hat{\xi}_\mu$ tel que la
perturbation métrique sur la brane en $\tilde{y} = 0$ soit : $\tilde{h}_{\mu\nu}
= \bar{h}^{(m)}_{\mu\nu}+{2\over\ell}\eta_{\mu\nu}\hat{\xi}^y$.

Ainsi, pour une source de matière statique à symétrie sphérique sur la brane
\ba\label{qqq:sourcegt}
T_{\mu\nu} & = & \rho(r) u_\mu u_\nu,
\ea
où la quadrivitesse est $u^\mu = (1,0,0,0)$ et $u_\mu u^\mu = -1$, on peut
résoudre exactement les équations. En statique, l'équation de déplacement
(\ref{qqq:bending}) s'écrit $\nabla^2 \hat{\xi}^{y} = (\kappa^2/6)T$, elle
s'intègre aisément de sorte qu'on trouve
\ba\label{qqq:hxi}
\bar{h}^{(\xi)} & = & {4\over 3}\int_0^r{dr'\over
  {r'}^2}\int_0^{r'}dr''{r''}^2 V(r''),\cr
V(r) & = & {\kappa^2\over 2}\int d^3\mathbf{x'}G(\mathbf{x},\mathbf{x'})\rho(\mathbf{x'}).
\ea
On déduit aussi facilement que
$h_{00} = -(8/3)V(r)$ et $h_{rr} = -[8/(3r^3)]\int_0^r dr'{r'}^2 V(r')$. À
l'extérieur de la source on peut supposer que $G(\mathbf{x},\mathbf{x'}) \approx
G(\mathbf{x})\equiv G(r)$ puis, en définissant la masse de la source $M = \int
d^3\mathbf{x} \rho$, on trouve à partir de l'expression (\ref{qqq:greencorr})
de la fonction de Green que 
\ba
V(r) & \approx & {-\kappa^2 M\over 8\pi\ell r}\left(1+{\ell^2\over
    2 r^2}\right)
\ea
 et donc que le déplacement de la brane due à la matière
vaut
\ba\label{qqq:solbending}
\hat{\xi}^y & \approx & {\kappa^2 M \over 24\pi r}.
\ea
Finalement le \emph{potentiel gravitationnel Newtonien} créé par la source sur la
brane est obtenu dans la limite de champ faible par la perturbation métrique
sur la brane $\tilde{h}_{00}/2$ en jauge GN :
\ba\label{qqq:newton}
{\tilde{h}_{00}\over 2} & \approx & {G M\over r}\left(1+{2\ell^2 \over 3 r^2}+
...\right)
\ea
où $G = G_5/\ell = M_5^3/\ell$ est la constante de gravitation
quadridimensionnelle. On constate qu'aux grandes distances $r \gg \ell$ le
potentiel gravitationnel créé par la source de matière correspond au potentiel
Newtonien standard quadridimensionnel. La signature d'une dimension
supplémentaire se manisfeste donc à très courte distance $r \lesssim \ell$,
indiquant que l'effondrement gravitationnel est à cette échelle très différent
de celui du trou noir de Schwarzchild à quatre dimension.  Les
corrections à la gravité quadridimensionnelle sont très faible à basse
énergie comme c'était prévu à cause de la suppression des modes KK massifs par la barrière de
potentiel (voir paragraphe \ref{subsec:spectrers}). Les plus récents tests de la loi
de Newton ont mesuré un comportement standard quadridimensionnel en $1/r^2$ de la
force de gravitation jusqu'à des distances de $0.1$ millimètres \cite{exp}. Les
expérimentateurs ne sont pas encore en mesure de pouvoir tester la force de
gravitation à des distances plus courtes que le dixième de millimètre pour
éventuellemnt détecter un comportement de plus haute dimension de la gravitation dans
l'ultra-violet. Cette borne expérimentale indique qu'on ne peut exclure, pour
une dimension supplémentaire ayant une structure anti-de Sitter, une taille
aussi \emph{grande} qu'un dixième de millimètre : $\ell \lesssim 0.1$ mm.

Si les deux points de la fonction de Green (\ref{qqq:greengt}) sont choisis sur la brane et comme
le mode zéro est dominant à basse énergie on peut tronquer la fonction de
Green au mode zéro et
obtenir que $G_R(x^\mu,\tilde{y} = 0, {x'}^\mu, \tilde{y}' = 0) \approx
\delta^{(4)}\left(x^\mu-{x'}^\mu\right)/\left(\ell~\Box_4\right)$. On en déduit
que la perturbation métrique sur la brane $\tilde{h}_{\mu\nu}
= \bar{h}^{(m)}_{\mu\nu}+{2\over\ell}\eta_{\mu\nu}\hat{\xi}^y$ est similaire à
la solution standard de la théorie d'Einstein 4D linéarisée :
\ba
\tilde{h}_{\mu\nu} & \approx & -{16\pi G\over \Box_4}\left(T_{\mu\nu}-{1\over
  2}\eta_{\mu\nu}T\right).
\ea


\section{Cosmologie branaire dans le modèle RS}\label{sec:cosmors}

Le Modèle de Randall-Sundrum reproduit donc la théorie de la relativité
générale à quatre dimensions sur la brane à basse énergie grâce à la courbure
du bulk $AdS^5$ qui entraine l'exitence
conjointe d'un état lié du graviton et d'une barrière de potentiel s'opposant
à l'interaction des gravitons massifs du bulk avec la brane. Des modifications à
la loi de gravitation de Newton, dues à la présence d'une cinquième dimension
infinie et non-compacte, se manifestent à courte distance sur la brane
 $r \lesssim \ell$, où $\ell$ est le rayon de courbure du bulk anti-de
Sitter. Si le cas d'une brane Minkowskienne (plate et sans contenu de
matière) permet d'appréhender les conséquences gravitationnelles dans le
scénario branaire RS, ce cas n'est pas réaliste du point de vue de la
cosmologie. L'Univers primordial a subi en réalité divers types d'expansion cosmologique au cours de
son histoire thermique et contenu divers types de matière : une fraction de
secondes après le Big-Bang l'Univers connait d'abord une phase d'Inflation, où son
expansion est exponentielle, puis son expansion décélère durant l'ère dominée
par la radiation et continue de décélérer durant l'ère ultérieure dominée par
la matière. Il est donc nécessaire d'étudier le cas plus réaliste d'une brane
homogène et isotrope de géométrie Friedmann-Robertson-Walker (FRW), ayant un
tenseur énergie-impulsion non nul dépendant du temps et
une histoire d'expansion cosmologique arbitraire, plongée dans le bulk $AdS^5$
du modèle de Randall-Sundrum. C'est l'objet de cette section, où l'on commence
par généraliser le modèle RS au cas d'une brane de géométrie purement de
Sitter pour ensuite énoncer les solutions cosmologiques homogènes et isotropes
dans le modèle RS.

\subsection{Brane de Sitter}\label{subsec:dsrs}

Le cas particulier d'une brane de géométrie de Sitter ($dS^4$) peut aider à
modéliser un univers branaire dans sa phase inflationnaire en première
approximation puisque lors de l'Inflation l'univers subit une expansion
quasi-de Sitter. Les calculs perturbatifs effectués dans la section
(\ref{sec:bound}) dans le cas d'une brane Minkowskienne peuvent être
généralisés au cas d'une brane $dS$ (voir la référence \cite{lands} pour le
détail des calculs ainsi que les références \cite{gs,llg,fk} pour des
discussions similaires).

\begin{figure}
  \begin{center}
\includegraphics[width=3.5cm]{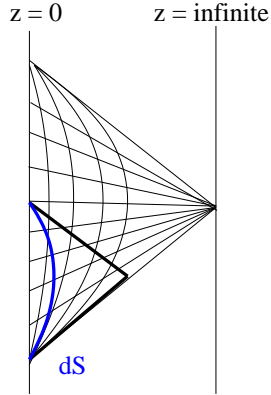}
 \end{center}
\caption{
Diagramme de Carter-Penrose pour une cosmologie de Sitter (dS) plongée dans $AdS^5$. }
\label{fig:penrosedS}
\end{figure}

Un espace-temps de Sitter subit une expansion exponentielle
uniforme caractérisée par le facteur d'échelle $a(t) = e^{H t}$ parce qu'il ne contient pas de matière mais seulement une constante
cosmologique positive $\Lambda_4 = H^2 > 0$, où $H$ est le facteur de Hubble,
constant dans ce cas. Un Univers à $(3+1)$ dimensions en expansion de Sitter
peut être contenu dans une portion de l'espace $AdS^5$, décrite dans les coordonnées statiques de Poincaré
\ba\label{qqq:poincare2}
ds^2 & = & \ell^2 {-dt^2+d\mathbf{x}_{3}^2+dz^2 \over z^2},
\ea
au moyen du plongement explicite ($-\infty < \tau < +\infty$, $-\infty < \chi
< +\infty$) :
\ba\label{qqq:dsads}
t & = & -{1\over H}e^{-H\tau}\cosh \left(H\chi\right)\cr
z & = & {1\over H}e^{-H\tau}\sinh \left(H\chi\right).
\ea
De cette transformation de coordonnées on obtient la métrique $AdS_5$ écrite sous la forme
\ba\label{qqq:metricdsads}
ds^2 & = & {H^2\ell^2\over\sinh^2\left(H\vert\chi\vert\right)}\left[-d\tau^2+e^{2H\tau}d\mathbf{x}_3^2+d\chi^2\right]
\ea
qui décrit une brane de géométrie $dS^4$ fixée en $\chi = \chi_b =
(1/H)\sinh^{-1}\left(H\ell\right)$ dans l'espace $AdS_5$. On a
rajouté une valeur absolue sur la dimension supplémentaire $\chi$ pour mettre
en évidence la symétrie $Z_2$ propre au modèle de Randall-Sundrum. L'expansion uniforme
quadridimensionnelle de l'Univers $dS^4$ est équivalente du point de vue à
cinq dimensions au mouvement de la brane dans le bulk statique $AdS^5$ selon
la trajectoire rectiligne
\ba\label{qqq:trajdsads}
t_b(\tau) & = & -{1\over H}e^{-H\tau}\sqrt{1+H^2\ell^2}\cr
z_b(\tau) & = & \ell e^{-H\tau}.
\ea
Les géodésiques ne sont pas des lignes droites dans l'espace courbe $AdS$, de 
telle sorte que l'expansion uniforme ($H = constante$) équivaut en fait à une trajectoire
uniformément accélérée de la brane $dS^4$ dans le bulk. Notons que l'infini
futur $\tau\rightarrow +\infty$ de l'espace $dS^4$ correspond à un temps fini
(t = 0) dans le bulk $AdS^5$ (Fig. \ref{fig:penrosedS}).

Au lieu d'étudier l'évolution des perturbations de manière covariante comme
dans la section \ref{sec:bound}, on se focalise sur les perturbations purement
tensorielles, c'est-à-dire les ondes gravitationnelles au sens
quadridimensionnel (deux degrés de liberté). Elles correspondent aux fluctuations
transverses et sans trace de la métrique ($\partial_i h^i_j = h_i^i = 0$
où les indices sont élevés avec la métrique plate Euclidienne
tridimensionnelle $\delta_{ij}$) : \ba\label{qqq:metricdsadspert}
d\tilde{s}^2 & = &
{H^2\ell^2\over\sinh^2\left(H\vert\chi\vert\right)}\left[-d\tau^2+e^{2H\tau}\left(\delta_{ij}+h_{ij}\right)d\mathbf{x}_3^2+d\chi^2\right]. 
\ea
Par définition les perturbations tensorielles sont invariantes de jauge
puisqu'elles ne sont pas affectées par une reparamétrisation des
coordonnées. Elles ne couplent pas non plus à la matière ni aux perturbations de
matière\footnote{Si on néglige les perturbations tensorielles anisotropes de matière.}, en ce sens elles évoluent librement dans le bulk $AdS$ et peuvent
être décrite par un champ scalaire canonique $\Psi(\tau,\chi)$ minimalement
couplé à la gravité après transformation de Fourier dans les trois directions spatiales transverses
selon $h_{ij}(\tau,\mathbf{x},\chi) = \int d^3\mathbf{x}
e^{i\mathbf{p}\cdot\mathbf{x}} \Psi(\tau,\chi)e_i\otimes e_j$, où le tenseur
$e_i\otimes e_j$ est transverse et sans trace. On peut vérifier que la
linéarisation des équations d'Einstein réalisée à la section \ref{sec:bound}
conduit pour les perturbations tensorielles à une équation de Klein-Gordon
selon laquelle évolue le champ scalaire $\Psi$. Dans les coordonnées
(\ref{qqq:metricdsads}) pour le fond $AdS$, l'équation de Klein-Gordon s'écrit
\ba\label{qqq:eomds}
\left[-\left(\partial^2_\tau+3H\partial_\tau+{p^2\over
    a^2(\tau)}\right)+\left(\partial^2_\chi-{3H\over
    \tanh\left(H\chi\right)}\partial_\chi\right)\right]\Psi & = & 0,
\ea
où $a(\tau) = e^{H\tau}$ est le facteur d'échelle traduisant l'expansion de Sitter de l'Univers.
Cette équation du mouvement est séparable si $H = constante$. On peut donc
rechercher les modes de spin $2$ massifs $m$ du graviton tels que $\Psi_m
(\chi,\tau) = u_m(\tau)\psi_m(\chi)$ et
\ba\label{qqq:separationds}
-\left(\partial^2_\tau+3H\partial_\tau+{p^2\over a^2(\tau)}\right) u_m(\tau) & = & m^2 u_m(\tau).
\ea
L'équation (\ref{qqq:separationds}) est celle d'un champ scalaire massif dans
l'espace $dS^4$. En utilisant le temps conforme $\eta = -(1/H)e^{-H\tau}$, on
peut remarquer cette équation est de type Bessel pour le champ $v_m(\eta) =
a(\eta)u_m(\eta)$  . Les modes de fréquence \emph{positive}/ \emph{negative}
sont donc donnés par \ba\label{qqq:soltds}
u_m^{\pm}(\tau) & = & {N_m\over H^{3/2}} e^{-{3\over 2}H\tau}
H^{(1)/(2)}_\nu(\tau),
\qquad \nu = {\sqrt{9/4 - m^2/H^2}},
\ea
où $H^{(1)/(2)}_\nu$ sont les fonctions de Hankel. Le mode $u_m^{\pm}$ est dit
de fréquence \emph{positive/negative} parce qu'il se comporte asymptotiquement
dans le passé ($\tau,\eta\rightarrow -\infty$) comme une onde plane de
fréquence positive/négative, comme dans l'espace plat : $u_m^+\left(\eta =
-(1/H)e^{-H\tau}\right)\sim \eta e^{\pm i\eta}$. Cette asymptote correspond de
plus aux modes aux échelles subhorizon, $p\eta = p/(aH) \gg 1$, c'est-à-dire dont la
longueur d'onde est plus petite que le rayon de Hubble (ou l'horizon des
évènements) de l'espace $dS^4$. Ce n'est pas étonnant puisque qu'à petite échelle
la courbure de l'espace de Sitter ne se fait plus sentir et ce dernier
correspond quasiment à l'espace plat de Minkowski.

Le profil des modes du graviton dans la dimension supplémentaire est donné par
la forme des fonctions d'onde $\psi_m(\chi)$. Il est pratique de reformuler le
problème sous la forme d'un problème de mécanique quantique comme cela a été
fait pour le cas de la brane Minkowskienne à la section
\ref{subsec:spectrers}. Le champ défini par $\phi_m(\chi)= {\left(H\ell\right)^{3/2}\over\sinh^{3/2}\left(H\vert\xi\vert\right)}
 \psi_m(\chi)$ satisfait l'équation de type Schrödinger suivante dans le bulk
 $AdS^5$ :
\ba\label{qqq:schrods}
\left[-\partial^2_\chi+V(\chi)\right]\phi_m(\chi) & = & m^2\phi_m(\chi),
\ea
où la brane est située en $\chi = \chi_b = (1/H)\sinh^{-1}(H\ell)$ et le potentiel de Schrödinger vaut
\ba\label{qqq:potentialds}
V(\chi) & = & {15H^2\over 4\sinh^2\left(H\chi\right)}+{9\over 4}H^2-{3\sqrt{1+H^2\ell^2}\over\ell}\delta(\chi-\chi_b).
\ea
\begin{figure}
  \begin{center}
\includegraphics[width=6.5cm]{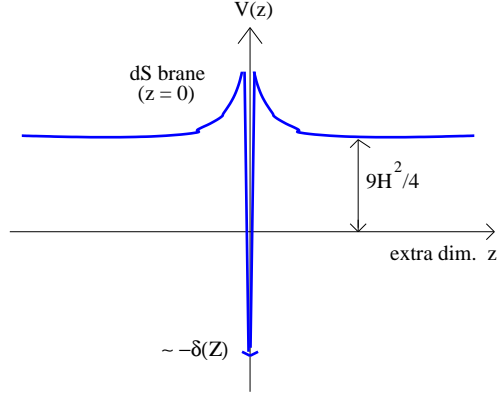}
 \end{center}
\caption{
Potentiel-volcan pour une brane de Sitter plongée dans Anti-de Sitter. On a
translaté $z = \chi-\chi_b$.}
\label{fig:volcanodS}
\end{figure}

La fonction $\delta$ encode les conditions de bord suivantes sur la brane :
\ba\label{qqq:cbrsds}
\left[\partial_\chi+{3H\over 2\tanh\left(H\chi_b\right)}\right]\phi_m(\chi)\vert_{\chi
  = \chi_b} & = & 0.
\ea
Ce potentiel-volcan pour la brane de Sitter (Fig. \ref{fig:volcanodS}) peut être comparé au potentiel-volcan du
modèle RS initial avec la brane Minkowskienne (section
\ref{subsec:spectrers}). On remarque cette fois la présence d'un
gap d'énergie ${9H^2/4}$, le potentiel tendant asymptotiquement, loin de la
brane ($\chi \rightarrow\infty$), vers cette valeur. Bien-sûr, dans la limite
$\Lambda_4 = H^2 = 0$, on retrouve les résultats du modèle RS initial avec
une brane Minkowskienne.
La forme du potentiel (\ref{qqq:potentialds}) indique que le spectre des gravitons
dans le modèle RS avec une brane de Sitter inclue un unique état lié, le mode
zéro $\phi_0(\chi) =
{\left(H\ell\right)^{3/2}\over\sinh^{3/2}\left(H\vert\xi\vert\right)}$, discret et
normalisable, dont la fonction d'onde a son support localisé près de la
brane. Le spectre inclue également un continuum d'états libres qui démarre à
$m > 3H/2$ à cause du gap. Le mode zéro s'identifie au graviton standard sans
masse de la gravité linéarisée à $4d$ et les modes $m > 3H/2$ s'identifient à
des gravitons massifs de KK du point de vue quadridimensionnel. Ce modèle
branaire, avec une brane de Sitter plongée dans le bulk $AdS_5$, est aussi
viable que le modèle RS initial du
point de vue cosmologique puisque la
localisation de l'état lié permet de reproduire la gravité standard à $4d$ sur
la brane tout en réalisant une expansion de Sitter de l'Univers. L'interaction
des gravitons massifs avec la brane est d'autant plus supprimée à basse
énergie à cause du gap $m > 3H/2$.

\subsection{Brane cosmologique}\label{subsec:bdel}

L'exploration de signatures cosmologiques de la présence de dimensions
supplémentaires dans le modèle de Randall-Sundrum doit tenir compte d'un
scénario cosmologique plus réaliste où l'Univers $4d$, homogène et isotrope, subit une expansion
cosmologique arbitraire de type Friedmann-Robertson-Walker (FRW)
\ba\label{qqq:frwmetric}
ds_{(4)}^2 & = & -d\tau^2+a^2(\tau)\delta_{ij}dx^idx^j
\ea
et dont
l'évolution du facteur d'échelle $a(\tau) = e^{\int H(t\tau) d\tau}$, où
$H(\tau)$ est le facteur de Hubble dependant du temps sur la brane, est dictée par le type de matière (le tenseur
d'énergie-impulsion) présent
dans l'Univers. On supposera que l'Univers est plat dans les trois directions
spatiales, comme semblent le confirmer les observations cosmologiques actuelles, telles que
les cartes d'anisotropies du Fond Diffus Cosmologique (CMB en Anglais pour Cosmic
Microwave Background). Donc la métrique tridimensionnelle est Euclidienne :  
$\delta_{ij}$ ($i,j = 1,...,3$).  Binetruy, Deffayet, Ellwanger et Langlois
ont calculé en 2000 les
solutions cosmologiques branaires homogènes et isotropes plongées dans
l'espace $AdS^5$ et montré que le facteur d'échelle $a(\tau)$ ne suit
pas exactement l'évolution cosmologique conventionnelle\footnote{On entend par
  ``conventionnelle'' l'évolution du facteur d'échelle dans le modèle standard
  de la cosmologie $4D$ pour un univers FRW.} \cite{bdel}. Un calcul similaire,
mais de fa\c{c}on covariante, a été effectué la même année par Shiromizu, Maeda
et Sasaki \cite{smz}.

Un univers à $(3+1)$ dimensions du type Friedmann-Robertson-Walker,
spatialement plat, homogène et isotrope, avec une expansion cosmologique
arbitraire, caractérisée par le facteur d'échelle $a(\tau)$, où $\tau$ est le
temps propre, peut être plongé dans une portion de l'espace $AdS^5$ de rayon de
courbure $\ell$, décrit en fonction des coordonnées de Poincaré statiques par
\ba
ds^2 & = & \ell^2 {-dt^2+d\mathbf{x}_{3}^2+dz^2 \over z^2},
\ea
au moyen de la trajectoire suivante pour la brane
\ba
z_b(\tau ) =
e^{-\int H(\tau)d\tau},\qquad
t_b(\tau )=
\int {d\tau\over\ell}\sqrt{\ell^2\dot{z_b}^2(\tau)+z_b^2(\tau)}.
\ea
$H(\tau)$ est le facteur de Hubble sur la brane. On peut construire
explicitement des coordonnées Gaussiennes normales (GN) en calculant les
courbes géodésiques dans le plan $(z,t)$, normales à la trajectoire de
la brane de temps propre $\tau$ et qui s'étendent à la distance propre $\xi$
de la brane. On obtient ainsi la transformation de coordonnées suivantes
entre les coordonnées GN et les coordonnées de Poincaré :
\ba
z(\xi,\tau) & = & \frac{e^{-\int d\tau ~H(\tau )}}
{ \cosh(\xi /\ell)
-\sqrt{1+\ell^2H^2}\sinh(\xi/ \ell)},\cr
t(\xi,\tau) & = &
\int{d\tau\over\ell}\left[e^{-\int d\tau ~H(\tau )}\sqrt{1+\ell^2H^2}\right]-{\ell He^{-\int
   d\tau ~H(\tau )}\sinh(\xi/ \ell )\over
\cosh(\xi/ \ell)-\sqrt{1+\ell^2H^2(\tau )}\sinh(\xi /\ell)}.\qquad
\ea
On déduit de cette transformation la métrique GN à cinq dimensions de $AdS^5$ :
\ba\label{eqnarray:exactmetric0} ds^2 & = &
-\left(\sqrt{1+\ell^2
H^2}\sinh(\xi/\ell)-\cosh(\xi/\ell)+{\ell^2\dot{H}\over\sqrt{1+\ell^2H^2}}\sinh(\xi/\ell)\right)^2
d\tau^2\cr & &
+\left(\sqrt{1+\ell^2H^2}\sinh(\xi/\ell)-\cosh(\xi/\ell)\right)^2
e^{2\int H(\tau)d\tau} d\mathbf{x}_{3}^2\cr & & +d\xi^2. \ea
Dans ces coordonnées la brane FRW est stationnaire par rapport au bulk, située en
$\xi = 0$, et $\xi$
mesure la distance propre à la brane. Il reste à connaître l'évolution du
facteur d'échelle $a(\tau)$ (ou du facteur de Hubble $H(\tau)$) en fonction de
l'évolution de la matière sur la brane.

Il s'agit donc de retrouver la solution métrique $g_{AB}$ ($A,B = 0,...4)$ des
équations d'Einstein à cinq dimensions suivantes :
\ba\label{qqq:einstein}
G_{AB}\equiv R_{AB}-{1\over 2}g_{AB}R & = & -\Lambda_5 g_{AB}+\kappa^2\delta(\xi)T_{AB},
\ea
où $\Lambda_5$ est la constante cosmologique de l'espace-temps total à cinq
dimensions (le bulk), $\kappa^2 = 8\pi/M^3_5$ si $M_5$ est la masse de Planck
à 5D. Dans le cadre du modèle de Randall-Sundrum on recherche une solution de
fond anti-de Sitter $AdS^5$, donc de constante cosmologique négative
$\Lambda_5 = -6/\ell^2 < 0$, où $\ell$ dénote le rayon de courbure du
bulk. $T_{AB}$ est le tenseur énergie-impulsion dépendant du temps, confiné sur la brane à $(3+1)$ dimensions en $\xi = 0$ si $x^4 \equiv \xi$ dénote la cinquième coordonnée d'espace-temps (ou dimension supplémentaire) :
\ba\label{qqq:stress}
T_{AB} & = & diag\left(-\rho,P,P,P,0\right).
\ea
Tout au long de ce mémoire on notera $\rho$ et $P$ respectivement la densité d'énergie (ou de masse) et la pression de la matière \emph{effective} sur la brane. Ces composantes sont reliées à la matière \emph{réelle} sur la brane, constituée d'un fluide parfait sur la brane ayant pour équation d'état
$P_M = w\rho_M$, où $P_M$ et $\rho_M$ sont repectivement la pression et la
densité d'énergie du fluide, selon
\ba\label{qqq:effreel}
\rho = \rho_M+\sigma, \qquad P = P_M-\sigma,
\ea
si $\sigma$ est la tension critique de la brane, ajustée dans le modèle de Randall-Sundrum de telle sorte que la constante cosmologique effective sur la brane plongée dans le bulk $AdS$ (de constante cosmologique négative $\Lambda_5$) soit zéro en l'absence de matière sur la brane. Son expression a été donnée en (\ref{qqq:finetuning}). Il est commode d'utiliser des coordonnées gaussiennes normales pour résoudre (\ref{qqq:einstein}) et de considérer l'Ansatz homogène et isotrope suivant pour la métrique :
\ba\label{qqq:typicmetric}
ds^2 = g_{AB}dx^A dx^B & = &
d\xi^2-N^2(\xi,\tau)d\tau^2+A^2(\xi,\tau)d\mathbf{x}_{3}^2.
\ea
Dans ces coordonnées la brane reste fixée en $\xi = 0$ et la coordonnée supplémentaire $\xi$ mesure la distance propre à la brane. On peut toujours reparamétriser le temps propre sur la brane $\tau$ pour que $N(\tau,\xi = 0) = 1$ sur la brane. En notant de plus $a(\tau) = A(\tau,\xi = 0)$, on voit que la métrique induite sur la brane en $\xi = 0$ est celle d'un Univers FRW à $(3+1)$ dimensions en expansion avec le facteur d'échelle $a(\tau)$ (voir métrique (\ref{qqq:frwmetric})). En insérant l'Anstaz (\ref{qqq:typicmetric}) dans les équations d'Einstein (\ref{qqq:einstein}) et dans les conditions de raccordement d'Israel sur la brane ($\mu,\nu = 0,...,3$)
\ba
K_{\mu\nu}(\xi =
 0^+) & = & {-\kappa^2\over 2}\left[T_{\mu\nu}-\frac{1}{3}Tg^{(4)}_{\mu\nu}\right],
\ea
où
 $g^{(4)}_{\mu\nu}$ est la métrique FRW induite sur la brane et $K_{\mu\nu} = (1/2) \partial_\xi g^{(4)}_{\mu\nu}$ est
 la courbure extrinsèque de la brane, on obtient que
\ba
A(\tau,\xi) & = & a(\tau)\left[\cosh(\xi/\ell)-\left(1+{\rho(\tau)\over \sigma}\right)\sinh(\xi/\ell)\right].
\ea
Tout au long de ce mémoire, un point désigne une dérivée temporelle et un prime une dérivée selon la dimension supplémentaire. De l'égalité $T_{04} = 0$, on obtient de plus que $N(\tau,\xi) = \dot{A}(\tau,\xi)/\dot{a}(\tau)$, de telle sorte que la solution métrique $AdS^5$ contenant une brane $FRW^4$ est finalement
\ba\label{qqq:bdelmetric}
ds^2 & = & -\left[\cosh(\xi/\ell)-\left(1+{\rho(\tau)\over \sigma}\right)\sinh(\xi/\ell)-{\dot{a(\tau)}\over a(\tau)}{\dot{\rho(\tau)}\over \sigma}\sinh(\xi/\ell)\right]^2d\tau^2\cr
& & +\left[\cosh(\xi/\ell)-\left(1+{\rho(\tau)\over \sigma}\right)\sinh(\xi/\ell)\right]^2
a^2(\tau) d\mathbf{x}_{3}^2\cr
& & +d\xi^2. \ea
L'équation de conservation du tenseur-énergie impulsion $\nabla_A T^{AB} = 0$ (de fa\c{c}on équivalente l'équation de Bianchi sur le tenseur d'Einstein $\nabla_A G^{AB} = 0$), où $\nabla_A$ est la dérivée covariante associée à la métrique ($\nabla_A g^{AB} = 0$), entraine de plus l'équation de conservation
\ba\label{qqq:cons}
\dot{\rho} & = & -3H(P+\rho).
\ea
Cette équation a la même forme que l'équation de conservation standard à quatre dimensions, où le facteur de Hubble sur la brane, caractérisant l'expansion cosmologique de la brane, est défini par $H(\tau) = \dot{a}(\tau)/a(\tau)$.
Cependant, on peut trouver une intégrale première des équations d'Einstein, conduisant à l'équation de Friedmann modifiée suivante sur la brane : \ba\label{qqq:fried}
H^2 \equiv \left({\dot{a}\over a}\right)^2 & = & {\Lambda_5\over 6}+{\kappa^4\over 36}\rho^2+{\mathcal{C}\over a^4}
\ea
qui implique une évolution cosmologique non-standard du facteur
d'échelle. $\mathcal{C}$ est une constante d'intégration qui, si elle est
non-nulle, décrit le paramètre de masse d'un espace Schwarzchild-anti-de
Sitter $SAdS^5$. Comme on se place dans le cadre du modèle de Randall-Sundrum,
avec un bulk $AdS$ pur, on fixera $\mathcal{C} = 0$ dans la suite. Réexprimons
l'équation de Friedmann (\ref{qqq:fried}) et l'équation de conservation
(\ref{qqq:cons}) en fonction de la densité d'énergie du fluide parfait sur la
brane :
\ba\label{qqq:bwdyn}
H^2 & = & {\kappa^2\over 3\ell}\rho_M\left(1+{\kappa^2\ell\over
    12}\rho_M\right),\cr
\dot{\rho} & = & -3H(1+w)\rho_M.
\ea
On constate que l'équation de Friedmann standard à quatre dimensions ($H^2 =
\kappa_4^2\rho_M/3$) est effectivement retrouvée à basse énergie ($\rho_M \ll
\sigma$) sur la brane, si $\kappa_4^2 = \kappa^2/\ell$. Par contre des
modifications à la cosmologie standard interviennent à haute énergie de
fa\c{c}on quadratique en la densité d'énergie sur la brane. C'est donc dans
l'Univers primordial (temps court ou haute énergie) que se manifeste la
signature de dimensions supplémentaires éventuelles. On peut obtenir
l'évolution cosmologique du facteur d'échelle sur la brane en résolvant les
équations (\ref{qqq:bwdyn}) selon :
\ba\label{qqq:amod}
a(\tau) & \propto & \left[t\left(t+t_\sigma\right)\right]^{1\over 3(1+w)}, \qquad \mbox{avec $t_\sigma = {2\ell\over 3(1+w)}$},
\ea
et $\rho_M = a^{-3(1+w)}$, où on rappelle que $w = P_M/\rho_M$. À grand temps
$t\gg t_\sigma$, ou basse énergie, on retrouve l'évolution cosmologique
standard d'un univers FRW 4D contenant un tenseur-énergie impulsion de type
fluide parfait : $a(\tau) \propto t^{2/3(1+w)}$. Par contre dans l'Univers
primordial, à temps court $t\ll t_\sigma$, ou haute énergie, l'évolution du
facteur d'échelle n'est plus conventionnelle : $a(\tau) \propto
t^{1/3(1+w)}$. Notons que cette solution n'a pas de sens dans le cas $w = -1$
mais en fait, dans ce cas, la solution de l'équation de conservation est de
toute fa\c{c}on $\rho = constante$, et donc l'équation de Friedmann modifiée
entraine que la brane est de géométrie de Sitter pur, sauf que le facteur de
Hubble constant (ou constante cosmologique) diffère de la constante de Hubble d'un univers de Sitter dans le cas standard à quatre dimensions.

Terminons en réécrivant la métrique (\ref{qqq:bdelmetric}) de Binetruy,
Deffayet, Ellwanger et Langlois en fonction du
facteur de Hubble $H(\tau)$ seulement car c'est cette forme que nous
utiliserons par la suite, notamment au chapitre \ref{chapter:dissip} où nous 
expose nos résultats. Au moyen de la relation (\ref{qqq:fried}), avec
$\mathcal{C} = 0$, on obtient que la métrique (\ref{qqq:bdelmetric}) se
réécrit :
\ba\label{qqq:apexactmetric} ds^2 & = &
-\left(\sqrt{1+\ell^2
H^2}\sinh(\xi/\ell)-\cosh(\xi/\ell)+{\ell^2\dot{H}\over\sqrt{1+\ell^2H^2}}\sinh(\xi/\ell)\right)^2
d\tau^2\cr & &
+\left(\sqrt{1+\ell^2H^2}\sinh(\xi/\ell)-\cosh(\xi/\ell)\right)^2
e^{2\int H(\tau)d\tau} d\mathbf{x}_{3}^2\cr & & +d\xi^2, \ea
telle que nous l'avons déjà calculée en (\ref{eqnarray:exactmetric0}).

Peut-on calculer l'évolution des perturbations cosmologiques autour de cette
métrique de fond décrivant un univers FRW homogène et isotrope (dans les trois
directions spatiales transverses), en expansion arbitraire, contenant
un tenseur énergie-impulsion dépendant du temps et plongé dans l'espace $AdS^5$,
tel qu'on l'a fait pour des branes ``vides'', Minkowski et de Sitter ? C'est le
but principal de ce travail de recherche, exposé aux chapitres
\ref{chapter:pert} et \ref{chapter:dissip} : essayer de résoudre l'évolution
des perturbations cosmologiques dans un Univers branaire en expansion quelconque.


\chapter{La théorie des perturbations cosmologiques dans les Univers
  branaires}\label{chapter:pert}

La métrique FRW à quatre dimensions (\ref{qqq:frwmetric}) décrit l'Univers
observable à $(3+1)$ dimensions à condition que celui-ci soit effectivement
homogène et isotrope dans les trois dimensions spatiales. En réalité notre
Univers n'apparait homogène et isotrope qu'à grande échelle, tel que le
confirme les relevés astronomiques, mais comporte des
inhomogénéités et des anisotropies  à plus
petite échelle, telles que les amas de galaxies ou les galaxies
elles-même. Depuis les années 80 \cite{peebles} on pense que la formation des grandes
structures de l'Univers (amas de galaxies,...) resulte de l'attraction
gravitationnelle de fluctuations de densité initiales à petite échelle, inhomogènes et
anisotropes. En ce sens la compréhension de la formation des structures astrophysiques de
l'Univers demande d'étudier un espace-temps légèrement différent de l'espace
FRW, c'est-à-dire légèrement inhomogène et anisotrope, en perturbant tout
simplement la métrique de fond FRW décrivant l'espace-temps ainsi que le tenseur énergie-impulsion
décrivant la matière. La théorie des perturbations cosmologiques linéaire standard à
quatre dimensions, dans le cadre relativiste, est bien connue depuis les
travaux pionniers de Lifshitz et Khalatnikov en 1963 \cite{lifshitz} et le formalisme invariant de jauge
développé par Bardeen en 1980 \cite{bardeen}. Il s'agit de linéariser\footnote{C'est-à-dire tronquer les
  perturbations de métrique et de matière au premier ordre.} les équations
d'Einstein et de calculer l'évolution des perturbations de métrique et de
matière. De la même manière que les équations d'Einstein relient
inextricablement la géométrie de l'espace-temps (métrique) au contenu matériel
de ce même espace-temps (tenseur énergie-impulsion), les équations linéarisées
d'Einstein relient les perturbations de métrique (qui incluent les ondes
gravitationnelles) aux fluctuations de densité de matière. Ces deux types de degrés de liberté se propagent dans
l'espace-temps de fond non-perturbé. Puisque l'espace-temps de fond
non-perturbé est homogène et isotrope, on peut effectuer une transformation de
Fourier des variables de perturbations dans les trois directions spatiales,
de telle sorte que les équations d'évolution des perturbations, issues des
équations d'Einstein linéarisées, se réduisent simplement à des équations aux
dérivées ordinaires (EDO), linéaires et d'ordre deux en temps. Ainsi il est facile
de les résoudre complètement et trouver le comportement des perturbations à
différentes échelles. Notons de plus qu'on est capable d'exprimer les
anisotropies de température et de polarisation du Fond Diffus Cosmologique
(CMB) en fonction des perturbations inhomogènes et anisotropes de métrique et
de matière à travers la formule de Sachs-Wolfe par exemple \cite{sw}. Le rayonnement du
Fond Diffus Cosmologique est de type corps noir à $2.7$ K mais présente un
spectre d'anisotropies de température et de polarisation de l'ordre de
$3\times 10^{-5}$ K. Ce rayonnement provient de la surface de dernière diffusion, au moment du
découplage des photons avec la matière (redshift $z \approx 1100$), et
représente donc l'image la plus ancienne qu'on ait de l'Univers.
L'extrème précision des cartes du CMB provenant de WMAP (et 
Planck dans un futur proche) nous pousse donc à étudier la théorie des perturbations cosmologiques
dans des modèles non-standard, tels que les modèles branaires, afin de
comparer les résultats théoriques aux observations du CMB, et de pouvoir
valider, contraindre ou invalider ces modèles. La théorie
 des perturbations cosmologiques demeure un outil indispensable quand on veut
 comparer les résultats théoriques aux observations cosmologiques.

En cosmologie branaire, la théorie des perturbations cosmologiques est bien
plus compliquée, et le calcul de l'évolution des perturbations cosmologiques
dans un univers branaire n'a pas encore complètement abouti. Cela est du à la presence de
la brane dans l'espace-temps $AdS^5$ qui, en tant que défaut topologique brise
l 'homogénéité et l'isotropie spatiale à cinq dimensions. En ce sens les
équations d'évolution des variables de perturbation issues de la linéarisation
des équations d'Einstein ne sont plus des EDO en temps mais des équations aux
dérivées partielles (EDP) d'ordre deux en temps et en la dimension
supplémentaire. En fait, dans le cas de branes vides de matière et à symétrie
maximale, telles que des branes Minkowski ou de Sitter, on a vu aux sections
\ref{subsec:spectrers} et \ref{subsec:dsrs} que les EDP pour les perturbations
pouvaient être complètement intégrées parce qu'elles étaient séparables
 et qu'on pouvait calculer le spectre des
gravitons (c'est-à-dire les perturbations de métrique) analytiquement. Mais
hormis dans ces cas de haute symétrie, le mouvement de
la brane FRW dans le bulk $AdS$ est arbitraire et compliqué. Par conséquent,
la forme des composantes de la métrique de fond,
donnée en (\ref{qqq:bdelmetric}) ou (\ref{qqq:apexactmetric}), est
compliquée et non-séparable. Il s'ensuit que les EDP pour les perturbations ne sont plus
séparables et donc difficile à résoudre au moins analytiquement. Notons que, outre cette
difficulté d'ordre technique, la théorie des perturbations linéaires en cosmologie
branaire fait également face à une difficulté d'ordre plus fondamental : le
problème des conditions initiales. Pour résoudre complètement le problème des
perturbations cosmologique branaires il faut spécifier des conditions
initiales sur un nombre infini de degrés de liberté. Afin de tester les
predictions des scenarios branaires et en amont les théories de cordes à
l'aide des observations cosmologiques, il est
indispensable de calculer au moins en partie l'évolution des perturbations
cosmologiques dans les univers branaires. C'est un défi encore non résolu pour
la cosmologie branaire, et un enjeu majeur pour la cosmologie théorique moderne.





\section{Théorie des perturbations invariantes de jauge en cosmologie
  branaire}\label{sec:invjaugepert}

De nombreux formalismes existent concernant la théorie des perturbations
cosmologiques en cosmologie branaire (on pourra
par exemple lire les revues \cite{langloisreview, durrer}) : dans
\cite{vandebruck} les auteurs introduisent la jauge dite ``longitudinale 5D'', qui est la jauge
utilisée dans le formalisme de Mukohyama (voir section \ref{sec:muko}), pour
exprimer les perturbations. Dans cette jauge les perturbations invariantes de
jauge coincident avec les perturbations naturelles de la métrique. Les
conditions de jonctions linéarisées pour les perturbations sont par contre plus maniables
dans la jauge dite Gaussienne normale, où la brane perturbée reste à sa
position fixe. Dans \cite{bridgman} les auteurs ont montré que les deux choix
de jauge étaient équivalents pour les équations des perturbations. Pour
faciliter le passage d'une jauge à l'autre il peut être utile d'utiliser un
formalisme invariant de jauge. Ici nous présentons le formalisme
invariant de jauge, couramment
utilisée dans la littérature \cite{langloisreview,vandebruck,durrer,deff1,bridgman,pert2,jap}). Cette partie concernant la théorie des perturbations est certes technique mais c'est le pilier de la cosmologie théorique moderne.

On peut toujours décomposer les perturbations d'un tenseur à deux indices en
trois catégories : les perturbations
purement "scalaires", les perturbations purement "vectorielles" et les perturbations
purement "tensorielles".

\subsection{Perturbations de métrique}\label{subsec:metricpert}

En perturbant à l'ordre linéaire la métrique générale $g_{AB}$ ($A,B = 0,...,4$) donnée en (\ref{qqq:typicmetric}) dans les coordonnées Gaussiennes normales, on obtient que la métrique perturbée $g_{AB}+\delta g_{AB}$ s'écrit
$$
g_{AB}+\delta g_{AB}  =
\left[
\begin{array}{ccc}
-N^2\left(1+2\bar{\mathcal{A}}\right) & A^2\left(\partial_i \bar{B}-\bar{S}_i\right) & N\bar{\mathcal{A}}_\xi \\
A^2\left(\partial_j \bar{B}-\bar{S}_j\right) & A^2\left[\left(1+2\bar{\mathcal{R}}\right)\delta_{ij}+2\partial_i\partial_j \bar{E}+\partial_{(i} \bar{F}_{j)}+\bar{E}_{ij}\right] & A^2\left(\partial_i \bar{B}_\xi-\bar{S}_{\xi i}\right) \\
N\bar{\mathcal{A}}_\xi & A^2\left(\partial_i \bar{B}_\xi-\bar{S}_{\xi i}\right) & 1+2\bar{\mathcal{A}}_{\xi\xi} \\
\end{array}
\right],
$$
où $i,j = 1,...,3$ dénotent les trois directions spatiales sur la brane.
$\bar{\mathcal{A}}$, $\bar{B}$, $\bar{\mathcal{A}}_\xi$, $\bar{B}_\xi$, $\bar{\mathcal{A}}_{\xi\xi}$, $\bar{\mathcal{R}}$ et $\bar{E}$ sont les perturbations scalaires de métrique, $\bar{S}_i$, $\bar{F}_i$ et $\bar{S}_{\xi i}$ sont les perturbations vectorielles ($3$-vecteurs sans divergence), et $\bar{E}_{ij}$ représente les perturbations tensorielles ($3$-tenseur transverse et sans trace)\footnote{On choisit de mettre une barre sur chaque perturbation lorsqu'on ne specifie pas de jauge particulière.}. L'avantage de cette décomposition est que les perturbations scalaires, vectorielles et tensorielles évoluent indépendamment à l'ordre linéaire, selon des équations d'onde découplées l'une de l'autre.

La transformation de jauge (reparamétrisation des coordonnées d'espace-temps à l'ordre linéaire) $x^A\rightarrow x^A+\delta x^A$ se décompose en transformations scalaires et vectorielles seulement, selon
\ba\label{qqq:rep}
\tau & \rightarrow & \tau+\delta\tau,\cr
x^i & \rightarrow & x^i+\partial^i \delta x+\delta x^i,\cr
\xi & \rightarrow & \xi+\delta\xi.
\ea
$\delta_\tau$, $\delta x$ et $\delta\xi$ sont des scalaires et $\delta x^i$ est un $3$-vecteur sans divergence. Sous la transformation de jauge (\ref{qqq:rep}), les perturbations scalaires de métrique se transforment selon ($\delta g_{AB}\rightarrow \delta g_{AB}-\nabla_{(A} x_{B)}$)
\ba\label{qqq:scaltrans}
\bar{\mathcal{A}} & \rightarrow & \bar{\mathcal{A}}-\dot{\delta\tau}-{\dot{N}\over N}\delta\tau-{N'\over N}\delta\xi,\cr
\bar{\mathcal{R}} & \rightarrow & \bar{\mathcal{R}}-{\dot{A}\over A}\delta\tau-{A'\over A}\delta\xi,\cr
\bar{B} & \rightarrow & \bar{B}+{N^2\over A^2}\delta\tau-\dot{\delta x},\cr
\bar{B}_\xi & \rightarrow & \bar{B}_\xi-\delta x'-{1\over A^2}\delta\xi,\cr
\bar{E} & \rightarrow & \bar{E}-\delta x,\cr
\bar{\mathcal{A}}_\xi & \rightarrow & \bar{\mathcal{A}}_\xi+N\delta\tau'-{1\over N}\dot{\delta\xi},\cr
\bar{\mathcal{A}}_{\xi\xi} & \rightarrow & \bar{\mathcal{A}}_{\xi\xi}-\delta\xi'.
\ea
On peut construire deux variables de perturbation invariantes sous les transformations de jauge 3D spatiales :
\ba\label{qqq:longvar}
\bar{\sigma} = -\bar{B}+\dot{\bar{E}},\qquad
\bar{\sigma}_\xi = -\bar{B}_\xi+\bar{E},
\ea
qui se transforment sous (\ref{qqq:rep}) selon $\bar{\sigma} \rightarrow \bar{\sigma}-(N^2/ A^2)\delta\tau$, $\bar{\sigma}_\xi \rightarrow \bar{\sigma}_\xi+(1/A^2)\delta\xi$. Par suite on peut aussi construire deux variables de perturbation invariantes sous les transformations de jauge 4D : $\bar{\Phi} =  \bar{\mathcal{A}}-(1/N)\left(A^2\bar{\sigma}/N\right)^{.}$ et $\bar{\Psi} = -\bar{\mathcal{R}}+\dot{A}A\bar{\sigma}/N^2$. Puisqu'il y a trois fonctions scalaires inconnues pour les transformations de jauge  5D (\ref{qqq:rep}), on en déduit qu'on peut définir quatre perturbations de métrique scalaires invariantes de jauge au sens 5D parmi les sept variables scalaires de départ :
\ba\label{qqq:invjaugescal}
\mathcal{A} & = & \bar{\mathcal{A}}-{1\over N}\left({A^2\bar{\sigma}\over N}\right)^{.}+{N'\over N}A^2\bar{\sigma}_\xi,\cr
\mathcal{A}_\xi & = & \bar{\mathcal{A}}_\xi+{\left(A^2\bar{\sigma}_\xi\right)^{.}\over N}+{\left(A^2\bar{\sigma}\right)'\over N}-2{N'\over N^2}A^2\bar{\sigma},\cr
\mathcal{A}_{\xi\xi} & = & \bar{\mathcal{A}}_{\xi\xi}+\left(A^2\bar{\sigma}_\xi\right)',\cr
\mathcal{R} & = & \bar{\mathcal{R}}+AA'\bar{\sigma}_\xi-{\dot{A}A\over N^2}\bar{\sigma}.
\ea
Sous la transformation de jauge (\ref{qqq:rep}), les perturbations vectorielles de métrique se transforment selon
\ba\label{qqq:vectrans}
\bar{S}_i & \rightarrow & \bar{S}_i+\dot{\delta x_i},\cr
\bar{S}_{\xi i} & \rightarrow & \bar{S}_{\xi i}+\delta x_i',\cr
\bar{F}_i & \rightarrow & \bar{F}_i-\delta x_i.
\ea
On peut écrire les perturbations vectorielles de métrique selon $\bar{S}_i = \bar{S}(\tau,\xi)e_i(\mathbf{x})$, $\bar{S}_{\xi i} = \bar{S}_\xi(\tau,\xi)e_i(\mathbf{x})$ et $\bar{F}_i = \bar{F}(\tau,\xi)e_i(\mathbf{x})$, où $e_i(\mathbf{x})$ est un $3$-vecteur sans divergence ($\partial_i e^i(\mathbf{x}) = 0$).
Puisqu'il y a une fonction vectorielle inconnue pour les transformations de
jauge 5D (\ref{qqq:rep}), on en déduit qu'on peut définir deux perturbations
de métrique vectorielles invariantes de jauge au sens 5D parmi les trois
variables vectorielles de départ :
\ba\label{qqq:invjaugevec}
v & = & \bar{S}+\dot{\bar{F}},\cr
v_{\xi} & = & \bar{S}_\xi+F'.
\ea
Il n'y a pas de fonction tensorielle dans la transformation de jauge
(\ref{qqq:rep}), de telle sorte que les perturbations tensorielles de métrique $E_{ij} \equiv \bar{E}_{ij}$ sont automatiquement invariantes de jauge, $E_{ij} \rightarrow E_{ij}$.

\subsection{Perturbations de matière}\label{subsec:matterpert}

Les perturbations du tenseur énergie-impulsion $T_{AB}$ donné en (\ref{qqq:stress}), décrivant la matière présente sur la brane\footnote{Notons que la brane ne reste plus nécessairement fixée en $\xi = 0$ après perturbation sauf dans le choix de jauge Gaussienne normale, par définition. On a déjà discuté le phénomène de "brane bending" au chapitre \ref{chapter:rs}.}, en $\xi = \xi_b(\tau,\mathbf{x})$,
$$
T_A^B+\delta T_A^B  =
\delta(\xi-\xi_b)\left[
\begin{array}{ccc}
-(\rho+\delta\rho) & \delta p^j & 0 \\
-A^{-2}\delta p_{i} & \left(P+\delta P\right)\delta_i^j+\delta\tilde{\pi}_i^j & 0\\
0 & 0 & 0
\end{array}
\right]
$$
peuvent également se décomposer en perturbations scalaires, vectorielles et tensorielles grâce à la symétrie $SO(3)$ de l'espace-temps non-perturbé, comme pour les perturbations de métrique, selon
\ba\label{qqq:decompmatter}
\delta p_{i} & = & \partial_i\delta q+\delta q_i,\cr
\delta\tilde{\pi}_{ij} & = & \left(\nabla_i\nabla_j-{1\over 3}\delta_{ij}\nabla^2\right)\delta\pi+\partial_{(i}\delta\pi_{j)}+\delta\pi_{ij}.
\ea
$\delta\rho$, $\delta P$, $\delta q$ et $\delta\pi$ sont les perturbations
scalaires de matière. $\delta q^i$ et $\delta\pi^i$ décrivent les
perturbations vectorielles de matière, sans divergence, et $\delta\pi_i^j$
représente la perturbation tensorielle de la matière, transverse et sans
trace. Notons que $\delta\tilde{\pi}_{ij}$ est le tenseur anisotrope
de perturbation du fluide $T_{AB}$.

Sous la transformation de jauge temporelle $\tau\rightarrow \tau+\delta\tau$
les perturbations de matière se transforment selon
\ba\label{qqq:mattrans}
\delta\rho & \rightarrow & \delta\rho -\dot{\rho}\delta\tau,\cr
\delta P & \rightarrow & \delta P -\dot{P}\delta\tau,\cr
\delta q & \rightarrow & \delta q +\left(\rho+P\right)\delta\tau,
\ea
alors que $\delta q^i$ et $\delta\tilde{\pi}_{ij}$ sont invariants. À partir
de (\ref{qqq:mattrans}) et en utilisant l'équation de conservation
(\ref{qqq:cons}) on constate qu'on peut construire une perturbation de matière
invariante de jauge $\Delta$, appelée contraste de densité :
\ba\label{qqq:delta}
\rho\Delta & = & \delta\rho-3H\delta q.
\ea
Également on peut construire une autre perturbation scalaire de matière
invariante de jauge $\Gamma_s$, appelée perturbation d'entropie :
\ba\label{qqq:entropie}
\Gamma_s & = & \delta P-c_s^2\delta \rho,
\ea
où $c_s^2 = \dot{P}/\dot{\rho}$ est la vitesse du son du fluide.

Une fois identifiées les perturbations de métrique et de
matière invariantes de jauge, on peut calculer leur évolution cosmologique au
moyen des équations d'Einstein linéarisées, de forme générale : 
\ba\label{qqq:linearizedeinstein}
\delta G_{\mu\nu}\left(\delta g_{AB}\right) \equiv \delta R_{AB}-{1\over
  2}\left(\delta g_{AB}R+g_{AB}\delta R\right) & = & -\Lambda_5 \delta
g_{AB}+\kappa^2\delta(\xi-\xi_b)\delta T_{AB}.
\ea
Plus précisément on peut séparer l'évolution dans le bulk et l'évolution sur
la brane : les gravitons (perturbations de métrique $\delta g_{AB}$) évoluent
dans le bulk $AdS^5$ selon l'équation d'Einstein linéarisée dans $AdS^5$ :
\ba\label{qqq:linearizedeinsteinbulk}
\delta G_{\mu\nu}\left(\delta g_{AB}\right) & = & -\Lambda_5 \delta
g_{AB},
\ea
et les perturbations de matière évoluent sur la brane et couplent aux
perturbations métriques selon les conditions de raccordement d'Israel
linéarisées et $Z_2$-symétriques, sur la brane :
\ba
\delta K_\mu^\nu(\xi =
 \xi_b) & = & {-\kappa^2\over 2}\left[\delta
   T_\mu^\nu-\frac{1}{3}\delta_\mu^\nu\delta T\right],
\ea
où $\mu,\nu = 0,...,3$. Les équations linéarisées d'évolution des
perturbations invariantes de jauge et les conditions de bord linéarisées
reliant les perturbations métriques et les perturbations de matière, invariantes de jauge, peuvent
donc être obtenues. Mais leurs expressions sont lourdes et nous préférons réécrire
ces équations dans le formalisme de Mukohyama (section \ref{sec:muko}), qui
permet de réduire ces équations de perturbations en des formes plus ``compactes''.

L'intérêt de travailler avec des quantités indépendantes du choix de jauge est
qu'on peut les faire évoluer dans la jauge qui nous arrange le plus suivant le problème considéré et on peut jongler d'une jauge à l'autre si nécessaire. Bien-sûr, une
fois qu'on a identifié les perturbations indépendantes de jauge, on peut
simplifier leur expression en choisissant la jauge la plus commode. Dans la
prochaine section \ref{sec:muko}, nous choisissons la ``jauge longitudinale
5D''  pour les perturbations de type scalaires qui est la jauge utilisée dans
le formalisme de Mukohyama.


\section{Formalisme de Mukohyama}\label{sec:muko}

On choisit de se placer dans la ``jauge longitudinale 5D'' \cite{vandebruck}, définie par
\ba\label{qqq:longgauge}
\bar{\sigma} = \bar{\sigma}_\xi & = & 0,
\ea
où $\bar{\sigma}$ et $\bar{\sigma}_\xi$ ont été définis en
(\ref{qqq:longvar}). Cette jauge est appelée comme telle parce que c'est une
sorte de généralisation, dans un modèle branaire, de la jauge longitudinale (ou
de Newton), utilisée en théorie des perturbations standards sur l'espace FRW à
$(3+1)$ dimensions. Dans cette jauge (\ref{qqq:longgauge}), les perturbations
scalaires de métriques indépendantes de jauge (\ref{qqq:invjaugescal})
coincident avec les quatres perturbations scalaires de métrique naturellement
introduites au début de la section \ref{subsec:metricpert} :
$\bar{\mathcal{A}}$, $\bar{\mathcal{A}}_\xi$, $\bar{\mathcal{A}}_{\xi\xi}$ et
$\bar{\mathcal{R}}$. Dans la suite nous omettons les barres sur ces quantités
invariantes de jauge exprimées dans la ``jauge longitudinale 5D''.

Mukohyama a montré le premier que, \emph{en l'absence de perturbations de
  matière dans le bulk}, les perturbations scalaires de métrique dans la
``jauge longitudinale 5D'' ($\mathcal{A}$, $\mathcal{A}_\xi$,
$\mathcal{A}_{\xi\xi}$ et $\mathcal{R}$) peuvent toutes être exprimées en
fonction d'une seule et unique variable scalaire maîtresse, $\Omega$, qui se
propage suivant une simple équation d'onde à cinq dimensions
\cite{pert2} (un travail similaire a été effectué dans l'article \cite{jap}).
De fa\c{c}on similaire, les perturbations vectorielles invariantes de jauge
 de la métrique ($v$ et $v_\xi$) peuvent être exprimées en fonction d'une
 seule et unique variable scalaire maîtresse, $\Omega^{(V)}$ , qui se
 propagent suivant une simple équation d'onde dans cinq dimensions. Les
 perturbations purement tensorielles de la métrique sont déjà invariantes de
 jauge et peuvent directement être considérées comme une variable
 maîtresse. Nous ne chercherons pas à refaire les calculs de Mukohyama mais
 seulement à énoncer les équations compactes qu'il a obtenues. Ces équations
 nous seront utiles pour le prochain chapitre \ref{chapter:dissip}. De manière
 générale, les équations de perturbations dans le formalisme de Mukohyama sont
 couramment utilisées dans la littérature concernant l'étude des perturbations
 cosmologiques en cosmologie branaire parce que ce formalisme simplifie
 élégamment l'attirail assez lourd des équations de perturbations exprimées
 dans d'autres formalismes.

\subsection{Perturbations tensorielles}\label{subsec:mukotens}

Les perturbations tensorielles, transverses et sans trace, sont les plus
simples à étudier : elles décrivent des ondes gravitationnelles (au sens
quadridimensionnel) et se propagent librement dans le bulk $AdS^5$,
indépendamment de la présence de matière sur
la brane (au moins lorsqu'on fait l'hypothèse simplificatrice qu'il n'y a pas de
perturbation tensorielle anisotrope de matière sur la brane,
$\delta\pi_i^j \equiv 0$). Les fluctuations tensorielles, transverses et sans trace, $E_{ij}$, de la
métrique
\ba
d\tilde{s}^2 =  \left(g_{\mu\nu}+h_{\mu\nu}\right)dx^\mu dx^\nu & = &
d\xi^2-N^2(\xi,\tau)d\tau^2+A^2(\xi,\tau)\left(\delta_{ij}+E_{ij}\right)dx^idx^j
\ea
sont, de plus, automatiquement invariantes de jauge. On peut ainsi les décrire à
l'aide d'un champ scalaire canonique sans masse minimalement couplé à
la gravité , $\Psi(t,\xi; p)$, et transformé de Fourier dans les trois
dimensions spatiales transverses, selon
\ba
E_{ij}(t,\xi,{\bf x}) & = & \int dp\Psi(t,\xi; p)e^{i{\bf p}\cdot{\bf x}}e_i\otimes e_j,
\ea
qui se propage dans le bulk $AdS^5$ selon l'équation de Klein-Gordon
\ba\label{eqnarray:eomt} -{1\over N^2}\left[\ddot{\Psi}+\left(3{\dot{A}\over A}-{\dot{N}\over
      N}\right)\dot{\Psi}\right]+\left[\Psi''+\left(3{A'\over A}+{N'\over
      N}\right)\Psi'\right]-{p^2\over A^2}\Psi & = & 0.\ea
$\mathbf{p}$ est le $3$-moment (tri-impulsion) dans les trois dimensions
spatiales transverses. Les conditions de jonction d'Israel linéarisées se
réduisent à des conditions de bord de type Neumann sur la brane pour les
perturbation tensorielles :
\ba\label{eqnarray:bct}
\partial_\xi \Psi\bigr\vert_{\xi = 0} & = & 0.
\ea
Ces conditions de bord sont homogènes sous l'hypothèse qu'il n'y a pas de
perturbation anisotrope de matière ($\delta\pi_i^j \equiv 0$).

Le champ scalaire $\Psi$ est le champ scalaire ``maître'' de Mukohyama pour les
perturbations tensorielles quoique sa définition soit triviale dans ce cas.

\subsection{Perturbations vectorielles}\label{subsec:mukovec}

On a vu que les perturbations vectorielles de la métrique
 \ba
d\tilde{s}^2 & = &
d\xi^2-N^2(\xi,\tau)d\tau^2+A^2(\xi,\tau)\left(\delta_{ij}+\partial_{(i} \bar{F}_{j)}\right)dx^idx^j\cr
             &   &-A^2(\xi,\tau)\bar{S}_{\xi i}d\xi dx^i-A^2(\xi,\tau)\bar{S}_i dt dx^i.
\ea
pouvaient être décrites par les deux perturbations indépendantes de jauge $v$, $v_\xi$ définies en (\ref{qqq:invjaugevec}). En l'absence de perturbations de matière
dans le bulk, Mukohyama a montré que les perturbations vectorielles de la
métrique peuvent être générées à partir d'un unique champ scalaire
``maître'' $\Omega^{(V)}(t,\xi)$, selon
\ba
v & = & {N\over A^3}\Omega^{(V)'},\cr
v_\xi & = & {1\over N A^3}\dot{\Omega}^{(V)}.
\ea
La nouvelle variable maîtresse $\Omega^{(V)}$, décrivant les perturbations
vectorielles de métrique, satisfait l'équation d'onde à cinq dimensions
suivante :
\ba\label{qqq:eoms} -{1\over N^2}\left[\ddot{\Omega}^{(V)}-\left(3{\dot{A}\over A}+{\dot{N}\over
      N}\right)\dot{\Omega}^{(V)}\right]+\left[\Omega^{(V)''}-\left(3{A'\over A}-{N'\over
      N}\right)\Omega^{(V)'}\right]-{p^2\over A^2}\Omega^{(V)} & = & 0.\qquad\ea
On constatera que celle-ci diffère de l'équation de Klein-Gordon, satisfaite
par les perturbations tensorielles. Les conditions de jonction d'Israel linéarisées se
réduisent à des conditions de bord de type Dirichlet sur la brane pour les
perturbation vectorielles :
\ba\label{eqnarray:bcv}
\dot{\Omega}^{(V)}\bigr\vert_{\xi = 0} & = & 0.
\ea
Ces conditions de bord sont homogènes sous l'hypothèse qu'il n'y a pas de
perturbation anisotrope de matière.

\subsection{Perturbations scalaires}\label{subsec:mukoscal}

Les quatre perturbations scalaires indépendantes de jauge,
exprimées dans la ``jauge longitudinale 5D'', transforment la métrique selon
 \ba
d\tilde{s}^2 & = &
\left(1+2\mathcal{A}_{\xi\xi}\right)d\xi^2-N^2(\xi,\tau)\left(1+2\mathcal{A}\right)d\tau^2+N(\xi,\tau)\mathcal{A}_\xi
d\xi dt\cr
             &   &+A^2(\xi,\tau)\left(1+2\mathcal{R}\right)\delta_{ij}dx^idx^j.
\ea
Les perturbations scalaires 5D de métrique couplent aux perturbations
scalaires 4D de matière sur la brane, définies par le tenseur énergie-impulsion
perturbé
$$
T_\mu^\nu+\delta T_\mu^\nu  =
\left[
\begin{array}{cc}
-(\rho+\delta\rho) & \delta q^{,j} \\
-A^{-2}\delta q_{,i} & \left(P+\delta P\right)\delta_i^j
\end{array}
\right].
$$
Les indices grecs dénotent les $(3+1)$ coordonnées sur la brane et les indices
latins en minuscules dénotent les trois dimensions spatiales sur la brane.
On a négligé les perturbations anisotropes scalaires du tenseur
énergie-impulsion parce qu'elles n'ont pas lieu d'être présentes lorsqu'on
considère un fluide parfait ou un champ scalaire sur la brane \cite{deff1}. 
Rappelons que les composantes de matière effective sur la brane,
$\rho$ et $P$, sont reliées aux composantes du fluide réel sur la brane,
$\rho_M$ et $P_M = w \rho_M$, selon $\rho = \rho_M+\sigma$, $P = P_M-\sigma$, où $\sigma$
est la tension de la brane dans le modèle de Randall-Sundrum. Nous avons
discuté cela à la section \ref{subsec:bdel}. En l'absence de perturbations de matière
dans le bulk, Mukohyama a trouvé que les perturbations scalaires de la
métrique peuvent toutes être générées à partir d'un unique champ scalaire
``maître'' $\Omega(t,\xi)$, selon
\ba
\mathcal{A} & = & -{1\over 6A}\left\{\left(2\Omega''-{N'\over N}\Omega'\right)+{1\over
    N^2}\left(\ddot{\Omega}-{\dot{N}\over
      N}\dot{\Omega}\right)-{1\over\ell^2}\Omega\right\},\cr
\mathcal{A}_\xi & = & {1\over N A}\left(\dot{\Omega}'-{N'\over N}\dot{\Omega}\right),\cr
\mathcal{A}_{\xi\xi} & = & {1\over 6A}\left\{\left(\Omega''-2{N'\over N}\Omega'\right)+{2\over
    N^2}\left(\ddot{\Omega}-{\dot{N}\over
      N}\dot{\Omega}\right)+{1\over\ell^2}\Omega\right\},\cr
\mathcal{R} & = & {1\over 6A}\left\{\left(\Omega''+{N'\over N}\Omega'\right)-{1\over
    N^2}\left(\ddot{\Omega}-{\dot{N}\over
      N}\dot{\Omega}\right)-{2\over\ell^2}\Omega\right\}.\label{eqnarray:R}
\ea
On remarquera que le champ scalaire ``maître'' $\Omega$ n'est pas sans
dimension mais possède la dimension d'une longueur au carré. La variable
maîtresse satisfait l'équation d'onde à cinq dimensions suivante : 
\ba\label{qqq:eoms} -{1\over N^2}\left[\ddot{\Omega}-\left(3{\dot{A}\over A}+{\dot{N}\over
      N}\right)\dot{\Omega}\right]+\left[\Omega''-\left(3{A'\over A}-{N'\over
      N}\right)\Omega'\right]-\left({p^2\over A^2}-{1\over
    \ell^2}\right)\Omega & = & 0.\ea
On constatera que celle-ci diffère de l'équation de Klein-Gordon, satisfaite
par les perturbations tensorielles. La linéarisation des conditions de
jonction d'Israel se fait plus aisément dans la jauge Gaussienne normale parce
que la brane reste fixée en $\xi = 0$ dans cette jauge. On peut exprimer ces
conditions de jonction liant les perturbations de métrique et de matière
indépendantes de jauge dans la jauge Gaussienne normale, puisqu' on sait
passer de la jauge Gaussienne normale à la jauge longitudinale 5D. On peut dès
lors re-exprimer les conditions de jonction calculées dans la jauge gaussienne
normale en fonction de la variable maîtresse $\Omega$. Les conditions de
jonction d'Israel se réduisent aux conditions de bord
``non-locales''\footnote{Le terme "non-local", utilisé pour les conditions de
  bord des perturbations scalaires, n'a rien à voir avec la non-localité
  étudiée au chapitre \ref{chapter:dissip} qui est due à l'information
  provenant du bulk. Comme Deffayet l'a expliqué dans son article
  \cite{deff2}, la terminologie "non-locale" pour les conditions de bord n'est
  pas très adaptée, bien que fréquemment utilisée dans la littérature, parce
  que les conditions de bord contiennent ici seulement un nombre fini de dérivées.} suivantes sur la brane \cite{scalbc}
\ba\label{qqq:sbc0}
\kappa^2 A\delta\rho & = & -3{\dot{A}\over A}\left(\dot{\Omega}'-{N'\over N}\dot{\Omega}\right)-{p^2\over
A^2}\left(\Omega'-{A'\over A}\Omega\right)\Bigr\vert_{\xi = 0},\cr
\kappa^2 A\delta
q & = & -\left(\dot{\Omega}'-{N'\over N}\dot{\Omega}\right)\Bigr\vert_{\xi = 0},\cr
\kappa^2 A\delta
P & = &
\left(\ddot{\Omega}'-{A'\over A}\ddot{\Omega}\right)+2{\dot{A}\over
  A}\left(\dot{\Omega}'-{N'\over
    N}\dot{\Omega}\right)-\left(\left(2{\dot{A}\over A}\right)'+\left({\dot{N}\over N}\right)'\right)\dot{\Omega}\cr
                 &   & +\left({A'\over A}-{N'\over N}\right)\left({1\over\ell^2}-{2\over 3}{p^2\over A^2}\right)\Omega-\left({A'\over A}-{N'\over N}\right)\left(2{A'\over A}-{N'\over N}\right)\Omega'\Bigr\vert_{\xi = 0}.
\ea

Les perturbations de type vectorielles ne survivent pas dans un modèle $Z_2$-symétrique tel que le modèle de Randall-Sundrum. C'est pourquoi nous nous concentrerons désormais sur les perturbations de type tensorielles (ondes gravitationnelles) et les perturbations de type scalaires (ondes gravitationnelles et matière).

On comprendra que l'équation du mouvement (\ref{qqq:eoms}) et les conditions
de bord (\ref{qqq:sbc0}), pour prendre l'exemple des perturbations scalaires,
ne seront pas triviales à résoudre : l'équation du mouvement n'est plus
séparable dans une géométrie de fond du type (\ref{qqq:apexactmetric}),
décrivant une brane FRW en expansion quelconque dans le bulk $AdS^5$. Dans
l'article \cite{deff2}, Deffayet a montré cependant que le problème de
l'évolution des perturbations scalaires était un problème bien posé au sens
mathématique dans les cas particuliers d'un fluide parfait de perturbations adiabatiques ou d'un
champ scalaire sur la brane : l'équation du mouvement (\ref{qqq:eoms}) combinée
aux conditions de bord (\ref{qqq:sbc0}) forment alors un système clos d'équations.


\chapter{Dissipation et non-localité dans un Univers branaire en expansion
  (article)}\label{chapter:dissip}

Ce chapitre présente nos résultats sur les processus de dissipation et de
non-localité dans un Univers branaire en expansion, qui ont donné lieu à la
publication de l'article \cite{moi}: \emph{Dissipation and
  nonlocality in a general expanding braneworld universe}, Mathieu
Remazeilles, Phys. Rev. D79:043523 (arXiv:0807.4238 [hep-th]).

\section{Introduction}\label{sec:intropaper}

Dans ce chapitre nous étudions l'évolution à la fois des perturbations
tensorielles et des perturbations scalaires  dans un Univers branaire du type
Randall-Sundrum ayant une histoire d'expansion cosmologique arbitraire. Au lieu de
rechercher des solutions exactes aux équations de perturbations présentées à
la section \ref{sec:muko}, nous sommes intéressés en priorité par le rôle des
gravitons du bulk (perturbations de métrique) dans leur interaction avec les
degrés de liberté localisés sur la brane.  Nous faisons le choix d'adopter un
point de vue quadridimensionnel, en considérant les degrés de liberté
localisés sur la brane comme un système quantique \emph{ouvert} couplé à un grand environnement
composé du continuum des gravitons du bulk. Lorsque l'expansion cosmologique
de la brane est non-uniforme, les degrés de liberté sur la brane et les degrés de liberté du
bulk interagissent quantiquement au cours du temps. Les excitations quantiques
des degrés de liberté sur la brane peuvent se désintégrer en émettant des gravitons dans le bulk qui
peuvent s'échapper vers l'infini futur dans la dimension supplémentaire, ce qui conduit à une forme de
\emph{dissipation} du point de vue quadridimensionnel d'un observateur confiné
sur la brane. Les gravitons du bulk peuvent également être réfléchis (ou
diffractés) dans le bulk (dimension supplémentaire) à cause de la
courbure du bulk $AdS$, puis ré-absorbés par la brane, et ainsi transformés de
nouveau en
quanta sur la brane, ce qui conduit à une forme de \emph{non-localité} du
point de vue quadridimensionnel. La dissipation et la non-localité sont
inscrits dans le propagateur retardé du bulk $AdS$, qui peut être systématiquement
inséré dans le propagateur retardé effectif de la brane à quatre dimensions par
resommation des effets de rétroaction du bulk (``backreaction'' en Anglais) à tous les
ordres dans le couplage brane-bulk. Dans ce
travail nous estimons, au moyen du propagateur retardé effectif sur la brane, les taux
de dissipation du mode lié du graviton (tenseur) ainsi que des degrés de
liberté de matière sur la brane (scalaires) dans différents régimes cosmologiques et pour différentes sources de matière sur la brane.

Nous avons vu dans les chapitres précédents comment des processus physiques
peuvent être interprétés différemment entre un point de vue à cinq dimensions
et le point de vue à quatre dimensions d'un observateur confiné sur une brane
: par exemple l'impulsion d'un graviton 5D dans la cinquième dimension
correspond de fa\c{c}on effective à une masse pour le graviton sur la brane
d'un point de vue quadridimensionnel. De même, dans cette représentation de plus
haute dimension, l'expansion de l'Univers quadridimensionnel peut être
interprété comme le mouvement accéléré de la brane 4D dans le bulk 5D décrit
dans des coordonnées statiques (coordonnées de Poincaré). Enfin dans ce
chapitre nous allons voir comment la réduction dimensionnelle donne aux
transitions 5D entre les modes quantiques sur la brane et dans le bulk
l'apparence d'une forme de dissipation du point de vue quadridimensionnel d'un
observateur confiné sur la brane. Nous utiliserons le formalisme de Mukohyama
exposé à la section \ref{sec:muko}. La principale difficulté dans le calcul de
l'évolution des perturbations est que, excepté dans le cas d'une brane ayant
une accélération uniforme (brane plate ou brane $dS$) où des solutions
analytiques ont pu être obtenues (voir chapitre \ref{chapter:rs}), les
équations du mouvement ne sont généralement pas séparables à cause du
mouvement compliqué dans le bulk de la brane FRW en expansion. Une courte
revue sur le problème d'évolution des perturbations cosmologiques dans les
Univers branaires en expansion est donnée dans la référence \cite{der2}. Ce
problème a été examiné numériquement et un certain nombre de résultats
intéressants ont été obtenus 
\cite{num0,num1,scalinf,num2,num3,num4,num5,scal,scalinf2}. Le problème des
perturbations cosmologiques dans les Univers branaires est en principe
entièrement soluble numériquement une fois que le vide initial dans $AdS$ est
connu. Cependant la spécification de conditions initiales dans le bulk pose un
problème de nature plus fondamentale. Il n'y a pas de choix unique,
physiquement motivé, de conditions initiales dans un espace de fond $AdS$
\cite{trodden,bucher} (contrairement à la situation dans un espace $dS$). Par
conséquent les résultats de ces études numériques sont sujets à certaines
suppositions concernant les conditions initiales.
Ici nous réalisons une approche
analytique basée sur le calcul du propagateur retardé effectif sur la brane
par resommation  à partir des deux propagateurs retardés "nus" sur la brane et dans le bulk
$AdS$, l'objectif étant de fournir plus d'intuition concernant l'évolution des
perturbations cosmologiques dans un Univers branaire en expansion. 
Dans cet article nous avons exploré l'évolution
des perturbations cosmologiques dans un scénario du type Randall-Sundrum
avec une brane FRW ayant un mouvement arbitraire dans le bulk $AdS^5$
$Z_2$-symétrique. Le but de ce travail est d'estimer la magnitude des effets
physiques ressentis sur la brane et dûs à la présence d'une dimension supplémentaire infinie mais courbée. Nous nous inspirons des idées développées par Binétruy, Bucher et Carvalho \cite{bbc}.

\begin{figure}
  \begin{center}
\includegraphics[width=8.5cm]{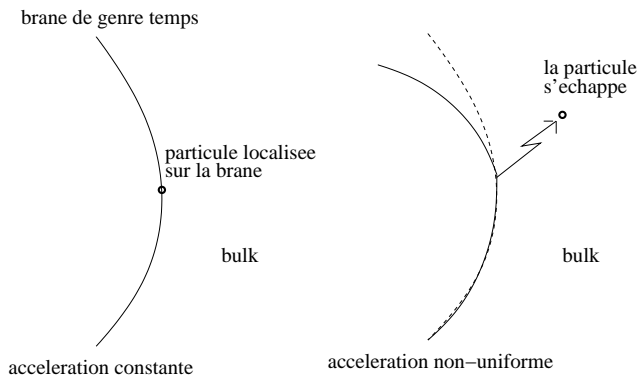}
 \end{center}
\caption{
Représentation classique.}
\label{fig:picturethese}
\end{figure}

Lorsque le paramètre de Hubble sur la brane, $H(\tau)$, varie dans le temps,
l'accélération de la brane dans le bulk varie également et des gravitons sont
émis dans le bulk. On peut illustrer cet effet par le problème
\emph{classique} équivalent suivant : considérons les particules localisées
sur la brane (tel que l'état lié du graviton par exemple) comme des particules
classiques, un changement d'accélération de la brane dans le bulk peut
entrainer que cette particule se délocalise de la brane et s'échappe dans la
dimension supplémentaire (Fig. \ref{fig:picturethese})
On peut se demander quel est le taux de dissipation effectif ou la probabilité
qu'une particule s'échappe de la brane quand l'accélération de la brane change de fa\c{c}on \emph{adiabatique} ?

Les gravitons émis dans le bulk peuvent soit s'échapper vers l'infini futur ou
bien être ré-absorbés par la brane à cause de réflections de ces gravitons
dans le bulk $AdS$ courbe. Du point de vue quadridimensionnel d'un observateur
sur la brane, ces processus donne l'apparence de générer respectivement de la
dissipation et de la non-localité \cite{bbc}. Ces effets sont "codés" dans le
propagateur retardé du graviton du bulk $AdS$ et peuvent être insérés
systématiquement dans le propagateur effectif de la brane, en resommant la
série perturbative prenant en compte les effets de rétroaction à tous les
ordres. L'approche utilisée dans ce travail est une perspective
quadridimensionnelle qui considère les degrés de liberté localisés sur la
brane comme un système quantique ouvert couplé à un grand environnement
composé des gravitons du bulk. Dans le langage la théorie quantique des champs
hors-équilibre, le  propagateur du bulk joue le même rôle qu'une
"self-énergie" dans le sens où elle "habille" les champs "nus" (composés des
degrés de libertés discrets localisés sur la brane et sans interaction
avec le bulk).

Dans cet article nous estimons d'abord le taux de dissipation de l'état lié du
graviton en étudiant l'évolution des perturbations \emph{tensorielles} dans le
cadre d'une brane FRW plongée dans le bulk $AdS$. Gorbunov, Rubakov et
Sibiryakov avaient obtenu en 2001 des résultats analytiques pour l'ordre de
magnitude des modifications au spectre de puissance standard 4D pour les
perturbations tensorielles, dans le cas où le facteur de Hubble sur la brane
change \emph{instantanément} \cite{pert3} (voir aussi \cite{pert33}). Ici nous
considérons la situation plus réaliste où le facteur de Hubble sur la brane
change \emph{continûment} et \emph{adiabatiquement}, dans le sens où $\dot{H}
\ll H^2$. Nous explorons également la dissipation de degrés de liberté
purement localisés sur la brane (c'est-à-dire confinés), tels qu'un fluide
parfait (de perturbation adiabatique) ou un champ scalaire en roulement lent
(inflaton) sur la brane, en étudiant l'évolution des perturbations
\emph{scalaires} et la nature du couplage entre les perturbations de métrique
et les perturbations de matière. Nous utilisons les coordonnées Gaussiennes
normales (GN) pour couvrir l'espace-temps $AdS$ et plonger la brane FRW dans
une tranche du bulk $AdS$ ; ainsi la coordonnée de la dimension supplémentaire
mesure la distance propre à la brane, et la position de la brane dans le bulk
reste fixe par définition. Les coordonnées GN ont aussi l'avantage de
simplifier la forme des conditions de jonction linéarisées sur la brane. Le
principal désavantage des coordonnées GN est la présence de singularités de
coordonnées dans le bulk à une distance finie de la brane bien que le bulk
$AdS$ soit en fait régulier et extensible au-delà de la singularité en
choisissant un autre système de coordonnées. À cause du mouvement arbitraire
de la brane FRW dans le l'espace $AdS$, les composantes de la métrique ont une
forme compliquée, ce qui rend non-séparables les équations du mouvement pour
les perturbations. C'est pourquoi, \emph{hormis} dans le cas de l'inflaton sur la
brane (section \ref{sec:inflaton}), nous utilisons une géométrie approchée
comme étant la limite de la métrique exacte près de la brane afin de réaliser
une séparation des équations. Cette limite peut être légitimée dans le modèle
de Randall-Sundrum par le fait que le support de la fonction d'onde décrivant
l'état lié est localisée près de la brane. Nous pensons que la physique proche
de la brane est pertinente pour décrire des effets de dissipation dans les
Univers branaires. Afin de se concentrer sur les effets dissipatifs, on peut
approximer à haute énergie ($H\ell\gg 1$) l'inhomogénéité du bulk courbe
$AdS$, responsable des réflections des gravitons dans le bulk et par conséquent d'effets non-locaux sur la brane. Cette approximation est discutée ultérieurement dans ce chapitre.

\section{Processus de dissipation et de non-localité dans un Univers branaire en expansion}\label{sec:secone}

\begin{figure}[htbp]
\begin{center}
\includegraphics[width=10cm]{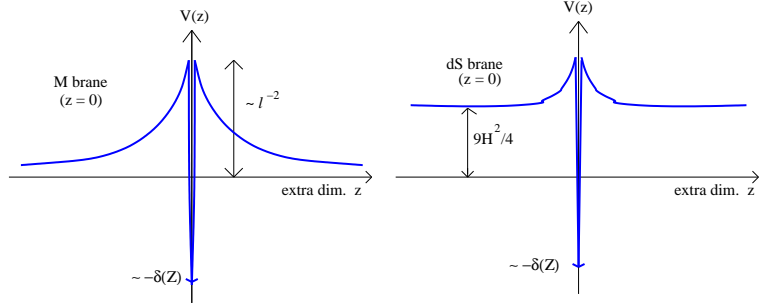}
\end{center}
\caption{\small{{\bf Gauche: potentiel-volcan pour une brane Minkowski.} La
  hauteur du potentiel est $\mathcal{O}(\ell^{-2})$ et le
  potentiel décroît comme $1/z^2$ si $z$ est la dimension supplémentaire. Le
  mode zéro du graviton est lié à la brane et entouré d'un continuum de
  gravitons massifs libres $m > 0$. {\bf Droite: potentiel-volcan pour une
  brane de Sitter.} La hauteur du potentiel est $\mathcal{O}(H^2+\ell^{-2})$ et le
  potentiel décroît comme $1/\sinh^2(z)$. Le mode zéro du graviton est lié à
  la brane et entouré d'un continuum de
  gravitons massifs libres $m > 3H/2$, caractérisé par un gap
  $\mathcal{O}(H^2)$ entre le mode zéro et le continuum.}} \label{fig:volcano}
\end{figure}

Nous avons vu au chapitre \ref{chapter:rs} que l'évolution des perturbations
cosmologiques dans le modèle de Randall-Sundrum a été résolu analytiquement
pour des Univers branaires hautement symétriques, où le facteur de Hubble sur la
brane $H$ était constant. C'était le cas d'une brane Minkowski ($H = 0$),
statique, ou d'une brane de Sitter pur ($H = const.>0$), suivant une
expansion uniforme, ou une trajectoire uniformément accélérée dans le bulk
$AdS$. Les équations linéarisées pour les perturbations étaient séparables par
symétrie et
on pouvait réduire le problème à la résolution d'une équation de type
Schrödinger, où le potentiel obtenu avait l'aspect d'un potentiel-volcan : la
présence de la brane créant un puit de potentiel du type fonction $\delta$ et
la courbure du bulk $AdS$ créant une barrière de potentiel décroissante
(Fig. \ref{fig:volcano}).

La forme du potentiel entraîne l'existence d'un unique état lié à la brane, le
mode de masse zéro $m = 0$, dans le spectre de gravitons des équations
d'Einstein linéarisées, ainsi qu'un continuum d'états de diffusion,
c'est-à-dire de gravitons libres et massifs de Kaluza-Klein de masse $m \geq 3H/2$ dans le cas
de la brane $dS$ \cite{lands}. Le mode zéro s'identifie naturellement au graviton standard
sans masse et reproduit la gravitation quadridimensionnelle sur la brane à
basse énergie. Il représente un degré de liberté branaire dans le sens où il
reste localisé près de la brane. L'amplitude d'interaction des gravitons
massifs avec la brane est supprimée près de la
brane à basse énergie à cause de la barrière de potentiel. Dans le cas d'un
contenu de matière dépendant du temps dans l'Univers, l'expansion cosmologique n'est plus
uniforme : le facteur de Hubble change avec le temps, ce qui entraine un
changement dans la forme des potentiels-volcans. Par conséquent, les degrés de
liberté sur la brane et les degrés de liberté du bulk interagissent et donc
génère des transitions entre les modes.

Le système brane-bulk est un système quantique Hamiltonien, nécessairement
conservatif puisque la densité de l'espace des phases doit être préservée par
translation dans le temps. En ce sens, le système brane-bulk est
intrinsèquement non-dissipatif, et l'apparence de dissipation ne peut se
produire que par le résultat d'un ``coarse graining'' (\emph{i.e} une moyenne
sur l'information manquante). L'interaction entre les quanta, dûe à l'expansion
de la brane, peut être caractérisisée par une transformation de Bogoliubov
(matrice $S$ pour un système linéaire) reliant les modes de fréquence
\emph{positive} et de fréquence \emph{négative} entre le vide ``in'' initial
et le vide ``out'' résultant. Cependant, du point de vue d'un observateur
confiné sur la brane, l'Univers quadridimensionnel, en tant que sous-variété,
est un système quantique ouvert. Par conséquent le mélange entre les modes
localisés sur la brane et les modes délocalisés dans le bulk créé l'apparence
d'une dissipation pour l'observateur quadridimensionnel incapable d'accéder aux
modes du bulk. L'observateur sur la brane n'est sensible qu'à une partie du vide
complet de l'espace $AdS$, partie composée seulement des modes discrets localisés sur
la brane. L'espace de Hilbert du sytème branaire est ``tronqué'' à cause de la réduction
dimensionnelle, et le vide du bulk, composé du continuum des modes du bulk,
contient l'information manquante. Lorsqu'on observe les perturbations
cosmologiques aujourd'hui, on mesure les valeurs moyennes dans le vide
d'observables quadratiques en les opérateurs de création et d'annihilation
localisés sur la brane aujourd'hui, à savoir $a_{brane,out}$ et
$a^\dagger_{brane,out}$. La matrice $S$ exprime les opérateurs  ``out''
comme des combinaisons linéaires de $a_{brane,in}$ et $a^\dagger_{brane,in}$
d'un côté, et de $a_{bulk,in}(k)$ et $a^\dagger_{bulk,in}(k)$ de l'autre. Une
paramétrisation utile de cette transformation a été proposée dans
\cite{bbc}. Les auteurs définissent $A_{brane,in}$ et
$A_{bulk,in}$ comme étant entièrement sur la brane et dans le bulk
respectivement, et normalisés tel que $\left[A_{brane,in},A^\dagger_{brane,in}\right] =
\left[A_{bulk,in},A^\dagger_{bulk,in}\right] = 1$. Dès lors $a_{brane,out}$
peut être exprimé en fonction de ceux-ci selon une des trois possibilités
suivantes : soit
\ba
a_{brane,out} & = & \cos\theta A_{brane,in}+\sin \theta A_{bulk,in}
\ea
où $0\leq\theta\leq\pi/2$; ou bien
\ba
a_{brane,out} & = & \cosh u A_{brane,in}+\sinh u A^\dagger_{bulk,in}
\ea
où $0\leq u\leq +\infty$; ou bien
\ba\label{qqq:parametrization}
a_{brane,out} & = & \sinh u A^\dagger_{brane,in}+\cosh u A_{bulk,in}
\ea
où $0\leq u\leq +\infty$. $A_{brane,in}$ peut être construit
entièrement comme une combinaison linéaire de $a_{brane,in}$ et
$a^\dagger_{brane,in}$, et de même $A_{bulk,in}$  peut être construit
entièrement comme une combinaison linéaire de $a_{bulk,in}(k)$ and the
$a^\dagger_{bulk,in}(k)$, $k$ indexant les modes du continuum. Nous remarquons 
en (\ref{qqq:parametrization}) que l'état initial dans le bulk peut
avoir un rôle considérable, voire dominant, dans la détermination de ce qui
est observé sur la brane aujourd'hui. 

\begin{figure}[htbp]
\begin{center}
\includegraphics[width=9cm]{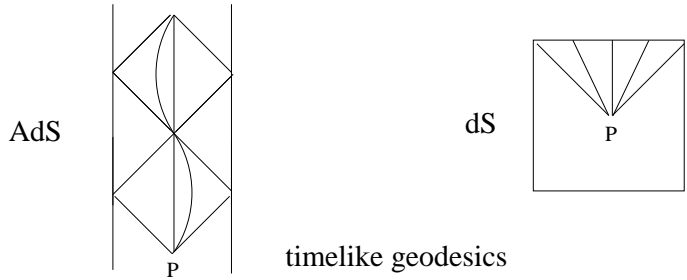}
\end{center}
\caption{\small{Géodésiques de genre temps dans $AdS$ (gauche) et dans $dS$ (droite)}}\label{fig:hair}
\end{figure}

En cosmologie standard la théorie inflationnaire permet de résoudre le
problème des conditions initiales et ce succès repose sur la structure causale
de la géométrie de fond de Sitter ($dS$) : l'espace $dS$ "n'a pas de cheveux"
("no hair" en Anglais) dans le sens où les irrégularités initiales sont
effacées au cours du temps puisque les géodésiques de genre temps divergent
dans $dS$, perdant toute causalité (Fig. \ref{fig:hair}). Cela entraine
l'homogénéité et l'isotropie observée dans l'Univers aujourd'hui et justifie
l'existence de conditions initiales naturelles dans $dS$. En revanche dans les
scenarios branaires à la Randall-Sundrum, la géométrie de fond est Anti-de
Sitter ($AdS$), dont la structure causale entraine que les amplitudes des
perturbations initiales sont conservées au cours du temps parce que les
géodésiques de genre temps d'abord divergent puis re-convergent et cela
pertpétuellement (Fig. \ref{fig:hair}) : il n'y a donc pas de conditions
initiales naturelles dans $AdS$ \cite{trodden,bucher}. Les méthodes numériques employées
dans la littérature \cite{num0,num1,scalinf,num2,num3,num4,num5,scal,scalinf2} nécessitent un choix particulier de conditions
initiales dans le bulk $AdS$ pour faire évoluer les perturbations et rendre
compte de ce qui observé aujourd'hui sur la brane. Le vide initial souvent
choisi dans les méthodes numériques est le vide invariant sous de Sitter (voir
par exemple \cite{num4}), c'est-à-dire que les conditions initiales sont
spécifiées à partir d'une combinaison des modes définis dans un découpage
("slicing" en Anglais) $dS^4$ de $AdS^5$. Les conditions initiales ne sont donc
définies que sur l'horizon de Cauchy du bulk créé par le "slicing" $dS$. Ce
choix n'est pas forcément légitime pour $AdS$ d'autant plus que, comme on le
décrit à la section \ref{sec:sectwo}, des gravitons du bulk initiaux provenant
de l'horizon de Poincaré de $AdS$ mais situés en dehors de l'Horizon de Cauchy
du slicing $dS$ peuvent affecter causalement les modes de la brane dans le
futur, lorsque l'expansion de la brane décélère. Ajoutons qu'il a été observé
numériquement \cite{num4,scal} que la
contribution initiale dans le bulk était sous-dominante voire négligeable
(quelques pourcent) par rapport à la contribution initiale sur la brane dans
la détermination des spectres de puissance observés aujourd'hui sur la brane. 
Cependant ces conclusions dépendent très certainement du choix
particulier du vide initial spécifié dans ces schémas numériques. Comme on
l'a vu au paragraphe précédent, avec la paramétrisation explicite (\ref{qqq:parametrization}), les
contributions relatives des conditions initiales dans le bulk et sur la brane
peuvent varier considérablement suivant sur quelle base est défini le vide
initial. Nous adoptons ici une
approche effective à quatre dimensions analytique et basée sur les
propagateurs retardés de la brane $FRW$ et du bulk $AdS$, ce qui évite la
spécification d'un vide initial dans le bulk. À partir des fonctions de Green
retardées du bulk et de la brane "nue", nous construisons le propagateur
retardé effectif sur la brane tenant compte de l'interaction avec le bulk.

\begin{figure}[htbp]
\begin{center}
\includegraphics[width=10cm]{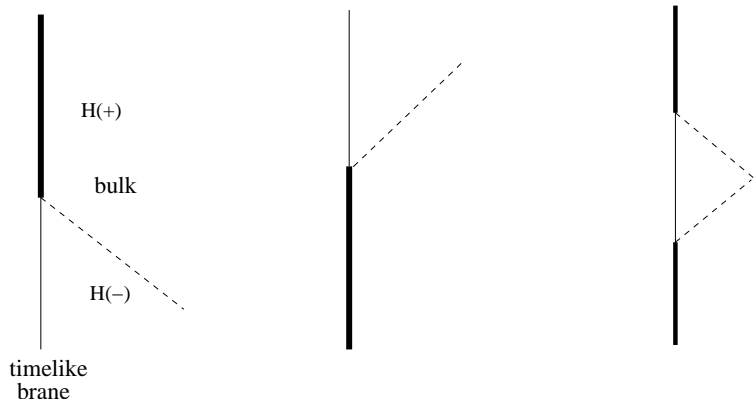}
\end{center}
\caption{\small{{\bf Processus fondamentaux de l'interaction brane-bulk dans 
  une perspective quadridimensionnelle.}  De la gauche vers la droite: absorption,
  dissipation, non-localité. La non-localité se manifeste à cause de la
  diffraction des gravitons dans le bulk $AdS$ courbe.}}\label{fig:interaction}
\end{figure}

L'interaction brane-bulk peut être résumée par les processus suivants, fondamentaux
du point de vue quadridimensionnel, et  illustrés sur la figure
Fig. \ref{fig:interaction}. Un état du vide initial peut être complètement
caractérisé en spécifiant l'état quantique des gravitons incidents sur
l'horizon de Cauchy du passé $H_{(-)}$ dans le bulk et l'état quantique des degrés de liberté
sur la brane à l'intersection de la brane avec $H_{(-)}$. Ultérieurement les
degrés de liberté de la brane et ceux du bulk interagissent au cours du
temps. Les gravitons du bulk peuvent être absorbés et transformés en quanta
sur la brane. De fa\c{c}on similaire, les excitations quantiques sur la brane
peuvent se désintégrer en éméttant des gravitons dans le bulk. Ceux-ci peuvent
soit s'échapper vers l'infini futur, conduisant à de la \emph{dissipation} du
point de vue quadridimensionnel, ou bien peuvent être ré-absorbés par la brane
à cause de la courbure du bulk, conduisant à de la \emph{non-localité} du
point de vue quadridimensionnel.

On peut considérer l'interaction brane-bulk dans une perspective
quadridimensionnelle en regardant les degrés de liberté de la brane comme un
système quantique ouvert couplé à un grand environnement, le bulk, ayant un grand nombre
de degrés de liberté. Schématiquement, on peut diagonaliser par bloc en
faisant une transformation de Fourier dans les trois dimensions spatiales
transverses. Pour un univers branaire en expansion, avec un fluide parfait ou
un champ scalaire sur la brane, l'interaction entre un degré de liberté
scalaire de la brane $q(t)$ et un degré de liberté scalaire du bulk $u(t,x)$,
où $x$ représente la dimension supplémentaire, est régie par un système
d'équations couplées qui, sous des approximations justifiées plus loin dans ce
chapitre, peuvent être écrites dans la forme générale suivante
\ba
\left[-\partial_t a(t,x) \partial_t + \partial_x b(t,x) \partial_x + c(t,x)\right] u(t,x) & = & 0,\cr
\left[\partial_x +\lambda(t)\right] u(t,x)\vert_{x = 0} & = & P_1(\partial_t)q(t),\cr
\left[\partial_t^2+\omega^2_0(t)\right]q(t) & = & P_2(\partial_t)u(t,x)\vert_{x = 0},
\ea
où $P_1$ et $P_2$ sont des polynômes en $\partial_t$ de degré un et de degré
deux respectivement, avec des coefficients dépendant du temps. La première
équation est l'équation du mouvement dans le bulk, la seconde est la condition
de bord sur la brane, de type Neumann, et la troisième l'équation du mouvement pour le degré de
liberté sur la brane. Les degrés de liberté de la brane et du bulk sont reliés
selon
\ba
q(t) & = & \int dt' G_{brane}(t,t') P_2(\partial_{t'}) u(t',0),\cr
u (t,x) & = & \int dt' G_{bulk}(t,t',x,0) P_1(\partial_{t'}) q(t'),
\ea
où $G_{brane} = \left[\partial_t^2+\omega^2_0(t)\right]^{-1}$ est le
propagateur retardé nu\footnote{Par ``nu'' on entend le propagateur des champs
  libres de la brane, c'est-à-dire excluant l'interaction avec le bulk.} de la
brane et $G_{bulk}$ le propagateur retardé du bulk dans sa forme
Neumann. Après interaction, le propagateur retardé effectif de la brane peut
être obtenu en resommant la série géométrique à tous les ordres de la
perturbation (ou des effets de rétroaction) :
\ba
\hat{G}_{brane} & = & G_{brane}+G_{brane}P_2(D_t)G_{bulk}P_1(D_t)G_{brane}\cr
                &  & +G_{brane}P_2(D_t)G_{bulk}P_1(D_t)G_{brane}P_2(D_t)G_{bulk}P_1(D_t)G_{brane}+ ...\cr
                & = & {1\over G_{brane}^{-1}-P_2(D_t)G_{bulk}P_1(D_t)},
\ea
où les deux points du propagateur du bulk résident sur la brane. Ainsi le
degré de liberté sur la brane se propage selon l'équation
intégro-différentielle
\ba
\hat{G}^{-1}_{brane}\circ q & = & \partial^2_t q+\omega_0^2(t)q+\int dt'
K_{bulk}(t-t')q(t') = 0.
\ea
Ici le noyau du bulk est donné par
\ba
K_{bulk}(t,t') = -P_2(\partial_t)G_{bulk}(t,t')P_1(\partial_{t'})
\ea
et joue le rôle d'une ``self-énergie'' qui habille le champ nu de la
brane. Les effets de dissipation locale et les effets non-locaux sont tous
contenus dans ce noyau d'interaction. Le propagateur retardé nu de la brane
$G_{brane} = \left[\partial_t^2+\omega^2_0(t)\right]^{-1}$ est le propagateur
d'un oscillateur localisé sur la brane et contribue uniquement aux transitions
entre modes 4D dues à l'expansion de l'Univers.

On peut considérer pour illustration les couplages suivants
\ba\label{qqq:ex}
P_1(\partial_t) & = & -\gamma_{1}(t) \partial_t, \qquad P_2(\partial_t)  =  \gamma_{2}(t) \partial_t.
\ea
Puisque la forme Neumann du propagateur retardé du bulk projeté sur la brane s'écrit
de fa\c{c}on générale
\ba
G_{bulk}(t,t',x = 0, x' = 0) = \theta(t-t')G(t,t',0,0),
\ea
le noyau d'interaction avec le bulk peut se décomposer en parties singulières
et régulières
\ba
K_{bulk}(t,t') & = & K^{sing}_{bulk}(t,t')+K^{reg}_{bulk}(t,t'),
\ea
où
\ba
K^{sing}_{bulk}(t,t') & = & \delta(t-t')\gamma_1(t)\gamma_2(t)G(t,t,0,0)\partial_t\cr
K^{reg}_{bulk}(t,t') & = & \theta(t-t')\gamma_2(t)\partial_t G(t,t',0,0)\gamma_1(t')\partial_{t'}.
\ea
La partie singulière est responsable de processus de dissipation locaux et la
partie régulière décrit des processus non-locaux, tels que ceux résultant de
réflections des gravitons dans le bulk $AdS$ courbe. La somme infinie sur la
partie singulière 
\ba
\tilde{G}_{brane} & = &
G_{brane}+G_{brane}\left(-K^{sing}_{bulk}\right)G_{brane}+G_{brane}\left(-K^{sing}_{bulk}\right)G_{brane}\left(-K^{sing}_{bulk}\right)G_{brane}+ ...,\cr
                & = & {1\over \partial_t^2+\Gamma(t)\partial_t+\omega^2_0(t)}
\ea
équivaut en effet à introduire un terme de dissipation local $\Gamma(t) =
\gamma_1(t)\gamma_2(t)G(t,t,0,0)$ dans l'équation effective du mouvement pour
le degré de liberté sur la brane. En resommant ensuite sur la partie
régulière 
\ba
\hat{G}_{brane} & = &
\tilde{G}_{brane}+\tilde{G}_{brane}\left(-K^{reg}_{bulk}\right)\tilde{G}_{brane}\cr
                 & & +\tilde{G}_{brane}\left(-K^{reg}_{bulk}\right)\tilde{G}_{brane}\left(-K^{reg}_{bulk}\right)\tilde{G}_{brane}+ ...,
\ea
on rajoute la non-localité dans l'équation effective du mouvement de telle sorte
que le degré de liberté sur la brane se propage selon
\ba
\hat{G}^{-1}_{brane}\circ q & = & \ddot{q}(t)+\Gamma(t)\dot{q}(t)+\omega^2_0(t)q(t)+\gamma_2(t)\int_{-\infty}^t dt' G(t,t',0,0)\gamma_1(t')\dot{q}(t')  =  0.\qquad
\ea
Puisqu'une partie singulière locale ne dépend pas de la courbure du fond, il s'ensuit
que le terme $G(t,t,0,0)$, qui apparaît dans le taux de dissipation locale,
peut être remplacé par le propagateur de \emph{Minkowski} dans sa forme
Neumann. Nous rappelons que le propagateur retardé de Minkowski est
\ba
G^{ret}_{Mink}(t,t',x,x') & = &
\theta(t-t')\theta\left((t-t')^2-(x-x')^2\right)J_0\left(p\sqrt{(t-t')^2-(x-x')^2}\right),
\ea
où $p$ est le moment dans les trois dimensions spatiales transverses. Donc
$G(t,t^+,0,0) = G_{Mink}(t,t^+,0,0) = 1$ et le taux de dissipation locale dépend
seulement des couplages
\ba
\Gamma(t) & = &\gamma_1(t)\gamma_2(t).
\ea

Notons que nous n'avons fait aucune approximation pour calculer ce taux de
dissipation locale. Pour des couplages autres que (\ref{qqq:ex}), des termes locaux
supplémentaires peuvent apparaître contribuant dans l'équation effective du
mouvement par example à un décalage de
fréquence. Il se peut que la dissipation soit non-locale (c'est-à-dire qu'elle
se manisfeste avec une sorte de mémoire dans le temps) lorsqu'il n'y a pas de
dérivées dans les couplages. Ce formalisme à partir des fonctions de Green est pratique
pour discriminer les transitions entre les modes 4D sur la brane dues à
l'expansion de l'Univers des transitions entre les modes de la brane et les
modes du bulk dûes à la présence d'une dimension supplémentaire. De plus ce formalisme
permet de distinguer les processus de dissipation locaux des processus
non-locaux. Le but de notre travail est d'estimer par ces méthodes les taux de
dissipation de certains degrés de liberté confinés sur la brane (ou bien
localisés près de la brane) dans le cas d'une expansion arbitraire de la brane
et sans symétrie particulière.

\section{Perturbations scalaires : Inflaton en roulement lent sur la brane}\label{sec:inflaton}

Les perturbations scalaires sont d'un intérêt particulier parce qu'elles couplent aux
perturbations de matière sur la brane et donc peuvent affecter le contenu du
tenseur énergie-impulsion de l'Univers. Elles sont toutes décrite par l'unique
champ "maître" $\Omega$ tel qu'on l'a déjà discuté à la section
\ref{sec:muko}. Nous prenons la transformée de Fourier dans les trois
directions spatiales transverses à cause de l'homogénéité et l'isotropie, et
évoluons séparément chaque mode de Fourier. Dans cette section nous étudions le cas des perturbations scalaires, où le
degré de liberté sur la brane est un champ scalaire avec un potentiel en
roulement lent ("slow-roll" en anglais), $V(\phi)$, de telle sorte que la
géométrie induite sur la brane est quasi-de Sitter. 
Durant l'Inflation en roulement lent, l'expansion de l'Univers est donc adiabatique dans le sens où $\dot{H} \ll
H^2$. Le champ scalaire de l'inflaton, $\phi(\tau)$, peut être caractérisé par une densité d'énergie $\rho_M$ et une pression $P_M$
\ba
\rho_M  & = & {1\over 2}\dot{\phi}^2+V(\phi),\cr
P_M & = & {1\over 2}\dot{\phi}^2-V(\phi).
\ea
On peut combiner les équations de Mukohyama, exposées à la section
\ref{subsec:mukoscal} pour les perturbations scalaires,  pour obtenir des équations couplées simples pour le système brane-bulk. Suivant les calculs effectués dans les références \cite{scalinf,scalinf2}, on introduit la variable de
Mukhanov-Sasaki indépendante de jauge, définie dans \cite{mukhanov}, pour
décrire le degré de liberté scalaire sur la brane : 
\ba
Q  & = & \delta\phi-{\dot{\phi}\over H}\mathcal{R}_b,
\ea
où $\delta\phi$ est la perturbation de l'inflaton et $\mathcal{R}_b$ est une des perturbations scalaires de métrique, à savoir la perturbation de courbure (\ref{eqnarray:R}),
projetée sur la brane. Les équations du système brane-bulk dans le système de
coordonnées Gaussiennes normales général (\ref{qqq:typicmetric}) est alors
donné par \cite{scalinf,scalinf2} : 
\ba\label{qqq:braneinf}
 -{1\over N^2}\left[\ddot{\Omega}-\left(3{\dot{A}\over A}+{\dot{N}\over
      N}\right)\dot{\Omega}\right]+\left[\Omega''-\left(3{A'\over A}-{N'\over
      N}\right)\Omega'\right]-\left({p^2\over A^2}-{1\over \ell^2}\right)\Omega & = & 0,\cr
-{p^2\over
  a^2}\left[H\left(\Omega'-{A'\over A}\Omega\right)+{\kappa^2\dot{\phi}^2\over 6}\left(\dot{\Omega}-H\Omega\right)\right]\Biggr\vert_{\xi = 0} & = &
\kappa^2 a \dot{\phi}^2\left({H\over\dot{\phi}}Q\right)^{.},\cr
\ddot{Q}+3H\dot{Q}+ \left({p^2\over
a^2}+V''(\phi)\right)Q +\left\{{\ddot{H}\over H}-2{\dot{H}\over H}{V'(\phi)\over\dot{\phi}}-2\left({\dot{H}\over H}\right)^2\right\}Q& = & J(\Omega)\vert_{\xi = 0},
\ea
où $p$ est le moment transverse et la source $J(\Omega)$ est donnée par
\ba\label{qqq:braneinf2}
J(\Omega) & = & -{\dot{\phi}\over H}\Biggl[\left({-\dot{H}\over H}+{\ddot{H}\over 2\dot{H}}\right){p^2\over 3 a^3}\left(\dot{\Omega}-H\Omega\right)+\left(1-{\dot{H}\over 2\mathcal{H}^2}\right){p^4\Omega\over 9a^5}\cr
          &   & +{p^2\over 6a^3}\left(\ddot{\Omega}-H\dot{\Omega}+{p^2\over 3a^2}\Omega-\left({N'\over N}-{A'\over A}\right)\Omega'\right)\cr
          &   & +{\dot{H}\over \mathcal{H}^2}{p^2\over 6a^3}\left(\mathcal{H}\Omega'-H\dot{\Omega}-{1\over\ell^2}\Omega+{p^2\over 3a^2}\Omega\right)\Biggr].
\ea
Ici les notations sont $a(\tau) = A(\tau,\xi = 0)$ et $\mathcal{H} =
(A'/A)\vert_{\xi = 0}$. De l'équation de Friedmann (\ref{qqq:bwdyn}), on a de
plus que
\ba
\kappa^2\dot{\phi}^2/2 & = & -\ell\dot{H}/\sqrt{1+\ell^2H^2}.
\ea

Les équations (\ref{qqq:braneinf}),
(\ref{qqq:braneinf2}) peuvent être simplifiées dans le cas de l'Inflation à roulement lent, comme suit : on néglige toutes les corrections adiabatiques de l'expansion, comme des termes du type $\dot{H}/H^2$, excepté
pour les termes impliqués dans les couplages entre $Q$ et $\Omega$. D'après
ces approximations et après changement d'échelle du champ "maître" selon
$A\Omega \rightarrow \Omega$, nous soutenons que le système des équations couplées se simplifie comme
\ba\label{eqnarray:lastsyst}
G_{bulk}^{-1}(\tau,\xi,\tau',\xi')\circ\Omega(\tau',\xi')  & = & 0,\cr
\Omega'\vert_{\xi = 0} & = &P_1(\partial_\tau)Q(\tau),\cr
G_{brane}^{-1}(\tau,\tau')\circ Q(\tau') & = & P_2(\partial_\tau)\Omega(\tau,\xi)\vert_{\xi = 0},
\ea
où $G_{bulk}$ est la fonction de Green de l'équation du mouvement exacte dans
le bulk,\\
\noindent
 $G_{brane}~=~\left[\partial_\tau^2+3H(\tau)\partial_\tau+p^2/a^2\right]^{-1}$ est la
fonction de Green ``nue'' de l'Inflaton sur la brane, $P_1(\partial_\tau)~=~-(\kappa^2 a^2\dot{\phi}/p^2)\partial_\tau$ et $P_2(\partial_\tau) =
-(p^2\dot{\phi}/(6a^2 H))\left[\partial_\tau^2+H(\tau)\partial_\tau+p^2/a^2\right]$.
La première équation dans (\ref{eqnarray:lastsyst}) est l'équation d'onde du bulk.
La présence de dérivées en temps des champs dans les couplages brane-bulk induit de la dissipation \emph{locale} pour le champ scalaire sur la brane. Le propagateur effectif sur la brane, prenant en compte l'interaction
avec le bulk, est obtenu, comme expliqué à la section \ref{sec:secone}, en
resommant la série perturbative géométrique à tous les ordres de rétroaction
\ba\label{qqq:effpropinf}
& & \hat{G}_{brane}\cr
& = & G_{brane} + G_{brane}\left[{p^2\dot{\phi}\over
    6a^2H}\left( \ddot{G}^N_{bulk}+H\dot{G}^N_{bulk}+{p^2\over
        a^2}G^N_{bulk}\right){\kappa^2 a^2 \dot{\phi}\over p^2}D_\tau\right] G_{brane} + ...\cr
                & = & {1\over G^{-1}_{brane}-{p^2\dot{\phi}\over
    6a^2H}\left(\ddot{G}^N_{bulk}+H\dot{G}^N_{bulk}+{p^2\over
        a^2}G^N_{bulk}\right){\kappa^2 a^2 \dot{\phi}\over p^2}D_\tau}
\ea
où $G^{-1}_{brane} = \left[D_\tau^2+3H D_\tau+p^2/a^2\right]$ décrit la propagation nue sur la brane. $G^N_{bulk}$ est le propagateur retardé du bulk dans sa forme Neumann et projeté sur la brane, et a la forme générale
\ba\label{qqq:gbulkform}
G^N_{bulk}(\tau,\tau,\xi = 0,\xi' = 0) & = & \theta(\tau-\tau')G^N(\tau,\tau',0,0).
\ea
Les dérivées par rapport au temps du propagateur du bulk (\ref{qqq:gbulkform}) apparaissant dans le propagateur effectif de la brane (\ref{qqq:effpropinf}) produisent des termes singuliers et réguliers.
Cela suggère que la propagation de l'inflaton ``habillé'' contient des
corrections adiabatiques locales et non-locales :
\ba\label{qqq:geqscal}
 &  & \left[1-{\kappa^2\dot{\phi}^2\over 6H}G^N(\tau,\tau,0,0)\right]\ddot{Q}+\left[3H-{\kappa^2\dot{\phi}^2\over 6H}\left(2HG^N(\tau,\tau,0,0)+\dot{G}^N(\tau,\tau,0,0)\right)\right]\dot{Q}\cr
& & +{p^2\over a^2}Q+\mbox{ (terme non-local)} = 0.
\ea
Ici le terme non-local dépend de la courbure $\ell$ de l'espace $AdS$ et de la courbure intrinsèque $H$ de la brane, et est donné par les dérivées par rapport au temps de la partie régulière $G^N$ de la fonction de Green $G^N_{bulk}$ de l'équation d'onde dans le bulk $AdS$ (\ref{eqnarray:lastsyst}):
\ba
\mbox{(terme non-local)} & = & -{\kappa^2\dot{\phi}^2\over 6H}\int_0^{+\infty}
ds \left[\ddot{G}^N(s)+2H\dot{G}^N(s)+{p^2\over
        a^2}G^N(s)\right]\dot{Q}(\tau-s).\quad
\ea
Les termes locaux ne dépendent pas de la courbure, ce qui signifie que la partie régulière $G^N(\tau,\tau,0,0)$ de la fonction de Green du bulk est égale au propagateur de Minkowski à l'origine dans
(\ref{qqq:geqscal}). Puisque la forme Neumann du propagateur de Minkowski $G^N_{Mink}(\tau,\tau',\xi,\xi')$ est donné par
\ba
G^N_{Mink}(\tau,\tau',\xi,\xi') &  = & G_M(\tau,\tau',\xi,\xi')+G_M(\tau,\tau',\xi,-\xi')
\ea
où
\ba
G_{Mink}(\tau,\tau',\xi,\xi') & = &
\theta\left((\tau-\tau')^2-(\xi-\xi')^2\right)J_0\left(p\sqrt{(\tau-\tau')^2-(\xi-\xi')^2}\right),
\ea
on a $G^N(\tau,\tau^+,0,0) = G^N_{Mink}(\tau,\tau^+,0,0) = 1$ et $\dot{G}^N(\tau,\tau^+,0,0) =
\dot{G}^N_{Mink}(\tau,\tau^+,0,0) = 0$, de telle sorte que l'équation du mouvement effective pour l'inflaton se réduit à
\ba
\left[1-{\kappa^2\dot{\phi}^2\over
    6H}\right]\ddot{Q}+\left[3H-{\kappa^2\dot{\phi}^2\over
    6H}H\right]\dot{Q}+{p^2\over a^2}Q+\mbox{(terme non-local)} & = & 0.
\ea
On peut renormaliser le terme cinétique à un en divisant cette équation par le coefficient du terme cinétique et obtenir
\ba\label{qqq:effeominf}
\ddot{Q}+\left[3H+\Gamma(\tau)\right]\dot{Q}+\left({p^2\over
    a^2}+\Lambda(\tau)\right)Q+\mbox{ (terme non-local)}+\mathcal{O}(\kappa^4\dot{\phi}^4) & = & 0.\quad
\ea
De l'équation (\ref{qqq:effeominf}) nous observons que l'interaction de
l'inflaton avec les gravitons du bulk conduit à de la dissipation
\emph{locale} dans la dimension supplémentaire à travers le terme de friction
locale apparaissant dans l'équation effective sous la forme d'une dérivée
première par rapport au temps $\Gamma(\tau)\dot{Q}$. Il y a de plus un
décalage de phase donné par $\Lambda(\tau) =
(p^2/a^2)\kappa^2\dot{\phi}^2/(6H)$. Nous trouvons que le décalage de phase
local de la fréquence nue de l'Inflaton induit par la dimension supplémentaire
est : 
\ba
\Lambda(\tau) & = & \mathcal{O}(1){\dot{H}\over
  H^2}{H\ell\over\sqrt{1+H^2\ell^2}}{p^2\over a^2}.
\ea
Nous trouvons que le taux de dissipation local de l'inflaton dû à la dimension supplémentaire est :
\ba
\Gamma(\tau) & = & {\kappa^2\dot{\phi}^2\over 3} \cr
             & = & \mathcal{O}(1){\dot{H}\over H^2}{H\ell\over\sqrt{1+H^2\ell^2}}H.
\ea
On observe que le terme de dissipation $\Gamma\dot{Q}$ domine la
correction de phase $\Lambda Q$ aux échelles superhorizon alors que c'est le contraire aux
échelles subhorizon. De plus cette correction peut être du même ordre de
grandeur que les corrections standard à l'Inflation (Stewart-Lydth corrections
$\sim (\dot{H}/H^2)H^2 Q$ \cite{scalbc}).

Si l'Inflation a lieu dans le régime où le rayon de courbure du bulk est beaucoup plus grand que le rayon de Hubble
 ($H\ell \gg 1$), le taux de dissipation local de l'inflaton se comporte comme
\ba
\Gamma(\tau) & \stackrel{H\ell \gg 1}{\sim} & \mathcal{O}(1){\dot{H}\over H^2}H.
\ea
Dans le régime quasi-quadridimensionnel ($H\ell \ll 1$) le taux de dissipation local se comporte comme
\ba
\Gamma(\tau) & \stackrel{H\ell \ll 1}{\sim} & \mathcal{O}(1){\dot{H}\over H^2}(H\ell)H.
\ea
Le taux de dissipation est supprimé dans le régime quasi-quadridimensionnel
par un facteur $\left(H\ell\right)$. Le champ scalaire sur la brane se dissipe
\emph{linéairement} par rapport au paramètre de roulement lent à n'importe
quelle échelle. La dissipation de l'inflaton dans la dimension supplémentaire
est donc dominante aux échelles superhorizon et à haute énergie. 

Dans le cas des perturbations tensorielles et dans le cas du fluide parfait
pour les perturbations scalaires, les équations de Mukohyama n'auront pas de
couplage dérivatif, ce qui fait que les corrections effectives sont
non-locales. Il est donc nécessaire dans ces cas de connaître la forme du
propagateur du bulk. On est donc amené à simplifier les équations des
perturbations en approximant la géométrie de fond. C'est l'objet des sections
qui suivent. 

\section{Métrique de fond}\label{sec:sectwo}

Un Univers homogène et isotrope Friedmann-Robertson-Walker (FRW) à $(3+1)$
dimensions avec une histoire d'expansion arbitraire, caractérisée par le
facteur d'échelle $a(\tau)$ fonction du temps propre $\tau$, peut être plongé
dans une portion de l'espace $AdS^5$ de rayon de courbure $\ell$, décrit dans
des coordonnées de bulk statique (coordonnées de Poincaré) par
\ba
 ds^2 & = & \ell^2 {-dt^2+d\mathbf{x}_{3}^2+dz^2 \over z^2},
\ea
au moyen du plongement explicite suivant
\ba
z_b(\tau ) =
e^{-\int H(\tau)d\tau},\qquad
t_b(\tau )=
\int {d\tau\over\ell}\sqrt{\ell^2\dot{z_b}^2(\tau)+z_b^2(\tau)},
\ea
où $H(\tau)$ est le facteur de Hubble sur la brane. Nous pouvons
construire explicitement des coordonnées Gaussiennes normales en calculant la
courbe géodésique dans le plan $(z,t)$, normale à la trajectoire de la brane
de temps propre $\tau$ et s'étendant à la distance propre $\xi$ à partir de la
brane. Nous obtenons donc la projection suivante entre les coordonnées
Gaussiennes normales et les coordonnées de Poincaré :
\ba
z(\xi,\tau) & = & \frac{e^{-\int d\tau ~H(\tau )}}
{ \cosh(\xi /\ell)
-\sqrt{1+\ell^2H^2}\sinh(\xi/ \ell)},\cr
t(\xi,\tau) & = &
\int{d\tau\over\ell}\left[e^{-\int d\tau ~H(\tau )}\sqrt{1+\ell^2H^2}\right]-{\ell He^{-\int
    d\tau ~H(\tau )}\sinh(\xi/ \ell )\over
\cosh(\xi/ \ell)-\sqrt{1+\ell^2H^2(\tau )}\sinh(\xi /\ell)},\quad
\ea
d'où il s'ensuit l'élément de longueur suivant
\ba\label{eqnarray:exactmetric} ds^2 & = &
-\left(\sqrt{1+\ell^2
H^2}\sinh(\xi/\ell)-\cosh(\xi/\ell)+{\ell^2\dot{H}\over\sqrt{1+\ell^2H^2}}\sinh(\xi/\ell)\right)^2
d\tau^2\cr & &
+\left(\sqrt{1+\ell^2H^2}\sinh(\xi/\ell)-\cosh(\xi/\ell)\right)^2
e^{2\int H(\tau)d\tau} d\mathbf{x}_{3}^2\cr & & +d\xi^2. \ea
Dans ces coordonnées la brane est stationnaire par rapport au bulk, et $\xi$ measure
la distance propre à la brane. Notons que nous avons déjà calculé cette
métrique à la section \ref{subsec:bdel} du chapitre \ref{chapter:rs} en
présentant les solutions cosmologiques homogènes et isotropes dans le modèle
de Randall-Sundrum. La seule différence est que nous étions partis des
perturbations de matière pour exprimer la métrique alors qu'ici l'élaboration
de la métrique est purement géométrique.

\begin{figure}[htbp]
\begin{center}
\includegraphics[width=10cm]{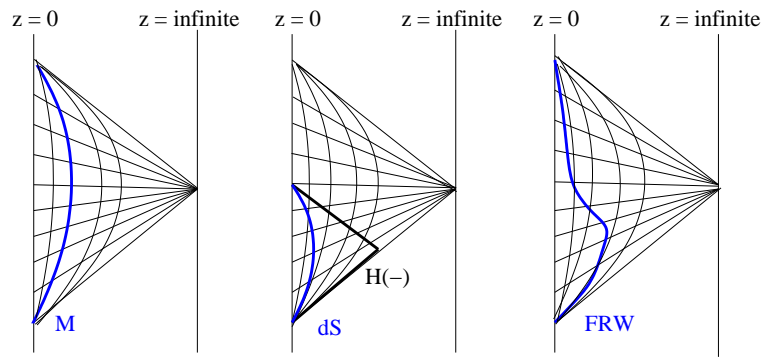}
\end{center}
\caption{\small{{\bf Diagrammes de Carter-Penrose pour différentes cosmologies
  branaires plongées dans $AdS$.} La ligne en gras indique la trajectoire de
  la brane de genre temps. La direction horizontale représente la dimension
  supplémentaire. L'horizon de Cauchy passé est noté $H(-)$. De la gauche vers
  la droite: brane Minkowski, brane de Sitter, brane
  Friedmann-Robertson-Walker. Le diagramme le plus à droite décrit un Univers
  dont l'expansion est initialement inflationnaire puis ralentit.}} \label{fig:penrose}
\end{figure}

Bien que les coordonnées Gaussiennes normales soient commodes pour décrire une
brane en expansion arbitraire, elles souffrent néanmoins d'un certain nombre
de désavantages. En particulier une description du bulk avec ces coordonnées
s'effondre quand les géodésiques spatiales normales à la brane développe des
caustiques, focalisant soit dans le temps à $\xi_h  = \ell{\coth}^{-1}\left(\sqrt{1+\ell^2
    H^2}+\ell^2\dot{H}/\sqrt{1+\ell^2H^2}\right)$, soit dans les dimensions
spatiales transverses à $\xi_s =  \ell{\coth}^{-1}\left(\sqrt{1+\ell^2
    H^2}\right)$. Même si ces singularités peuvent donner l'illusion d'un
horizon, l'espace $AdS$ du bulk est régulier en ces points et peut être étendu
au-delà en utilisant un autre ensemble de coordonnées. Considérons le problème
des conditions initiales dans ces coordonnées en examinant le diagramme de
Carter-Penrose pour des cosmologies du type Randall-Sundrum
(Fig. \ref{fig:penrose}). Il faut spécifier des données initiales sur une surface de
temps Gaussien normal constant, dans le bulk aussi bien que sur la
brane. Supposons qu'on limite notre ambition à prédire ce qui se passe sur la
brane dans le futur. L'évolution d'un Univers branaire dans sa phase dominée
par la matière (fluide sans pression $P = 0$), qui initialement était dans sa phase
inflationnaire, peut être causalement affecté dans le futur par de
l'information inconnue provenant de l'extérieur de l'horizon de Cauchy
initial du passé, parce que la taille de l'horizon du bulk a augmenté durant
la décélération de l'expansion de la brane.

Le mouvement arbitraire et compliqué de la brane dans le bulk brise la symétrie
de translation dans le temps du bulk, de telle sorte que les composantes de la
métrique dans l'équation (\ref{eqnarray:exactmetric}) exhibent une forme
non-séparable. Il s'ensuit que les équations de Mukohyama régissant la
propagation des perturbations et présentées au chapitre \ref{chapter:pert} sont
également non-séparables. L'évolution des perturbations du bulk dans la métrique de fond
 (\ref{eqnarray:exactmetric}) n'est donc pas maniable par des méthodes analytiques
 sans une certaine forme d'approximation. Par conséquent nous allons modifier
 la métrique en ne retenant que ses propriétés les plus importantes et
 supposer que la solution du bulk est quasi-séparable. Du point de vue d'un
 observateur sur la brane, la plupart de l'action a lieu dans un volume de
 quelques longueurs de courbure de $AdS$ (ou longueur de courbure 
 apparente), où virtuellement tout le quadri-volume est concentré. Les modes
 du bulk liés à la brane ont presque tout leur support localisé là, donc une
 approximation médiocre de la métrique loin de la brane risque seulement de
 fournir une approximation pauvre de la queue de la fonction d'onde des états
 liés, mais la queue représente une probabilité quasiment nulle. Nous pouvons
 aussi espérer que chaque quanta qui s'échappe de la brane, à cause d'effets
 non-adiabatiques de l'expansion cosmologique, ne revient pas à
 cause de réflections ou diffractions dans le bulk courbe\footnote{En fait nous allons
 voir comment approximer l'inhomogénéité du bulk $AdS$ et omettre ainsi les effets non-locaux
 dûs aux réflections afin de se concentrer sur la dissipation à la section \ref{subsec:plateau}, où on introduira l'approximation par un
 ``potentiel-plateau'' valable à haute énergie $H\ell\gg 1$.}. Par conséquent un observateur sur la brane sera peu
 sensible à la fa\c{c}on dont ces quanta s'échappent, qui va dépendre de la
 forme de la métrique loin de la brane. De plus, dans l'approximation WKB, la
 plupart des quanta deviennent classiques à une distance courte de la
 brane. Pour toutes ces raisons nous pouvons approcher la géométrie de fond
 (\ref{eqnarray:exactmetric}) par sa géométrie proche de la brane ($\xi \ll
 \ell$), à l'aide de l'élément de longueur suivant
\ba\label{eqnarray:approxmetric} ds^2 \approx
d\xi^2-e^{-2\alpha_1(\tau){\vert\xi\vert\over
\ell}}d\tau^2+e^{-2\alpha_2(\tau){\vert\xi\vert\over
\ell}}a^2(\tau)d\mathbf{x}_{3}^2, \ea où on a $Z_2$-symétrisé en $\xi$
et où les facteurs de gauchissement (``warp factors'') dépendant du temps et le
facteur d'échelle sont respectivement donnés par
\ba \alpha_1(\tau) & = &
\sqrt{1+\ell^2H^2}+{\ell^2\dot{H}\over\sqrt{1+\ell^2H^2}},\cr
\alpha_2(\tau) & = & \sqrt{1+\ell^2H^2},\cr a(\tau) & = & e^{\int
H(\tau) d\tau}. \ea
Dans cette approximation (\ref{eqnarray:approxmetric}) les singularités de
coordonnées se trouvent rejetées à l'infini.


\section{Perturbations tensorielles dans le bulk approximé}\label{sec:secthree}

Nous étudions le cas le plus simple des perturbations
tensorielles, décrivant les ondes gravitationnelles, qui évoluent indépendamment du contenu de matière sur la
brane. Nous utilisons la géométrie de fond approchée
(\ref{eqnarray:approxmetric}), qui est fiable près de la brane. Chaque
polarisation des perturbations tensorielles est décrite par un champ scalaire
canonique sans masse, minimalement couplé à la gravité $\Psi$, comme on l'a déjà discuté à la
section \ref{subsec:mukotens} du chapitre \ref{chapter:pert}. Nous prenons la
transformée de Fourier dans les trois directions spatiales transverses et
évoluons séparément chaque mode de Fourier. Comme nous allons le voir dans la
section qui suit, même dans la géométrie approchée
(\ref{eqnarray:approxmetric}), les équations du mouvement ne sont toujours pas
séparables. On peut réaliser la séparation
des équations en approximant l'inhomogénéité du bulk $AdS$, mais on se concentre en même temps seulement sur les processus
de dissipation en rejetant les processus non-locaux dus à la diffraction des
gravitons dans le bulk.

\subsection{Le potentiel-plateau}\label{subsec:plateau}

\begin{figure}[htbp]
\begin{center}
\includegraphics[width=15cm]{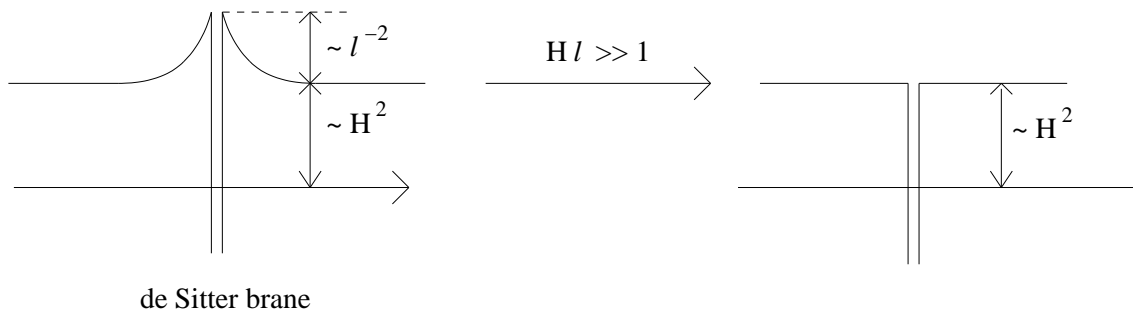}
\end{center}
\caption{\small{{\bf Approximation du potentiel-volcan par le
      potentiel-plateau à haute énergie $H\ell\gg 1$.}} } \label{fig:volcanolim}
\end{figure}

L'équation de Klein-Gordon du mouvement du champ scalaire sans masse $\Psi$, donnée en
(\ref{eqnarray:eomt}), est dans la métrique
de fond approchée (\ref{eqnarray:approxmetric})
\ba
& &\Biggl[-\partial_\tau^2-\left(3{\dot{a}\over
a}+\left({\dot{\alpha_1}\over\ell}-3{\dot{\alpha_2}\over\ell}\right)\vert\xi\vert\right)\partial_\tau+e^{-2\alpha_1{\vert\xi\vert\over
\ell}}\left(\partial_\xi^2-sgn(\xi)\left({\alpha_1\over\ell}+{3\alpha_2\over\ell}\right)\partial_\xi\right)\cr
& &-{p^2\over a^2}e^{-2\vert\xi\vert\left({\alpha_1\over
\ell}-{\alpha_2\over \ell}\right)}\Biggr]\Psi =  0,
\ea
où $\mathbf{p}$ est le moment dans les trois dimensions transverses. À partir
de cette équation, valide dans l'intervalle  $-\infty < \xi <+\infty$, on peut
extraire la condition de bord de Neumann homogène (\ref{eqnarray:bct}) en $\xi
= 0$ pour les modes pairs (\emph{i.e.} $Z_2$-symétriques). En changeant
d'échelle le champ physique selon \ba\label{eqnarray:recale} \Phi(\tau,\xi) =
e^{-\left(\alpha_1(\tau)+3\alpha_2(\tau)\right){\vert\xi\vert\over
2\ell}}a(\tau)^{3\over
2}\Psi(\tau,\xi), \ea
l'équation du mouvement devient
\ba\label{eqnarray:rescaledeom} & &
\Biggl[-\partial_\tau^2+\mathcal{O}\left(\vert\xi\vert\right)\partial_\tau+\left({9\over 4}H^2+{3\over
2}\dot{H}-{p^2\over a^2(\tau)}e^{-2\vert\xi\vert\left({\alpha_1\over
\ell}-{\alpha_2\over \ell}\right)}+\mathcal{O}\left(\vert\xi\vert,\vert\xi\vert^2\right)\right)\cr
& &+e^{-2\alpha_1{\vert\xi\vert\over
\ell}}\left(\partial_\xi^2+\delta(\xi)\left({\alpha_1\over\ell}+{3\alpha_2\over\ell}\right)-{1\over 4}\left({\alpha_1\over\ell}+{3\alpha_2\over\ell}\right)^2\right)\Biggr]\Phi =  0.
\ea
L'équation du mouvement n'est pas encore séparable. Cependant, comme l'état
lié est localisé près de la brane, une approximation adéquate pour l'évolution
de l'état lié du graviton peut être obtenue en retenant seulement le
comportement dominant autour de $\xi = 0$ des coefficients dépendant de $\xi$,
ce qui simplifie l'équation. Sans doute c'est une approximation grossière pour
les bouts de la fonction d'onde de l'état lié mais nous soutenons que les
bouts contribuent de fa\c{c}on négligeable. Par conséquent on pose
\ba
e^{-2\alpha_1{\vert\xi\vert\over
\ell}} \approx 1, \qquad  e^{-2\vert\xi\vert\left({\alpha_1\over
\ell}-{\alpha_2\over \ell}\right)} \approx 1, \qquad
\mathcal{O}\left(\vert\xi\vert\right),\mathcal{O}\left(\vert\xi\vert^2\right)
\approx 0,
\ea
réduisant l'équation (\ref{eqnarray:rescaledeom}) à une équation de type Schrödinger
\ba\label{eqnarray:schro}
-\partial_\xi^2\Phi+V(\xi,\tau)\Phi =
-\partial_\tau^2\Phi+\left[{9\over 4}H^2+{3\over
2}\dot{H}-{p^2\over a^2(\tau)}\right]\Phi \ea avec
le ``potentiel-plateau'' effectif \ba\label{eqnarray:potH} V(\xi,\tau) & = &
-V_0(\tau)\delta(\xi)+{V_0(\tau)^2\over 4}\ea
où
\ba
V_0(\tau)  & = & {1\over\ell}\left(4\sqrt{1+\ell^2H^2}+{\ell^2\dot{H}\over\sqrt{1+\ell^2H^2}}\right).
\ea
La condition de bord pour le champ renormalisé $\Phi$ est inscrite dans dans
la fonction $\delta$ du potentiel et peut être obtenue en intégrant l'équation 
$Z_2$-symétrique (\ref{eqnarray:schro}). De manière équivalente on
aurait pu imposer la condition de bord séparément
\ba \left[\partial_\xi+{V_0(\tau)\over
    2}\right]\Phi\vert_{\xi = 0^+} = 0
\ea
en $\xi = 0$ et restreindre le domaine de (\ref{eqnarray:schro}) à $\xi > 0$,
supprimant ainsi la fonction $\delta$. Sous cette approximation, le cratère du
potentiel-volcan habituel \cite{rs2,lands} est fidèlement conservé mais le
paysage autour du sommet reste à élevation constante, tel qu'indiqué à la
figure Fig. \ref{fig:volcanolim} à droite.

Lorsque la brane possède une géométrie de Sitter pur, les équations d'Einstein
linéarisées dans $AdS$ pour les perturbations tensorielles peuvent être réduite à une équation de type
``Schrödinger'' pour un champ scalaire dans des coordonnées du bulk conformes
et une fois que le champ a été changé d'échelle. L'expression exacte du
potentiel effectif a été calculée en \cite{lands} et a été donnée à la section
\ref{subsec:dsrs}. Le potentiel a dans ce cas la forme habituelle d'un volcan
avec une barrière de potentiel décroissante résultant de la courbure du bulk
$AdS$ (Fig. \ref{fig:volcanolim} à gauche). La hauteur de la barrière de
potentiel est $\mathcal{O}(\ell^{-2})$ et le gap (écart) d'énergie entre le mode zéro
et le continuum est $\mathcal{O}(H^2)$. Donc l'approximation du
potentiel-plateau que nous avons utilisée précédemment consiste à négliger
l'inhomogénéité du bulk $AdS$ et par conséquent la diffraction des gravitons par le
bulk courbe. Cette approximation est légitime dans le régime cosmologique où
le rayon de courbure du bulk est beaucoup plus grand que l'horizon de Hubble,
$H\ell\gg 1$, c'est-à-dire à haute énergie (ou tôt dans l'histoire de
l'Univers). C'est parce que dans ce régime, l'intervalle d'énergie de la
barrière de potentiel est de l'ordre $\ell^{-2}$, ce qui est beaucoup plus
faible que l'échelle d'énergie de Hubble $H^2$ caractérisant le gap. Dans ce
régime le potentiel-volcan ressemble au potentiel-plateau
(Fig. \ref{fig:volcanolim}), ce qui légitime l'approximation. Dans le régime opposé où le rayon de courbure de
la dimension supplémentaire est beaucoup plus petit que l'horizon de Hubble,
$H\ell \leq 1$, notre approximation échoue parce que l'intervalle d'énergie de
l'inhomogénéité du bulk est maintenant non-négligeable. Donc dans ce régime le
potentiel-plateau donnerait seulement des bornes supérieures et inférieures
grossières au vrai potentiel-volcan.

\subsection{Dissipation de l'état lié du graviton pour une brane
inflationnaire à haute énergie}\label{subsec:bounddissip}

Nous constatons que la forme du potentiel-plateau (\ref{eqnarray:potH})
implique, pour le spectre 5D de gravitons, la présence d'un unique état lié à la brane, le mode zéro du graviton, et d'un continuum de modes libres de diffusion dont les valeurs propres de masse démarrent au-dessus du plateau. Nous développons la fonction d'onde sur la base des modes normaux
\ba\label{eqnarray:modekg} \Phi(\tau,\xi) = c_b(\tau)\phi_b(\xi;\tau)+\int_0^{+\infty} dk
c_k(\tau)\phi_k(\xi;\tau) \ea où le $b$ indexe le mode lié et les $k$ indexent
les modes du continuum.  Les fonctions propres des modes $\phi_n(\xi;\tau)$
($n = b,k$) sont dépendantes du temps et satisfont \ba\label{eqnarray:eqproprekg}
\left(-\partial^2_\xi-V_0(\tau)\delta(\xi)+{V_0(\tau)^2\over
4}\right)\phi_n(\xi;\tau) = m_n^2(\tau)\phi_n(\xi;\tau). \ea
Le spectre $Z_2$-symétrique de l'équation (\ref{eqnarray:eqproprekg}) consiste
effectivement en un unique état lié avec $m_b^2(\tau) = 0$ :
\ba\label{eqnarray:zeromode} \phi_{b}(\xi;\tau) = N_b(\tau)
e^{-{V_0(\tau)\over 2}\vert\xi\vert},~~~~N_b(\tau) = \sqrt{V_0(\tau)/ 2} \ea
et en un continuum d'états libres $k$ du bulk avec $m_k^2(\tau) =
k^2+V_0^2(\tau)/ 4$ :
\ba\label{eqnarray:kmode} \phi_{k}(\xi;\tau) =
N_k(\tau)\left[\cos\left(k\xi\right)-{V_0(\tau)\over
2k}\sin\left(k\vert\xi\vert\right)\right],~~~~N_k(\tau) =
{k\over\sqrt{\pi}\sqrt{k^2+{V_0^2(\tau)\over 4}}}. \ea
Un gap d'énergie de $V_0(\tau)^2/ 4$ sépare l'état lié du continuum.

\begin{figure}[htbp]
\begin{center}
\includegraphics[width=5cm]{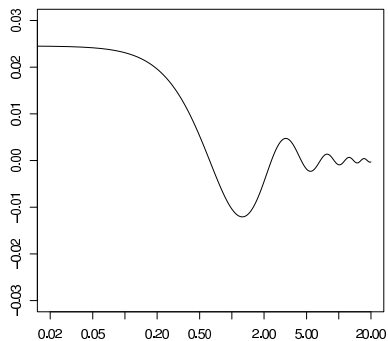}
\end{center}
\caption{\small{{\bf Évolution du noyau d'interaction avec le bulk, $K(s)$, en
      fonction du temps $s$, aux échelles superhorizon $p << aH$ et dans la
      limite $H\ell\gg 1$}.
Sur la figure $H = 1$.}}\label{fig:kernel}
\end{figure}

Nous considérons l'Univers pendant l'époque de l'Inflation, lorsque la
géométrie de la brane est quasi-de Sitter et l'expansion \emph{adiabatique},
dans le sens où $\dot{H} \ll H^2$. Dans ce régime l'interaction brane-bulk
peut être calculée de fa\c{c}on perturbative dans les couplages faibles. Dans le cas
des perturbations tensorielles, le seul degré de liberté sur la brane est
l'état lié discret du graviton (mode zéro). Nous pouvons calculer le taux de
dissipation de l'état lié du graviton dû au mouvement inflationnaire de la
brane dans le bulk. Dans l'approximation adiabatique on a que 
\ba\label{eqnarray:potadiab}
V_0(\tau) & \approx & 4\sqrt{1+\ell^2 H^2(\tau)}/\ell, \qquad \dot{V_0}
\approx 4\dot{H}{H\ell\over\sqrt{1+\left(H\ell\right)^2}}.
\ea
On obtient l'équation d'évolution des coefficients $c_n(\tau)$ ($n = b,k$) dépendant du temps du développement en modes
 dans le bulk approximé décrit en (\ref{eqnarray:approxmetric}) :

\ba\label{eqnarray:unkgtau}
\ddot{c}_n+\left[{p^2\over a^2(\tau)}-{9\over 4}H^2+m_n^2(\tau)\right] c_n & \approx &
-2\dot{V}_0\left\langle\phi_n\Bigg\vert{\partial\phi_n\over\partial
  V_0}\right\rangle\dot{c}_n-\sum_{m\neq n}2\dot{V}_0\left\langle\phi_n\Bigg\vert{\partial\phi_m\over\partial
  V_0}\right\rangle\dot{c}_m\cr
& & +\mbox{ (termes d'ordre supérieur) },
\ea
où les termes d'ordre supérieur incluent deux dérivées temporelles du potentiel (\emph{e.g.} $\ddot{H}, \dot{H}^2$). Les termes $\left\langle\phi_n\vert\partial\phi_n/\partial
  V_0\right\rangle$ s'annulent puisque les modes propres sont réels. Les termes  $\left\langle\phi_k\vert\partial\phi_{k'}/\partial
  V_0\right\rangle$ connectant deux états du continuum sont également nuls à cause de l'orthogonalité\footnote{L'orthogonalité est le résultat ici de la forme simplifiée du potentiel-plateau.}. Les seuls éléments matriciels non nuls sont ceux qui connectent le mode lié à un mode du continuum, \ba
\left\langle\phi_{k}\Bigg\vert{\partial\phi_{b}\over\partial V_0}\right\rangle =
2N_k^{*}N_b\int_0^{+\infty}d\xi\left[\cos\left(k
\xi\right)-{V_0(\tau)\over
2k}\sin\left(k\xi\right)\right]\left(-{\xi\over
  2}\right)e^{-{V_0(\tau)\over 2}\xi}
 =  {N_k^{*}(\tau)N_b(\tau)\over
  k^2+{V_0^2(\tau)\over 4}}.\qquad
\ea
L'équation (\ref{eqnarray:unkgtau}) se réduit donc à
\ba
\ddot{c}_k+\Omega_k^2(\tau)c_k & \approx & -\gamma_k(\tau)\dot{c}_b(\tau)+\mathcal{O}\left(\dot{V_0}^2,\ddot{V_0}\right),\label{eqnarray:system2} \\
\ddot{c}_b+\Omega_b^2(\tau)c_b & \approx & \int dk \gamma_k(\tau)\dot{c}_k(\tau)+\mathcal{O}\left(\dot{V_0}^2,\ddot{V_0}\right),\label{eqnarray:system}
\ea
couplant ainsi l'état lié $c_b$ aux états $c_k$ du continuum avec $k > 0$. Dans l'approximation adiabatique le facteur de couplage au premier ordre $\gamma_k(\tau)$ et les fréquences $\Omega_b$, $\Omega_k$ sont
\ba
\gamma_k(\tau) & = &
\dot{V_0}(\tau)\sqrt{2V_0(\tau)\over\pi}{k\over
  \left(k^2+{V_0^2(\tau)\over 4}\right)^{3/2}},\cr
\Omega_b^2(\tau) & \approx
&{p^2\over a^2(\tau)}-{9\over
    4}H^2(\tau),\cr
\Omega_k^2(\tau) & \approx & {p^2\over a^2(\tau)}-{9\over
    4}H^2(\tau)+k^2+{V_0^2(\tau)\over 4}.
\ea
Le couplage dépendant du temps dans (\ref{eqnarray:system2}) génère des transitions entre le mode lié et le continuum, parce que les modes nus à l'ordre zéro agissent comme une source pour les modes au prochain ordre. Dès que l'accélération de la brane change dans le bulk, à travers la dérivée du facteur de Hubble $\dot{H}(\tau)$ (ou de manière équivalente
$\dot{V_0}(\tau)$), le mode \emph{in} agit comme une source à travers le facteur de couplage et génère différents modes au premier ordre. Par conséquent des transitions du mode lié vers le continuum se produisent au premier ordre dans les couplages, et les modes continus du premier ordre agissent ensuite eux-même comme une source pour le mode lié, comme un effet de rétroaction au deuxième ordre. Le mode lié à la brane et les modes continus du bulk sont reliés selon
\ba
c_k(\tau)  & = & \int_{-\infty}^\tau d\tau' G^k_{bulk}(\tau,\tau')\left(-\gamma_k(\tau')\right)\dot{c}_b(\tau'),\cr
c_b(\tau) & = & \int_{-\infty}^\tau d\tau' G_{brane}(\tau,\tau')\int dk \gamma_k(\tau')\dot{c}_k(\tau'),
\ea
où $G^k_{bulk}(\tau,\tau')$ et $G_{brane}(\tau,\tau')$ sont les fonction de Green retardées nues satisfaisant
\ba\label{eqnarray:related}
\left(\partial_\tau^2 + \Omega_k^2(\tau)\right)G^k_{bulk}\left(\tau,\tau'\right) & = & \delta\left(\tau-\tau'\right),\cr
\left(\partial_\tau^2 + \Omega_b^2(\tau)\right)G_{brane}\left(\tau,\tau'\right) & = & \delta\left(\tau-\tau'\right).
\ea
Nous pouvons exprimer la fonction de Green retardée du mode $k$ du bulk dans l'approximation WKB
\ba
G^k_{bulk}(\tau,\tau') & \approx & \theta(\tau-\tau'){\sin\left(\int_{\tau'}^\tau
    \Omega_k(\tau)d\tau\right)\over \sqrt{\Omega_k(\tau)\Omega_k(\tau')}}.
\ea
Cette approximation WKB est raisonnable parce que les modes propres du bulk oscillent tout le temps. Il n'y a pas de point de rebroussement puisque $\Omega_k^2 \approx
p^2/a^2-9H^2/4+k^2+V_0^2/4 > 0$ est toujours vérifié. La condition
$\dot{\Omega}_k \ll \Omega_k^2$ est satisfaite à la fois pour des modes subhorizon, $p \gg aH$, et des modes superhorizon, $p \ll aH$.
Le propagateur effectif sur la brane, avec l'interaction avec les modes du
bulk prise en compte, est obtenue en resommant la série perturbative
géométrique à tous les ordres : 
\ba\label{eqnarray:effgreentensor}
\hat{G}_{brane} & = &  G_{brane} + G_{brane}\left[\int dk \gamma_k D_\tau G^k_{bulk}\left(-\gamma_k\right)D_\tau\right] G_{brane}\cr
& + & G_{brane}\left[\int dk \gamma_k D_\tau G^k_{bulk}\left(-\gamma_k\right)D_\tau\right] G_{brane}\left[\int dk \gamma_k D_\tau G^k_{bulk}\left(-\gamma_k\right)D_\tau\right] G_{brane}
 + ...\cr
              & = & {1\over G^{-1}_{brane}+\int dk \gamma_k D_\tau G^k_{bulk}\gamma_k D_\tau}.
\ea
L'interaction quadridimensionnelle entre les modes de la brane, dûe à
l'expansion de l'Univers, est contenue dans $G^{-1}_{brane}$, alors que
l'interaction entre le mode lié et les modes du continuum est contenue dans le
noyau du bulk :
\ba\label{eqnarray:self}
K(\tau,\tau') & = & \int dk \gamma_k(\tau) \partial_\tau G^k_{bulk}(\tau,\tau')\gamma_k(\tau'),\cr
               & = & \theta(\tau-\tau')\int dk\gamma_k(\tau)\sqrt{{\Omega_k(\tau)\over \Omega_k(\tau')}}\cos\left(\int_{\tau'}^\tau
    \Omega_k(\tau)d\tau\right)\gamma_k(\tau'),
\ea
de telle sorte que l'évolution du mode lié à la brane est gouverné par
l'équation intégro-différentielle :
\ba\label{eqnarray:exacteq}
 &  & \hat{G}_{brane}^{-1}\circ c_b\cr
  & = & \partial_\tau^2c_b(\tau)+\Omega_b^2(\tau)c_b(\tau)+\int_{-\infty}^\tau
  d\tau' K(\tau,\tau')\partial_{\tau'}c_b(\tau') = 0,
\ea
où le noyau d'interaction avec le bulk $K(\tau,\tau')$ "habille" le champ nu de l'état lié.
La dissipation éventuelle de l'état lié apparait donc ici comme
\emph{non-locale} (avec mémoire) à un observateur sur la brane.
Ce caractère dissipatif se manifeste de fa\c{c}on non-locale parce que l'état lié n'est pas un degré de liberté purement local sur la brane mais localisé près de la brane, avec une extension typique dans le bulk donnée par  $\ell_{att}
= (V_0/2)^{-1}$. L'existence de l'état lié dépend intrinsèquement de la
courbure du bulk $\ell^{-1}$. Cette condition est incompatible avec une
interaction locale avec le bulk. Dans la limite de haute énergie $H\ell \gg 1$ (où le potentiel-plateau est une approximation fiable), on a $V_0\approx 4H$, $\dot{V_0} = 4\dot{H}$. Le noyau d'interaction avec le bulk  (\ref{eqnarray:self}) se simplifie dans la limite $H\ell
\gg 1$ comme
\ba\label{eqnarray:superself}
K(s\equiv \tau-\tau') & \stackrel{H\ell\gg 1, p\ll aH}{\approx} &
\theta(s)\dot{H}^2\int_0^\infty dk {128k^2 H\over
  \pi\left(k^2+4H^2\right)^3} \cos\left(\sqrt{k^2+{7\over
      4}H^2}~s\right)
\ea
aux échelles superhorizon,
et comme
\ba\label{eqnarray:subself}
K(\tau,\tau') & \stackrel{H\ell\gg 1, p\gg aH}{\approx} &
\theta(\tau-\tau')\dot{H}^2\int_0^\infty dk {128k^2 H\over
  \pi\left(k^2+4H^2\right)^3} \left({k^2+{p^2\over a(\tau)^2}\over k^2+{p^2\over
    a(\tau')^2}}\right)^{1/4}\cos\left(\int_{\tau'}^\tau \sqrt{k^2+{p^2\over
    a(\bar{\tau})^2}}d\bar{\tau}\right)\cr
              & \approx & \mathcal{O}(1)H^2\left({\dot{H}\over
H^2}\right)^2e^{-{H\over 2}(\tau-\tau')}\cos\left({p\over H}\left(e^{-H\tau}-e^{-H \tau'}\right)\right)\theta(\tau-\tau')
\ea
aux échelles subhorizon, où dans ce dernier cas l'intégrale sur $k$ est supprimée pour $k \gg H$.
Nous avons utilisé le fait que $H$, $\dot{H}$ sont quasiment constant au cours du temps dans l'approximation adiabatique. Le noyau superhorizon (\ref{eqnarray:superself}) est tracé sur la figure Fig. \ref{fig:kernel}.

Nous observons que le noyau est non-local, variant typiquement sur un échelle de temps de Hubble $H^{-1}$. L'interaction avec le bulk contient donc une mémoire (ou un délai en temps). Par conséquent une approximation locale du noyau ne serait possible que si le mode lié variait avec une échelle de temps caractéristique plus grande que le temps de Hubble.

Au cours de l'Inflation, le rayon de Hubble de l'Univers reste quasiment
constant dans le temps, alors que la longueur d'onde physique d'un mode
augmente de fa\c{c}on exponentielle à cause de l'expansion. Au début de
l'Inflation un mode de nombre d'onde donné est initialement subhorizon de
sorte que $p\gg a(\tau)H$. Au fur et à mesure de l'expansion de l'Univers, ce
mode traverse l'horizon pour devenir superhorizon. Alors que les modes
oscillent aux échelles subhorizon, ils deviennent "gelés" après le croisement
de l'horizon de Hubble. Un observateur est sensible seulement aux modes
superhorizon sur la brane qui sont les modes accessibles aux observations
cosmologiques aujourd'hui et qui constituent les "conditions initiales" de
l'Univers à la fin de l'Inflation. Pour ces raisons nous nous concentrerons
sur les modes aux échelles superhorizon (ou, de fa\c{c}on équivalente, aux asymptotes à grand temps des modes)
\ba
p\ll a(\tau)H.
\ea
L'évolution des modes de la brane aux échelles superhorizon, prenant en compte
l'interaction avec le bulk à travers le noyau integral $K(\tau,\tau')$, est
non-locale. Par conséquent l'évolution (et la dissipation éventuelle) des
modes superhorizon sur la brane dépend en fait de leur évolution passée aux
échelles subhorizon, à travers l'interaction du bulk avec les modes aux échelles subhorizon. Aux échelles superhorizon, la fréquence nue du mode lié est imaginaire
\ba
\Omega_b^2(\tau) = -{9\over 4}H^2(\tau).
\ea
Le coefficient du mode lié, nu, $c_b^{(0)}(\tau)$, excluant l'interaction avec le bulk, est la
superposition d'un mode croissant et d'un mode décroissant
\ba\label{qqq:cbare}
c_b^{(0)}(\tau)  & = & A_+e^{+{3\over 2}\int H(\tau)d\tau} +A_-e^{-{3\over
    2}\int H(\tau)d\tau},\cr
                 & \approx & A_+e^{+{3\over 2}H\tau} +A_-e^{-{3\over
    2}H\tau},
\ea
 aux échelles superhorizon, dans l'approximation WKB. Puisque nous avions changé d'échelle le champ physique en (\ref{eqnarray:recale}) par le facteur $a^{3/2}(\tau)$, le vrai
 mode lié est en fait la superposition d'un mode constant et d'un mode décroissant. Les solutions nues (\ref{qqq:cbare}) correspondent également aux asymptotes superhorizon de la solution quadridimensionnelle donnée par la fonction de Hankel dans la limite d'une géométrie purement de Sitter à quatre dimensions :
\ba
H^{(1),(2)}_{3/2}\left(-p\eta\equiv {p\over a(\tau)H} \right)& \stackrel{p/aH \ll
  1}{\longrightarrow}& \sim A_+a^{3/2}(\tau)+A_-a^{-3/2}(\tau),\cr
                     & \stackrel{p/aH \gg
  1}{\longrightarrow}& \sim e^{\pm i {p\over a(\tau)H}}
\ea
où $\eta$ est le temps propre conforme et $a(\tau) = e^{H\tau}$.
Le mode décroissant n'est pas observable et a peu d'intérêt cosmologique pour la formation des structures à grande échelle dans l'Univers. Dans la suite nous considérons seulement la solution dominante composée du mode croissant.

Nous pouvons maintenant rechercher le mode croissant superhorizon de l'état
lié "habillé", c'est-à-dire tenant compte de l'interaction avec le bulk, en
utilisant l'Ansatz superhorizon :
\ba\label{eqnarray:ansatzbound}
c_b(\tau) & = & \exp[+\gamma \tau],
\ea
où
\ba
\gamma & = & {3\over 2}H+\Delta\gamma,\qquad \Delta\gamma\ll H.
\ea
Comme le noyau d'interaction avec le bulk est non-local, nous devrions prendre des précautions en insérant cet Ansatz superhorizon  dans le noyau integral de l'équation (\ref{eqnarray:exacteq}). L'intégration complète et rigoureuse devrait bien-entendu prendre en compte la contribution du régime subhorizon, où le mode lié a une forme différente
de l'Ansatz superhorizon (\ref{eqnarray:ansatzbound}). Cependant, nous conjecturons que, lorsqu'on observe la solution "habillée" aux échelles superhorizon aujourd'hui, les effets intégrés des modes subhorizon dans leur interaction avec le bulk sont sous-dominant en comparaison de la contribution des modes superhorizon. C'est parce que l'amplitude des modes subhorizon est supprimée par rapport à l'amplitude des modes
superhorizon. On peut également remarquer en (\ref{eqnarray:subself}) que le noyau integral aux échelles subhorizon est diminué par le facteur $\exp\left[-\int H(\tau)/2~d\tau\right]$ par rapport à sa limite superhorizon. Utiliser l'Ansatz (\ref{eqnarray:ansatzbound}) à tout temps pour résoudre l'équation
intégro-differentielle (\ref{eqnarray:exacteq}) pour l'état lié est en fait équivalent à lisser les oscillations subhorizon de faible amplitude dans la solution exacte du mode lié.
Pour ces raisons nous considérons l'Ansatz superhorizon
(\ref{eqnarray:ansatzbound}) suffisant pour résoudre l'équation
intégro-differentielle (\ref{eqnarray:exacteq}) et calculer l'ordre de magnitude du taux de dissipation de l'état lié aux échelles superhorizon.

Nous procédons donc maintenant au calcul du facteur d'atténuation du mode lié
croissant dans le régime de haute énergie $H\ell \gg 1$.
Dans ce régime le rayon de courbure de la dimension supplémentaire est beaucoup plus grand que le rayon de Hubble de sorte que taux de dissipation de l'état lié dans la dimension supplémentaire doit atteindre sa valeur maximale. Comme nous l'avons discuté au paragraphe \ref{subsec:plateau}, le potentiel-plateau est une approximation fiable dans ce régime pour appréhender le comportement et la magnitude de l'interaction brane-bulk. La hauteur du potentiel-plateau vaut alors $V_0^2(\tau)/ 4  \approx
4H^2(\tau)$ et le coefficient du mode lié "habillé" évolue aux échelles superhorizon selon l'équation
\ba\label{eqnarray:effeom2}
\ddot{c}_b(\tau) -{9\over 4}H^2(\tau)c_b(\tau)+\int_{0}^{+\infty}
  ds K(s)\dot{c}_b(\tau-s) = 0,
\ea
où $K(s)$ est le noyau d'interaction avec le bulk pris dans la limite superhorizon, éqn. (\ref{eqnarray:superself}).

L'insertion de l'Ansatz (\ref{eqnarray:ansatzbound}) dans (\ref{eqnarray:effeom2}) conduit à l'équation algébrique
\ba
\gamma^2-{9\over 4}H^2 & = & -{128\over\pi}\dot{H}^2H\int_0^\infty dk {k^2\gamma^2\over
  \left(k^2+4H^2\right)^3\left(k^2+{7\over 4}H^2+\gamma^2\right)},
\ea
qui, à l'ordre linéaire en $\Delta\gamma$, devient
\ba
3H\Delta\gamma & \approx & -{128\over\pi}\dot{H}^2H\left[{9\over 4}H^2\int_0^\infty dk {k^2\over
      \left(k^2+4H^2\right)^4}\right]
\ea
de telle sorte que
\ba
\Delta\gamma & \approx & -{3\over 2^{5}}H\left({\dot{H}\over
    H^2}\right)^2.
\ea

Le signe moins trouvé dans $\Delta\gamma$ signifie une atténuation du mode
croissant dans la dimension supplémentaire. Le potentiel-plateau possède
l'avantage d'éliminer les processus de diffraction (ou réflection) dans le
bulk courbe en approximant l'inhomogénéité du bulk, ne préservant ainsi que
les processus de dissipation due à la dimension supplémentaire. Même si l'état
lié ne se dissipait pas dans le régime subhorizon et subissait seulement un
décalage de fréquence, on observe que l'état lié subit de la dissipation
classiquement dans le régime superhorizon selon
\ba
c_b(\tau) & = & e^{-\int\Gamma(\tau)d\tau}c_b^{(0)}(\tau)
\ea
où $c_b^{(0)}(\tau)$ est le mode croissant nu. Le taux de dissipation $\Gamma(\tau) \equiv -\Delta\gamma$ de l'état lié dans le régime superhorizon est donc donné par
\ba
\Gamma(\tau) & = & \mathcal{O}(1)\left({\dot{H}\over
    H^2}\right)^2 H.
\ea
Les gravitons émis dans la dimension supplémentaire sont oscillant à tout
temps puisque leur fréquence reste au dessus du plateau du
potentiel. Terminons cette section en ré-exprimant le facteur d'atténuation en
fonction du paramètre de slow-roll de l'Inflation $\epsilon_H = -\dot{H}/H^2$,
caractérisant l'adiabaticité de l'expansion de l'Univers, et du nombre
d'e-folds $N$ :
\ba
e^{-\int\Gamma(\tau)d\tau} & = & e^{-\mathcal{O}(1)\epsilon_H^2N}.
\ea

L'état lié du graviton (mode zéro des perturbations tensorielles) se dissipe
de fa\c{c}on non-locale et \emph{quadratiquement} dans le paramètre de
slow-roll. La dissipation de l'état lié est donc sous-dominante par rapport à
la dissipation de l'Inflaton (perturbation scalaire, section \ref{sec:inflaton}).

\section{Perturbations scalaires dans le bulk approximé : fluide parfait adiabatique aux échelles subhorizon}\label{sec:secfour}

Dans cette section nous explorons l'évolution des perturbations scalaires de
métrique, dans la géométrie de fond approchée (\ref{eqnarray:approxmetric}),
couplant cette fois aux perturbations adiabatiques d'un fluide parfait sur la
brane dans le régime subhorizon.  Dans la géométrie de fond approchée (\ref{eqnarray:approxmetric}), fiable près de la brane, l'équation du mouvement (\ref{qqq:eoms}) du champ "maitre" de Mukohyama dans l'intervalle $ 0 < \xi
<+\infty$ s'écrit :
\ba\label{eqnarray:eomss} & &
\Biggl[-\partial_\tau^2+\left(3{\dot{a}\over
a}-\left({\dot{\alpha_1}\over\ell}+3{\dot{\alpha_2}\over\ell}\right)\xi\right)\partial_\tau+e^{-2\alpha_1{\xi\over
\ell}}\left(\partial_\xi^2-\left({\alpha_1\over\ell}-{3\alpha_2\over\ell}\right)\partial_\xi\right)\cr
& &-{p^2\over a^2}e^{-2\xi\left({\alpha_1\over
\ell}-{\alpha_2\over \ell}\right)}+{e^{-2\alpha_1{\xi\over
\ell}}\over\ell^2}\Biggr]\Omega =  0,
\ea
où $p$ est le tri-moment transverse. De fa\c{c}on similaire au cas tensoriel de la section \ref{sec:secthree} précédente, nous réduisons l'équation du mouvement (\ref{eqnarray:eomss}), en faisant le changement de variable
\ba\label{qqq:rescalemaster}
e^{-\left({\alpha_1-3\alpha_2\over\ell}\right){\xi\over
    2}}a(\tau)^{-3/2}\Omega & \rightarrow & \Omega.
\ea
L'équation dans le nouveau champ s'écrit :
\ba\label{eqnarray:rescaledeoms} & &
\Biggl[-\partial_\tau^2+\mathcal{O}\left(\xi\right)\partial_\tau+\left({9\over 4}H^2-{3\over
2}\dot{H}-{p^2\over a^2(\tau)}e^{-2\xi\left({\alpha_1\over
\ell}-{\alpha_2\over \ell}\right)}+{e^{-2\alpha_1{\xi\over
\ell}}\over\ell^2}+\mathcal{O}\left(\xi,\xi^2\right)\right)\cr
& &+e^{-2\alpha_1{\xi\over
\ell}}\left(\partial_\xi^2-{1\over 4}\left({\alpha_1\over\ell}-{3\alpha_2\over\ell}\right)^2\right)\Biggr]\Omega =  0.
\ea
Elle contient des termes dépendant de $\xi$ représentant l'inhomogeneité du bulk $AdS$.
L'équation du mouvement n'est malgré tout pas encore séparable. Cependant, comme à la section
\ref{subsec:plateau} pour le cas tensoriel, lorsque nous avons introduit le potentiel-plateau, nous simplifions de nouveau la dépendance en $\xi$ en gardant seulement le comportement dominant des coefficients en $\xi$ autour de $\xi = 0$. On pose
\ba
e^{-2\alpha_1{\xi\over
\ell}} \approx 1, \qquad  e^{-2\xi\left({\alpha_1\over
\ell}-{\alpha_2\over \ell}\right)} \approx 1, \qquad
\mathcal{O}\left(\xi\right),\mathcal{O}\left(\xi^2\right)
\approx 0,
\ea
réduisant l'équation (\ref{eqnarray:rescaledeoms}) à
\ba\label{qqq:schroscalar}
\left[\partial_\tau^2-\partial_\xi^2+\bar{\omega}^2(\tau)+\lambda^2(\tau)\right]\Omega & = & 0
\ea
où
\ba
\lambda(\tau) & = &{\alpha_1-\alpha_2\over 2\ell}
= {\ell\dot{H}\over 2\sqrt{1+\ell^2H^2}} ~\leq 0,\label{eqnarray:v1}\cr
\bar{\omega}^2(\tau) & = & {p^2\over a^2}-{5\over 4}H^2+{1\over
2}\dot{H}.\label{eqnarray:omtilde}
\ea
La courbure du bulk $\ell$ est contenue dans $\lambda(\tau)$ et la courbure intrinsèque de la brane $H$ est contenue dans  $\bar{\omega}$.
Soulignons le "mauvais" signe du potentiel $\lambda(\tau)$ pour les
perturbations scalaires puisque $\dot{H}(\tau)<0$. En fait la brane va agir
dans ce cas comme un potentiel $(+\delta)$ répulsif pour les perturbations
scalaires de métrique dans un cadre $Z_2$-symétrique, de telle sorte qu'il ne
peut exister d'état lié gravitationnel scalaire près de la brane, au moins
tant qu'il n'y a pas de couplage avec la matière sur la brane (voir équation
de bord (\ref{eqnarray:bc})).


Le cas d'un fluide parfait sur la brane (d'équation d'état $P_M =
w\rho_M$) entraîne quelques difficultés dans le calcul de l'évolution des
perturbations parce que l'expansion de la brane n'est plus
adiabatique. $\dot{H}/H^2$ n'est pas petit excepté lorsque le fluide est
sous-dominant par rapport à la tension de la brane $\sigma$ (constante
cosmologique). Dans cette section nous explorons la dissipation éventuelle des
perturbations du fluide dans le régime subhorizon parce que, dans ce régime, l'approximation adiabatique de l'expansion est valide. En effet aux échelles subhorizon la geometrie de la brane apparait quasiment Minkowskienne. 

Les conditions de bord (\ref{qqq:sbc0}), reliant le champ "maitre" des
perturbations de métrique aux perturbations du fluide sur la brane, deviennent
dans la géométrie de fond approximée (\ref{eqnarray:approxmetric}) valide près
de la brane :
\ba\label{eqnarray:sbc}
\kappa^2 a\delta\rho & = & -3H\left(\dot{\Omega}'+{\alpha_1\over\ell}\dot{\Omega}\right)-{p^2\over
a^2}\left(\Omega'+{\alpha_2\over\ell}\Omega\right)\Bigr\vert_{\xi = 0},\cr
\kappa^2 a\delta
q & = & -\dot{\Omega}'-{\alpha_1\over\ell}\dot{\Omega}\Bigr\vert_{\xi = 0},\cr
\kappa^2 a\delta
P & = &
\left(\ddot{\Omega}'+{\alpha_2\over\ell}\ddot{\Omega}\right)+2H\left(\dot{\Omega}'+{\alpha_1\over\ell}\dot{\Omega}\right)+\left({\dot{\alpha_1}\over\ell}+2{\dot{\alpha_2}\over\ell}\right)\dot{\Omega}\cr
                 &   & +{\dot{\alpha_2}\over\ell
H}\left({1\over\ell^2}-{2\over 3}{p^2\over a^2}\right)\Omega-{\dot{\alpha_2}\over\ell
H}\left({\alpha_1\over\ell}-2{\alpha_2\over\ell}\right)\Omega'\Bigr\vert_{\xi = 0}.
\ea
Il est possible de simplifier les conditions de bord (\ref{eqnarray:sbc}) pour les perturbations scalaires de métrique en exprimant celles-ci en fonction de la perturbation de matière indépendante de jauge $\left(\rho-\sigma\right)\Delta = \delta\rho-3H\delta
q$ (contraste de densité). On obtient alors les conditions de bord suivantes :
\ba\label{eqnarray:bc}
\left(\Omega'+{\alpha_2(\tau)\over\ell}\Omega\right)\Bigr\vert_{\xi = 0} & = &
-\kappa^2{\left(\rho-\sigma\right) a^3\over p^2}\Delta(\tau)
\ea
qui, en fonction du champ "maitre" renormalisé (\ref{qqq:rescalemaster}),
conduisent aux conditions de bord 
\ba
\left(\partial_\xi+\lambda(\tau)\right)\Omega\vert_{\xi = 0} = -\kappa^2{\left(\rho-\sigma\right) a^{3/2}\over p^2}\Delta(\tau),
\ea
où $\lambda$ a été donné en (\ref{eqnarray:v1}).

L'équation du mouvement pour les perturbations de matière sur la brane peut
également être obtenue à partir des conditions de bord (\ref{eqnarray:sbc}) en
imposant une équation d'état sur les perturbations de matière. Nous
considérons des perturbations adiabatiques de matière (\emph{i.e} sans entropie)
\ba
\delta P & = & c_s^2 \delta\rho,
\ea
où $c_s^2 = \dot{P}/\dot{\rho}$ est la vitesse du son du fluide. On obtient l'équation du mouvement pour le contraste de densité $\Delta$, similaire à l'équation
obtenue dans
 \cite{scal},
\ba\label{eqnarray:eqm}
& \ddot{\Delta}+ & \left(8+3c_s^2-6\epsilon\right)H\dot{\Delta}+\left[{p^2c_s^2\over
  a^2}+\left(10+6c_s^2-14\epsilon+3\epsilon^2\right){\kappa^2\rho_M\over
  2\ell}+\left(5+3c_s^2-9\epsilon\right){\kappa^4\rho_M^2\over 12}\right]\Delta\cr
& = & \epsilon{p^4\over 3 a^5}a^{3/2}\Omega\vert_{\xi = 0},
\ea
où, selon la dynamique des Univers branaires \cite{bdel},\cite{deff1}, présentée à la section \ref{subsec:bdel},
\ba\label{qqq:bwdyn0}
H^2 & = & {\kappa^2\over 3\ell}\rho_M\left(1+{\kappa^2\ell\over
    12}\rho_M\right) = -{1\over\ell^2}+{\kappa^4\over 36}\rho^2,\cr
\dot{\rho} & = & -3H(P+\rho) = -3H(1+w)\rho_M,\cr
 \epsilon & = & 1+w, \qquad \rho_M = \rho-\sigma, \qquad P_M = P+\sigma.
\ea
L'équation d'état du fluide parfait sur la brane est $P_M = w\rho_M$. Nous pouvons aussi changer d'échelle le degré de liberté sur la brane comme $e^{(1/2)\int
  \left(8+3c_s^2-6\epsilon\right)H d\tau}\Delta \rightarrow \Delta$ pour obtenir une équation d'oscillateur. Finalement les équations couplant les perturbations scalaires de la brane aux perturbations scalaires du bulk peuvent se résumer sous la forme générique
 \ba
\ddot{\Omega}-\Omega''+\bar{\omega}^2(\tau)\Omega+\lambda^2(\tau)\Omega & = & 0, \cr
\Omega'\vert_{\xi = 0}+\lambda(\tau)\Omega\vert_{\xi = 0} & = &
-\gamma_1(\tau)\Delta,\cr
\ddot{\Delta}+\omega_0^2(\tau)\Delta & = & \gamma_2(\tau) \Omega\vert_{\xi =
0}\label{eqnarray:genm},
\ea
où $\lambda$ et $\bar{\omega}$ ont été donné en (\ref{eqnarray:omtilde}) et
\ba\label{eqnarray:param}
\omega^2_0(\tau) & = & {p^2c_s^2\over
  a^2}+\left(10+6c_s^2-14\epsilon+3\epsilon^2\right){\kappa^2\rho_M\over
  2\ell}+\left(5+3c_s^2-9\epsilon\right){\kappa^4\rho_M^2\over 12}\cr
  & & -{1\over 4}\left(8+3c_s^2-6\epsilon\right)^2H^2-{1\over 2}\left(8+3c_s^2-6\epsilon\right)\dot{H}+9H^2\epsilon(\epsilon-c_s^2-1),\cr
\gamma_1(\tau) & = & \kappa^2{\rho_M a^{3/2}\over p^2}e^{-(1/2)\int \left(8+3c_s^2-6\epsilon\right)H d\tau},\cr
\gamma_2(\tau) & = & \epsilon{p^4\over 3 a^5}a^{3/2}e^{+(1/2)\int \left(8+3c_s^2-6\epsilon\right)H d\tau}.
\ea
L'observation immédiate, en référence à la section \ref{sec:secone}, est l'absence de couplages dérivatifs en temps pour le système brane-bulk dans (\ref{eqnarray:genm}). Par conséquent la dissipation éventuelle du fluide sur la brane vers le bulk ne peut être que non-locale malgré que le fluide parfait soit local (\emph{i.e.} vraiment localisé sur la  brane). Dit autrement, aucun terme de friction locale de la forme $\Gamma(\tau)\partial_\tau$
ne peut apparaitre dans l'équation effective sur la brane, cela signifie que la dissipation éventuelle du fluide sur la brane contient nécessairement une mémoire ou un délai en temps. La non-localité de l'interaction brane-bulk provient des effets de courbure.

\begin{figure}[htbp]
\begin{center}
\includegraphics[width=9cm]{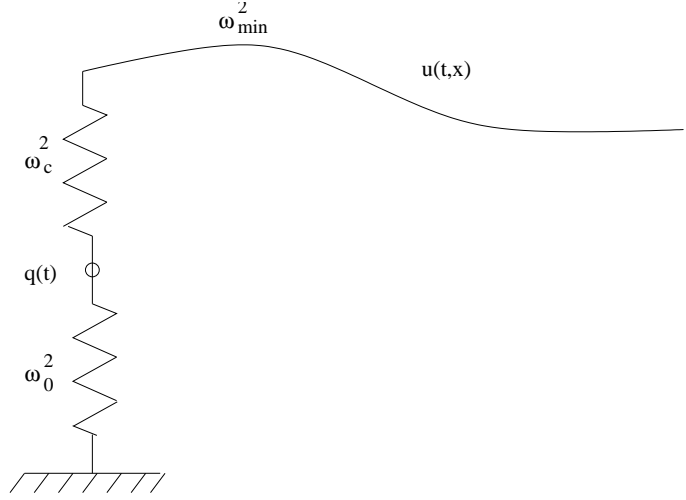}
\end{center}
\caption{\small{{\bf Modèle mécanique pour les perturbations branaires
scalaires} dans le cas de perturbations adiabatiques d'un fluide parfait. 
La corde, représentant les gravitons du bulk, est couplée, par un ressort
intermédiaire $\omega^2_c$, à un oscillateur harmonique représentant les perturbations du fluide sur la brane.}} \label{fig:spring}
\end{figure}

La physique sous-jacente peut être illustrée en utilisant le modèle mécanique suivant. Considérons une corde couplée par son extrémité à un oscillateur harmonique avec deux ressorts, comme indiqué sur la figure Fig. \ref{fig:spring}.
Les équations pour ce modèle sont similaires à celles obtenues pour les
perturbations scalaires du système brane-bulk : 
\ba
\ddot{u}-u''+\omega^2_{min}(t) u & = & 0,\label{eqnarray:string} \cr
\gamma(t) u'\vert_{x = 0}-\omega^2_c(t) u\vert_{x = 0} & = &
-\omega^2_c(t) q(t),\label{eqnarray:link}\cr
\ddot{q}+\left(\omega_0^2(t)+\omega_c^2(t)\right) q & = & \omega^2_c(t) u\vert_{x = 0},\label{eqnarray:osc}
\ea
où $\gamma $ est la tension de la corde exprimée comme une fréquence, et
$\omega _0^2$ et $\omega _c^2$ sont les constantes de raideur des deux ressorts exprimées comme des fréquences au carré.
Le ressort intermédiaire entre la masse et la corde joue le rôle d'un
tampon. En l'absence de ce tampon, le couplage avec la corde introduirait un
terme local de dérivée première en temps (terme de friction) dans l'équation
effective pour l'oscillateur harmonique, indiquant alors une dissipation
locale de l'oscillateur vers la corde. Ce n'est pas le cas ici.

En l'absence du terme de masse (force de rappel) sur la corde, les équations seraient
\ba
\ddot u(x,t)-u^{\prime \prime }(x,t)&=&0,\cr
\gamma u^{\prime }(x=0,t)-{\omega _c}^2u(x=0,t)&=&-{\omega _c}^2q(t),\cr
\ddot q(t)+
({\omega _0}^2+{\omega _c}^2)q(t)-
{\omega _c}^2u(x=0,t)
&=&
f(t),
\ea
où
$f(t)$ est un terme de for\c{c}age. Pour des coefficients indépendant du temps, les Ansätze
$u(x,t)=u\cdot \exp [i\omega (x-t)],$
$q(t)=q\cdot \exp [-i\omega t],$ et
$f(t)=f\cdot \exp [-i\omega t]$
entrainent
\ba
-\omega ^2q(\omega )+
({\omega _0}^2+{\omega _c}^2)q(\omega )-
\frac{{\omega _c}^4}{\omega _c^2-i\gamma \omega }q(\omega )=f(\omega ),
\ea
qui équivaut à
\ba
\ddot q(t)+
({\omega _0}^2+{\omega _c}^2)q(t)+
\int _0^\infty ds~K(s)~q(t-s)
&=&
f(t),
\label{qqq:a}
\ea
où
\ba\label{qqq:kno}
K(t)&=&-\int _{-\infty }^{+\infty }
\frac{d\omega }{(2\pi )}
\frac{{\omega _c}^4~\exp [-i\omega t]}{{\omega _c}^2-i\omega \gamma }\cr
&=&-\frac{{\omega _c}^4}{\gamma }~\theta (t)~\exp [-{\omega _c}^2t/\gamma ]\cr
&=&-({\omega _c}^2\lambda )~\theta (t)~\exp [-\lambda t]
\ea
et $\lambda ={\omega _c}^2/\gamma .$ Par conséquent, l'équation (\ref{qqq:a})
devient
\ba
\ddot q(t)+
({\omega _0}^2+{\omega _c}^2)q(t)-
{\omega _c}^2\lambda
\int _0^\infty ds~\exp [-\lambda s]~q(t-s)
&=&
f(t).
\label{qqq:b}
\ea
Dans le cas limite où $\lambda \gg \omega _0, \omega _c,$ on peut faire le développement
\ba
\lambda\int _0^\infty ds~\exp [-\lambda s]~q(t-s)\approx q(t)-\lambda^{-1}\dot q(t),
\ea
tel que l'équation(\ref{qqq:b}) peut être approximée par
\ba
\ddot q(t)+\gamma
\dot q(t)
+{\omega _0}^2q(t)=f(t).
\label{qqq:c}
\ea

\begin{figure}[htbp]
\begin{center}
\includegraphics[width=15cm]{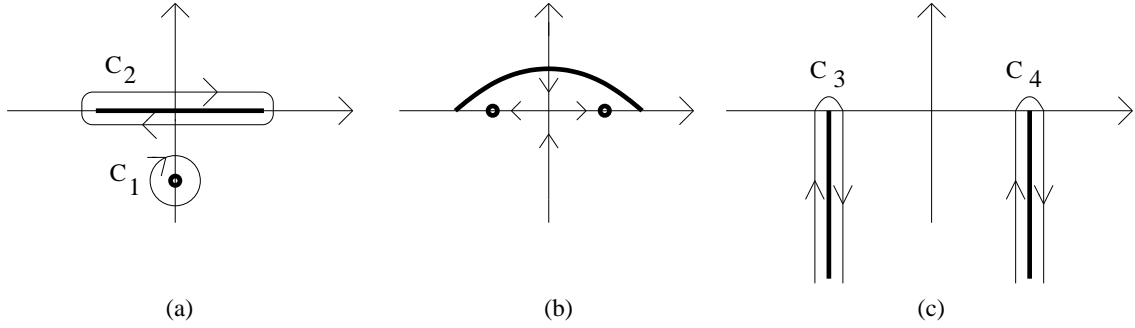}
\end{center}
\caption{\small{{\bf Propriétés d'analyticité du noyau dans le plan complexe}. Schéma (a):
si $\omega_{min} < \lambda=\omega_c^2/\gamma$ alors il y a un pôle isolé sur
l'axe imaginaire négatif et une coupure $[-\omega_{min},+\omega_{min}]$
sur l'axe réel. Schémas (b) et (c): si $\omega_{min} >
\lambda=\omega_c^2/\gamma$ alors le pôle se déplace vers l'axe réel, sous la
coupure, ainsi que son image mirroir qui vient du second feuillet de Riemann. 
On peut déformer la coupure en la repoussant vers l'infini négatif (schéma (c)).}  } \label{fig:pole}
\end{figure}

Nous considérons ensuite l'équation d'onde pour la corde incluant le terme de masse (force de rappel), de sorte qu'il y a maintenant une fréquence minimum $\omega _{min}$
pour exciter les modes de la corde (ou les modes se propageant dans le bulk). On a
\ba
\ddot u-u^{\prime \prime }+{\omega_{min}^2}u=0.
\ea
Dans ce cas le noyau est modifié pour devenir
\ba
K(t)=-\int _{-\infty }^{+\infty }
\frac{d\omega }{(2\pi )}
\frac{{\omega _c}^4~\exp [-i\omega t]}{{\omega _c}^2
-i\gamma \sqrt{\omega ^2-{\omega_{min}^2}}}.
\ea
Lorsque $\omega _{min}$ n'est pas trop grand, $\omega _{min} < \lambda =
\omega_c^2/\gamma$, il y a un pôle isolé en $\omega =-i\Lambda $, où
\ba
\Lambda & = & \sqrt{\frac{{\omega _c}^4}{\gamma ^2}-{\omega_{min}^2}} =  \sqrt{\lambda^2-\omega _{min}^2},
\ea
et une coupure sur l'axe réel s'étendant de
$\omega =-\omega _{min}$ à $\omega =+\omega _{min}.$
Pour obtenir un noyau causal, on doit prendre un contour passant au-dessus de la coupure sur l'axe réel. Pour $t>0,$ où le noyau a son support, nous évaluons le contour en le déformant tel qu'il y ait deux contributions
$K(t)=K_1(t)+K_2(t)$ (voir Fig. \ref{fig:pole}(a)).
Le premier terme a pour origine le contour encerclant le pôle dans le sens des
aiguilles d'une montre :
\ba
K_1(t)=
-(\omega _c^2\lambda)\theta(t) \exp [-\Lambda t]
\frac {\sqrt{\Lambda ^2+{\omega_{min}^2} }}{\Lambda }.
\ea
Le second a pour origine le contour encerclant la coupure dans le sens des
aiguilles d'une montre, comme indiqué à la figure Fig.\ref{fig:pole}(a) :
\ba
K_2(t)=
-\omega_c^2\lambda \int _{-\omega _{min}}^{+\omega _{min}}
\frac{d\omega }{(2\pi )}
\exp [-i\omega t]\left[\frac{1}
{\sqrt{{\omega_{min}^2}-\omega ^2}+\lambda }-\frac{1}
{-\sqrt{{\omega_{min}^2}-\omega ^2}+\lambda }\right].
\ea

Quand $\omega_{min}$ approche $\lambda = \omega_c^2/\gamma$, le pôle sur le demi-plan inférieur s'approche de $\omega = 0$. Il y a un second pôle, image miroir du premier, $\omega = +i\Lambda$. Cependant ce pôle réside sur l'autre feuillet de Riemann, caché derrière la coupure. Lorsque $\omega_{min}$ excède $\lambda =
\omega_c^2/\gamma$, ces deux pôles entrent en collision et diffusent à angle droit en résidant sur la ligne réelle de la coupure (ou juste en dessous après la déformation faite à la figure Fig. \ref{fig:pole}(b)). On peut faire disparaître ces deux pôles en déformant la coupure vers le bas comme indiqué à la figure Fig. \ref{fig:pole}(c). La coupure a été repoussée vers l'infini négatif. Désormais le noyau intégral peut être exprimé comme
\ba
K(t) & = & -\omega^2_c\lambda\int_{C_3}\frac{d\omega }{(2\pi )}
\frac{\exp [-i\omega t]}{\lambda
-i\sqrt{\omega ^2-{\omega_{min}^2}}}-\omega^2_c\lambda\int_{C_4}\frac{d\omega }{(2\pi )}
\frac{\exp [-i\omega t]}{\lambda
-i \sqrt{\omega ^2-{\omega_{min}^2}}},
\ea
où les contours $C_3$ and $C_4$ sont indiqués à la figure Fig. \ref{fig:pole}(c). On peut réécrire cela comme
\ba\label{qqq:kyes}
K(t) & = & -i \omega^2_c\lambda\int_0^\infty \frac{dy}{2\pi} \left( \frac{e^{-yt}e^{-i\omega_{min} t}}{\lambda-i\sqrt{y(-y-i2\omega_{min})}}-\frac{e^{-yt}e^{-i\omega_{min} t}}{\lambda+i\sqrt{y(-y-i2\omega_{min})}}\right)\cr
     & & - i \omega^2_c\lambda\int_0^\infty \frac{dy}{2\pi} \left( \frac{e^{-yt}e^{+i\omega_{min} t}}{\lambda-i\sqrt{y(-y+i2\omega_{min})}}-\frac{e^{-yt}e^{+i\omega_{min} t}}{\lambda+i\sqrt{y(-y+i2\omega_{min})}}\right).
   \ea
En fait cette expression tient toujours pour $\omega_{min} < \lambda =
\omega_c^2/\gamma$.
À des temps très grands, le noyau
 (\ref{qqq:kyes}) se comporte asymptotiquement comme
\ba
K(t) &\stackrel{t\rightarrow\infty}{\approx}&
\mathcal{O}(1)\gamma\sqrt{\omega_{min}}{\cos\left(\omega_{min}t\right)-\sin\left(\omega_{min}t\right)\over
  t^{3/2}}.
\ea

Il semble peu aisé de manipuler les expressions du noyau en espace réel. C'est pourquoi nous calculons aussi le noyau dans l'espace de Fourier pour une étude complémentaire. Avec le terme de for\c{c}age $f(t) =
f\cdot \exp [-i\omega t]$, l'équation de l'oscillateur harmonique est
\ba
-\omega ^2q(\omega )+
({\omega _0}^2+{\omega _c}^2)q(\omega )-
\frac{{\omega _c}^2\lambda}{\lambda-i\sqrt{\omega^2-\omega^2_{min}} }q(\omega )
& = & f(\omega ),\cr
\hat{G}^{-1}(\omega)q(\omega) & = & f(\omega).
\ea
La moyenne dans le temps de la puissance transmise à l'oscillateur (qui est une mesure de la dissipation) est
\ba
\bar{P} & = & {1\over T}\int_0^T f(t)\dot{q}(t)dt,\cr
        & = & \omega\left(\Re [f] \Im [q] - \Im [f] \Re [q]\right),\cr
        & = & \vert f\vert^2 \omega \Im \left[\hat{G}(\omega)\right].
\ea
Si le noyau n'a pas de partie imaginaire, le système échange de l'énergie mais ne se dissipe pas puisque la puissance moyenne s'annule. Pour $\omega < \omega_{min}$ la partie imaginaire du noyau s'annule justement. De l'expression de  $\hat{G}(\omega)$, on déduit la puissance dissipée moyenne
\ba
\bar{P} & = & \vert f\vert^2  {\omega^2_c\lambda\omega
  \sqrt{\omega^2-\omega^2_{min}}\over
  \lambda^2\left(\omega_0^2-\omega^2\right)^2+\left(\omega^2-\omega^2_{min}\right)\left(\omega_0^2+\omega^2_c-\omega^2\right)^2}~~\qquad (\omega^2 > \omega_{min}^2).
\ea
La puissance dissipée est nulle lorsque
$\omega^2 \leq \omega_{min}^2$. On peut comprendre qualitativement comment le comportement change quand $\omega _{min}$ varie. La corde peut être vue comme un filtre passe-haut avec le seuil $\omega _{min}$. Seule la radiation de plus haute fréquence peut s'échapper dans la corde vers l'infini, conduisant ainsi à un flux d'énergie loin de l'oscillateur.  À plus basse fréquence la corde est excitée autour de l'oscillateur mais de fa\c{c}on localisée avec une amplitude décroissant de manière exponentielle à distance de l'oscillateur puisque le moment $k = \sqrt{\omega^2-\omega^2_{min}}$ de l'équation d'onde de la corde devient imaginaire.
 Dans ce cas il n'y a pas de dissipation à long terme parce que la corde vibre en phase avec l'oscillateur. Bien-sûr le couplage décale la fréquence de l'oscillateur parce que la corde ajoute de l'inertie et une force de rappel sur l'oscillateur.

Pour revenir au système brane-bulk, le seuil est donné par
\ba
 \omega^2_{min} = \lambda^2+\bar{\omega}^2
 \ea
où $\lambda$, $\bar{\omega}$ ont été calculés en
(\ref{eqnarray:omtilde}). Dans le cas d'un couplage faible avec le bulk ($\dot{H}\ll H^2$),
la fréquence du degré de liberté sur la brane est seulement légèrement décalée de la valeur nue $\omega_0$ donnée à l'équation (\ref{eqnarray:param}). Aux échelles subhorizon ($p\gg aH$),
\ba
0 < \omega_0^2 \approx {c_s^2 p^2\over a^2}+\mathcal{O}\left(\dot{H}\right) & < &
\omega^2_{min}\approx {p^2\over a^2}+\mathcal{O}\left(\dot{H}\right)
\ea
puisque $c_s^2\leq 1$. Donc la puissance dissipée est exactement zéro. Dans
l'approximation adiabatique, valide aux échelles subhorizon, le fluide ne peut
pas se dissiper dans le bulk parce que la fréquence naturelle des modes du
bulk,  à nombre d'onde égal par ailleurs, est trop haute, empêchant les
perturbations du fluide d'entrer en résonance avec les gravitons du bulk.

\section{Discussion}\label{sec:discuss}

Dans ce chapitre nous avons presenté nos résultats \cite{moi}. Nous avons
proposé des approches analytiques pour estimer l'ordre de
magnitude des effets dissipatifs en cosmologie branaire. Nous avons supposé
que l'expansion de l'Univers était adiabatique, dans le sens où $\dot{H}\ll
H^2$. Nous avons exploré l'évolution des perturbations scalaires et des
perturbations tensorielles. Dans le cas des perturbations scalaires, avec un
champ scalaire en roulement lent (Inflaton) sur la brane, sans aucune autre
approximation nous avons obtenu le taux de dissipation local de l'Inflaton dû
à l'interaction avec les gravitons du bulk: $\Gamma\sim
H\left(\dot{H}/H^2\right)\left(H\ell/\sqrt{1+H^2\ell^2}\right)$, qui est
linéaire dans le facteur de slow-roll. C'est notre principal résultat et il
s'accorde avec les résultats numériques obtenus par Koyama \&
al. dans \cite{scalinf3}: dans le régime de haute énergie de l'Inflation
branaire ($H\ell\gg 1$), les corrections à l'Inflation standard dûes au
couplage avec les perturbations de métrique du bulk sont également trouvées
linéaires dans le facteur de slow-roll. De plus la dépendance en $H\ell$ du
taux de dissipation que nous avons trouvé correspond exactement au profil de
la correction slow-roll tracée numériquement dans \cite{scalinf3}. Dans le cas des perturbations
scalaires avec un fluide parfait sur la brane ainsi que dans le cas des perturbations
tensorielles, nous avons fait face à des processus de dissipation non-locaux,
cela nous a amené à simplifier la géométrie de fond dans ces cas : nous avons considéré
que la géométrie de fond proche de la brane est une approximation fiable pour
décrire l'interaction brane-bulk, au moins les processus dissipatifs. Nous
avons explicitement éliminé les effets non-locaux dûs à la diffraction des
gravitons émis dans le bulk inhomogène $AdS$ en approximant cette inhomogénéité
(potentiel-plateau). On s'est concentré de cette fa\c{c}on sur les processus de
dissipation. Cette approximation est bien légitime dans le régime de haute
énergie de l'Inflation branaire ($H\ell\gg 1$) parce que la courbure du bulk
devient négligeable comparée à l'accélération de la brane. 
En utilisant une approche de ``resommation quantique'' nous avons été capable
de séparer les transitions entre les quanta de la brane dûes à l'expansion
non-uniforme de l'Univers des transitions entre les quanta de la brane et les
quanta du bulk dûes à l'interaction avec la dimension supplémentaire. 

À cause de l'expansion non-uniforme de la brane, les degrés de liberté
localisés sur la brane interagissent avec les gravitons du bulk, ce qui
conduit éventuellement à de la dissipation sur la brane d'un point de vue
quadridimensionnel. Ces effets sont inscrits dans le propagateur du bulk
$AdS$. Nous avons calculé l'atténuation du mode croissant de l'état lié du
graviton aux échelles superhorizon, due à l'interaction avec les modes KK du
bulk. Nous avons trouvé que le taux de dissipation de l'état lié aux échelles
superhorizon est de l'ordre de $\Gamma\sim
H(\dot{H}/H^2)^2$ lorsque $H\ell\gg 1$.
Nous avons aussi montré que les perturbations adiabatiques d'un fluide parfait
sur la brane ne peuvent pas se dissiper dans la dimension supplémentaire aux
échelles subhorizon parce que la fréquence minimum d'excitation des modes KK du bulk, à
nombre d'onde égal à celui du fluide, est trop haute pour pouvoir obtenir une
résonance.  Nous avons rencontré des difficultés pour appliquer des méthodes
analytiques dans le cas du fluide parfait aux échelles superhorizon, parce
qu'à de telles échelles l'expansion de l'Univers n'est plus adiabatique, au
moins tant que le fluide domine la tension de la brane. Nous avons également
montré qu'un champ scalaire en roulement lent sur la brane (``slow-roll
inflaton'') se dissipe de fa\c{c}on locale à n'importe quelle échelle, alors
que l'état lié du graviton se dissipe de fa\c{c}on non-locale. De plus
l'inflaton dissipe avec un taux plus important $\Gamma\sim H(\dot{H}/
H^2)$, linéaire dans le paramètre adiabatique de roulement lent, que le taux
de dissipation de l'état lié du graviton $\Gamma\sim
H(\dot{H}/H^2)^2$, quadratique dans le paramètre
adiabatique de roulement lent. Cette différence est la conséquence du
caractère local de l'interaction de l'inflaton avec les modes du bulk
contrairement au caractère non-local de l'interaction de l'état lié du
graviton avec les modes du bulk.

L'approche analytique développée ici repose sur le calcul du propagateur
retardé effectif sur la brane. Comme nous avons pris en compte les effets de rétroaction à la fois pour les
perturbations tensorielles et pour les perturbations scalaires sur la brane
inflationnaire, nous sommes désormais en mesure de calculer les corrections
branaires au rapport tenseur-scalaire $T/S$ standard de l'Inflation à un
champ. Ce sera l'objet de nos futures recherches. Il serait également 
intéressant d'aller au-delà de l'approximation du potentiel-plateau
pour prendre en compte l'inhomogénéité du bulk $AdS$ et la diffraction des
gravitons par le bulk courbe qui entraine des effets non-locaux sur la
brane. La non-localité est inscrite dans le propagateur du bulk $AdS$ et
dépend fortement de la courbure du bulk. En ce sens l'expression exacte du
propagateur du graviton dans $AdS$ est nécessaire pour calculer les processus
non-locaux. La géométrie courbe du bulk entraine également l'existence de
gravitons métastables (c'est-à-dire des états quasi-liés sur la brane), même
si la brane est statique, comme l'a montré Seahra dans \cite{seahra}.
On pourrait comparer le flux des gravitons du bulk qui
se propagent par tunnel quantique avec le flux des états quasi-liés. Ce sera
l'objet d'une future publication. Notons enfin que les méthodes employées ici
peuvent s'appliquer à tout autre modèle de théorie des champs en interaction
avec un bord ou une sous-variété localisant une partie des champs.


\chapter{Le propagateur retardé covariant du graviton dans l'espace Anti-de
  Sitter}\label{chapter:adsprop}

Nous avons vu au chapitre \ref{chapter:dissip} que, pour un Univers branaire
en expansion non-uniforme, des gravitons pouvaient être émis dans le bulk
$AdS$. Nous avons montré que cela apparaissait comme de la dissipation du
point de vue d'un observateur confiné sur la brane et nous avons estimé les
taux de dissipation de différentes perturbations de matière sur la brane en
utilisant une approche de "resommation quantique", où le propagateur du bulk
$AdS$ est systématiquement inséré à tous les ordres de perturbation en les
couplages pour obtenir le propagateur effectif sur la brane.  À cause de la
courbure du bulk $AdS$, les gravitons qui ont été émis de la brane peuvent
aussi être diffractés dans le bulk courbe puis ré-absorbés sur la brane et
transformés de nouveau en quanta sur la brane, conduisant ainsi à de la non-localité du point de vue quadridimensionnel d'un observateur sur la brane. Cette non-localité est inscrite dans la forme du propagateur du bulk $AdS$ pour le graviton. Comme des effets non-locaux dépendent forcément de la courbure, il s'avère nécessaire de connaître la forme exacte du propagateur dans $AdS$ si l'on veut estimer correctement la magnitude des processus non-locaux sur la brane. Les calculs explicites  du propagateur retardé du graviton dans $AdS$, effectués dans ce chapitre, étaient motivés par le problème du calcul des perturbations cosmologiques dans un Univers branaire en expansion.

Dans ce chapitre nous trouvons la forme explicite du propagateur retardé du
graviton dans $AdS^{d+1}$, où $d$ est arbitraire, de fa\c{c}on complètement
covariante et classique. La méthode utilisée pour obtenir les fonctions de
Green dans la première partie du chapitre est essentiellement celle employée
par D'Hoker, Freedman, Mathur, Matusis et Rastelli \cite{ads}, où les mêmes
techniques  ont été appliquées pour calculer la version Euclidienne de la
fonction à deux points, sans fixer de jauge et en séparant explicitement
parties physiques et parties de pure jauge. Dans la seconde partie du chapitre
nous essayons de déterminer la forme \emph{retardée} du propagateur du
graviton par des considérations purement classiques. En fait nous souhaitons
montrer, sans avoir recours à des arguments quantiques, que la prescription pour calculer la forme retardée du propagateur du graviton dans $AdS$ est exactement
\ba\label{qqq:prescription}
G^{ret}_{\mu\nu\mu'\nu'}(I) = \theta(t-t')\left[G_{\mu\nu\mu'\nu'}(I+i0)-G_{\mu\nu\mu'\nu'}(I-i0)\right],
\ea
 si $G_{\mu\nu\mu'\nu'}(I)$ est la fonction à deux points du graviton dans
 $AdS$ et $I(P,P')$ est la séparation invariante relativiste entre deux
 points $P,P'$ sur la variété $AdS$. En effet, la littérature ne nous semble pas très claire pour justifier la prescription (\ref{qqq:prescription}) pour construire la forme retardée du propagateur dans un espace courbe.  Plusieurs tentatives s'appuient sur les relations de commutations canoniques ou le choix d'un vide quantique \cite{allen}, alors que le propagateur retardé est un objet \emph{classique}.
 Les notations utilisées dans ce chapitre sont $A_{\mu,\nu}\equiv \partial_\nu
 A_\mu$, $A_{\mu;\nu}\equiv \nabla_\nu A_\mu$ et $\Box = g^{\mu\nu
 ~(0)}\nabla _\mu \nabla _\nu$, où $\nabla_\mu$ est la dérivée covariante
 associée à la métrique du fond $AdS$ non-perturbée $g^{\mu \nu  ~(0)}$. Le
 travail que nous exposons dans ce chapitre est informel, il n'a pas été soumis à publication.

\section{Fonction de Green covariante du graviton dans $\boldsymbol{AdS^{d+1}}$}\label{sec:deuxpoints}

\subsection{Équation d'Einstein linéarisée covariante dans $\boldsymbol{AdS^{d+1}}$}\label{subsec:coveom}

La métrique $g_{\mu \nu
}^{(0)}$ de l'espace-temps $AdS^{d+1}$ de rayon de courbure $\ell^2$ est la solution des équations d'Einstein à $(d+1)$ dimensions
\ba
R^{(0)}_{\mu\nu}-{1\over 2}g^{(0)}_{\mu \nu} R^{(0)} & = & -\Lambda_5 g^{(0)}_{\mu \nu},
\ea
où la constante cosmologique négative vaut $\Lambda_5 = -d(d-1)/(2\ell^2)$. L'espace $AdS^{d+1}$ étant maximalement symétrique, on en déduit aisément respectivement le tenseur de Riemann, le tenseur de Ricci et le scalaire de courbure :
\ba\label{qqq:zerotenseurs}
R^{(0)}_{\mu\nu\rho\sigma} = -{1\over\ell^2}\left(g^{(0)}_{\mu \rho}g^{(0)}_{\nu \sigma}-g^{(0)}_{\mu \sigma}g^{(0)}_{\nu\rho}\right),\qquad R^{(0)}_{\mu\nu} = -{d\over\ell^2}g^{(0)}_{\mu \nu},\qquad R^{(0)} = -{d(d+1)\over\ell^2}.
\ea
Sous l'influence d'une perturbation de matière $T_{\mu\nu}^{(1)}$ apparaissant à l'ordre un, la nouvelle métrique perturbée
$g_{\mu \nu
}$ de $AdS^{d+1}$
\ba g_{\mu \nu } & = & g_{\mu \nu
}^{(0)}+h_{\mu \nu },
\ea
diffère de la métrique de $AdS^{d+1}$ par la perturbation $h_{\mu\nu}$ à l'ordre un, et est solution des équations d'Einstein\footnote{Le couplage avec la matière est posé egal à $\kappa^2 \equiv 1$.}
\ba
R_{\mu\nu}-{1\over 2}g_{\mu \nu
} R & = & -\Lambda_5g_{\mu \nu
}+T_{\mu\nu}^{(1)}
\ea
qui peuvent se ré-écrire sous la forme
\ba\label{qqq:covlineins}
R_{\mu\nu}+d g_{\mu \nu
} & = & \bar{T}_{\mu\nu}^{(1)}
\ea
où on a défini le tenseur énergie-impulsion ``trace-modifié'' ($T \equiv
T_\sigma^\sigma$) :
\ba\label{qqq:tracemod}
\bar{T}_{\mu \nu } =T_{\mu \nu
}-{1\over d-1}g_{\mu \nu }^{(0)}T.
\ea
En gravité linéarisée dans $AdS^{d+1}$, les gravitons correspondent aux perturbations $h_{\mu\nu}$ de la métrique de fond $g_{\mu \nu
}^{(0)}$ et satisfont les équations d'Einstein (\ref{qqq:covlineins}) à l'ordre linéaire :
\ba\label{qqq:covlineins2}
R^{(1)}_{\mu\nu}+dh_{\mu \nu
} & = & \bar{T}_{\mu\nu}^{(1)}.
\ea
Les symboles de Christoffel se scindent en une partie d'ordre zéro et une partie d'ordre un
\ba
\Gamma^a_{bc} & = & {1\over 2}g^{ad}\left(g_{dc,b}+g_{bd,c}-g_{bc,d}\right)\cr
              & = & \Gamma^{(0)a}_{bc}+\Gamma^{[h]a}_{bc},
\ea
où
\ba\label{qqq:gamma1}
\Gamma^{[h]a}_{bc} & = & {1\over 2}g^{(0)ad}\left(h_{dc,b}+h_{bd,c}-h_{bc,d}\right)-{1\over 2}h^{ad}\left(g^{(0)}_{dc,b}+g^{(0)}_{bd,c}-g^{(0)}_{bc,d}\right)\cr
& = & {1\over 2}g^{(0)ad}\left(h_{dc,b}+h_{bd,c}-h_{bc,d}\right)-{1\over 2}g^{(0)ad}h^{ed}g^{(0)ef}\left(g^{(0)}_{fc,b}+g^{(0)}_{bf,c}-g^{(0)}_{bc,f}\right)\cr
& = & {1\over 2}g^{(0)ad}\left(h_{dc,b}+h_{bd,c}-h_{bc,d}\right)-g^{(0)ad}\Gamma^{(0)e}_{bc}h^{ed}\cr
& = & {1\over 2}g^{(0)ad}\left(h_{dc,b}+h_{bd,c}-h_{bc,d}\right)\cr
&   & +{1\over 2}g^{(0)ad}\left(-\Gamma^{(0)e}_{bc}h^{de}-\Gamma^{(0)e}_{cb}h^{ed}-\Gamma^{(0)e}_{bd}h^{ec}
+\Gamma^{(0)e}_{db}h^{ec}-\Gamma^{(0)e}_{cd}h^{be}+\Gamma^{(0)e}_{dc}h^{be}\right)\cr
& = & {1\over 2}g^{(0)ad}\left(h_{dc;b}+h_{bd;c}-h_{bc;d}\right).
\ea
Le tenseur de Riemann à l'ordre un vaut quant à lui
\ba\label{qqq:riemann1}
R^{(1)a}_{bcd} & = & \Gamma^{[h]a}_{bd,c}-\Gamma^{[h]a}_{bc,d}+\Gamma^{[h]a}_{ec}\Gamma^{(0)e}_{bd}+\Gamma^{[h]e}_{bd}\Gamma^{(0)a}_{ec}- \Gamma^{[h]a}_{ed}\Gamma^{(0)e}_{bc}-\Gamma^{[h]e}_{bc}\Gamma^{(0)a}_{ed}\cr
& = & \Gamma^{[h]a}_{bd;c}-\Gamma^{[h]a}_{bc;d}.
\ea
On déduit de (\ref{qqq:riemann1}) et (\ref{qqq:gamma1}) que le tenseur de Ricci à l'ordre un $R^{(1)}_{ab} = R^{c(1)}_{acb}$ apparaissant dans les équations d'Einstein linéarisées (\ref{qqq:covlineins2}) vaut
\ba
R^{(1)}_{ab} & = & -{1\over 2}\left(h_{ab;cc}+h_{;ab}-h_{ac;bc}-h_{bc;ac}\right).
\ea
En notant que $h_{ac;bc}+h_{bc;ac} = h_{ac;cb}+h_{bc;ca}+h_{ac;[bc]}+h_{bc;[ac]}$ et utilisant la relation $[\nabla_d,\nabla_c]h_{ab} = R^{(0)}_{aedc}h^e_b+R^{(0)}_{bedc}h_a^e$, on peut ré-écrire
\ba\label{qqq:ricci1}
R^{(1)}_{ab} & = & -{1\over 2}\left(h_{ab;cc}+h_{;ab}-h_{ac;cb}-h_{bc;ca}+2R^{(0)}_{acbd}h^{cd}-2R^{(0)}_{ac}h_b^c\right).
\ea
En insérant (\ref{qqq:ricci1}) dans (\ref{qqq:covlineins2}), connaissant les expressions (\ref{qqq:zerotenseurs}), on obtient l'équation de propagation complètement covariante du graviton dans $AdS^{d+1}$ avec source $T^{(1)}_{\mu\nu}$ :
\ba
\Box h_{\mu\nu}+\nabla _{\mu}\nabla _{\nu}h-\nabla _{\mu}\nabla
_{\sigma}h_{\sigma\nu}-\nabla _{\nu}\nabla
_{\sigma}h_{\mu\sigma}+{2\over\ell^2}(h_{\mu\nu}-g^{(0)}_{\mu\nu}h) = -2\bar{T}^{(1)}_{\mu\nu},
\label{qqq:general}
\ea
où $\bar{T}_{\mu\nu}$ est défini en (\ref{qqq:tracemod}). Bien-sûr on pourrait choisir la jauge covariante de Lorentz $\nabla_\nu\left(h_{\mu \nu}-{1\over 2}g_{\mu \nu }^{(0)}h\right) = 0$ de fa\c{c}on à simplifier l'équation du mouvement selon
\ba
\Box h_{\mu\nu}+{2\over\ell^2}(h_{\mu\nu}-g_{\mu\nu}h) = -2\bar{T}^{(1)}_{\mu\nu},
\ea
mais on préfèrera travailler avec l'équation (\ref{qqq:general}) parce qu'elle
est complètement covariante, sans choix de jauge particulier.

\subsection{Fonction de Green}\label{subsec:green}

La solution $h_{\mu\nu}$ de l'équation (\ref{qqq:general}) est connectée à la source $T^{(1)}_{\mu\nu}$ selon
\ba\label{eqnarray:ht} h_{\mu\nu}(\underline{x}) & = & \int
d^{d+1}\underline{x}'\sqrt{g'}G_{\mu\nu\mu'\nu'}(\underline{x},\underline{x}')T^{(1)}_{\mu'\nu'}(\underline{x}'),
\ea
où la fonction de Green $G_{\mu\nu\mu'\nu'}(\underline{x},\underline{x}')$ satisfait
\ba\label{eqnarray:eomg}
& & \Box G_{\mu \nu \mu '\nu '}+\nabla _{\mu}\nabla _{\nu}G_{\sigma \sigma \mu '\nu '}-\nabla _{\mu}\nabla
_{\sigma}G_{\sigma \nu \mu '\nu '}-\nabla _{\nu}\nabla
_{\sigma}G_{\mu \sigma \mu '\nu '}+{2\over \ell^2}[
G_{\mu \nu\mu
  '\nu '} -g_{\mu \nu}^{(0)}G_{\sigma \sigma \mu '\nu '}]\cr
& = & \left( g_{\mu \mu '}g_{\nu \nu '}+g_{\mu \nu '}g_{\nu \mu
    '}-{2\over d-1}g_{\mu \nu}g_{\mu' \nu '}\right) ~
{\delta ^{(d+1)} ({\bf x},{\bf x'})\over\sqrt{g}}.
\ea
Le dernier terme du membre de droite de
(\ref{eqnarray:eomg}) est un terme "trace-modifiant". Ce terme est nécessaire puisque nous avons défini $G_{\mu \nu \mu '\nu '}$ connectant
$ h_{\mu \nu }$ avec $T_{\mu \nu }^{(1)}$ (au lieu du membre de droite de
(\ref{qqq:general}) : $\bar T_{\mu \nu }^{(1)}$).

\begin{figure}
  \begin{center}
\includegraphics[width=8.5cm]{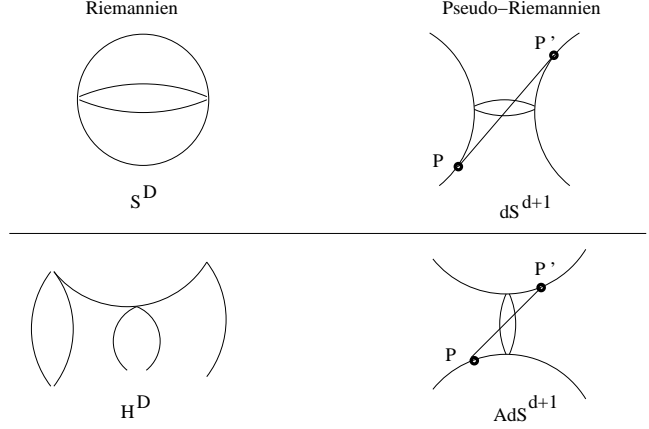}
 \end{center}
\caption{
Distances cordales sur des variétés maximalement symétriques
pseudo-riemanniennes. $D = d+1$.}
\label{fig:allen}
\end{figure}

\begin{figure}
  \begin{center}
\includegraphics[width=6.5cm]{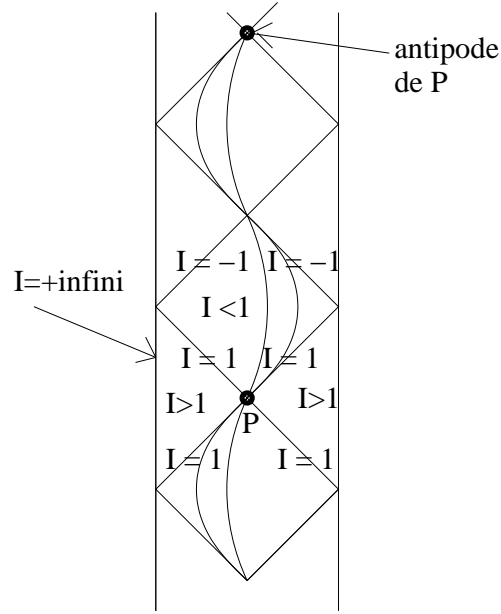}
 \end{center}
\caption{
Diagramme de Carter-Penrose de l'espace $CAdS$ global. On a représenté les
géodésiques de genre temps, ainsi que la valeur prise par l'invariant de
distance $I$ sur les zones de genre temps et les zones de genre espace.}
\label{fig:cads}
\end{figure}

L'espace-temps $AdS^{d+1}$ de rayon de courbure $\ell$ est une variété pseudo-Riemanienne (ou Lorentzienne) qui correspond à l'hypersurface à $(d+1)$ dimensions d'un hyperboloide  plongé dans un espace plat à $(d+2)$ dimensions de signature $(-,+,...,+,-)$, selon
\ba
-X_0^2-X_{d+1}^2+\sum_{i=1}^d X_i^2 & = & -\ell^2.
\ea
La métrique de l'espace enveloppant à $(d+2)$ dimensions est 
\ba
ds^2 & = & -(dX_0^2+dX_{d+1}^2)+\sum_{i=1}^d X_i^2.
\ea
Le groupe d'isométrie de $AdS^{d+1}$ est donc $SO(2,d)$. Il s'avère commode
pour faire des calculs explicites d'utiliser les coordonnées
\ba
X_0  =  \cos \left(t/\ell\right) \cosh\left(\xi/\ell\right),\qquad X_{d+1} = \sin \left(t/\ell\right) \cosh\left(\xi/\ell\right),\qquad X_i = \sinh\left(\xi/\ell\right) \hat{n}_i,
\ea
où le vecteur $\hat{n}(\theta^1,...,\theta^{d-1})$ décrit l'hypersphère
$S^{d-1}$ et $t\in[-\pi,\pi]$. Ces coordonnées couvrent tout l'espace
$AdS^{d+1}$ avec la métrique\footnote{Notons que cette métrique globale de
$AdS$ correspond à celle qu'on a calculée en (\ref{eqnarray:exactmetric}) au chapitre \ref{chapter:dissip} dans le cas $H = 0$.} de $AdS^{d+1}$ :
\ba
ds^2 & = & -\cosh ^2[\xi/\ell ]dt^2+d\xi ^2+\sinh ^2[\xi/\ell ]d\Omega^2_{d-1},
\ea
où $d\Omega^2_{d-1}$ est la métrique de la sphère $S^{d-1}$. Il s'avère de plus commode d'utiliser la séparation
invariante relativiste entre deux points $P,P'$: $I(P,P') =
-X_0{X'}_0-X_{d+1}{X'}_{d+1}+\sum_{i=1}^d X_i{X'}_i = \sigma(P,P')+1$, où
$\sigma(P,P') =
(1/2)\eta_{\alpha\beta}(X^\alpha-{X'}^\alpha)(X^\beta-{X'}^\beta)$ est la
demi-distance entre $P,P'$ dans l'espace plat de plongement de signature
$(-,+,...,+,-)$, dite distance cordale (Fig. \ref{fig:allen}). 
Dans les coordonnées globales, la séparation invariante s'écrit
\ba
I(P,P')=
\cosh [\xi/\ell ] \cosh [\xi '/\ell] \cos \left[(t-t')/\ell\right]-\sinh [\xi/\ell ] \sinh [\xi '/\ell](\hat{n}\cdot\hat{n}')_{S^{d-1}}.
\ea
Lorsque $P$ coincide avec $P'$ ou réside dans le cône de lumière passé ou futur de $P'$, on a  $I(P,P')=+1.$ Si $P$ et $P'$ peuvent être reliés par une géodésique de genre temps de longueur propre $\tau$ (en unité de la longueur de courbure de $AdS$), $I(P,P')=\cos [\tau ].$ Lorsque $P$
coincide avec le point antipodal de $P'$ (noté $\bar P'$) ou
réside sur le cône de lumière passé ou futur de $\bar P',$ $I(P,P')=-1.$ Si
$P$ et $P'$ sont connectés par une géodésique de genre espace de longueur propre
$\sigma $ (de nouveau en unité de la longueur de courbure de $AdS$),
$I(P,P')=\cosh [\sigma ],$ et si $P$ peut être connecté à $\bar
P'$ par une géodésique de genre espace et de longueur propre $\sigma ,$
$I(P,P')=-\cosh [\sigma ]$ (Fig. \ref{fig:cads}).

On peut exprimer $G_{\mu\nu\mu '\nu '}$ en séparant explicitement les parties invariantes de jauge des parties artefacts de jauge \cite{ads}
\ba
\label{eqnarray:ansatz}
G_{\mu\nu\mu '\nu '}(x, x')
&=&B^{(1)}(I)g_{\mu\nu}g_{\mu '\nu '}\cr
&+&B^{(2)}(I)\left[
(\nabla _\mu \nabla _{\mu '}I)(\nabla _\nu \nabla _{\nu '}I)
+(\nabla _\mu \nabla _{\nu '}I)(\nabla _\nu \nabla _{\mu '}I)
\right] \cr
&+&\nabla _{(\mu}\left[\nabla _{\nu)}\nabla _{\mu'} I\nabla _{\nu'} I
    B^{(3)}(I)\right]
+\nabla _{(\mu'}\left[\nabla _{\nu')}\nabla _{\mu} I\nabla _{\nu} I
    B^{(3)}(I)\right]\cr
&+&\nabla _{(\mu}\left[\nabla _{\nu)} I \nabla _{\mu'} I\nabla _{\nu'} I
    B^{(4)}(I)\right]
+\nabla _{(\mu'}\left[\nabla _{\nu')} I \nabla _{\mu} I\nabla _{\nu} I
    B^{(4)}(I)\right]\cr
&+&\nabla _{\mu}\left[\nabla _{\nu} I g_{\mu '\nu '} B^{(5)}(I)\right]
+\nabla _{\mu'}\left[\nabla _{\nu'} I g_{\mu\nu} B^{(5)}(I)\right],
\label{eqn:greenstensor}
\ea
ce qui est bien équivalent à la forme habituelle
\ba
G_{\mu \nu \mu '\nu '}({\bf x}, {\bf x}')
&=&A^{(1)}(I) g_{\mu \nu }g_{\mu '\nu '}\cr
&+&A^{(2)}(I)\left[
(\nabla _\mu \nabla _{\mu '}I)(\nabla _\nu \nabla _{\nu '}I)
+
(\nabla _\mu \nabla _{\nu '}I)(\nabla _\nu \nabla _{\mu '}I)
\right] \cr
&+&A^{(3)}(I)
(\nabla _\mu I)(\nabla _{\mu '}I)(\nabla _\nu I)(\nabla _{\nu '}I)\cr
&+&A^{(4)}(I)\Bigl[
(\nabla _\mu \nabla _{\mu '}I)(\nabla _\nu I)(\nabla _{\nu '}I)
+
(\nabla _\mu \nabla _{\nu '}I)(\nabla _\nu I)(\nabla _{\mu '}I)\cr
&&~~~~~~
+
(\nabla _\nu \nabla _{\mu '}I)(\nabla _\mu I)(\nabla _{\nu '}I)
+
(\nabla _\nu \nabla _{\nu '}I)(\nabla _\mu I)(\nabla _{\mu '}I)
\Bigr] \cr
&+&A^{(5)}(I)\left[
(\nabla _\mu I)(\nabla _{\nu }I)g_{\mu '\nu '}+
g_{\mu \nu }(\nabla _{\mu '}I)(\nabla _{\nu '}I)\right].
\ea
décomposée sur la base des cinq bi-tenseurs de $AdS$, symétriques sous le
groupe d'isométrie $SO(2,d)$. Nous noterons $T^{(i)}_{\mu\nu\mu'\nu'}(I)$
chacun des cinq bi-tenseurs de base accompagnant les cinq fonctions scalaires
$A^{(i)}(I)$. 

Les termes impliquant $B^{(3)}$, $B^{(4)}$ and $B^{(5)}$ dans (\ref{eqnarray:ansatz}) sont des gradients par rapport à ${\bf x}$ ou ${\bf x'}$. Les gradients par rapport à ${\bf x'}$
ne contribuent pas à la propagation des composantes \emph{physiques} de
$h_{\mu\nu}$ parce qu'on intègre sur des courants conservés : $\nabla _{\mu}T^{(1)}_{\mu\nu} = 0$. Les gradients par rapport à ${\bf
  x}$ donnent des modifications non-pertinentes de $h_{\mu\nu}$ dans (\ref{eqnarray:ht}) puisqu'ils correspondent aux diffeomorphismes ou transformations de jauge
$h_{\mu\nu} \rightarrow h_{\mu\nu}+\nabla _{(\mu}\xi_{\nu)}$. Donc $B^{(3)}$, $B^{(4)}$ et $B^{(5)}$ sont des parties de pure jauge, alors que $B^{(1)}$ et
$B^{(2)}$ sont les parties physiques du propagateur du graviton. On posera donc $B^{(3)} = B^{(4)} = B^{(5)} \equiv 0$ pour la suite. La forme de $G_{\mu\nu\mu '\nu '}$, avec ses parties physiques seulement, connecte donc les gravitons physiques à la source de matière selon
\ba\label{qqq:solphys}
 h_{\mu\nu}
&=&\int d^{(d+1)}x \left[
B^{(1)}T^{(1)}_{\mu\nu\mu '\nu '}
+B^{(2)}T^{(2)}_{\mu\nu\mu '\nu '}\right]T_{\mu '\nu '}.
\ea

En insérant la partie physique de la fonction de Green $G_{\mu \nu \mu '\nu
'}(I)$ dans l'équation d'Einstein linéarisée covariante (\ref{eqnarray:eomg}),
et en utilisant les formules algébriques données dans l'appendice
\ref{sec:formulae} de ce chapitre, on trouve, après un calcul trivial mais
laborieux, l'équation du mouvement pour la fonction de Green physique du
graviton dans $AdS^{d+1}$ : 
\ba
&&\Box G_{\mu\nu\mu '\nu'} +\nabla _{\mu}\nabla _{\nu}G_{\sigma\sigma\mu '\nu'}-\nabla _{\mu}\nabla
_{\sigma}G_{\sigma\nu\mu '\nu'}-\nabla _{\nu}\nabla
_{\sigma}G_{\mu\sigma\mu '\nu'}+2\left[
G_{\mu\nu\mu '\nu '}
-g_{\mu\nu} G_{\sigma\sigma\mu '\nu '}\right]=\cr
&=&
\left[ (I^2-1)(B^{(1)})''+2dI(B^{(1)})' -2dB^{(1)} -4B^{(2)}+2I(B^{(2)})'\right]
 T^{(1)}_{\mu\nu\mu '\nu'}\cr
&+&\left[
  (I^2-1)(B^{(2)})''+(d-1)I(B^{(2)})'-2(d-1)B^{(2)}\right]T^{(2)}_{\mu\nu\mu
  '\nu'}\cr
&+&\left[2(B^{(2)})''\right]T^{(3)}_{\mu\nu\mu
  '\nu'}\cr
&+&\left[-I(B^{(2)})''-(d-1)(B^{(2)})'\right]T^{(4)}_{\mu\nu\mu '\nu'}\cr
&+&\left[2(B^{(2)})''+(d-1)(B^{(1)})''\right]g_{\mu '\nu
  '}(\nabla_{\mu}I)(\nabla_{\nu}I)\cr
&+&\left[-2I(B^{(2)})'-4(d+1)B^{(2)}\right]g_{\mu \nu
  }(\nabla_{\mu'}I)(\nabla_{\nu'}I),
\ea
si $P,P'$ ne sont pas confondus ($I \neq 1$). Ensuite on peut manipuler les différents termes tensoriels de fa\c{c}on à isoler des dérivées totales du type $\nabla_{\mu'}$, qui s'annulent par intégration par partie puisque
$\nabla_{\mu'}T_{\mu'\nu'} = 0$. Il restera trois bi-tenseurs de base indépendants qui donneront trois équations pour $B^{(1)}$ et
$B^{(2)}$. Nous donnons les trois manipulations : on peut transformer le terme
en $g_{\mu \nu}(\nabla_{\mu'}I)(\nabla_{\nu'}I)$ en fonction de $T^{(1)}$ pour trouver que
\ba
\left[-2I(B^{(2)})'-4(d+1)B^{(2)}\right]g_{\mu \nu
  }(\nabla_{\mu'}I)(\nabla_{\nu'}I)
&=&\nabla_{\mu '} \left[\left(-2IB^{(2)}-2(2d+1)\int B^{(2)}\right)g_{\mu \nu
  }\nabla_{\nu'}I\right]\cr
&+&\left[2I^2B^{(2)}+2(2d+1)I\int B^{(2)}\right]T^{(1)}_{\mu\nu\mu '\nu'},
\ea
on peut aussi combiner le terme en $T^{(4)}$ avec un terme en $T^{(2)}$
pour trouver que
\ba
-(d-1)(B^{(2)})'T^{(4)}_{\mu\nu\mu '\nu'}-2(d-1)B^{(2)}T^{(2)}_{\mu\nu\mu '\nu'}
&=&\nabla_{(\mu '} \left[-(d-1)
B^{(2)}(\nabla_{\mu}I)(\nabla_{\nu}\nabla_{\nu ')}I)\right]\cr
&+&\nabla_{(\mu '} \left[-(d-1)B^{(2)}(\nabla_{\nu}I)(\nabla_{\mu}\nabla_{\nu ')}I)\right]\cr
&+&4(d-1)B^{(2)}(\nabla_{\mu}I)(\nabla_{\nu}I)g_{\mu '\nu '},
\ea
et enfin on peut combiner le terme en $T^{(3)}$ avec un terme en $T^{(2)}$ et un terme en $T^{(4)}$
pour trouver que
\ba
2(B^{(2)})''T^{(3)}_{\mu\nu\mu '\nu'}-2I(B^{(2)})'T^{(2)}_{\mu\nu\mu '\nu'}-I(B^{(2)})''T^{(4)}_{\mu\nu\mu '\nu'}
&=&\nabla_{(\mu '} \left[(B^{(2)})'\nabla_{\nu
    ')}I\nabla_{\mu}I\nabla_{\nu}I\right]\cr
&+&\nabla_{(\mu '} \left[-I
(B^{(2)})'(\nabla_{\mu}I)(\nabla_{\nu}\nabla_{\nu ')}I)\right]\cr
&+&\nabla_{(\mu '} \left[-(d-1)B^{(2)}(\nabla_{\nu}I)(\nabla_{\mu}\nabla_{\nu ')}I)\right]\cr
&+&2I(B^{(2)})')g_{\mu' \nu'}\nabla_{\mu}I\nabla_{\nu}I.
\ea
De ces trois manipulations il résulte que l'équation pour le propagateur se
simplifie selon : 
\ba
&&\Box G_{\mu\nu\mu '\nu'} +\nabla _{\mu}\nabla _{\nu}G_{\sigma\sigma\mu '\nu'}-\nabla _{\mu}\nabla
_{\sigma}G_{\sigma\nu\mu '\nu'}-\nabla _{\nu}\nabla
_{\sigma}G_{\mu\sigma\mu '\nu'}+2\left[
G_{\mu\nu\mu '\nu '}
-g_{\mu\nu} G_{\sigma\sigma\mu '\nu '}\right]=\cr
&=&\left[ (I^2-1)(B^{(1)})''+2dI(B^{(1)})' -2dB^{(1)} -4B^{(2)}+2I(B^{(2)})'+2I^2B^{(2)}+2(2d+1)I\int B^{(2)}\right]
T^{(1)}_{\mu\nu\mu '\nu'} \cr
&+&\left[ (I^2-1)(B^{(2)})''+(d+1)I(B^{(2)})'\right]T^{(2)}_{\mu\nu\mu '\nu'}\cr
&+&\left[ 2(B^{(2)})''+2I(B^{(2)})'+4(d-1)B^{(2)}+(d-1)(B^{(1)})''\right]g_{\mu '\nu '}(\nabla_{\mu}I)(\nabla_{\nu}I),
\ea
de telle sorte que les composantes scalaires physiques $B^{(1)}$, $B^{(2)}$ du
propagateur du graviton satisfont le système d'équations : 
\ba
&& (I^2-1)(B^{(2)})''+(d+1)I(B^{(2)})' = 0,\label{eqn:one}\\
&& (d-1)(B^{(1)})''+2(B^{(2)})''+2I(B^{(2)})'+4(d-1)B^{(2)} = 0\label{eqn:two},\\
&& (I^2-1)(B^{(1)})''+2dI(B^{(1)})' -2dB^{(1)}
-4B^{(2)}+2I(B^{(2)})'+2I^2B^{(2)}\cr
& &+2(2d+1)I\int B^{(2)} = 0\label{eqn:three}.
\ea

On pourra observer que ce sont des équations similaires à celles trouvées dans \cite{ads}.
L'équation (\ref{eqn:one}) est l'équation du propagateur d'un champ scalaire sans masse dans $AdS^{d+1}$ dont les deux solutions indépendantes sont connues. La solution de l'équation
(\ref{eqn:two}) peut s'écrire ensuite :
\ba
\label{eqnarray:solH}
B^{(1)} = {-1\over d-1}\left[2B^{(2)}+2I\int B^{(2)}+4(d-2)\int \int
  B^{(2)}\right],
\ea
et on peut vérifier que cette solution est consistante avec la troisième
équation (\ref{eqn:three}). Notons que ces mêmes équations auraient pu être
trouvées plus rapidement en utilisant l'opérateur du mouvement dans la jauge
de Lorentz, $B^{(1)}$ et $B^{(2)}$ étant indépendants de jauge.


On arrive au point délicat. Un couple de solutions indépendantes à l'équation de Klein-Gordon (\ref{eqn:one}) est donné au moyen de fonctions hypergéométriques selon
\ba\label{qqq:b2}
B^{(2)}(I) & = & C_1F\left(0,-{d-1\over 2};-\left(d-1\right);{-2\over I-1}\right)\cr
& + &{C_2\over \left(I-1\right)^d}F\left(d,{d+1\over 2};d+1;{-2\over I-1}\right).
\ea
Ces solutions pour la fonction à deux points ne sont bien définies que sur un espace $AdS^{d+1}$ Euclidien ou pour une séparation de genre espace entre les deux points sur un espace $AdS^{d+1}$ Lorentzien, c'est-à-dire que ces solutions sont analytiques pour $I > 1$. La première solution est régulière et naturellement prolongeable en $I \leq 1$ dans les zones de genre temps, par contre la seconde solution est singulière sur le cône de lumière en $I = 1$ et possède une coupure $I \leq 1$ dans les zones de genre temps.
 La prescription habituelle pour la fonction à deux points dans $AdS$ qu'on
 peut trouver dans \cite{ads,allen} sélectionne le vide quantique dit
 "Euclidien" qui exige tout d'abord que la fonction à deux points corresponde,
 à l'origine ($I = 1$), à la fonction à deux points de l'espace plat, ce qui
 semble naturel puisque deux points infiniment proches ne doivent pas
 ressentir les effets de courbure. La prescription du vide ``Euclidien'' exige
 de plus que la fonction à deux points décroisse le plus vite possible à
 l'infini spatial sur les bords de $AdS$, en $I = +\infty$. Cette seconde
 condition de la prescription sélectionne donc la solution singulière pour la
 fonction à deux points : $B^{(2)}(I) = C_2(1/I-1)^d
 F\left(d,(d+1)/2;d+1;-2/(I-1)\right)$, où la normalisation est obtenue par la
 première condition $N = \Gamma[(d+1)/2]/((4\pi)^{(d+1)/2}d\ell^{d-1})$. Par
 exemple pour $(d+1) = 5$ dimensions, la fonction à deux points Euclidienne
 est donc \cite{ads}
\ba
B^{(2)}(I) & = & -{1\over 8\pi^2}\left[{I(2I^2-3)\over
(I^2-1)^{3/2}}-2\right],\cr
B^{(1)}(I) & = & {1\over 12\pi^2}\left[{I\left(6I^4-9I^2+2\right)\over (I^2-1)^{3/2}}-6I^2\right].
\ea

 La fonction de Green retardée est ensuite obtenue dans la littérature
 \cite{allen} selon la prescription suivante : elle est proportionnelle au
 commutateur et donc correspond au saut à travers la coupure
 $G(I-1+i0)-G(I-1-i0)$, où on a omis les indices tensoriels pour ne pas
 alourdir les notations. La question est pourquoi ce choix particulier de vide
 quantique ? En fait ce choix de vide est relié aux conditions au bord de
 $AdS$, nécessaires pour construire une théorie quantique de champs cohérente dans $AdS$. L'espace $AdS$ n'est pas globalement hyperbolique (contrairement à l'espace Minkowski ou de Sitter), cela peut être admis en remarquant que l'infini spatial de $AdS$ (bord de $AdS$) peut être atteint en un temps fini, ainsi de l'information inconnue venant du bord peut affecter le cône de lumière futur. Dans \cite{quantads} les auteurs ont montré qu'une théorie quantique des champs n'était bien définie sur l'espace $AdS$ qu'à condition de préciser des conditions au bord de $AdS$, notamment celles qui conduisent à la prescription énoncée plus haut.
 Cependant la fonction de Green \emph{retardée} est un objet \emph{classique} qui ne dépend pas d'un choix de vide quantique (c'est le commutateur dans un cadre quantique). On doit donc pouvoir redémontrer quelle est la prescription à adopter pour la fonction de Green retardée sans faire appel à ces arguments quantiques. Notamment on va essayer de trouver des arguments classiques dans la section \ref{sec:retarde} montrant pourquoi le propagateur retardé du graviton dans $AdS$ ne sélectionne que la solution singulière dans (\ref{qqq:b2}).

\section{Propagateur retardé (ébauche)}\label{sec:retarde}

Cette section est une ébauche de démonstration classique (non quantique) de la
prescription à adopter pour construire la forme \emph{retardée} du
propagateur covariant du graviton dans $AdS^{d+1}$, qui un objet classique.

\subsection{Méthode de construction}\label{subsec:method}

Chaque étape pour construire le propagateur \emph{retardé} est donnée dans
cette section pour le cas scalaire. En effet le propagateur tensoriel complet
du graviton étant essentiellement déterminé à partir de la fonction $B^{(2)}$
qui n'est rien d'autre qu'un champ scalaire canonique sans
masse dont l'équation du mouvement est l équation de Klein-Gordon (\ref{eqn:one}), la généralisation au cas tensoriel est immédiate. Nous pouvons résumer la méthode en trois étapes :
\begin{itemize}
\item L'équation de Klein-Gordon \emph{homogène} pour le propagateur scalaire dans $AdS$ a deux solutions indépendantes. L'une, $R(I)$, est régulière sur le cône de lumière ($I = 1$ ou $I =
-1$), l'autre, $S(I)$, est singulière sur le cône de lumière
\ba G(I) & = & A
.S(I)+B .R(I). \ea
De plus la solution singulière est bien définie seulement dans les zones de
genre espace (c'est-à-dire à l'extérieur des cônes de lumière passé et futur)
ou pour des distances Euclidiennes, \emph{i.e}
\ba I > 1, \qquad \mbox{équivalent
en espace plat à une distance de genre espace $\tau^2 > 0$}. \ea
Afin d'obtenir le propagateur \emph{Lorentzien} complet, on doit prolonger analytiquement la solution singulière dans la zone de genre temps ($I \leq 1$). Il y a deux possibilités. La continuation analytique peut être réalisée dans le demi-plan complexe supérieur ou dans le demi-plan complexe inférieur, de telle sorte que dans la zone de genre temps il y a trois solutions indépendantes de l'équation homogène : la solution régulière et les deux solutions singulières analytiquement prolongées de part et d'autre de la coupure $I \leq 1$. Ces deux différentes solutions singulières deviennent dégénérées dans la zone de genre espace :
\ba G(I) & = & A_1 .S(I+i\epsilon)+A_2 .S(I-i\epsilon)+B
.R(I)
~\mbox{ dans la zone de genre temps,}\qquad\\
G(I) & = & \left(A_1+A_2\right) .S(I)+B .R(I)
~~~~~~~~~~~~~~~~~\mbox{ dans la zone de genre espace}.\qquad \ea
\item En imposant ensuite que le propagateur retardé classique soit \emph{causal} (\emph{i.e} zéro dans les zones de genre espace), on trouve qu'une fa\c{c}on naturelle de répondre à la causalité est de conclure pour la solution singulière que \ba A_1+A_2 & = & 0, \ea et la causalité pour la solution régulière demande de la tronquer par une fonction $\theta$. Finalement
\ba\label{qqq:causal}
G_c(I) & = & A_1 \left\{S(I+i\epsilon)-
S(I-i\epsilon)\right\}+B .R(I)*\theta(1-I^2) 
\ea 
est le candidat le plus
     général qui soit maintenant \emph{causal} et \emph{bien défini dans la
     zone de genre temps}. Notons que cette prescription causale pour la
     partie singulière est une fa\c{c}on analytique de la tronquer en évitant
     l'utilisation d'une fonction $\theta$. Une fois que nous seront convaincus que
     la partie régulière ne participe pas à la construction de la forme
     retardée, nous comprendrons que cette prescription pour la partie
     singulière peut être généralisée par abus à la fonction à deux points complète, c'est-à-dire qu'elle peut être appliquée également à la partie régulière, faisant disparaître cette dernière de la forme retardée du propagateur. 
\item  La dernière étape est de déterminer les coefficients $A_1$ et
$B$ dans (\ref{qqq:causal}). On utilise pour cela les deux conditions
suivantes :
\begin{enumerate}
\item La source de la fonction de Green est localisée à l'origine (séparation nulle entre les deux points),
\item on doit retrouver le propagateur de Minkowski à l'origine.
\end{enumerate}
La seconde condition peut être utilisée pour la normalisation du
 propagateur. La première condition va déterminer la combinaison correcte de
 $A_1$ et $B$. Cette première condition nous dit que, \emph{excepté à l'origine}, l'opérateur $\Box$ appliqué au propagateur doit donner zéro sur le cône de lumière
 \ba <\Box G_c(u,v),\phi(u,v)> & = & 0,
\label{eqnarray:condd}\ea où $\phi$ est une fonction test et $u,v$
sont les coordonnées nulles du cône de lumière (voir appendice
\ref{sec:lightcoord}). L'intégration dans (\ref{eqnarray:condd}) est effectuée sur le cône de lumière, en dehors de l'origine, \emph{i.e} dans la boite infinitésimale $0<a<v<b$ et $-\epsilon < u < +\epsilon$ ($\epsilon\rightarrow 0$). En dimension paire d'espace-temps, la singularité en $u$ est isolée de sorte que le théorème des résidus sera utilisé dans le calcul de
(\ref{eqnarray:condd}). En dimension impaire, il y a une coupure en $u$ donc une approche différente doit être adoptée dans ce cas. Le résultat de ce calcul sera que seul le coefficient
$A_1$, devant la partie singulière de $G_c$, doit être non-nul pour que la
condition (\ref{eqnarray:condd}) soit vérifiée. La conclusion immédiate sera
que la prescription pour la forme retardée du propagateur est $G^{ret}(I)
=\theta(t-t')\left[ S(I+i0)-S(I-i0)\right]$, ou plus généralement,
puisque la partie régulière disparait de toute fa\c{c}on dans cette
prescription :
$$G^{ret}(I) = \theta(t-t')\left[ G(I+i0)-G(I-i0)\right],$$ avec
$G(I)$ la solution homogène Euclidienne bien définie dans la zone de genre
espace. Cela prouvera que la prescription dans l'espace $AdS$ est la même que dans l'espace Minkowski.
\end{itemize}

Dans la section \ref{subsec:method} nous rendons explicite la construction du
propagateur retardé en répétant les trois étapes décrites au-dessus et en
calculant (\ref{eqnarray:condd}) par intégration de contour dans le plan
complexe sur les coordonnées nulles du cône de lumière, achevant ainsi la
démonstration de la prescription à adopter pour la forme retardée du propagateur du graviton dans $AdS$.

\subsection{Démonstration explicite}\label{subsec:demo}

En toute dimension le propagateur scalaire sans masse dans $AdS$ satisfait l'équation
\ba \Box G(I)&=& (\underline{\nabla }I)^2{d^2G\over dI^2}+(\Box
I){dG\over dI}\cr &=& (I^2-1) {d^2G\over dI^2}+(d+1)I {dG\over dI} =
0 \ea
 en dehors de l'origine. C'est la même équation dans le cas tensoriel pour $B^{(2)}(I)$. On trouve plus commode de ré-écrire cette équation en fonction de la variable $\tilde{I} = (I-1)/ (I+1)$ comme une équation hypergéométrique
\ba \tilde{I}(1-\tilde{I}){d^2G\over d\tilde{I}^2}+\left({d-3\over
     2}\tilde{I}+{d+1\over 2}\right){dG\over d\tilde{I}} = 0,\ea
dont la solution générale, définie dans la zone de genre espace ou Euclidienne $\tilde{I} > 0$, peut s'écrire dans une forme adaptée à la parité de la dimension d'espace-temps
 \ba
G^{odd}(\tilde{I}) & = & C_1+C_2{1\over \tilde{I}^{d-1\over
2}}F\left({1-d\over 2},1-d;{3-d\over 2};\tilde{I}\right)~~\mbox{ pour $d+1$ impair,
 $\tilde{I} > 0$}, \label{eqnarray:oddsol}\\
G^{even}(\tilde{I}) & = & C_1+C_2{\left(1-\tilde{I}\right)^d \over \tilde{I}^{d-1\over
2}}F\left(1,{1+d\over 2};1+d;1-\tilde{I}\right)~~\mbox{ pour $d+1$ pair,
$\tilde{I} > 0$}\label{eqnarray:evsol}.\qquad \ea
La constante $C_1$ est la solution régulière et la solution sous le coefficient $C_2$ est singulière sur le cône de lumière et non-définie à l'intérieur (zones de genre temps) : la coupure est en effet $\tilde{I} \leq
0$, ou de fa\c{c}on équivalente $I \leq 1$, puisque, pour
$d+1$ pair, la solution singulière contient des fonctions logarithme et puissance entières négatives en $\tilde{I}$ et, pour $d+1$ impair, des fonctions puissances demi-entières négatives en $\tilde{I}$.

D'après la seconde étape de la section \ref{subsec:method}, nous devons proplonger analytiquement la solution singulière des $\tilde{I}$ Euclidiens aux $\tilde{I}$ Lorentziens, ou dit autrement des zones de genre espace (extérieur du cône de lumière $\tilde{I} > 0$) aux zones de genre temps
(intérieur du cône de lumière $\tilde{I} \leq 0$). Il y a deux continuations
analytiques : dans le demi-plan  $\tilde{I}$ complexe supérieur et dans le
demi-plan $\tilde{I}$ complexe inférieur. Nous avons vu que la causalité
implique d'écrire le \emph{propagateur complet Lorentzien et causal} comme
(par exemple pour $d+1$ impair) :
\ba
G_c^{odd}(\tilde{I}) & = &
C_1\theta(-\tilde{I})+C_2F\left({1-d\over 2},1-d;{3-d\over 2};\tilde{I}\right)\left[{1
      \over \left(\tilde{I}+i\epsilon\right)^{d-1\over
2}}-{1 \over \left(\tilde{I}-i\epsilon\right)^{d-1\over
2}}\right].\qquad
\ea

La dernière étape est de déterminer les coefficients $C_1$ et
$C_2$. Un moyen possible est d'appliquer la première condition de la section
\ref{subsec:method} : la source de la fonction de Green est localisée à
l'origine seulement. On a donc, excepté à l'origine, \ba <\Box G_c(u,v),\phi(u,v)> & = & 0,
\label{eqnarray:cond}\ea où $\phi$ est une fonction test et l'intégration est effectuée dans la boite infinitésimale  $0<a<v<b$ et $-\epsilon < u <+\epsilon$ sur le cône de lumière, hors origine. $u,v$ sont les coordonnées du cône de lumière définies dans l'appendice
\ref{sec:lightcoord} et elles correspondent à $\tilde{I} = -uv$. Dans ces coordonnées, l'opérateur d'Alembertien $\Box$ s'écrit \ba \sqrt{|g|}\Box &
= & {\partial\over\partial u} f(u,v) {\partial\over\partial
v}+{\partial\over\partial v} f(u,v) {\partial\over\partial u},\ea
 où $f(u,v)=\left({u-v\over 1+uv}\right)^{d-1}$ vient de la géométrie $AdS^{d+1}$.
 Montrons tout d'abord que la partie régulière $C_1\theta(u)\theta(v)$ de la fonction de Green retardée ne suffirait pas à elle-seule à satisfaire la condition exposée au-dessus puisque
 \ba
\sqrt{|g|}\Box \theta(u)\theta(v) &=& \{{\partial\over\partial u} f(u,v)
{\partial\over\partial v}+{\partial\over\partial v} f(u,v)
{\partial\over\partial u}\}\theta(u)\theta(v)\cr
                                  &=& {\partial\over\partial u} f(u,v)
                                  \delta(v)\theta(u)+{\partial\over\partial v}
                                  f(u,v) \delta(u)\theta(v)\cr
                                  &=& (d-1)(u-v)^{d-2}[\delta(u)\theta(v)+\delta(v)\theta(u)]
\ea
d'après l'expression de $f$. Donc $<\Box C_1\theta(u)\theta(v),\phi(uv)>\neq 0$ sur le cône de lumière lorsque $d+1 >
2$. Il y a une fonction $\delta$ avec son support sur le cône de lumière.

Nous allons voir que la partie singulière de $G_c$, \emph{i.e} la fonction causale sous $C_2$,
vérifie systématiquement la condition (\ref{eqnarray:cond}) en n'importe
quelle dimension, de sorte la conclusion immédiate sera que $C_1 = 0$
nécessairement. On distingue pour la preuve le cas en dimension paire et le
cas en dimension impaire :

\subsubsection{Cas $d+1$ pair}\label{subsubsec:even}

\ba
G_c^{even}(\tilde{I}) =
C_1\theta(-\tilde{I})+C_2\left[S(\tilde{I}+i\epsilon)-S(\tilde{I}-i\epsilon)\right]
\ea avec la solution singulière, éqn. (\ref{eqnarray:evsol}),
\ba
S(\tilde{I}) \equiv S(-uv) = {\left(1+uv\right)^d \over (-uv)^{d-1\over
2}}F\left(1,{1+d\over 2};1+d;1+uv\right).
\ea

Les singularités en $\tilde{I} = 0$ correspondent aux singularités en
$u = 0$ sur le cône de lumière dans la boite ($0<a<v<b$,$-\epsilon < u <+\epsilon$), puisque $v
> 0$. On vérifie la condition (\ref{eqnarray:cond}) pour la partie singulière de $G_c^{even}$, \emph{i.e}
$C_2\left[S(\tilde{I}+i\epsilon)-S(\tilde{I}-i\epsilon)\right]$, en calculant le résidu. De fa\c{c}on concrète, si on pose $\phi(-uv)\sim\phi(0)\equiv 1$, la vérification de
(\ref{eqnarray:cond}) équivaut à calculer dans les coordonnées $u,v$
 \ba \int_a^b dv \left(\int_{\mathcal{C}_+}du \sqrt{|g|}
\Box
     S(-uv)-\int_{\mathcal{C}_-}du \sqrt{|g|} \Box S(-uv)\right),\label{eqnarray:contours}
\ea
où $\mathcal{C}_+$ est le contour d'intégration qui évite la singularité en passant par le demi-plan $u$ complexe supérieur et $\mathcal{C}_-$ est le contour d'intégration qui évite la singularité en passant par le demi-plan $u$ complexe inférieur. L'intégrand, dans le cas $d+1$ pair, possède une singularité isolée $u = 0$,
tel qu'on le verra un peu plus bas, donc on peut refermer le contour et ré-écrire (\ref{eqnarray:contours}) comme
 \ba & & \int_a^b dv
\int_{\mathcal{C_\odot}}du \sqrt{|g|} \Box S(-uv)\cr & = &  \int_a^b
dv \int_{\mathcal{C_\odot}}du \left[{\partial\over\partial u} f(u,v)
{\partial\over\partial v}+{\partial\over\partial v} f(u,v)
{\partial\over\partial u}\right] S(-uv), \ea où
$\mathcal{C_\odot}$ est le contour d'intégration encerclant la singularité  $u
= 0$. Ensuite, par le théorème de Cauchy, l'expression se réduit à
 \ba \int_a^b dv
\int_{\mathcal{C_\odot}}du {\partial\over\partial v} f(u,v)
{\partial\over\partial u}S(-uv). \ea
En notant que  (formule 15.2.8 dans \cite{abram})
 \ba\label{qqq:handbook}
 {\partial\over\partial x}\left[S(x)\right] &
\equiv & {\partial\over\partial x}\left[{\left(1-x\right)^d \over
x^{d-1\over 2}}F\left(1,(1+d)/2;1+d;1-x\right)\right] \cr & = &
{d-1\over 2}{\left(1-x\right)^{d-1} \over x^{d+1\over 2}}, \ea l'expression devient
\ba\label{eqnarray:residue} & & {d-1\over 2}\int_a^b dv
\int_{\mathcal{C_\odot}}du {\partial\over\partial v} f(u,v) (-v)
{\left(1+uv\right)^{d-1} \over (-uv)^{d+1\over 2}} \cr & = &
{d-1\over 2}\int_a^b dv {\partial\over\partial v}\left(-{1\over
v}\right)^{d-1\over 2} \int_{\mathcal{C_\odot}}du {(u-v)^{d-1}\over
u^{d+1\over 2}}\cr & = &{d-1\over 2}\int_a^b dv
{\partial\over\partial v}\left(-{1\over v}\right)^{d-1\over 2}
(2i\pi){(d-1)!\over({d-1\over 2})!({d-1\over 2})!}(-v)^{d-1\over
2}\cr & = & 0 \ea parce que le résidu ne dépend pas de $v$.

On conclut que $C_1 = 0$ et que le propagateur retardé scalaire dans $AdS^{d+1}$ est, pour
$d+1$ pair,
\ba\label{qqq:retevenprop}
G^{ret}(\tilde{I}) & =  C_2\theta(t-t') & \Biggl[{\left(1-\tilde{I}-i0\right)^dF\left(1,{1+d\over 2};1+d;1-\tilde{I}-i0\right) \over \left(\tilde{I}+i0\right)^{d-1\over
2}}\cr
& & - {\left(1-\tilde{I}+i0\right)^d F\left(1,{1+d\over
      2};1+d;1-\tilde{I}+i0\right)\over \left(\tilde{I}-i0\right)^{d-1\over 2}}\Biggr].
      \ea
Par exemple, en $(3+1)$ dimensions,
\ba
G_{3+1}^{ret}(\tilde{I}) & = &
C_2\theta(t-t')\left[-2\log(\tilde{I}+i0)-{1\over\tilde{I}+i0}+(\tilde{I}+i0)\right]-\left[-2\log(\tilde{I}-i0)-{1\over\tilde{I}-i0}+(\tilde{I}-i0)\right]\cr
                        & =  &
2i\pi C_2\theta(t-t')\left[-2\theta(-\tilde{I})+\delta(\tilde{I})\right]\cr
                        & =  &
2i\pi C_2\theta(t-t')\left[-2\theta(1-I^2)+\delta(1-I^2)\right].
\ea
Le coefficient de normalisation $C_2$ est ensuite déterminé par la seconde condition en comparant le propagateur retardé dans $AdS$ au propagateur retardé dans Minkowski proche de l'origine ($\tilde{I}=0$ ou
$I=1$).
 La généralisation au cas tensoriel, c'est-à-dire le propagateur retardé physique du graviton dans $AdS^{d+1}$ ($d+1$ pair) est donc
 \ba
G^{ret}_{\mu\mu'\nu\nu'}(\tilde{I}) & = & B^{(1)ret}(I)g_{\mu\nu}g_{\mu '\nu '}\cr
&+&B^{(2)ret}(I)\left[
(\nabla _\mu \nabla _{\mu '}I)(\nabla _\nu \nabla _{\nu '}I)
+(\nabla _\mu \nabla _{\nu '}I)(\nabla _\nu \nabla _{\mu '}I)
\right]
\ea
avec $B^{(2)ret}(I)$ donné par (\ref{qqq:retevenprop}). Et $B^{(1)ret}(I)$ est déduit de $B^{(2)ret}(I)$ à partir de (\ref{eqnarray:solH}).

\subsubsection{Cas $d+1$ impair}\label{subsubsec:odd}
\ba
G_c^{odd}(\tilde{I}) =
C_1\theta(-\tilde{I})+C_2\left[S(\tilde{I}+i\epsilon)-S(\tilde{I}-i\epsilon)\right]
\ea avec la solution singulière (éqn. \ref{eqnarray:oddsol})
\ba
S(\tilde{I}) \equiv S(-uv) = {1\over (-uv)^{d-1\over
2}}F\left({1-d\over 2},1-d;{3-d\over 2};-uv\right).
\ea

De nouveau on veut vérifier  (\ref{eqnarray:cond}) pour la partie singulière de
$G_c^{odd}$, \emph{i.e} $C_2\left[S(\tilde{I}+i\epsilon)-S(\tilde{I}-i\epsilon)\right]$. En posant  $\phi(-uv)\sim\phi(0)\equiv 1$, la vérification de (\ref{eqnarray:cond}) équivaut à calculer,
dans les coordonnées $u,v$,
\ba
\int_a^b dv \left(\int_{\mathcal{C}_+}du \sqrt{|g|} \Box
     S(-uv)-\int_{\mathcal{C}_-}du \sqrt{|g|} \Box S(-uv)\right),\label{eqnarray:contours2}
\ea
où $\mathcal{C}_+$ est le contour d'intégration qui évite la singularité en passant par le demi-plan $u$ complexe supérieur et $\mathcal{C}_-$ est le contour d'intégration qui évite la singularité en passant par le demi-plan $u$ complexe inférieur. L'intégrand, dans le cas $d+1$ impair, possède une coupure en $u \geq 0$, à cause des puissances demi-entières négatives, donc on ne peut refermer le contour que sur la gauche de la singularité où $u < 0$, et ré-écrire
(\ref{eqnarray:contours2}) comme
\ba
& & \int_a^b dv \int_{\mathcal{C_\subset}}du \sqrt{|g|} \Box S(-uv)\cr
& = &  \int_a^b dv \int_{\mathcal{C_\subset}}du \left[{\partial\over\partial u} f(u,v) {\partial\over\partial v}+{\partial\over\partial v} f(u,v)
{\partial\over\partial u}\right] S(-uv),
\ea
où $\mathcal{C_\subset}$ est le contour en "trou de serrure" ("keyhole contour" en anglais) autour de la coupure $u  \geq
0$. De plus, par théorème de Cauchy, l'expression se réduit à
\ba
\int_a^b dv \int_{\mathcal{C_\subset}}du {\partial\over\partial v} f(u,v)
{\partial\over\partial u}S(-uv)
\ea
On a de plus que (formule 15.2.3 dans \cite{abram})
\ba
{\partial\over\partial x}\left[S(x)\right] & \equiv & {\partial\over\partial
  x}\left[{1 \over x^{d-1\over
2}}F\left({1-d\over 2},1-d;{3-d\over 2};x\right)\right] \cr
& = & {d-1\over 2}{\left(1-x\right)^{d-1} \over x^{d+1\over
2}}.
\ea
Donc on a affaire au même intégrand que pour le cas pair
(éqn. (\ref{qqq:handbook})), mais le contour d'intégration
$\mathcal{C_\subset}$ est différent :
\ba
{d-1\over 2}\int_a^b dv {\partial\over\partial v}\left(-{1\over v}\right)^{d-1\over 2}
\int_{\mathcal{C_\subset}}du {(u-v)^{d-1}\over u^{d+1\over
2}}.
\ea
La démonstration dans le cas impair est en progrès.

\begin{subappendices}

\section{Formules algébriques utiles}\label{sec:formulae}

Dans cet appendice, en utilisant la forme explicite de l'invariant de séparation
$$I = \cosh[\xi]\cosh[\xi']\cos[t-t']-\sinh[\xi]\sinh[\xi'](\hat{n}\cdot\hat{n}')_{S^{d-1}} $$
en fonction des coordonnées de $AdS^{d+1}$ avec la métrique globale
$ds^2=-\cosh ^2[\xi] dt^2 +d\xi ^2+\sinh ^2[\xi]d\Omega^2_{d-1},$
nous calculons quelques formules algébriques utilisées dans le chapitre.
\ba
\underline{\nabla}I
&=& \tilde \xi~(\sinh[\xi]\cosh[\xi']\cos[t-t'] -\cosh[\xi]\sinh[\xi'])
-\tilde t~\cosh[\xi']\sin [t-t']\cr
& & -\sinh[\xi]\sinh[\xi']\sum_{i = 1}^{d-1}\tilde \theta_i~\nabla_i(\hat{n}\cdot\hat{n}'),
\ea
\ba
(\underline{\nabla}I)^2 & = & (I^2-1),\label{eqnarray:4}\\
\underline{\nabla}\otimes \underline{\nabla}I
&=& I\underline{\underline g},\label{eqnarray:3}\\
\Box I &=& (d+1)I,\label{eqnarray:1}\\
\Box \underline{\nabla}I, &=& \underline{\nabla}I
\ea
puis
\ba
\underline{\nabla '}I
&=& \tilde \xi'~(\cosh[\xi]\sinh[\xi']\cos[t-t']-\sinh[\xi]\cosh[\xi'])
+\tilde t'~\cosh[\xi]\sin[t-t']\cr
& & -\sinh[\xi]\sinh[\xi']\sum_{i = 1}^{d-1}\tilde \theta'_i~\nabla'_i(\hat{n}\cdot\hat{n}'),
\ea
\ba
\Box \underline{\nabla'}I &=& (d+1)\underline{\nabla'}I,\\
(\underline{\nabla}I\cdot \underline{\nabla})(\nabla_{\mu}\nabla_{\nu'})I
&=&(\nabla_{\mu}I)(\nabla_{\nu'}I),\label{eqnarray:2}\\
\underline{\nabla}(\underline{\nabla}\otimes \underline{\nabla}' I)
&=&\underline{\underline{g}}\otimes \underline{\nabla}' I,\\
(\underline{\nabla}\otimes \underline{\nabla}' I)\cdot
 (\underline{\nabla}\otimes \underline{\nabla}' I)
&=&\underline{\nabla}' I\otimes \underline{\nabla}' I
+\underline{\underline g}'.
\ea

\section{Métrique $\boldsymbol{AdS}$ dans les coordonnées du cône de lumière}\label{sec:lightcoord}

On démarre avec la métrique globale
\ba
ds^2=-\cosh ^2[\xi ]dt^2+d\xi ^2+\sinh ^2[\xi ]d\Omega^2_{d-1}.
\ea
On calcule
\ba
r = \int dr \equiv \int {d\xi\over\cosh [\xi ]} = \int d\xi {\cosh ^2[\xi/2
  ]-\sinh ^2[\xi/2 ]\over\cosh ^2[\xi/2 ]+\sinh ^2[\xi/2 ]} & = &\int d\xi
  {1-\tanh ^2[\xi/2
  ]\over 1+\tanh ^2[\xi/2
  ]}\cr
                                                             & = & \int d\xi {d\over d\xi}\left(2\tan^{-1}\left(\tanh {\xi\over
        2}\right)\right)\cr
                                                             & = & 2\tan^{-1}\left(\tanh {\xi\over
        2}\right).
\ea
Donc on a $\tan {r\over 2}  = \tanh {\xi\over 2}$ et
\ba
\label{eqnarray:coord}
ds^2 & = &\cosh ^2[\xi ](-dt^2+dr^2)+\sinh ^2[\xi ]d\Omega^2_{d-1}.
\ea
De (\ref{eqnarray:coord}) on déduit facilement
\ba
\cosh ^2[\xi ] = {1\over \cos ^2[r ]}, \qquad \sinh ^2[\xi ] = \tan ^2[r ],
\ea
d'où
\ba
ds^2 & = &{-dt^2+dr^2 \over \cos ^2[r ]} +\tan ^2[r ]d\Omega^2_{d-1}.
\ea
En posant $r = (U-V)/2$ and $t = (U+V)/2$, on obtient
\ba
ds^2 & = &{-dUdV \over \cos ^2[(U-V)/2 ]} +\tan ^2[(U-V)/2 ]d\Omega^2_{d-1}.
\ea
En posant ensuite $u = \tan [U/2]$ and $v = \tan [V/2]$ tel que
\ba
-{dUdV \over \cos ^2[(U-V)/2 ]} = -4{dudv \over (1+uv)^2}, \qquad
\tan ^2[(U-V)/2 ]d\Omega^2_{d-1} = {(u-v)^2 \over (1+uv)^2}d\Omega^2_{d-1},
\ea
on obtient finalement
\ba
ds^2 = {4\over(1+uv)^2}\left(-dudv+\left({u-v\over 2}\right)^2d\Omega^2_{d-1}\right).
\ea

\end{subappendices}


\chapter{Reconstruction en espace réel des effets de lentille gravitationnelle
  sur le CMB }\label{chapter:cmb}

Dans ce dernier chapitre nous traitons un sujet en cosmologie standard plutôt
différent de la majeure partie de cette thèse, consacrée principalement à la
cosmologie branaire, mais auquel nous nous sommes intéressés en parallèle. Il
s'agit d'un travail effectué en collaboration avec Martin Bucher de
l'Université Paris-Sud en France et Kavilan Moodley de l'Université de
KwaZulu-Natal en Afrique du Sud. Ce travail est en progrès et devrait être publié
bientôt sur les archives :
\emph{CMB lensing reconstruction in real space}, Martin Bucher, Kavilan Moodley and Mathieu Remazeilles.

Nous utiliserons l'abréviation anglaise CMB (Cosmic Microwave Background) pour
désigner le rayonnement du Fond Diffus Cosmologique. Le CMB est un rayonnement
de photons qui nous parvient de la surface de dernière diffusion, c'est-à-dire
au moment du découplage entre les photons et les électrons lorsqu'ils ne
diffusent plus ensemble (époque de la recombinaison, redshift $z\sim 1100$,
température $T = 3000~K$). Les photons se propagent alors librement jusqu'à
nous ($z = 0$). Ce rayonnement du CMB est donc l'image la plus ancienne de
l'Univers observable dont on dispose. Il nous apparait aujourd'hui à des
fréquences micro-onde comme un rayonnement de type corps noir à la température
$T = 2.7~K$, le refroidissement et la basse fréquence du rayonnement
aujourd'hui étant dûs à l'expansion de l'Univers. Ce rayonnement du ciel a
d'abord été détecté par inadvertance comme un bruit micro-onde par les
ingénieurs Penzias et Wilson en 1965, puis détecté comme un corps noir par le
satellite COBE en 1992. En fait ce rayonnemnt
n'est pas totalement un corps noir mais présente des anisotropies de
température $\delta T/T \sim 10^{-5}$, observables lorsqu'on mesure le spectre
des corrélations angulaires de température entre deux points dans le ciel
séparés d'un angle $\boldsymbol{\theta}$ (bidimensionnel). Ces anisotropies de
température sont directement reliées aux perturbations cosmologiques au moment
du découplage \cite{sw}, et donc aux conditions initiales de l'Univers
produites à la fin de l'Inflation. En ce sens les spectres de puissance des
anisotropies de température s'avèrent être un outil observationnel puissant
pour sonder la physique de l'Univers primordial et notamment tester les
théories d'Inflation. Les anisotropies de température ont été détectées par le
satellite WMAP en 2003. Comme tout rayonnement
électromagnétique, le rayonnement du CMB est polarisé et présente donc de la
même fa\c{c}on des anisotropies de polarisation. Il est pratique et d'usage de
décomposer les anisotropies de polarisation sur une base de modes $E$ et $B$
(notés comme tel en référence aux propriétés de parité scalaire ("électrique") pour le mode E
et pseudo-scalaire ("magnétique") pour le mode B).

Les grandes structures dans l'Univers (amas de galaxies), présentes à un redshift $z \sim 3$, crééent un potentiel gravitationnel non-homogène (dit potentiel de lentille) et affectent donc le trajet des photons par un effet de lentille gravitationnelle faible (déviation de la lumière par un corps massif en relativité générale). Ce potentiel de lentille affecte donc les spectres primordiaux d'anisotropies de température T et de polarisation $E$ et $B$ et biaise donc l'image réelle de l'Univers primordial apparaissant aujourd'hui. Il est donc indispensable de "nettoyer" les spectres d'anisotropies \emph{observés} et \emph{contaminés} par les effets de lentille gravitationnelle faible ("weak gravitationel lensing" en Anglais) afin de restituer les spectres \emph{primordiaux}. Un moyen de parvenir à cela est de reconstruire le champ de lentille gravitationnelle $\Phi(\boldsymbol{\ell})$ (où $\ell\sim 180/\theta$) du CMB à l'aide d'estimateurs statistiques. On peut en effet construire des estimateurs quadratiques de $\Phi(\boldsymbol{\ell})$ à partir des anisotropies de température et/ou de polarisation \emph{observés}, dans l'espace harmonique (équivalent à l'espace de Fourier si le ciel était plat au lieu d'être sphérique). L'objectif est bien-entendu de construire un estimateur statistique efficace et optimal malgré qu'on soit soumis à plusieurs limitations dûes aux capacités des outils d'observation (résolution, sensibilité, surface du ciel couverte,un seul point de vue d'observation, ...) et aux avant-plans dans l'Univers (points sources, Voie Lactée, ...).

Dans ce travail nous tentons de reconstruire le champ de lentille
gravitationnelle du CMB directement en \emph{espace réel}, en montrant qu'on
ne perd que très peu d'information statistique lorsqu'on utilise nos
estimateurs statistiques de courte portée angulaire sur la sphère céleste et en
espace réel à la place des estimateurs habituels construits dans l'espace harmonique. Ces
derniers sont non-locaux et nécessitent en principe une analyse couvrant tout
le ciel en entier, sans coupures ni excisions. Étant donné que toute
l'information pertinente des effets de lentille réside aux petites échelles
angulaires, proches de l'échelle de résolution des cartes CMB du ciel, les
champs de dilatation et de cisaillement (champs qui sont directement reliés
aux observations de manière locale, contrairement au champ de déflection
angulaire ou au potentiel de lentille lui-même), créés par le potentiel de
lentille, peuvent être reconstruits au moyen de combinaisons quadratiques
impliquant seulement des pixels faiblement séparés dans le ciel. Bien que la
reconstruction théorique des effets de lentille est naturellement plus
abordable en espace harmonique, les méthodes en espace réel développées ici
(section \ref{sec:real}) ont l'avantage d'être plus rapides à implémenter
numériquement et certainement plus utiles lorsqu'on veut analyser des cartes
réalistes du CMB contenant la coupure de la Voie Lactée ou d'éventuelles
petites excisions pour soustraire les différents points sources. Nous
commen\c{c}ons par rappeler brièvement le formalisme utilisé pour décrire les
anisotropies du CMB en section \ref{sec:peucmb}, puis nous expliquons les
effets de lentille gravitationnelle sur les spectres d'anisotropies du CMB en
espace harmonique à la section \ref{sec:peulens}, enfin nous présentons une
ébauche de notre travail sur la reconstruction des effets de lentille (dilatation et cisaillement) en
espace réel à la section \ref{sec:real}.

On notera abusivement $T(\boldsymbol{n})\equiv \delta T_{CMB}/T_{CMB}$ les
fluctuations anisotropes de température du CMB dans le ciel bidimensionnel
dans la direction $\boldsymbol{n}$. Les anisotropies du champ de polarisation
du CMB sont décrites par les modes $E(\boldsymbol{n})$ et
$B(\boldsymbol{n})$. Pour décrire les effets de lentille sur le CMB, on
utilisera souvent l'approximation de ciel plat qui reste une approximation assez précise pour toutes les échelles angulaires couvertes par les instruments de mesure sauf pour les plus grandes échelles angulaires, où la courbure de la sphère céleste ne joue plus un rôle négligeable.

\section{Un peu de formalisme du CMB}\label{sec:peucmb}

Les anisotropies du CMB sont liées aux perturbations cosmologiques provenant elles-mêmes des fluctuations quantiques de l'inflaton. En ce sens les anisotropies sont des variables stochastiques et ce sont donc leurs propriétés statistiques que l'on cherche à étudier. La plupart des modèles d'Inflation prédisent des fluctuations gaussiennes, dans ce cas toute l'information statistique est contenue dans la fonction de corrélation à deux points
\ba
C(\theta) & = & \langle T(\boldsymbol{n})T(\boldsymbol{n'})\rangle
\ea
par exemple pour les anisotropies de température du CMB. Ici $\theta$ est l'angle bidimensionnel entre les deux points du ciel observés dans les deux directions $\boldsymbol{n}$ et $\boldsymbol{n'}$ et défini par $\cos(\theta) = \boldsymbol{n}\cdot\boldsymbol{n'}$. L'hypothèse d'isotropie de l'Univers entraine l'isotropie statistique des variables fluctuantes donc que la fonction de corrélation ne dépend que de l'écart angulaire entre les deux points. On peut, de fa\c{c}on conjuguée, décrire les anisotropies de température sur la sphère céleste dans l'espace des harmoniques sphériques selon
\ba
T(\boldsymbol{n})  =  \sum_{l = 1}^{\infty} \sum_{m = -\ell}^{+\ell} a_{\ell m}^T Y_{lm}(\boldsymbol{n})\qquad \mbox{où } a_{\ell m}^T = \int_{S^2} T(\boldsymbol{n})Y_{lm}(\boldsymbol{n}) d^2 \boldsymbol{n}.
\ea
Le spectre de puissance des anisotropies de température est donc
\ba
\langle a_{\ell m}^Ta_{\ell'm'}^{T*}\rangle & = & \int \int C(\theta) Y_{lm}(\boldsymbol{n})Y_{l'm'}^*(\boldsymbol{n'})d^2 \boldsymbol{n} d^2 \boldsymbol{n'}.
\ea
Il est pratique de décomposer la fonction de corrélation angulaire sur la base des polynômes de Legendre selon $C(\theta) = (1/(4\pi))\sum_\ell (2\ell+1)C_\ell P_\ell(\cos \theta)$ afin de réduire
\ba
\langle a_{\ell m}^Ta_{\ell'm'}^{T*}\rangle & = & \int \int \sum_{l'' = 0}^{\infty} {2\ell''+1\over 4\pi}C_{\ell''} P_{\ell''}(\cos \theta) Y_{lm}(\boldsymbol{n})Y_{l'm'}^*(\boldsymbol{n'})d^2 \boldsymbol{n} d^2 \boldsymbol{n'}
\ea
à l'aide de la formule $P_\ell(\cos \theta) = (4\pi/ (2\ell+1))\sum_{m = -\ell}^{+\ell}Y_{lm}(\boldsymbol{n})Y_{lm}^*(\boldsymbol{n'})$ à
\ba
\langle a_{\ell m}^Ta_{\ell'm'}^{T*}\rangle & = & C_\ell\delta_{\ell\ell'}\delta_{mm'}.
\ea
Les coefficients $C_\ell$ constituent le spectre de puissance angulaire et ne
dépendent que du paramètre $\ell$ appelé multipôle et relié à l'ecart
angulaire entre deux points du ciel selon $\ell\sim 180/\theta$, si $\theta$
est en degrés. Le monopôle $\ell = 0$ correspond à la température moyenne du
CMB : $T_{CMB} = 2.7~K$. Le dipôle $\ell = 1$ est du à l'effet Doppler créé
par notre mouvement, ou plutôt par le mouvement de la Voie Lactée.  Sur les cartes d'anisotropies du CMB est représenté le spectre de puissance RMS ("root mean square" en Anglais), défini par $\mathcal{C}_\ell = \ell(\ell+1)C_\ell$.

Beaucoup de contraintes sur la physique de l'Univers primordial peuvent être
obtenues à partir de l'analyse de ces  spectres d'anisotropies du CMB. Nous ne
les détaillerons pas ici, mais par exemple on peut montrer que la position des
pics acoustiques dans le spectre d'anisotropies de température, notamment à
$\ell \sim 220$, exige que les conditions initiales de l'Univers issues de
l'Inflation soient adiabatiques. Les courbes des spectres d'anisotropies ont
également permis de contraindre la valeur des paramètres cosmologiques tels
que, par exemple, la densité baryonique, la densité d'énergie noire, de
neutrinos, le paramètre de Hubble, ... \cite{spergel}. 

Dans la suite nous utiliserons \emph{l'approximation de ciel plat}, qui est une approximation robuste pour la plupart des échelles angulaires couvertes par les instruments de mesure mais critique aux plus grandes échelles angulaires, où la courbure de la sphère céleste n'est plus négligeable. Dans l'approximation de ciel plat, on peut développer les anisotropies de température sur la base de Fourier selon
\ba
T(\boldsymbol{\theta })
=\int \frac{d^2\ell }{(2\pi )^2}T( \ellb )
\exp \left[ i\ellb \cdot {\boldsymbol \theta }\right] .
\ea
où l'angle $\boldsymbol{\theta }$ bidimensionnel a cette fois la dimension d'une longueur, et le moment $\boldsymbol{\ell}$ bidimensionnel a la dimension de l'inverse d'une longueur. Notons que, la fluctuation de température étant un champ réel, on a $T^{*}(\ellb) = T(-\ellb)$. Par isotropie statistique la fonction de corrélation à deux points ne dépend que de la séparation entre les deux points
\ba
\langle T(\boldsymbol{\theta })T(\boldsymbol{\theta' })\rangle & = & C\left(\vert \boldsymbol{\theta }-\boldsymbol{\theta' }\vert\right).
\ea
Il s'ensuit
\ba
\langle T(\ellb)T^*(\ellb)\rangle & = & \int d^2 \boldsymbol{\theta }\int d^2 \boldsymbol{\theta '} e^{-i\ellb \cdot {\boldsymbol \theta }}e^{i\ellb' \cdot {\boldsymbol \theta'}} C\left(\vert \boldsymbol{\theta }-\boldsymbol{\theta' }\vert\right),\cr
                                  & = & \int d^2 \boldsymbol{\theta }\int d^2 \boldsymbol{r} e^{i(\ellb'-\ellb)\cdot {\boldsymbol \theta}}e^{i\ellb' \cdot {\boldsymbol r}} C\left(r\right),\cr
                                  & = & (2\pi)^2\delta^2(\ellb-\ellb')\int d^2 \boldsymbol{r}e^{i\ellb \cdot {\boldsymbol r}} C\left(r\right),
\ea
où on a effectué le changement de variable $\boldsymbol{r} = \boldsymbol{\theta }-\boldsymbol{\theta' }$ et $r = \vert\boldsymbol{r}\vert$. En utilisant la décomposition en fonctions de Bessel, $\exp\left[ir\cos\phi\right] = \sum_{n = -\infty}^{\infty} i^n J_n(r)\exp\left[in\phi\right]$, on définit le spectre de puissance des anisotropies dans l'approximation ciel plat selon
\ba
C_\ell  =  \int d^2 \boldsymbol{r}e^{i\ellb \cdot {\boldsymbol r}} C\left(r\right) = \int rdr\int d\phi_r e^{i\ell r\cos(\phi_\ell-\phi_r)}C(r) = 2\pi \int rdr J_0(\ell r)C(r).
\ea
Le spectre de puissance des anisotropies du CMB est statistiquement isotrope et donc diagonal en $\ellb$ :
\ba
\left< T( \ellb )~T^*( \ellb ') \right> =
(2\pi )^2~\delta ^2 ( \ellb - \ellb ')~C(\ell ).
\label{flatcltt}
\ea
Remarquons que $\langle \vert T(\boldsymbol{\theta })\vert^2\rangle = \int
(d^2\ell/(2\pi )^2)\int (d^2\ell' /(2\pi )^2)\exp\left[i(\ellb-\ellb')\cdot
\boldsymbol{\theta }\right]\left< T( \ellb )~T^*( \ellb ') \right> =
\int_0^\infty (d\ell/\ell)\ell^2C_\ell/2\pi$. La quantité RMS sans dimension
pour le spectre est dans l'approximation ciel plat : $\mathcal{C}_\ell = d\langle T(\boldsymbol{\theta })^2\rangle/d[\ln \ell] = \ell^2C_\ell/2\pi$, à comparer avec l'expression en ciel sphérique.

En pratique, on n'observe qu'une seule réalisation statistique de notre
Univers observable, et donc dans une hypothèse d'ergodicité spatiale, les
spectres $C_\ell^{obs}$ observés sont obtenus à partir d'une moyenne spatiale
sur tout le ciel alors que les $C_\ell$ théoriques correspondent une moyenne
statistique (en fait c'est la variance cosmique) de variables gaussiennes
$a_{\ell m}$ distribuées selon une loi de probabilité gaussienne\\
\noindent
 $P(a_{\ell m}) = (1/\sqrt{2\pi} C_\ell)\exp\left[-a^2_{\ell
 m}/2C_\ell\right]$. On utilise donc généralement l'estimateur statistique
 sans biais ($\langle \hat{C}_\ell \rangle_{spatiale} = C_\ell$) suivant :
\ba
\hat{C}_\ell = {1\over 2\ell+1}\sum_m a^{obs}_{\ell m}a^{obs*}_{\ell m}
\ea
pour estimer la variance cosmique $C_\ell$.

Bien-entendu tout le formalisme présenté dans cette section pour décrire les
anisotropies de température $T$ est identique pour les anisotropies de
polarisation $E$ et $B$. Si l'on note $C_\ell(TT)$ le spectre d'anisotropies
de température, on peut également obtenir les spectres d'anisotropies
$C_\ell(EE)$ et $C_\ell(BB)$ ainsi que les corrélations croisées $C_\ell(TE)$,
$C_\ell(TB)$ et $C_\ell(EB)$. Notons que pour des raisons de parité, les
perturbations scalaires ne peuvent engendrer que des modes $E$ (qui possède la
symétrie scalaire), ce qui indique que la détection de modes $B$ (qui
possèdent la symétrie pseudo-scalaire), à travers l'observation du spectre
$C_\ell(BB)$ (non-encore détecté), sera le signe de l'existence d'ondes
gravitationnelles (perturbations tensorielles) produites lors de l'Inflation,
et donc un test décisif pour les modèles d'Inflation. On pourrait alors
mesurer le rapport tenseur-scalaire
$T/S$ et magnitude du potentiel de slow-roll dans l'Inflation à un champ
(échelle d'énergie de l'Inflation).

\section{Effets de lentille gravitationnelle sur le CMB dans l'approximation de ciel plat}\label{sec:peulens}

\subsection{Effets de lentille}

La présence de grandes structures, telles que les amas de galaxies dans
l'Univers, constitue un avant-plan à $z\sim 3$ sur la ligne de visée entre la
surface de dernière diffusion et nous. Ces structures crééent un potentiel
gravitationnel qui dévie les géodésiques nulles (ou la trajectoire des photons
du CMB). Par conséquent les spectres d'anisotropies du CMB qu'on observe
aujourd'hui ont subit des distorsions dûes au potentiel gravitationnel. C'est
l'effet de lentille gravitationnelle faible sur le CMB ("CMB weak
gravitational lensing" en Anglais). Une revue complète des effets de lentille
gravitationnelle sur le CMB a été faite par Challinor et Lewis \cite{challinor}.

Dans le formalisme de la relativité générale, on peut calculer la déflection
(vecteur de déplacement sur la sphère céleste) des géodésiques nulles
perturbées (trajectoires des photons) à l'ordre linéaire, engendrée par le
potentiel gravitationnel $\Psi$ \cite{challinor} :
\ba
\boldsymbol{\xi} & = & -2\int_0^{\chi_*} d\chi {f_K(\chi_*-\chi)\over f_K(\chi_*)f_K(\chi)}\nabla_{\boldsymbol{n}}\Psi(\chi\boldsymbol{n};\eta_0-\chi),
\ea
où $\chi$ est la coordonnée radiale et $\chi_*$ est la distance à la surface
de dernière diffusion, $\nabla_{\boldsymbol{n}}$ est dérivée covariante sur la
sphère céleste $S^2$ et $f_K(\chi) = \sin(\sqrt{K}\chi)/\sqrt{K}$ ou
$\sinh(\sqrt{\vert K\vert}\chi)/\sqrt{\vert K\vert}$ ou $\chi$ décrit la
géométrie spatiale, selon que l'Univers est respectivement fermé ($K > 0$) ou
ouvert ($K < 0$) ou plat ($K = 0$) spatialement. On définit le potentiel de
lentille comme $\Phi(\boldsymbol{n}) = -2\int_0^{\chi_*} d\chi
\left(f_K(\chi_*-\chi)/(
f_K(\chi_*)f_K(\chi))\right)\Psi(\chi\boldsymbol{n};\eta_0-\chi)$, de telle sorte que l'angle de déflection sur la sphère céleste s'écrit simplement
\ba
\boldsymbol{\xi} & = & \nabla_{\boldsymbol{n}}\Phi(\boldsymbol{n}).
\ea
Dans l'approximation de ciel plat, les fluctuations de température "lentillées" sont donc $T^L(\boldsymbol{\theta}+\boldsymbol{\xi})$ dont le développement à l'ordre linéaire entraine que le décalage d'anisotropie de température sans et avec lentille $(T^L-T)$ est
\ba
\delta T(\boldsymbol{\theta })=-(\boldsymbol{\nabla }\Phi )\cdot
(\boldsymbol{\nabla } T).
\ea
De fa\c{c}on conjuguée, la correction (dûe aux lentilles) à l'anisotropie de température observée est donnée dans l'espace de Fourier (approximation ciel plat) par
\ba
\delta T(\ellb _F)
=\int \frac{d^2{\ellb _L}}{(2\pi )^2}~
(+\ellb _L)\cdot ({ \ellb }_F-\ellb _L)~
\Phi ( \ellb _L)~
T( \ellb _F- \ellb _L).
\ea
Les anisotropies de polarisation peuvent être décomposées sur la base des
modes $E$ et $B$, dans l'approximation de ciel plat :
\ba
P_{ij}(\boldsymbol{\theta })=
\int \frac{d^2{\ell _L}}{(2\pi )^2}~
\exp \left[ i \ellb \cdot {\boldsymbol \theta }\right]~
\left\{
\left[
2\boldsymbol{\hat  \ellb }_i\boldsymbol{\hat  \ellb }_j-\delta _{ij}
\right] E( \ellb )
+
\left[ (\boldsymbol{\hat  \ellb }
\times \boldsymbol{\hat z})_i\boldsymbol{\hat \ellb }_j
+\boldsymbol{\hat  \ellb }_i(\boldsymbol{\hat  \ellb }\times \boldsymbol{\hat z})_j\right] ~
B( \ellb )
\right\},\qquad
\ea
où
$\boldsymbol{\hat \ellb }=\boldsymbol{\ell }/\ell $ et $T(\boldsymbol{\theta })
+ \boldsymbol{\hat n}_i \boldsymbol{\hat n}_j P_{ij}(\boldsymbol{\theta })$
 donne l'anisotropie de température de la composante polarisée linéairement le long du vecteur $\boldsymbol{\hat n}.$
Il s'ensuit
\ba
\begin{pmatrix}
&\delta E( \ellb _F)\cr
&\delta B( \ellb _F)\cr
\end{pmatrix}
=
\int \frac{d^2{\ell _L}}{(2\pi )^2}~
(+ \ellb _L)\cdot ( \ellb _F- \ellb _L)~
\begin{pmatrix}
+\cos [2\Theta _{FI}]&+\sin [2\Theta _{FI}]\cr
-\sin [2\Theta _{FI}]&+\cos [2\Theta _{FI}]\cr
\end{pmatrix}
\begin{pmatrix}
E( \ellb _I= \ellb _F- \ellb _L)\cr
B( \ellb _I= \ellb _F- \ellb _L)\cr
\end{pmatrix},\qquad
\ea
où $\Theta _{FI}$ est l'angle entre le vecteur $ \ellb _I$ et le vecteur $ \ellb _F.$
Il s'ensuit que le spectre de puissance "lentillé" des anisotropies de
température vaut 
\ba
C^{\delta T\delta T}( \ellb _{\delta T})=
\int \frac{d^2{\ell _L}}{(2\pi )^2}~
[ \ellb _\Phi \cdot ( \ellb _{\delta T}- \ellb _\Phi )]^2 ~
C^{\Phi \Phi }( \ellb _\Phi )~
C^{TT}( \ellb_T=\vert  \ellb _{\delta T}- \ellb _\Phi \vert ).
\label{qqq:twelve}
\ea
Ici on a supposé implicitement que le potentiel de lentille $\Phi $
et l'anisotropie de température intrinsèque (non "lentillée") $T$ ne sont pas corrélés.
De fa\c{c}on similaire pour la polarisation "lentillée", on a le spectre de
puissance suivant :
\ba
\begin{pmatrix}
&C^{\delta E\delta E}( \ellb _F)\cr
&C^{\delta B\delta B}( \ellb _F)\cr
\end{pmatrix}
&=&
\int \frac{d^2{ \ellb _\Phi }}{(2\pi )^2}~
[\ellb _\Phi \cdot ( \ellb _F- \ellb _\Phi )]^2~C^{\Phi \Phi }( \ellb _\Phi )\cr
&&\times
\begin{pmatrix}
 \cos ^2[2\theta _{FI}]&\sin ^2[2\theta _{FI}]\cr
 \sin ^2[2\theta _{FI}]&\cos ^2[2\theta _{FI}]\cr
\end{pmatrix}
\begin{pmatrix}
C^{EE}(\vert  \ellb _F- \ellb _\Phi \vert )\cr
C^{BB}(\vert  \ellb _F- \ellb _\Phi \vert )\cr
\end{pmatrix}.
\ea
Ici on a supposé que $C^{EB}=C^{E\Phi }=C^{B\Phi }=0.$ Au niveau des fonctions de corrélation à deux points, les effets de lentille gravitationnelle distordent donc le spectre de puissance des corrélations $TT$ (température-température) et mélangent les spectres de puissance des polarisations $EE$ et $BB$, en plus de les distordre.
Pour des bas $\ell $ (\emph{i.e.} dans la limite $\ell \to 0$), on peut approximer l'intégrale (\ref{qqq:twelve}) par
\ba
\int _0^\infty \frac{d\ell }{\ell }~\ell ^6~c^{TT}(\ell )c^{\Phi \Phi }(\ell ).
\ea
Le fait que cette intégrale ne diverge pas dans l'infrarouge (
$\ell ^6c^{TT}(\ell )c^{\Phi \Phi }(\ell )$
possède un pic bien à droite du quadrupôle $\ell =2$) implique que pour
$\ell \ll \ell _{peak}^{\delta T, \delta T},$
le spectre $c^{\delta T, \delta T}(\ell )$ est quasiment un spectre de bruit blanc.
Un léger changement de $\ell $ change radicalement l'intégrand; donc dans ce régime $c^{\delta T, \delta T}(\ell )$ est presque constant. Dit autrement, toute la magnitude à bas $\ell $ est dûe à de très faibles lentilles agissant sur les anisotropies du CMB aux grandes échelles angulaires. Le même argument s'applique à $c^{\delta B, \delta B}(\ell );$
cependant la position du pic
$\ell _{peak}^{\delta B, \delta B}$ est décalée par rapport à
$\ell _{peak}^{\delta T, \delta T}$
du fait de l'allure différente du spectre $c^{EE}(\ell ).$

\subsection{Reconstruction du potentiel de lentille en espace de Fourier}

Nous nous occupons maintenant de la reconstruction du potentiel de lentille $\Phi (\ellb )$ en construisant des estimateurs quadratiques optimaux en espace de Fourier. Soit
\ba
T_{tot}=T+\delta T.
\ea
Comme la théorie est gaussienne, les fonctions de corrélations à trois points du type $\langle\Phi T T\rangle$ s'annulent lorsqu'on considère le potentiel de lentille comme un champ stochastique gaussien, de fa\c{c}on identique à $T$. Cependant, si nous considèrons le potentiel de lentille du CMB comme fixé, nous trouvons que les fonctions de corrélations "lentillées" $\langle T_{tot} T_{tot}\rangle$ (qui impliquent à l'ordre linéaire en $\Phi$ des fonctions à trois points $\langle\Phi T T\rangle$) ne s'annulent plus, et cette propriété peut être exploitée pour reconstruire le champ de lentille en utilisant des estimateurs statistiques quadratiques en $T$ (ou $E$ et $B$).
On peut développer
\ba
\left< T_{tot}(\ellb _1)~T_{tot}(\ellb _2)\right> &=&
  \left< \delta T(\ellb _1)~T(\ellb _2)\right>
+ \left< T(\ellb _1)~\delta T(\ellb _2)\right> \cr
&=&\int \frac{d^2\ell _\Phi }{(2\pi )^2}
(+\ellb _\Phi)\cdot (\ellb _1-\ellb _\Phi )~\Phi (\ellb _\Phi)~
\left< T(\ellb _1-\ellb _\Phi )~T(\ellb _2)\right>
+(\ellb _1\leftrightarrow \ellb _2)\cr
&=&-(\ellb _1+\ellb _2)\cdot \left[ \ellb _1~c^{TT}(\ell _1)
                              +\ellb _2~c^{TT}(\ell _2)\right]
\Phi (\ellb _1+\ellb _2).
\ea
Par conséquent,
\ba
\hat \Phi (\ellb ;\ellb ')=
\frac{-1}%
{\ellb \cdot \left[ \ellb '~c^{TT}(\ell ')
+(\ellb -\ellb ')~c^{TT}(\vert \ellb -\ellb '\vert )\right] }~
T_{tot}(\ellb -\ellb ')~T_{tot}(\ellb ')
\ea
génère une famille continue bidimensionnelle d'estimateurs statistiques pour $\Phi (\ellb )$ indexée par $\ellb '.$
Notons que $\ellb '$ et $(\ellb -\ellb ')$ indexent le même estimateur. Pour éliminer cette redondance, on introduit les coordonnées $\ellb '=(\ell '_\parallel , \ell '_\perp ),$ où
$\ell '_\parallel $ et $\ell '_\perp $ sont les composantes parallèles et perpendiculaires par rapport au vecteur  $\ellb .$ Si on exige $\ell '_\perp \ge 0,$ on évite le double comptage.\footnote{Dans la limite continue on n'a pas besoin de s'inquiéter du bord de ce domaine $\ell ^\prime _\perp =0,$
qui est de mesure zéro.} Ci-dessous, ${\sum _{\ellb '}}^\prime $
dénotera la somme avec $\ell '_\perp \ge 0.$
Nous recherchons maintenant la combinaison linéaire optimale
\ba
\hat \Phi _{opt}(\ellb )={\sum _{\ellb '}}^\prime
w(\ellb ;\ellb ') \hat \Phi (\ellb ;\ellb '),
\ea
où les poids satisfont
\ba
{\sum _{\ellb '}}^\prime
w(\ellb ;\ellb ') =1.
\ea
La variance de l'estimateur ci-dessus est
\ba
{\sum _{\ellb '}}^\prime
{\sum _{\ellb ^{\prime \prime }}}^\prime
w(\ellb ;\ellb ')~
V(\ellb ',\ellb ^{\prime \prime };\ellb )~
w^*(\ellb ;\ellb ^{\prime \prime }),
\ea
où
\ba
&&V(\ellb ',\ellb ^{\prime \prime };\ellb )
=
\left<
\Bigl(\hat \Phi (\ellb ;\ellb ')-\Phi (\ellb )\Bigr) ~
\Bigl(\hat \Phi (\ellb ;\ellb ^{\prime \prime })-\Phi (\ellb )\Bigr) ^*
\right>\cr
&&=\quad
\frac{1}%
{\ellb \cdot \left[ \ellb '~c^{TT}(\ell ')
+(\ellb -\ellb ')~c^{TT}(\vert \ellb -\ellb '\vert )\right] }~
\frac{1}%
{\ellb \cdot \left[
\ellb ^{\prime \prime }~c^{TT}(\ell ^{\prime \prime })
+(\ellb -\ellb ^{\prime \prime })~
c^{TT}(\vert \ellb -\ellb ^{\prime \prime }\vert )\right] }~\cr
&&\quad \quad  \times \Biggl<
T(\ellb -\ellb ')~T(\ellb ')~
T^*(\ellb -\ellb ^{\prime \prime })~T^*(\ellb ^{\prime \prime })
\Biggr> \cr
&&=\quad\frac%
{(2\pi )^4\cdot
\left[ \delta ^2(\ellb '-\ellb ^{\prime \prime }\right] ^2
\cdot c_{TT}(\ell ')~c_{TT}(\vert \ellb -\ellb '\vert )}
{\left\{ \ellb \cdot \left[ \ellb ^{\prime }~c^{TT}(\ell ^{\prime })
+(\ellb -\ellb ^{\prime })~
c^{TT}(\vert \ellb -\ellb ^{\prime }\vert )\right] \right\} ^2}.
\label{corrEst}
\ea
Pour donner du sens à l'expression ci-dessus, on passe à la limite discrète
telle que (\ref{corrEst}) devient
\ba
&&V(\ellb ',\ellb ^{\prime \prime };\ellb )=L^4 \cdot \delta _{\ellb ',\ellb ^{\prime \prime }}
\cdot \quad\frac%
{c_{TT}(\ell ')~c_{TT}(\vert \ellb -\ellb '\vert )}
{\left\{ \ellb \cdot \left[ \ellb ^{\prime \prime }~c^{TT}(\ell ^{\prime \prime })
+(\ellb -\ellb ^{\prime })~
c^{TT}(\vert \ellb -\ellb ^{\prime }\vert )\right] \right\} ^2},
\ea
où $L$ est le coté du carré d'observation du ciel dans l'approximation de ciel plat. Soit
\ba
B(\ell _{max}; \ellb)=\{ \ellb '\vert
\ell '_\perp \ge 0 ~{\rm et}~
\vert \ell ' \vert < L_{max} ~{\rm et}~
\vert \ell -\ell ' \vert < L_{max}  \} .
\ea
On trouve que, pour l'estimateur optimal formé des combinaisons linéaires
avec\\
\noindent 
$\ellb '\in B(L_{max} ;\ellb),$
\ba
V^{-1}_{opt}(\ellb ; \ell _{max})
&=&\sum _{ \ellb '\in B(\ell _{max}; \ellb)}V^{-1}(\ellb ; \ellb ')\cr
&=&L^{-4} \sum _{ \ellb '\in B(\ell _{max}; \ellb )}
\frac%
{\left\{ \ellb \cdot \left[ \ellb ^{\prime }~c^{TT}(\ell ^{\prime })
+(\ellb -\ellb ^{\prime })~
c^{TT}(\vert \ellb -\ellb ^{\prime }\vert )\right] \right\} ^2}
{c_{TT}(\ell ')~c_{TT}(\vert \ellb -\ellb '\vert )}\cr
&\approx &L^{-2}\int _{ \ellb '\in B(\ell _{max}; \ellb )}
\frac{d^2\ell '}{(2\pi )^2}
\frac%
{\left\{ \ellb \cdot \left[ \ellb ^{\prime }~c^{TT}(\ell ^{\prime })
+(\ellb -\ellb ^{\prime })~
c^{TT}(\vert \ellb -\ellb ^{\prime }\vert )\right] \right\} ^2}
{c_{TT}(\ell ')~c_{TT}(\vert \ellb -\ellb '\vert )},\qquad
\label{varianceEqn}
\ea
où, à la dernière ligne, on a utilisé une approximation continue.
Les poids optimaux sont donnés par
\ba
w(\ellb ;\ellb ')=
\frac%
{V^{-1}(\ellb ',\ellb '; \ellb )}
{\sum _{\ellb ^{\prime \prime }}
V^{-1}(\ellb ^{\prime \prime },\ellb ^{\prime \prime }; \ellb )}.
\ea
On peut ré-exprimer le résultat (\ref{varianceEqn}) ci-dessus en fonction d'un
spectre de puissance continu pour le bruit de la reconstruction (variance de
l'estimateur) :
\ba
C^{\delta \hat \Phi ~\delta \hat \Phi }(\ell )
&=&L^{-2}~V_{opt}(\ell , \ell _{max})\cr
&=&\left[
\int _{ \ellb '\in B(\ell _{max}; \ellb )}
\frac{d^2\ell '}{(2\pi )^2}
\frac%
{\left\{ \ellb \cdot \left[ \ellb ^{\prime }~c^{TT}(\ell ^{\prime })
+(\ellb -\ellb ^{\prime })~
c^{TT}(\vert \ellb -\ellb ^{\prime }\vert )\right] \right\} ^2}
{c_{TT}(\ell ')~c_{TT}(\vert \ellb -\ellb '\vert )}
\right]^{-1}.\quad 
\label{ReconstructionNoise}
\ea

Lorsque $\vert \ellb \vert \ll \ell _{max}$ et $c_{TT}(\ell ')\equiv {\rm constant}$ (\emph{i.e.} un bruit blanc), l'expression ci-dessus se simplifie selon
\ba
C^{\delta \hat \Phi ~\delta \hat \Phi }
(\ellb ; \ell _{max})
\approx
8\pi \vert \ellb \vert ^{-4}\ell _{max}^{-2}.
\ea
Les différents facteurs sont compris comme suit. Le facteur $\vert \ellb \vert ^{-4}$
est présent parce que c'est la déformation de cisaillement (et également la dilatation) qui se mesure localement. Pour $C^{\delta \hat \kappa ~\delta \hat \kappa }$, on a essentiellement  un bruit blanc, parce que c'est surtout l'effet de lentille sur les plus petites échelles de fluctuations, juste en dessous du "cut-off" $\ell _{max}$, qui fournit l'essentiel de l'information. Qualitativement, on peut dire que $\kappa $ est détecté en mesurant le spectre de puissance local sur les échelles juste en dessous du "cut-off" et en comparant ce dernier avec le spectre de puissance moyen. Il faut remarquer que $\Phi $
et ${\bf \xi}$  ne sont pas directement mesurables. On ne peut pas mesurer le déplacement absolu
${\bf \xi}$ parce que, la statistique étant invariante par translation, nous ne savons pas à quoi devraient ressembler les anisotropies "non-lentillées" sous-jacentes.
Par conséquent, le rapport signal sur bruit carré de la carte reconstruite est donné par
\ba
(S/N)^2= O(1) c_{\Phi \Phi }(\ell ) \ell ^4\ell _{max}^2.
\ea
À l'aide d'un ``broad binning'' (\emph{i.e.} avec $(\Delta \ell )/\ell \approx \mathcal{O}(1)$),
on peut améliorer le rapport signal sur bruit au carré pour la détection des
effets de lentille selon
\ba
(S/N)^2= O(1) c_{\Phi \Phi }(\ell ) \ell ^6\ell _{max}^2.
\ea

Estimons maintenant dans quelles proportions le bruit d'un spectre générique
(de type loi de puissance) pour le fond ``non-lentillé'' diffère de l'estimation
précedente qui supposait un spectre de type bruit blanc. Bien-sûr, lorsque
le spectre contient des zéros, on peut s'attendre à ce que le dénominateur
s'annule et la fonction diverge. Mais supposons pour le moment un spectre en
loi de puissance. Si nous comparons la magnitude de
 $\ell c(\ell )$ (le seul terme qui apparait dans le cas du bruit blanc) avec
 le terme donnant la correction dominante au spectre de bruit blanc $\ell ' \ell \nabla c(\ell '),$
on trouve que c'est de l'ordre de $n$ pour le cas d'une loi de puissance de la
forme $c(\ell ')={\ell '}^n.$ Par conséquent, quand l'indice spectral local
$d\ln [c]/d\ln [\ell ]$ est d'ordre un, on s'attend à ce que le calcul de
bruit blanc effectué ci-dessus soit corrigé d'un facteur d'ordre  1.

Finalement on modifie l'équation (\ref{ReconstructionNoise}) pour inclure le
bruit du détecteur et la largeur finie du faisceau selon
\ba
C^{\delta \hat \Phi ~\delta \hat \Phi }_{[TT]}(\ell )
&=&\left[
{1\over 2}\int
\frac{d^2\ell '}{(2\pi )^2}
\frac%
{\left\{ \ellb \cdot \left[ \ellb ^{\prime }~c^{TT}(\ell ^{\prime })
+(\ellb -\ellb ^{\prime })~
c^{TT}(\vert \ellb -\ellb ^{\prime }\vert )\right] \right\} ^2}
{\Bigl( c_{TT}(\ell ')+n_{TT}^{eff}(\ell ')\Bigr) ~
\Bigl( c_{TT}(\vert \ellb -\ellb '\vert )+n_{TT}^{eff}(\vert \ellb -\ellb '\vert )}
\right]^{-1}.\qquad
\label{ReconstructionNoiseBis}
\ea
Ici, les $n_{TT}^{eff}(\ell ')$ incluent un facteur d'atténuation du profil du
faisceau de la forme $\exp [+(\ell /\ell _{beam})^2].$
Avec ces modifications, le cut-off $\ell _{max}$ est donné par le profil du
faisceau et le bruit du détecteur correspond à un cut-off régulier. Le facteur
${1/2}$ sert à compenser le double comptage.

De le même fa\c{c}on on peut construire un estimateur quadratique en espace de
Fourier basé sur l'effet de lentille transformant le mode
$E$ mode en mode $B$.
\ba
C^{\delta \hat \Phi ~\delta \hat \Phi }_{[EB]}(\ell )
&=&\left[
{1\over 2}\int
\frac{d^2\ell '}{(2\pi )^2}
\frac%
{\left\{ \ellb \cdot \ellb ^{\prime }~c^{EE}(\ell ^{\prime })
\right\} ^2~\sin ^2[2\Theta _{EB}]}
{\Bigl( c_{EE}(\ell ')+n_{EE}^{eff}(\ell ')\Bigr) ~
n_{BB}^{eff}(\vert \ellb -\ellb '\vert )}
\right]^{-1}.
\label{ReconstructionNoiseEB}
\ea
Avec une sensibilité suffisante des détecteurs
(\emph{i.e.} $n^{eff}_{BB}, n^{eff}_{EE}\ll c_{EE}$), cet estimateur devient
particulièrement puissant grâce au facteur
 $(n_{BB}^{eff}/c_{EE})$ par comparaison à un estimateur construit à partir des
 corrélations $TT$, avec approximativement la même résolution
 angulaire. Pratiquement tout le signal $B$ peut être attribué sans ambiguité
 aux effets de lentille, sans risque de confusion avec le signal $B$
 intrinsèque ``non-lentillé'' bien plus faible. Deux autres estimateurs
 peuvent être construits de la même fa\c{c}on en utilisant les combinaisons
$TE,$ $EE.$ Pour former l'estimateur quadratique optimal, pour chaque $\ellb
'$ on peut construire les estimateurs $\hat \Phi _A(\ellb ; \ellb ')$ où les
indices  $A,B$ décrivent les combinaisons $TT,$ $TE,$ $EE,$
et $EB.$  Il n'est pas nécessaire d'inclure la combinaison $BB$ parce que cette combinaison
n'intervient qu'au deuxième ordre en $\Phi .$

\section{Reconstruction en espace réel de la dilatation et du cisaillement des
 cartes du CMB (ébauche)}\label{sec:real}

Nous avons vu à la section \ref{sec:peulens} qu'au niveau des fonctions à deux
points, l'effet de lentille gravitationnelle déforme les spectres des
corrélation TT (température-température) et mélange les spectres de
corrélation EE et BB (polarisation) en plus de les déformer. Les effets de
lentille génère en fait des non-Gaussianités se manifestant dans les
fonctions de corrélations à plus de points. Au niveau des fonctions à trois
points par exemple, $
\left<
T (\boldsymbol{\theta})
T(\boldsymbol{\theta}')
\right>
$ contient la fonction à trois points $\left<
T^{NL} (\boldsymbol{\theta})
\nabla T^{NL}\nabla \Phi\right>$ ($T^{NL}$ désigne la
température non-lentillée) qui ne s'annule plus si l'on considère le potentiel de
lentille comme un champ fixe, et non plus aléatoire. Cette propriété a été exploitée à
la section \ref{sec:peulens} pour reconstruire le champ de lentille au moyen
d'estimateurs quadratiques en $T$ (ou $E$ et $B$) dans l'espace harmonique (ou
de Fourier dans l'approximation du ciel plat).

Beaucoup d'efforts ont été dépensés dans la littérature pour obtenir une reconstruction
\emph{optimale} du potentiel de lentille en espace harmonique. Mais cette
méthode présupposait implicitement une couverture totale du ciel sans la
coupure galactique, sans les mauvais pixels dûs aux points sources qui devraient
normalement être excisés, et sans pondération non-uniforme tenant compte d'une
couverture irrégulière mais réaliste du ciel. Pour une reconstruction basée sur
la température, il a été montré comment construire un optimateur quadratique
optimal dans ce contexte idéalisé, et que l'amélioration gagnée
 lorsqu'on utilise plutôt un estimateur du maximum de
vraisemblance (plus optimal) est marginale \cite{seljak-hirata}, parce que la déformation dûe à l'effet de
lentille est faible par rapport à la variance cosmique et au bruit de
l'expérience (excepté cependant aux très grands $\ell$). Pour l'exploitation
des anisotropies de polarisation, la situation est différente : à haute
sensibilité de l'expérience, le mode de polarisation $B$, dû
quasi-exclusivement aux effets de
lentille sur le mode $E$, domine le bruit de l'expérience et les corrections
d'ordre supérieur de l'estimateur quadratique présentes dans l'estimateur du
maximum de vraisemblance ne sont plus négligeables.

Ici nous adoptons une approche en \emph{espace réel} pour la
reconstruction des effets de lentille. Dans les conditions idéales souvent
supposées et décrites ci-dessus, cette approche donnerait naturellement le
même résultat que l'approche conventionnelle en espace harmonique. Cependant
notre intérêt est de construire des estimateurs légèrement non-optimaux, qui
sont modifiés de fa\c{c}on à être de courte portée (courtes échelles
angulaires), de telle sorte que les coupures, excisions de pixels et couverture
non-uniforme du ciel puissent être prises en compte d'une fa\c{c}on simple et
flexible. Nous pensons que de tels estimateurs locaux définis dans l'espace
réel, non-idéaux mais plus robustes, vont se montrer supérieurs dans la
confrontation avec les complications inhérentes aux données réelles.

Nous présentons maintenant les relations entre les différentes descriptions
de l'effet de lentille et nous calculons la déformation des anisotropies dans
l'espace réel. La déformation par effet de lentille des anisotropies du CMB
sur la surface de dernière diffusion peut être décrite de trois manières :
par le potentiel de lentille $\Phi$, par le champ de déflection
$\boldsymbol{\xi }=-\nabla \Phi ,$ ou par les trois composantes du tenseur de
cisaillement
\ba
\kappa =
\begin{pmatrix}
\kappa _0+\kappa _{+}& \kappa _{\times }\cr
\kappa _{\times }& \kappa _0-\kappa _{+}\cr
\end{pmatrix}
=-\partial _a\partial _b\Phi .
\ea
Même si on avait accès au ciel entier, les descriptions
 $\Phi $ et $\boldsymbol{\xi }$ souffrent d'une certaine ambiguité.
$\Phi $ ne peut pas être distingué de $\Phi +\textrm{(constante)}$ et le champ
vectoriel $\boldsymbol{\xi }$ ne peut être mesuré qu'à une constante de
translation près (ou une rotation près si on prend en compte la courbure du
ciel). Si on observe seulement le spectre de puissance du CMB, un morceau du
ciel et son translaté ont nécessairement la même probabilité à cause de la
propriété d'isotropie. Par contre, la dilatation et le cisaillement (qui sont
les gradients du champ vectoriel de translation) sont \emph{localement} bien
définis. Cela se voit facilement en considérant l'effet d'une déformation
constante décrit par la matrice de déformation $S$ reliant les coordonnées
angulaires $\boldsymbol{\theta },$ actuellement observées sur la sphère
céleste pour un point de la surface de dernière diffusion, aux coordonnées
$\boldsymbol{\theta }'$ qu'aurait ce même point en l'absence d'effet de 
lentille\footnote{Dans la suite, sauf indication contraire,  on emploiera
  l'approximation du ciel plat où le vecteur ${\thetab }$ représente un point
  sur la sphère céleste ``aplatie'' et ${\ellb }$ represente un nombre
  d'onde. Par moments des sommations sur $(l,m)$ seront aussi utilisées.}.
On a
$\boldsymbol{\theta '}= S\boldsymbol{\theta }$ avec  $S=\exp [\boldsymbol{\kappa }].$
 Nous employons l'approximation du ciel plat et supposons que les déformations
 sont petites et donc qu'une approximation à l'ordre linéaire est acceptable.
On notera de fa\c{c}on générique $P(\ell)$ le spectre de puissance primordial
  (c'est-à-dire sans les effets de lentille) dans l'espace de Fourier pour
  n'importe quelle fonction de correlation à deux points.

Calculons la déformation des spectres de puissance par dilatation et
cisaillement. On ne cherchera pas la rigueur dans la démonstration qui suit
mais plutôt les arguments principaux qui amènent au résultat attendu.
\begin{figure}
  \begin{center}
\includegraphics[width=6.5cm]{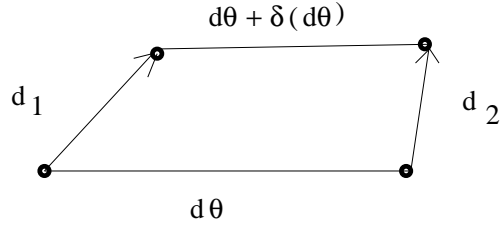}
 \end{center}
\caption{
Schema du déplacement de deux points dû aux effets de lentille.}
\label{fig:trapeze}
\end{figure}
Si deux points $\thetab$ et $\thetab'$ de la sphère céleste sont déplacés par
effet de lentille selon les vecteurs de déflection $\boldsymbol{d_1}$ et
$\boldsymbol{d_2}$ respectivement, alors, d'après le trapèze (Fig. \ref{fig:trapeze}), on a que la déformation $\delta(d\thetab)$ de
l'écart $d\thetab = \thetab-\thetab'$ entre les deux points vaut
$\delta(d\thetab) = \boldsymbol{d_2}-\boldsymbol{d_1}$. Dans la limite de
petites déflections et de points proches, et
puisque la déflection est
le gradient du potentiel de lentille ($\boldsymbol{d_1} = \nabla_{\thetab}
\Phi$, $\boldsymbol{d_2} = \nabla_{\thetab'}
\Phi$), on peut re-écrire
\ba
\delta(d\thetab)_a & = & {\partial^2 \Phi\over\partial\theta^a \partial\theta^b}
d\theta^b\cr
                 & = & -\kappa_{ab}(\theta) d\theta^b
\ea
où $\kappa_{ab}(\theta)$ est le tenseur de cisaillement
et où on somme sur les indices répetés ($\delta(d\thetab) = -\kappa\cdot
d\thetab$).
 La déformation du
spectre de puissance $P(d\thetab) = \langle T(\thetab)T(\thetab')\rangle$ dans l'espace réel vaut donc
\ba
\delta P(d\thetab) & = & P(d\thetab+\delta(d\thetab))-P(d\thetab)\cr
                               & = & \delta(d\thetab)_a
{\partial P(d\thetab)\over \partial d\theta_a}
\cr
                               & = & -\kappa_a^b(\theta) d\theta_b {\partial P(d\thetab)\over \partial d\theta_a}.
\ea
En supposant que la dilatation et le cisaillement varient lentement et en
remarquant que la transformée de Fourier de $x_b dP/dx_a $ vaut
$-(d/d\ell^b) [\ell^a P(\vert\ellb\vert)] = -\delta^a_b P(\ell)-\ell^a d P/d(\ell^b)$, on a dans l'espace de
Fourier la déformation suivante pour le spectre de puissance
\ba
P(\vert {\ellb }\vert )
& \to & P(\ell ) + \kappa_a^b \left(\delta^a_b P(\ell)+\ell^a {d P\over
  d\ell^b}\right)\cr
& \to & P(\ell ) + 2\kappa_0 P(\ell)+\kappa_a^b\ell^a{\ell_b\over \vert\ellb\vert} {d P\over
  d\ell}\cr
& \to & P(\ell ) + 2\kappa_0 P(\ell)+\kappa_a^b\ell^a{\ell_b\over \ell^2} {d P\over
  d(\log[\ell])}.
\ea
À l'ordre linéaire le spectre de puissance est donc modifié de la fa\c{c}on
suivante, préservant l'homogénéité mais pas l'isotropie du processus
stochastique sous-jacent:
\ba
P(\vert {\ellb }\vert )
\to P(\ell ) \biggl[
1 + \kappa _0
\left( \frac{d(\log[P(\ell )])}{d(\log [\ell ])} + 2\right)
+\left(
\frac{\kappa _+(\ell _1^2-\ell _2^2)+\kappa _\times  (2\ell _1\ell _2)}{\ell ^2}
\right)
\frac{d(\log[P(\ell )])}{d(\log [\ell ])}
\biggr].\qquad
\ea
Pour le cas d'un spectre invariant d'échelle
(\emph{i.e.} une loi de puissance de la forme $P(\ell )\propto \ell ^{-2}$)
il n'y a pas de modifications dans les corrélations qui soit dûe à la
composante de dilatation $\kappa_0$ du tenseur de cisaillement. De fa\c{c}on
similaire pour un spectre de bruit blanc parfait
(\emph{i.e.} une loi de puissance de la forme $P(\ell )\propto \ell ^0$),
 il n'y a aucune modification venant des composantes de cisaillement pur
 $\kappa _+$, $\kappa _\times$ dans les corrélations anisotropes à deux
 points.

Si les composantes de dilatation-cisaillement sont supposées lentement variables,
on peut donc construire les estimateurs de $\kappa_0$, $\kappa _+$ et $\kappa
_\times$, en repassant dans l'espace \emph{réel}, comme suit :
\ba
\hat \kappa _0 &=& N_0
\int _A d^2\theta ~
\int _A d^2\theta '~
\left[
T(\thetab )~T(\thetab ')-
\left< T({\thetab })~T({\thetab '})\right> _{\kappa =0}
\right]
\cr
&&\times \int \frac{d^2\ell}{(2\pi )^2}
\exp [i{\ellb }\cdot ({\thetab }-{\thetab '})]~
\frac{P(\ell )}{[P(\ell )+N(\ell )]^2}
\left( \frac{d(\log[P(\ell )])}{d(\log [\ell ])} + 2\right)\cr
&=& N_0
\int _A d^2\theta ~
\int _A d^2\theta '~
\left[
T({\thetab })~T({\thetab '})-
\left< T({\thetab })~T({\thetab '})\right> _{\kappa =0}
\right]
K_0({\thetab }-{\thetab '}),
\ea
et de fa\c{c}on similaire, en posant $\ell_1 = \ell\cos (\vartheta )$ et $\ell_2 = \ell\sin (\vartheta )$, 
\ba
\begin{pmatrix}
\hat \kappa _{+}\cr
\hat \kappa _{\times }\cr
\end{pmatrix}
&=& N_{+,\times }
\int _A d^2\theta ~
\int _A d^2\theta '~T(\thetab )~T({\thetab '})~\cr
&&\times \int \frac{d^2\ell }{(2\pi )^2}
\exp [i{\ellb }\cdot ({\thetab }-{\thetab '})]~
\frac{P(\ell )}{[P(\ell )+N(\ell )]^2}
\frac{d(\log[P(\ell )])}{d(\log [\ell ])}
\begin{pmatrix}
\cos (2\vartheta )\cr
\sin (2\vartheta )\cr
\end{pmatrix}\cr
&=& N_{+,\times }
\int _A d^2\theta ~
\int _A d^2\theta '~T({\thetab })~T({\thetab '})~
\begin{pmatrix}
K_+({\thetab }-{\thetab '})\cr
K_\times ({\thetab }-{\thetab '})\cr
\end{pmatrix},
\ea
où les facteurs de normalisation sont donnés par
\ba
N_0&=&\left[ {\cal A}
\int \frac{d^2\ell }{(2\pi )^2}
\frac{[P(\ell )]^2}{[P(\ell )+N(\ell )]^2}
\left( \frac{d(\log[P(\ell )])}{d(\log [\ell ])} + 2\right)^2
\right] ^{-1},\cr
N_+&=&N_\times =\left[ \frac{\cal A}{2}
\int \frac{d^2\ell }{(2\pi )^2}
\frac{[P(\ell )]^2}{[P(\ell )+N(\ell )]^2}
\left( \frac{d(\log[P(\ell )])}{d(\log [\ell ])}\right)^2
\right] ^{-1}
\ea
et ${\cal A}$ est l'aire du domaine du ciel considéré.

Les expressions ci-dessus supposent un domaine du ciel spatialement plat,
bidimensionnel, d'aire étendue mais finie, et sujet à une déformation linéaire
spatialement uniforme. On pourrait considérer un domaine toroidal dans la
limite où la période du tore devient extrèmement grande. Il
est indispensable de considérer tout d'abord la relation entre ce problème simplifié et le problème
réel de reconstruction du champ de lentille pour des cartes du CMB de
résolution finie. À cause de la petitesse des déformations des anisotropies du CMB par les lentilles, la situation simplifiée considérée ci-dessus est moins différente qu'on pourrait le penser de la situation réelle où l'on a affaire à un ciel courbe et des champs de dilatation/cisaillement non-uniformes. 

\begin{figure}
  \begin{center}
     \includegraphics[width=7.5cm]{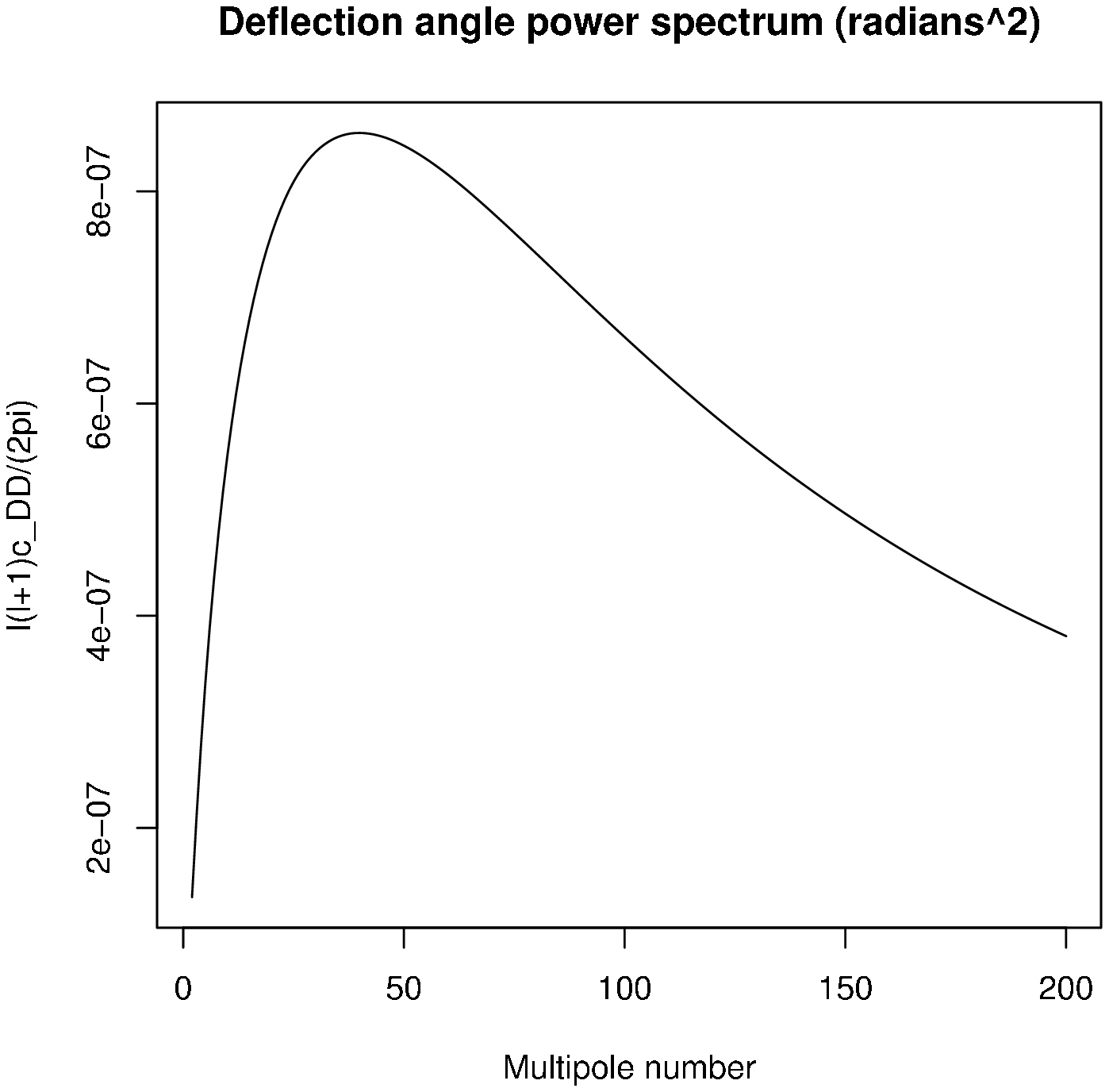}
     \includegraphics[width=7.5cm]{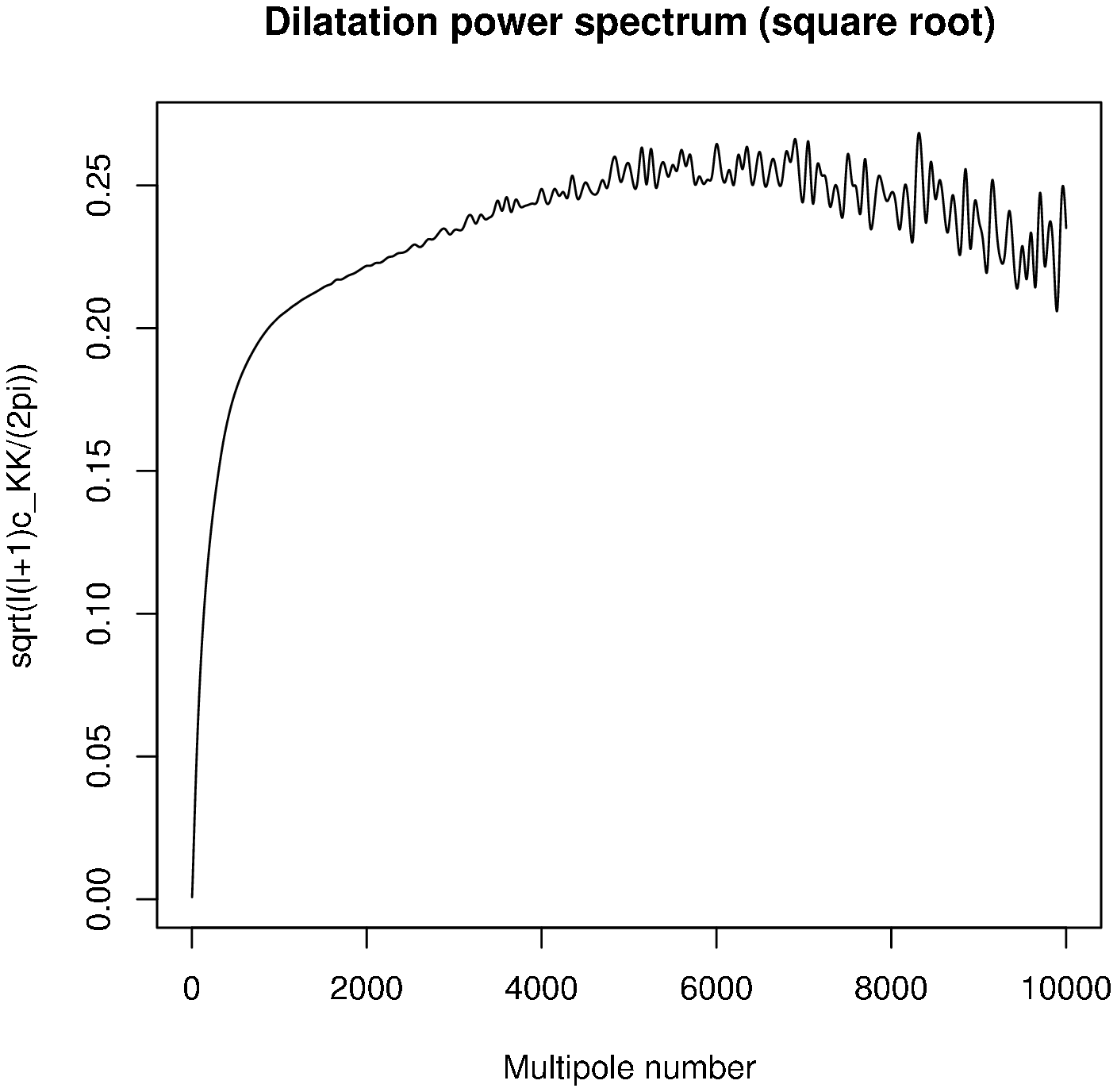}
  \end{center}
\caption{Le spectre de puissance du champ de lentille est représenté de deux
  manières. Les deux  tracés (de la gauche à la droite) illustrent le champ de lentille exprimé par le champ de déflection et le champ de dilatation/cisaillement, respectivement. On a tracé  
$C_{\ell }^{\boldsymbol{\xi \xi }}=\ell ^2 ~C_{\ell }^{\Phi \Phi },$
et 
$C_{\ell }^{\kappa \kappa }=\ell ^4 ~C_{\ell }^{\Phi \Phi },$
multipliés par $\ell (\ell +1)/(2\pi )$ selon la convention.
De ces deux tracés, le deuxième spectre de puissance (dilatation/cisaillement)
est plus directement relié à la déformation observée localement sur les cartes du CMB.
}
\label{Fig:LensingFields}
\end{figure}

Le tracé le plus à droite de la figure Fig. \ref{Fig:LensingFields} montre le spectre de puissance RMS de la dilatation/cisaillement ("Root Mean Square" en anglais, \emph{i.e.} $[\ell(\ell+1)]^2C_\ell^{\kappa\kappa}$) en fonction du nombre de multipôle, et on observe que pour $\ell < 3000$, la déformation est toujours plus faible qu'environ $3\% .$ Cela illustre que les effets de la dilatation/cisaillement se manisfestent en particulier aux très petites échelles angulaires sur les cartes d'anisotropies du CMB.

Pour l'estimateur linéaire idéal (couverture totale du ciel), l'inverse de la variance est
\ba
\frac{1}{\sigma ^2_{\hat \kappa _{ideal}}}&=&
\sum _{\ell , m}
\frac{c_\ell ^2}
{2(c_\ell +n_\ell )^2}
\left[
\frac{d(\ln [c_\ell ])}{d(\ln [\ell ])}+2
\right] ^2\cr
&\approx &
\int _0^\infty d\ell ~\ell ~
\frac{c_\ell ^2}
{(c_\ell +n_\ell )^2}
\left[
\frac{d(\ln [c_\ell ])}{d(\ln [\ell ])}+2
\right] ^2,
\ea
et pour un estimateur non-idéal avec le vecteur de poids $\{ w_\ell \},$
on définit
\ba
\hat \kappa _{\{ w_\ell \} }=
\frac{\lambda \left(N_0^{ideal}\right)^{1/2}}{\left< w, w \right> ^{1/2}}
\sum _{\ell , m}
\frac{1}{(c_\ell +n_\ell )^2}~w_{\ell }~\delta c_\ell ^{obs},
\ea
où $N_0^{ideal} = \sigma ^2_{\hat \kappa _{ideal}}$ et avec  
\ba
\lambda =\frac%
{
\left(
\sum _{\ell , m}
\frac{1}{(c_\ell +n_\ell )^2}~{w_{\ell }}^2
\right) ^{1/2}
\left(
\sum _{\ell , m}
\frac{c_\ell ^2}{(c_\ell +n_\ell )^2}~\left[ \frac{d(\ln [c_\ell ])}{d(\ln [\ell ])}+2 \right] ^2
\right) ^{1/2}
}
{\left(
\sum _{\ell , m}
\frac{c_\ell }{(c_\ell +n_\ell )^2}~w_{\ell }~
\left[ \frac{d(\ln [c_\ell ])}{d(\ln [\ell ])}+2 \right]
\right) }
=\frac{1}{\cos (\theta )}.
\ea
Le produit scalaire a été défini ici comme suit
\ba
\left< a, b\right> =\frac{1}{2}\int \ell ~d\ell ~
\frac{a_\ell b_\ell }{(c_\ell +n_\ell )^2}.
\ea
Ainsi, l'estimateur non-idéal n'est pas biaisé mais sa variance est donnée par
\ba
\sigma ^2_{\hat \kappa _{\{ w_\ell \} }}
=
\frac{1}{\cos ^2(\theta )}
\sigma ^2_{\hat \kappa _{ideal}}.
\ea
La variance de l'estimateur non-idéal (de support fini) est donc toujours plus grande que la variance de l'estimateur idéal excepté dans le cas particulier où 
les vecteurs\\
\noindent
$c_\ell[d(\ln [c_\ell ])/d(\ln [\ell ])+2]$
et $w_\ell $ sont liés (linéairement dépendants).
Des formules analogues peuvent être obtenues facilement pour $\kappa _+$ et $\kappa _\times $.

\begin{figure}
  \begin{center}
     \includegraphics[width=7.5cm]{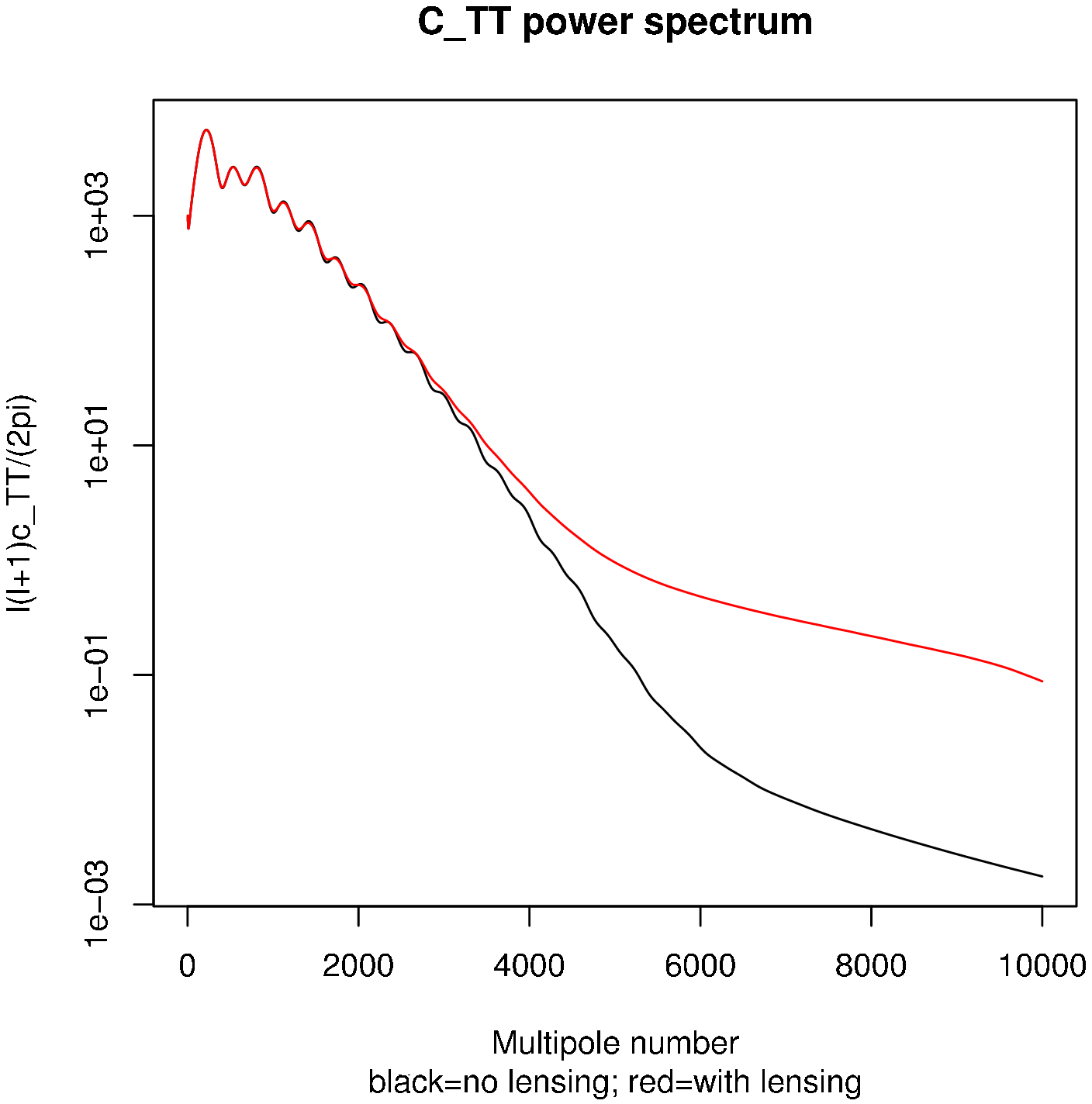}
     \includegraphics[width=8.5cm]{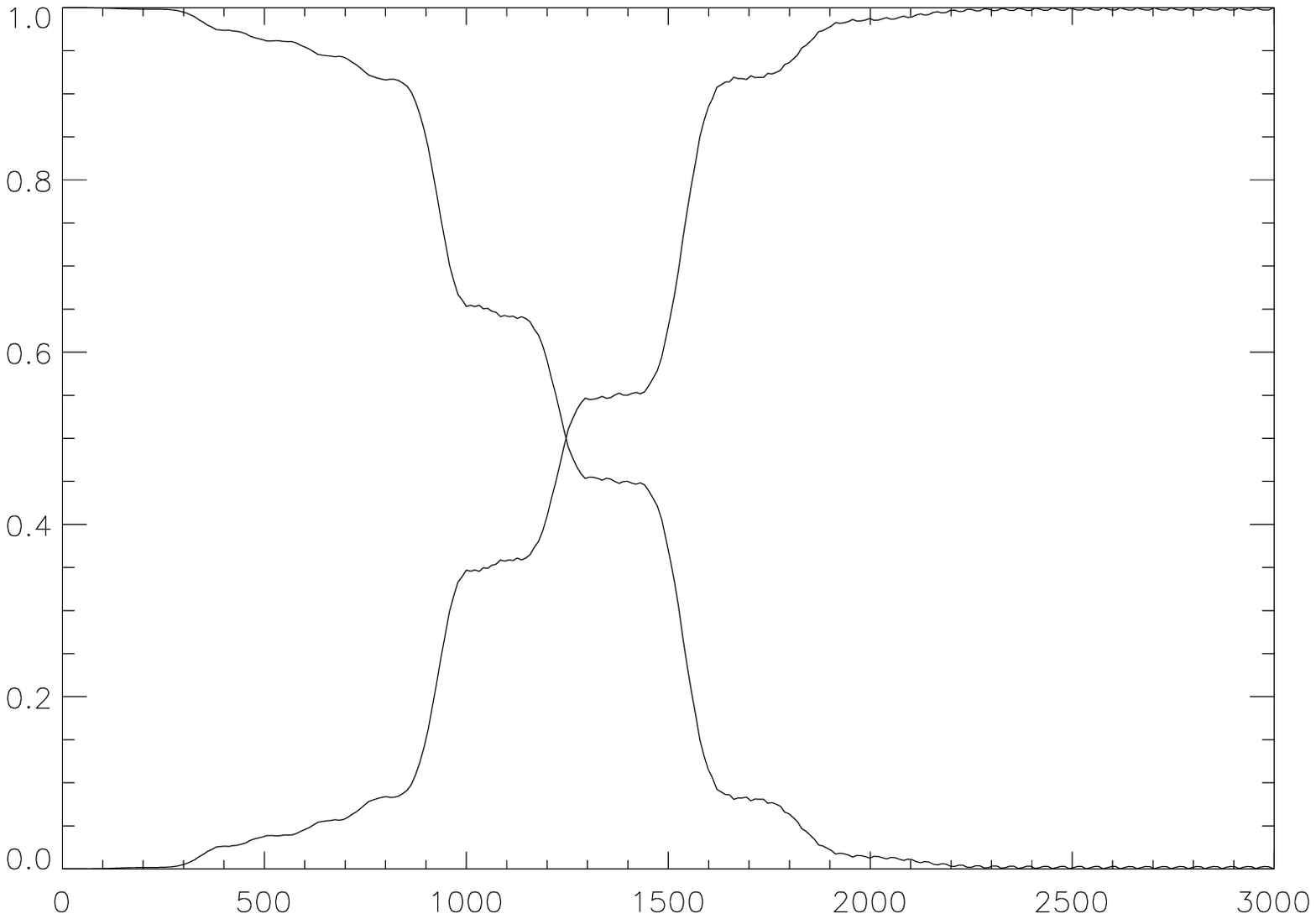}
  \end{center}
\caption{La figure de droite montre l'information cumulative normalisée 
$\chi ^2$ en fonction de $\ell $, intégrée des petits $\ell$ aux grands $\ell$ d'une part et intégrée des grands $\ell$ aux petits $\ell$ d'autre part, où on a utilisé les paramètres de sensibilité et de résolution de l'expérience PLANCK.
 On remarque que 80\% de l'information est concentrée dans l'intervalle $\ell~=~800$~-~$1600.$
Les plus petits $\ell $ ne contribuent presque pas à l'information parce que
le spectre angulaire dans le ciel est quasiment invariant d'échelle. Aux plus grands $\ell $, le bruit de l'instrument et l'étendue du faisceau contaminent le signal disponible. Dans l'intervalle intermédiaire, on observe une structure de plateaux connectés par des montées abruptes. Cette structure est la conséquence directe
des oscillations Doppler. Autour de ces oscillations, le spectre est presque invariant d'échelle et donc ne contient aucune information pour déterminer la dilatation.
}
\label{Fig:Info}
\end{figure}

\begin{figure}
  \begin{center}
\includegraphics[width=7.5cm]{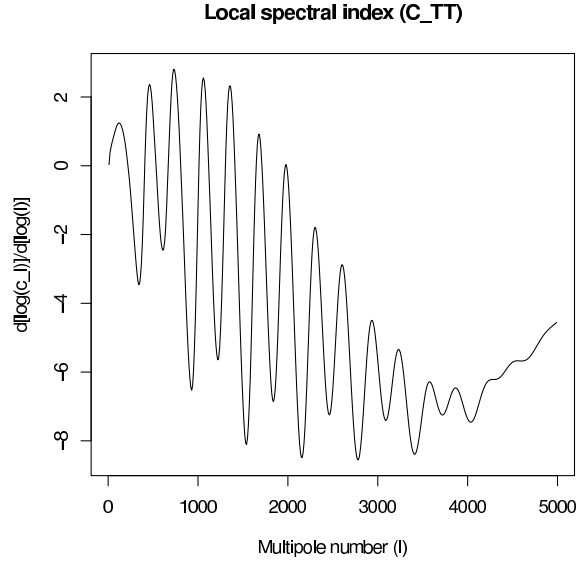} 
 \end{center}
\caption{
Index spectral local $d(\ln [c_\ell])/d(\ln [\ell])$ en fonction de $\ell $
pour la température ($c_\ell(TT)$ est tracé sur la figure Fig.~\ref{Fig:Info} à gauche) et pour la cosmologie standard (meilleur fit de WMAP). }
\label{Fig:SpectralIndex}
\end{figure}

Considérons comment l'information contenue dans l'estimateur idéal est répartie sur les différents multipôles. Sur la figure Fig.~\ref{Fig:Info}
on a tracé les quantités
\ba
F_<(\bar \ell )=
\frac{
\int _0^{\bar \ell }\ell ~d\ell ~
\frac{ {c_\ell }^2}{({c_\ell }+{n_\ell })^2}
\left[ \frac{d(\ln [c_\ell ])}{d(\ln [\ell ])}+2 \right] ^2
}{
\int _0^{\infty }\ell ~d\ell ~
\frac{ {c_\ell }^2}{({c_\ell }+{n_\ell })^2}
\left[ \frac{d(\ln [c_\ell ])}{d(\ln [\ell ])}+2 \right] ^2
}
\ea
et $F>(\bar \ell )=1-F_<(\bar \ell )$, où le bruit de l'expérience est
\ba
n_\ell
=n_0 ~\exp \left[ +\ell ^2\theta ^2_{beam}\right]
=n_0 ~\exp \left[ +\ell ^2/\ell ^2_{beam}\right]
\ea
et $\ell _{beam}=(810)(10'/\theta _{beam}^{fwhm}).$

Les noyaux idéaux de variance minimum pour les estimateurs 
$\hat \kappa _0,$
$\hat \kappa _+,$
et
$\hat \kappa _\times $
ont leur support nettement piqué sur les petites séparations angulaires, mais leur support s'étend néanmoins jusqu'aux grandes séparations. Nous recherchons maintenant combien d'information est perdue si le support aux grandes séparations est complètement coupé, comme c'est le cas pour notre estimateur non-idéal. La fa\c{c}on quantitative de caractériser cette perte d'information est de se demander de quel facteur la variance de l'estimateur est amplifiée par rapport à la variance de l'estimateur optimal, une fois qu'on a renormalisé notre estimateur "rogné" pour le rendre non-biaisé.  


\def\half{\frac{1}{2}}

Nous définissons notre estimateur au support tronqué en utilisant la fonction "fenêtre" :
\ba
W(\theta ;\theta _a,\theta _b)
\left\{
\begin{array}{ll}
1,& \theta \le \theta _a\cr
\half +\half \tanh \left[
\frac{1}{\theta -\theta _b}
+
\frac{1}{\theta -\theta _a}
\right] ,& \theta _a<\theta <\theta _b\cr
0,& \theta \ge \theta _b\cr
\end{array}
\right.
\ea
qui a la bonne propriété d'être une fonction $C^\infty $.
Le noyau de l'estimateur de variance minimum couvrant tout le ciel, défini par 
\ba
K_{ideal}(\theta )
=N_{ideal}
\int _0^\infty \ell ~d\ell ~
\frac{J_0(\ell \theta )}{(c_\ell +n_\ell )^2}~
c_\ell \left[ \frac{d[\ln (c_\ell )]}{d[\ln (\ell )]} +2 \right]
\ea
est donc tronqué selon
\ba
K(\theta ;\theta _a, \theta _b)=
K_{ideal}(\theta )~
W(\theta ;\theta _a, \theta _b)
\ea
tel que
\ba
K(\ell ;\theta _a, \theta _b)=
\int _0^\infty \theta ~d\theta ~
J_0(\ell \theta )~
K(\theta ;\theta _a, \theta _b)
\ea
La variance est alors amplifiée par le facteur 
$1/\cos ^2[\chi (\theta _a, \theta _b)]$
où
\ba
\cos \chi (\theta _a, \theta _b)=
\frac{
\left< K_{ideal}(\ell ), K(\ell ;\theta _a, \theta _b)\right>
}{
\left<
K_{ideal}(\ell ),
K_{ideal}(\ell )
\right> ^{1/2}
\left<
K(\ell ;\theta _a, \theta _b),
K(\ell ;\theta _a, \theta _b)
\right> ^{1/2}
}
\ea
et où on a défini le produit scalaire cette fois selon
\ba
\left<
K_a(\ell ),
K_b(\ell )
\right> =
\int _0^\infty \ell ~d\ell ~
 (c_\ell +n_\ell )^2 K_a(\ell )~K_b(\ell ).
\ea

Nous devons maintenant trouver la fa\c{c}on optimale d'effectuer la troncation
$W_{opt}(\theta ;\theta _a,\theta _b)$ de manière à minimiser l'amplification
de variance: ce travail est en progrès.


\pagestyle{plain}
\chapter*{Conclusions}\label{chapter:end}

Dans la majeure partie de ce travail de thèse nous avons exploré le
problème de l'évolution des perturbations cosmologiques dans un Univers
branaire du type Randall-Sundrum ayant une expansion cosmologique
arbitraire. L'objectif était d'obtenir des signatures cosmologiques de la
présence éventuelle d'une dimension supplémentaire infinie au moyen de la
théorie des perturbations cosmologiques et d'estimer la magnitude des
interactions entre la brane et le bulk. Nous avons insisté sur les deux
difficultés majeures intervenant dans le problème des perturbations
cosmologiques dans un Univers branaire en expansion : d'une part c'est un
problème d'ordre technique puisque le mouvement arbitraire de la brane en
expansion non-uniforme dans le bulk brise les symétries du problème et rend
les équations d'Einstein linéarisées non-séparables pour les perturbations
cosmologiques. D'autre part on est confronté à un problème d'ordre plus
fondamental qui est celui des conditions initiales inconnues dans le bulk
Anti-de Sitter ($AdS$). Il n'y a pas de conditions initiales naturelles dans
$AdS$ pour générer homogénéité et isotropie. Si le premier problème technique
peut être en partie résolu de façon numérique, le problème des conditions
initiales pour les gravitons du bulk $AdS$ persiste puisque les schémas
numériques spécifient nécessairement des conditions initiales particulières dans
le bulk pour faire évoluer les perturbations dans le temps. Nous avons 
proposé une méthode analytique basée sur les fonctions de Green retardées de la
brane $FRW$ et du bulk $AdS$ et avons suivi une approche effective quadridimensionelle en  considérant les degrés de liberté localisés sur la brane comme un système
quantique ouvert couplé à l'environnement composé des gravitons du bulk. Nous
avons vu que dans cette perspective quadri-dimensionnelle, les transitions
quantiques entre les modes du bulk et ceux de la brane apparaissaient sous la
forme de processus de dissipation et de non-localité du point de vue
quadri-dimensionnel d'un observateur sur la brane. En partant des propagateurs
retardés "nus" (c'est-à-dire sans couplage brane-bulk) de la brane et du bulk
$AdS$, nous avons construit le propagateur retardé effectif sur la brane en
resommant les effets de rétroaction du bulk à tous les ordres dans le couplage
brane-bulk. Nous avons montré comment les processus de dissipation et de
non-localité étaient codés dans ce propagateur effectif, et nous avons reussir à
extraire les taux de dissipation effectifs de diverses perturbations de
matière sur la brane en appliquant ce calcul aux équations linéarisées
d'Einstein exactes dans le formalisme de Mukohyama et pour diverses
configurations de champs. Dans l'hypothèse d'une expansion adiabatique de
l'Univers ($\dot{H}/H^2\ll 1$), nous avons trouvé que la perturbation de
l'inflaton (champ scalaire en roulement lent sur la brane) se dissipait de
façon locale et linéaire dans le facteur de slow-roll, alors que le mode zéro
lié du graviton (perturbations tensorielles) se dissipe de façon non-locale et
quadratique dans le facteur de slow-roll aux échelles superhorizon et à haute
énergie, et donc que la dissipation de la perturbation scalaire domine par
rapport à celle de la perturbation tensorielle. Nous avons trouvé aussi que
les perturbations adiabatiques d'un fluide parfait sur la brane ne se dissipent
pas aux échelles subhorizon.  

Au chapitre \ref{chapter:adsprop} nous avons présenté le calcul explicite du propagateur retardé covariant du graviton dans $AdS$. Ce dernier
étant un objet classique, nous avons voulu le calculer 
en utilisant uniquement des considérations classiques basées sur la
structure causale de l'espace $AdS$. Nous avons réussi à faire la
démonstration dans le cas d'un espace $AdS$ de dimension paire. Le cas en
dimension impaire est en progrès. En dehors de l'intérêt mathématique et
technique, le calcul du propagateur retardé du graviton dans
$AdS$ est nécessaire pour parfaire la description des processus non-locaux
intervenant en cosmologie branaire et que nous avons présentés dans cette
thèse.  

Dans le dernier chapitre nous avons exposé notre travail sur la reconstruction
des effets de lentille gravitationnelle sur le CMB dans une cosmologie
standard. Ce travail est encore en progrès mais est bien avancé et devrait
déboucher bientôt sur la publication de nos résultats. À partir des spectres
d'anisotropies observés et contaminés par les lentilles, nous reconstruisons
le champ de dilatation/cisaillement directement en
espace réel à l'aide d'estimateurs quadratiques de courte portée angulaire,
puisque l'information sur l'effet de lentille se situe principalement aux
courtes échelles angulaires sur la sphère céleste. Les méthodes locales en
espace réel que nous développons dans ce travail présentent l'avantage d'être
plus 
efficaces pour traiter des cartes réalistes du CMB contenant la coupure
galactique et de nombreuses petites excisions dûes aux point-sources qu'il
faut exclure. La méthode employée ici est générale et peut s'appliquer à
d'autres non-gaussianities, comme celles provenant des avant-plans
("foregrounds") ou celles issues de scenarios d'Inflation. Dans un
futur proche, nous aimerions étendre ce travail de recherche à d'autres
non-gaussianités pouvant affecter le rayonnement du CMB. 

L'approche analytique, que nous avons développée au chapitre
\ref{chapter:dissip}, pour traiter l'évolution
des perturbations cosmologiques dans des univers 
branaires en expansion du type Randall-Sundrum est plus générale et peut être
appliquée à tout autre théorie de champs en interaction avec des champs
localisés sur un bord ou une sous-variété. Nous pensons notamment au modèle
branaire Dvali-Gabadadze-Porrati (DGP) \cite{dgp} qui propose une explication à
l'accélération tardive de l'Univers (problème de l'énergie noire) et au modèle
branaire ekpyrotique \cite{ekpy} de cosmologies à rebonds qui propose
une alternative à l'Inflation pour le problème des conditions initiales de
l'Univers. Nous aimerions mettre en évidence les signatures cosmologiques de
tels modèles cordistes au moyen de la théorie des perturbations cosmologiques
en se plaçant dans des géométries de fond dépendantes du temps pour la
brane. Nous espérons pouvoir mener à bien ces recherches dans un avenir
proche.


\bibliography{./references}
\bibliographystyle{./apsrev}


\end{document}